\documentstyle[12pt,epsfig]{article}
\newcommand{\chha}{$\tilde{\chi}_1^{\pm}$}
\newcommand{\chhb}{$\tilde{\chi}_2^{\pm}$}
\newcommand{\chna}{$\tilde{\chi}_1^{0}$}
\newcommand{\chnb}{$\tilde{\chi}_2^{0}$}
\newcommand{\chnc}{$\tilde{\chi}_3^{0}$}
\newcommand{\chnd}{$\tilde{\chi}_4^{0}$}
\newcommand{\etm}{$E_T^{miss}$}
\newcommand{\g}{$\tilde{g}$}
\newcommand{\q}{$\tilde{q}$}
\newcommand{\bt}{$t\bar t$}
\newcommand{\bb}{$b\bar b$}
\newcommand{\zbb}{$Z b\bar b$}
\newcommand{\nchd}{$N_{\tilde{\chi}_1^{\pm} \tilde{\chi}_2^{0}}^{dir}$}
\newcommand{\nall}{$N_{ALL}$}
 
\def\gsim{\mathrel{\rlap{\raise.4ex\hbox{$>$}} {\lower.6ex\hbox{$\sim$}}}}
\def\lsim{\mathrel{\rlap{\raise.4ex\hbox{$<$}} {\lower.6ex\hbox{$\sim$}}}}

\def\gappeq{\mathrel{\rlap {\raise.4ex\hbox{$>$}}
{\lower.6ex\hbox{$\sim$}}}}
 
\def\lappeq{\mathrel{\rlap{\raise.4ex\hbox{$<$}}
{\lower.6ex\hbox{$\sim$}}}}

\begin{document}

\begin{titlepage}

\hspace*{-0.6cm}
{\small Available on CMS information server} {\hfill\large\bf CMS NOTE
1998/006}
\vspace*{-3mm}
\begin{center}
\includegraphics{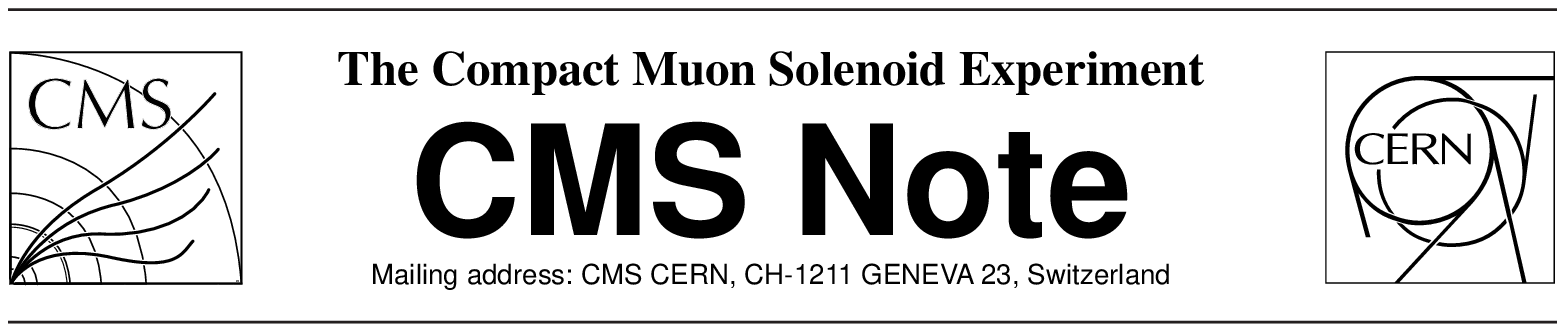}
\end{center}
\vspace*{-4mm}
{\hfill April 1998}

\vspace*{3mm}

\begin{center}

{\Huge{\bf{\sf Discovery potential for supersymmetry in CMS}}}

\end{center}

\vskip 1mm

\begin{center}
For the CMS Collaboration
\end{center}

\begin{center}
S.~Abdullin$^a$,  \v{Z}.~Antunovi\'{c}$^b$, F.~Charles$^c$,
D.~Denegri$^d$, U.~Dydak$^e$,
M.~D\v{z}elalija$^b$, \\
V.~Genchev$^f$, D.~Graham$^g$,
I.~Iashvili$^{h,i}$, A.~Kharchilava$^{h,i}$,
R.~Kinnunen$^j$, \\
S.~Kunori$^k$,
K.~Mazumdar$^l$,
C.~Racca$^i$, L.~Rurua$^{h,e}$,
N.~Stepanov$^a$, J.~Womersley$^m$
\end{center}

\vskip 1mm

\begin{center}
{\bf Abstract}
\end{center}

{\small
This work summarizes and puts in an overall perspective studies
done within CMS concerning the
discovery potential for squarks and gluinos,
sleptons, charginos and neutralinos, SUSY dark matter,
lightest Higgs, sparticle mass determination methods and the detector
design optimisation in view of SUSY searches.
It represents the status of our understanding of these subjects as of
Summer 1997.

As a benchmark model we used the minimal supergravity-inspired
supersymmetric standard model (mSUGRA) with a stable LSP. Discovery of
supersymmetry at the LHC should be relatively straightforward.
It may occur through the observation of a large excesses
of events in missing $E_T$ plus jets, or with one or more isolated
leptons. An excess of trilepton events or of isolated dileptons with
missing $E_T$, exhibiting a characteristic signature in the
$l^+ l^-$ invariant mass distribution could also be the first
manifestation of SUSY production.
Squarks and gluinos can be discovered for masses in
excess of 2~TeV.
Charginos and neutralinos can be discovered from an
excess of events in dilepton or trilepton final states.
Inclusive searches can give early indications
from their copious production in squark and gluino cascade decays.
Indirect evidence for sleptons can be obtained also from
inclusive dilepton studies.
Isolation requirements and a jet veto would
allow detection of both, the direct chargino/neutralino production
and of directly-produced sleptons.
Squark and gluino production may also represent a
copious source of Higgs bosons through cascade decays.
The lightest SUSY Higgs h~$\to b\overline b$ may
be reconstructed with a signal/background ratio of order 1
thanks to hard cuts on $E_T^{miss}$ justified by escaping LSP's.
The lightest supersymmetric particle of SUSY models
with conserved $R$-parity represents a very good candidate for the
cosmological dark matter. The region of parameter space where
this is true is well-covered by our searches, at least for
tan$\beta = 2$.

If supersymmetry exists at electroweak scale it could hardly escape
detection in CMS, and the study of supersymmetry will
form a central part of our physics program.
}

\end{titlepage}

\newpage

\vskip 10mm

\vskip 8mm

{\it
$^a$ Institute for Theoretical and Experimental Physics, Moscow, Russia

$^b$ University of Split, Split, Croatia

$^c$ Universit\'e de Haute Alsace, Mulhouse, France

$^d$ Centre d$`$Etudes Nucle\'aire de Saclay, Gif-sur-Yvette, France

$^e$ Institut f\"ur Hochenergiephysik, Wien, Austria

$^f$ Institute for Nuclear Research and Nuclear Energy, Sofia, Bulgaria

$^g$ Imperial College, University of London, London, United Kingdom

$^h$ Institute of Physics, Georgian Academy of Sciences, Tbilisi, Georgia

$^i$ Institut de Recherches Subatomiques, Strasbourg, France

$^j$ Helsinki Institute of Physics, Helsinki, Finland

$^k$ University of Maryland, College Park, USA

$^l$ Tata Institute of Fundamental Research-EHEP, Bombay, India

$^m$ Fermi National Accelerator Laboratory, Batavia, USA

}

\newpage


\pagestyle{empty}
\tableofcontents
\newpage


\pagenumbering{arabic}
\pagestyle{plain}

\section{Introduction}

There are strong arguments that the Standard Model (SM), despite
its phenomenological successes \cite{lepsldewwg}, is only a low energy
effective theory of spin-1/2 matter fermions interacting via spin-1
gauge bosons. A good candidate for ``new physics'' is the 
Supersymmetric (SUSY) extension of the SM which in its minimal
version (MSSM) doubles the number of known particles, introducing
scalar and fermion partners to ordinary fermions and bosons
and relating their couplings. This scheme provides the necessary
cancellation of quadratic divergences which appear in loop
corrections to the masses provided the masses of super-partners
are of the order of electroweak (EW) scale, i.e. $\sim$0.1
to 1 TeV.

One of the main motivations of the experiments at the Large
Hadron Collider (LHC) is to search for SUSY particles (``sparticles'').
Previous studies \cite{baer} have shown that at the LHC the
observation of an excess in event rates of specific final states over
the SM expectations would signal the production of SUSY particles.
The present study is based on a more precise description of the
Compact Muon Solenoid (CMS) detector and covers a wider spectrum of
physics channels. The primary goal is to understand the experimental
limiting factors for SUSY studies and contribute to the detector
optimisation while its design is not entirely frozen. The emphasis is
put on leptonic channels, which are as interesting in terms of discovery
potential, if not more, as the classical $multi$-$jets$ + \etm \
signature. The latter is more sensitive to instrumental backgrounds
which can be fully understood only once the cracks, dead areas and
volumes of diminished instrumental response due to detector
services, mechanical supports are fully specified and evaluated.
Another motivation for this study was the need to evaluate the discovery
potential of an LHC detector in terms of accessible sparticle spectrum,
sparticle mass reach, possibilities to determine the SUSY model
scenario at work and possibilities to determine sparticle masses and model
parameters in comparison and competition with a future $e^+e^-$ collider
and dedicated SUSY (WIMP) dark matter search experiments \cite{lhccsusy}.

This work summarizes and puts in an overall perspective specific studies
done within CMS concerning the observability and discovery potential for
squarks and gluinos \cite{gq, gqmp}, sleptons \cite{ll_excl, ll_incl},
charginos and neutralinos \cite{wz, inclchi2, chi12mass},
SUSY dark matter \cite{daniel1}, the lightest Higgs \cite{htobb},
sparticle mass determination methods \cite{inclchi2, chi12mass, ll_incl},
detector design optimisation in view of SUSY searches
\cite{calconf}, etc., and represents the status of our understanding
of these subjects as of summer 1997.

We discuss the specific SUSY model employed in section 2 and the
experimental signatures investigated in section 3. Detector issues and
simulations of detector response are described in section 4. 
Searches for SUSY in $lepton(s) + jets$ + \etm \ final states
from strongly interacting sparticle production are discussed
in section 5 followed by inclusive searches for the next-to-lightest
neutralino in section 6. Exclusive $2~leptons + no~jets +$ \etm \
and exclusive $3~leptons + no~jets +$ (\etm) channels are
discussed in sections 7 and 8.
Possibilities to observe h~$\rightarrow b \bar{b}$
in squark and gluino decays are discussed in section 9.
The methods to measure sparticle masses and
restrict model parameters are addressed in section 10.
Results and conclusions are summarized in section 11.

The potential of CMS to study the Higgs sector of SUSY has been
extensively discussed elsewhere \cite{higgs} and we do not address this
problem here.
Obtained results are based on two-loop level
calculations \cite{twoloop}.
To summarize, in the search of the various
MSSM Higgs bosons, most of the 
$m_A$, tan$\beta$ parameter space will be explored through at least 
one channel, provided an integrated luminosity
of $L_{int}  = 10^5$ pb$^{-1}$ is available. A difficulty still
exists in the
region of 120 GeV $\lappeq \ m_A \  \lappeq$ 240 GeV and
2 $\lappeq \ tan\beta \  \lappeq$ 10, which would require
$L_{int} = 6 \div 10 \times 10^5$ pb$^{-1}$ to be fully explored
(though section 9 indicates that the Higgs sector may also be 
detectable in the decays of other SUSY particles).


\section{SUSY model employed}

Since the phenomenological implications of SUSY are model-dependent, 
the discovery potential
of a detector has to be studied resorting to some
particular model, preferably with a limited number of free
parameters. This implies some loss of ``generality'', but ensures
tractable predictions. Once the possibilities and problems have been 
well understood within a defined scheme it is easier to
evaluate its generality.

We have chosen to make the investigation in the minimal
Super Gravity constrained version of the MSSM (mSUGRA) \cite{rev}
as implemented in ISAJET \cite{isajet}.
This mSUGRA scheme has a limited number of parameters,
has well established and simple mass relations, and is
implemented in event generators, besides having definite
phenomenological/theoretical attractiveness and plausibility.
In the mSUGRA model only five
extra parameters, which are not present in the SM,
need to be specified:
the universal scalar $m_{0}$ and gaugino $m_{1/2}$ masses,
the SUSY breaking universal trilinear coupling
$A_{0}$, the ratio of the vacuum expectation
values of the Higgs fields $tan \beta$ and the sign of the
Higgsino mixing parameter $sign(\mu)$. The first three
parameters are fixed at the gauge coupling unification
scale and the sparticle spectrum at the EW scale are then
obtained via renormalization group equations \cite{slm}.

Throughout this study except for cases specially mentioned
we largely limit ourselves to the set:
\begin{equation}
\mathrm{tan} \beta = 2, \ \ \ A_{0} = 0, \ \ \ \mu < 0,
\end{equation}
and also consider five representative parameter space points
suggested by theorists, and listed in Table 1.1.

\vspace{5mm}

Table 1.1: Representative mSUGRA points suggested by
theorists.

\begin{table}[htb]
\centering
\begin{tabular}{|l|c|c|c|c|c|}
\hline
Point & $m_0$ \ (GeV) & $m_{1/2}$ \ (GeV)& $A_0$ \ (GeV)&tan$\beta$&
\ \ \ $\mu$ \ \ \\
\hline
\hline
\ \ \ 1 & 100 & 300 & 300 & 2.1 & $>0$ \\
\hline
\ \ \ 2 & 400 & 400 &   0 & 2   & $>0$ \\
\hline
\ \ \ 3 & 400 & 400 &   0 & 10  & $>0$ \\
\hline
\ \ \ 4 & 200 & 100 & 0 & 2   & $<0$ \\
\hline
\ \ \ 5 & 800 & 200 & 0  & 10  & $>0$ \\
\hline
\end{tabular}
\end{table} 

\vspace{2mm}

In the following we discuss the mass relations among sparticles
in mSUGRA, display isomass curves in terms of model parameters,
discuss production mechanisms and decay modes accessible to
experimental observations.


\subsection{Charginos and neutralinos}

The mixing of the fermionic partners of the electroweak gauge and
Higgs bosons, the gauginos and the Higgsinos, gives rise to the
physical mass eigenstates called the charginos
($\tilde{\chi}^{\pm}_{1,2}$) and neutralinos 
($\tilde{\chi}^0_{1,2,3,4}$). They are among the lightest expected SUSY
particles and therefore present  particular interest.
The two lightest neutralinos and the
lightest chargino (\chna, \chnb, \chha)
have as their largest mixing component the gauginos, and
their masses are determined by the common gaugino mass, $m_{1/2}$.
Within mSUGRA:

\begin{equation}
M_{\tilde{\chi}^{0}_{1}} \approx 0.45 m_{1/2}
\end{equation}
\begin{equation}
M_{\tilde{\chi}^{0}_{2}} \approx
M_{\tilde{\chi}^{\pm}_{1}} \approx 2 M_{\tilde{\chi}^{0}_{1}}
\end{equation}
\begin{equation}
M_{\tilde{\chi}^{0}_{2}}
\approx (0.25 \div 0.35) M_{\tilde{g}}
\end{equation}

Figure 2.1 shows the isomass contours of (\chna, \chnb, \chha)
and  of (\chnc, \chnd, \chhb) in the ($m_0, m_{1/2}$) plane.
We remind the reader that in most of parameter space and in  the more
plausible scenarios 
the lightest neutralino, \chna, 
is the Lightest Supersimmetric
Particle (LSP) and as such is particularly interesting as it is one
of the most plausible cosmic dark matter candidates particle physics
provides.

The lightest chargino
\chha \ has several leptonic decay modes, giving an isolated lepton
and missing energy due to the undetectable neutrino
and LSP  (\chna \ in mSUGRA):
 
\vspace*{3mm}
 
\hspace*{5mm} $\bullet$ $\tilde{\chi}_1^{\pm}$ $\longrightarrow$  
$\tilde{\chi}_1^0$ $l^{\pm}$ $\nu$
\hspace*{25mm} three-body decay
 
\vspace{5mm}
 
\hspace*{5mm} $\bullet$ $\tilde{\chi}_1^{\pm}$ $\longrightarrow$
$\tilde{l}_{L}^{\pm}$ $\nu$
 
  \ \ \ \ \ \ \ \ \ \ \ \ \ \ \ \ \ \ \
  $\hookrightarrow$ $\tilde{\chi}_1^{0}$ $l^{\pm}$
 
\vspace{4mm}
 
\hspace*{5mm} $\bullet$ $\tilde{\chi}_1^{\pm}$ $\longrightarrow$  
$\tilde{\nu}_{L}$ $l^\pm$
\hspace*{29mm} two-body decays
 
 \ \ \ \ \ \ \ \ \ \ \ \ \ \ \ \ \ \ \
 $\hookrightarrow$ $\tilde{\chi}_1^{0}$ $\nu$
 
\vspace{4mm}
 
\hspace*{5mm} $\bullet$ $\tilde{\chi}_1^{\pm}$ $\longrightarrow$
$\tilde{\chi}_{1}^{0}$  $W^\pm$
 
 \ \ \ \ \ \ \ \ \ \ \ \ \ \ \ \ \ \ \ \ \ \ \
$\hookrightarrow$ $l^\pm$ $\nu$
 
\vspace{-42mm}
 
$$\hspace*{-50mm}\left\rbrace{
\begin{array}{l}
 
\\
 
\\
 
\vspace{4mm}
 
\\
 
\\
 
\vspace{4mm}
 
\\
 
\\
 
\end{array} }
\right.$$

\vspace*{1mm}
 
Figure 2.2 shows the branching ratios for the above listed \chha \ decay
channels as a function of ($m_0, m_{1/2}$).
One can see that in different regions these different leptonic decay modes
are complementary, amounting to a larger than
20$\%$ branching in the entire ($m_0, m_{1/2}$) plane (for $l=$ e, $\mu$).
Three-body decays are open in the
region $m_{1/2} \lappeq 200$ GeV and $m_{1/2} \gsim 0.5 m_{0}$,
whereas in the rest of parameter
space the two-body decays are dominant.
 
Leptonic decays of \chnb \ give two isolated leptons and
missing energy (\chna):

\vspace*{3mm}
 
\hspace*{5mm} $\bullet$ $\tilde{\chi}_2^{0}$ $\longrightarrow$
$\tilde{\chi}_1^0$ $l^+$ $l^-$

\vspace{4mm}

\hspace*{5mm}  $\bullet$ $\tilde{\chi}_2^{0}$ $\longrightarrow$
$\tilde{\chi}_1^{\pm}$ $l^{\mp}$ $\nu$
\hspace*{25mm} three-body decays
 
  \ \ \ \ \ \ \ \ \ \ \ \ \ \ \ \ \ \
$\hookrightarrow$ $\tilde{\chi}_1^{0}$ $l^\pm$ $\nu$

\vspace{5mm}
 
\hspace*{5mm} $\bullet$ $\tilde{\chi}_2^{0}$ $\longrightarrow$
$\tilde{l}_{L,R}^{\pm}$ $l^{\mp}$
\hspace*{27.5mm} two-body decay
 
  \ \ \ \ \ \ \ \ \ \ \ \ \ \ \ \ \ \ \
$\hookrightarrow$ $\tilde{\chi}_1^{0}$ $l^\pm$
 
\vspace{-39mm}
 
$$\hspace*{-50mm}\left\rbrace{
\begin{array}{l}

\\
 
\\
 
\\
 
\\

\end{array} }
\right.$$
  
\vspace{16mm}
 
Figure 2.3 shows the \chnb \ leptonic branching ratios
as a function of ($m_0, m_{1/2}$). Like the \chha \ case, the three-body
decay branching ratios are sizable at
$m_{1/2} \lsim 200$ GeV and $m_{1/2} \gsim 0.5 m_{0}$.
Beyond $m_{1/2} \simeq $ 200 GeV, the
three-body decay of \chnb \ is suppressed due to the opening up of
the channels \chnb $\rightarrow$\chna~h and \chnb $\rightarrow$
\chna~Z (``spoiler'' modes). The two-body decay  branchings are 
significant if $m_{0} \lsim 0.5 m_{1/2}$.

In hadronic collisions, the charginos and neutralinos
can be produced directly via a Drell-Yan mechanism or, more abundantly,
through the cascade decays of strongly interacting sparticles.
There is also the possibility of associated production
$\tilde{\chi}_i^{\pm}/\tilde{\chi}_j^{0}+\tilde{g}/\tilde{q}$.
There are 21 different reactions 
(8 $\tilde{\chi}_i^{\pm} \tilde{\chi}_j^{0}$,
3 $\tilde{\chi}_i^{\pm} \tilde{\chi}_j^{\pm}$ and
10 $\tilde{\chi}_i^{0} \tilde{\chi}_j^{0}$) for direct
chargino-neutralino pair production
among which \chha \chnb \ production followed by leptonic
decays of \chha \  and \chnb \ is most easy from the
detection point of view, yielding a 3 $leptons + no \ jets +$\etm \
event topology.
  
Figure 2.4 shows the cross-section times branching ratio,
$\sigma \times$ B$(3l  + invisible)$, where
$\sigma $ is the \chha \chnb \ direct 
production cross-section and B$(3l + invisible)$ is the
convolution of all the leptonic decays of \chha \ and \chnb \
listed above. Within a relatively small region of $m_{1/2}$ and for
$m_{0} \gappeq 400$ GeV, the $\sigma \times$ B$(3l + invisible)$ drops by
several orders of magnitude and vanishes at $m_{1/2} \simeq $ 200 GeV. For
$m_{0} \lappeq 400$ GeV the slope is less sharp thanks to the presence of 
two-body decay modes.

Probabilities for production of \chha \ and \chnb \ from gluinos
and squarks  are shown in Figs.~2.5a and 2.5b for
tan$\beta = 2$, $\mu < 0$.
One sees that these two
sources of charginos and neutralinos are complementary in ($m_{0},
m_{1/2}$) parameter space. This abundant production of \chnb \ from
strongly interacting sparticles over the whole ($m_{0}, m_{1/2}$)
plane has very useful experimental implications,
which will be discussed in detail later in this paper.
The convolution of the \chnb \ indirect production
cross-section with its leptonic decay branching ratio is
shown in Fig.~2.5d. Clearly, over a large portion of the
($m_{0}, m_{1/2}$) plane, one expects rather
large rates of same-flavor opposite-sign dileptons
originating from \chnb .

Similarly to the \chnb, the lightest chargino \chha \ is predominantly
produced from strongly interacting sparticles, as illustrated in
Figs.~2.6a,b. Moreover, the leptonic decay branching of the \chha \
always exceeds 0.1 per lepton flavor (Fig.~2.6c).
For low values of $m_0$, in the region where decay to sleptons are
kinematically allowed, the \chha \ decays to a lepton with a probability
close to 1.

Figure 2.7 shows the sum of
$\tilde{\chi}_2^{0} \to \tilde{\chi}_1^0 l^+ l^-$ and
$\tilde{\chi}_2^{0} \to \tilde{l} l \to \tilde{\chi}_1^0 l^+ l^-$
branching ratios for larger values of tan$\beta = 10$ and 35 for
both signs of $\mu$. These branchings to e$^+$e$^-$ or $\mu^+\mu^-$
are smaller than at tan$\beta = 2$ shown in Fig.~2.5c.


\subsection{Sleptons}

The SUSY partners of ordinary  leptons are scalars. 
Left and right-handed charged sleptons are not mass-degenerate.
The slepton masses are determined by $m_0$ and $m_{1/2}$ \cite{slm}:

\begin{equation}
\hspace{2cm} m^2_{\tilde {l}_R}=m_0^2+0.15m^2_{1/2}-
                                   sin^2{\theta}_W M_Z^2cos2\beta
\end{equation}

\begin{equation}         
\hspace{2cm} m^2_{\tilde {l}_L}=m_0^2+0.52m^2_{1/2}-
                                   1/2(1-2sin^2{\theta}_W)M_Z^2cos2\beta
\end{equation}

\begin{equation}
\hspace{2cm} m^2_{\tilde {\nu}}=m_0^2+0.52m^2_{1/2}+1/2M_Z^2cos2\beta
\end{equation}

Asymptotically, for $m_{1/2}\rightarrow 0$ the slepton mass is
determined by $m_0$. Charged left sleptons are the heaviest.
The mass dependence on $m_{1/2}$ for left (L) sleptons is 
stronger than for right ones (R), hence the left-right slepton mass 
splitting  increases  with $m_{1/2}$. Figure 2.8a shows slepton isomass
contours in the $(m_0,m_{1/2})$ plane. 

Since \chha, \chnb \ isomass contours (see Fig.~2.1)
are approximately straight lines of constant $m_{1/2}$, 
as one moves up a slepton isomass curve, the \chnb \ mass increases and 
eventually becomes higher than that of the slepton. Thus 
we may identify two domains in the $(m_{0},m_{1/2})$ parameter plane: one
is where left sleptons are more massive than \chha, \chnb \  
($m_0\gappeq 0.45\cdot m_{1/2}$, domain I), and the other, where they are
lighter ($m_0\lappeq 0.45\cdot m_{1/2}$, domain II).  
These mass relations, shown Fig.~2.8b are responsible for distinct
contributions to slepton production and decay mechanisms.
In domain I sleptons can  be produced only through a Drell-Yan
mechanism ($direct$ slepton production), via $q\bar q$ annihilation with
neutral or charged gauge boson exchange in the s-channel. 
Pairs with the same ``handedness'' can be produced, namely, 
$\tilde{l}_L\tilde{l}_L, \, 
\tilde{l}_R\tilde{l}_R, \, \tilde{\nu}\tilde{\nu}, 
\tilde{\nu}_l\tilde{l}_L$. 
In this part of parameter space
both direct and cascade decays  of sleptons to LSP are possible:

i) the left sleptons can decay to charginos and neutralinos 
via the following decays:

\vspace {0.15cm}
 
$\hspace{4cm} \tilde{l}_L^{\pm} \rightarrow l^{\pm} + \tilde{\chi}_{1,2}^0$ 

\vspace {0.15cm}
 
 $\hspace{4cm} \tilde{l}_L^{\pm} \rightarrow \nu_l + \tilde{\chi}_1^{\pm}$ 

\vspace {0.15cm}
  
 $\hspace{4cm} \tilde{\nu}_l \rightarrow \nu_l + \tilde{\chi}_{1,2}^0 $

\vspace {0.15cm}
 
 $\hspace{4cm} \tilde{\nu}_l \rightarrow l^{\pm} + \tilde{\chi}_1^{\mp} $ 

\vspace {0.15cm} 

ii) for right sleptons, only decays to neutralinos are 
possible; they  dominantly decay directly to the LSP:  

\vspace {0.15cm} 

 $\hspace{4cm} \tilde{l}_R^- \rightarrow l^- + \tilde{\chi}_1^0 $

\vspace {0.15cm}

If decays to the second neutralino or lightest chargino
are kinematically allowed, the decay to the LSP can proceed through 
several decay paths (cascade decays)
with missing energy and/or isolated leptons in final state: 

\vspace {0.15cm}
 
 $\hspace{4cm} \tilde{\chi}_2^0 \rightarrow  \tilde{\chi}_1^0 + l^+l^- $  

\vspace {0.15cm} 

 $\hspace{4cm} \tilde{\chi}_2^0 \rightarrow  \tilde{\chi}_1^0 + 
\nu \overline{\nu} $

\vspace {0.15cm} 

$\hspace{4cm} \tilde{\chi}_2^0 \rightarrow  \tilde{\chi}_1^0 + Z^0$

\vspace {0.15cm}
 
$\hspace{4cm} \tilde{\chi}_1^{\pm}\rightarrow \tilde{\chi}_1^0 + l^{\pm}+\nu$

\vspace {0.15cm}

$\hspace{4cm} \tilde{\chi}_1^{\pm} \rightarrow  \tilde{\chi}_1^0 + W^{\pm}$

\vspace {0.15cm}

In domain II $indirect$ slepton
production, from  directly or indirectly produced 

\vspace {0.15cm}

\hspace{-0.5cm} charginos/neutralinos is also  possible: 

\vspace {0.15cm}

$\hspace{4cm} \tilde{\chi}_2^0 \rightarrow \tilde{l}_{L,R}^{\pm}l^{\mp}$

\vspace {0.15cm} 

$\hspace{4cm} \tilde{\chi}_2^0 \rightarrow \tilde{\nu}_l \bar{\nu}_l$

\vspace {0.15cm} 

$\hspace{4cm} \tilde{\chi}_1^{\pm} \rightarrow \tilde{\nu}_ll^{\pm}$

\vspace {0.15cm} 

$\hspace{4cm} \tilde{\chi}_1^{\pm} \rightarrow \tilde{l}_L^{\pm}\nu_l$

\vspace {0.15cm}

When allowed, $indirect$ $\tilde{l}$ production arising from decays
of strongly produced $\tilde{g},\tilde{q}$ to \chnb \
or \chha, is the predominant production mode.
With increasing $m_{1/2}$ along a slepton isomass curve, the fraction of
direct decays into the LSP increases as the mass difference between
sleptons and charginos-neutralinos 
decreases, until  in domain II sleptons can only directly 
decay to the LSP.  
Sneutrinos in this domain thus decay totally invisibly, i.e.
no component ($\nu,\tilde{\chi}_1^0$) from sneutrino 
decay can be directly detected.
The slepton production and decay features described  above give rise to
several characteristic experimental signatures:

\begin{itemize}
\item $two\ leptons + E_T^{miss} + no\ jets$ event topology
arises from direct charged slepton pair
$\tilde{l}_L\tilde{l}_L,\tilde{l}_R\tilde{l}_R$ production,   
followed by direct  decays to the LSP.
The same signature is found in $\tilde{\nu}\tilde{\nu}$ production 
when one of the sneutrinos decays invisibly 
and the other has a cascade decay with two leptons in final state. 
This  signature can also be found when sleptons are produced among decay 
products of directly or indirectly produced charginos-neutralinos. 
In the case of indirectly produced sleptons there are the following
differences compared with direct production: 
i) single production is also possible, 
for example in $\tilde{\chi}_1^{\pm}\tilde{\chi}_2^0$ production followed by 
decay of the $\tilde{\chi}_2^0$ to a slepton; ii) a
$\tilde{\nu}_l\tilde{l}_L$ pair 
can be produced from direct $\tilde{\chi}_1^{\pm}
\tilde{\chi}_1^{\mp}$; iii) left handed sleptons production is 
favored; 
\item
$three\,\,leptons\,+E_T^{miss}+ no\,\,jets$ final states are 
expected only from  $\tilde{\nu}_l\tilde{l}_L$ production followed by
cascade decays to  LSP;
\item
$four\,\,leptons\,+E_T^{miss}+ no\,\,jets$ comes from
$\tilde{l}_L\tilde{l}_L,\,\tilde{\nu}\tilde{\nu}$, 
$\tilde{\nu}_l\tilde{l}_L$ production 
followed by cascade decays to LSP.

The last two topologies  can also be found when sleptons are
produced in the decay of directly produced charginos-neutralinos.

\item 
Indirect slepton production from gluinos and squarks
 through charginos, neutralinos (domain II), leads to 
a $two\,\,(or \,\,more)\,\,leptons \, + E_T^{miss} + jets$ signature.
This production has the largest cross-section because of 
the strongly interacting  $\tilde{g}$ and $\tilde{q}$, and 
again sleptons can be singly produced.
\end{itemize}

The largest contribution to the two leptons with no jets final state in
charged slepton pair production  comes from their direct decays to 
the LSP, $\tilde{l}_L \rightarrow l$ \chna.
However, with decreasing $m_{1/2}$ along an isomass curve the fraction
of slepton direct decays to LSP 
decreases as shown in  Fig.~2.9a. Figure 2.9b
The $\sigma(\tilde{l}_L\tilde{l}_L)\times$B$(2l+invisible)$ for directly 
produced left sleptons and
Fig.~2.9c shows the same
for right sleptons:
$\sigma(\tilde{l}_R\tilde{l}_R)\times$B$(2l+invisible)$.
As right sleptons decay directly into LSP's the  cross-section
times branching ratio
is constant along an isomass curve.
One can see that the topological cross-section
$\sigma(\tilde{l}_R\tilde{l}_R)\times$B$(2l+invisible)$ is larger 
than
$\sigma(\tilde{l}_L\tilde{l}_L)\times$B$(2l+invisible)$ in almost the
entire ($m_0,  m_{1/2}$) plane.

For sneutrino pair production, the main contribution to the 2 leptons 
and no jets topology comes from one sneutrino
decaying  directly to the LSP and  the second to
$\tilde\chi_1^{\pm}$ or $\tilde{\chi}_2^0$ followed by their leptonic
decays. The sneutrino  decay branching ratios are shown in Figs.~2.10a-c.
The $\sigma(\tilde{\nu}\tilde{\nu})\times$B$(2l+invisible)$,
shown in Fig.~2.10d,
is limited along the $m_{1/2}$ axis by the $\tilde\chi_1^{\pm}$, 
$\tilde{\chi}_2^0$ leptonic decay branching
ratios (see Figs.~2.1 and 2.2),
and along the $m_0$ axis by the sneutrino production cross-section. 

Figure 2.11 shows the total cross-section times branching ratio for 
$two\ lepton + E_T^{miss} + no\ jets$ event 
topology arising from all possible combinations of directly produced 
slepton pairs, $\tilde{l}_L\tilde{l}_L$, $\tilde{l}_R\tilde{l}_R$,
$\tilde{\nu}\tilde{\nu}$. 
It drops rapidly with increasing $m_0$ and $m_{1/2}$, from a maximum
value of the order of a few pb down to a few fb for $m_0 \sim m_{1/2}\sim
500$ GeV. 
In almost the entire $(m_0,m_{1/2})$ parameter 
plane slepton pair production is dominated by right sleptons. 

To illustrate the additional contribution from \chha
$\tilde{\chi}_1^{\mp}$ direct production  and the
potential contribution from \chha \chnb \  to
the {\it two same-flavor opposite-sign leptons}+ \etm \ + {\it no
jets} event topology, we show  
$\sigma (\tilde{\chi}_1^{\pm}
\tilde{\chi}_1^{\mp})\times$B$(\tilde{\chi}_1^{\pm}
\tilde{\chi}_1^{\mp} \rightarrow l^+ l^- + invisible)$,
$\sigma (\tilde{\chi}_1^{\pm} \tilde{\chi}_2^{0})\times$B$(\tilde{\chi}_2^{0}
\rightarrow \tilde{l}_L^{\pm} + l^{\mp})$ and 
$\sigma (\tilde{\chi}_1^{\pm} \tilde{\chi}_2^{0})\times$B$(\tilde{\chi}_2^{0}
\rightarrow \tilde{l}_R^{\pm} + l^{\mp})$
in Figs.~2.12a, b and c, respectively.
One can see that the contribution from electroweak indirectly produced 
sleptons is not negligible, and in some areas of the $(m_0,m_{1/2})$ 
parameters space is even higher than for direct production.
Indirect slepton production from strongly produced
$\tilde{g}, \tilde{q}$ is larger still, when allowed, but is (almost)
always accompanied by jets in the final state, so will not contribute
unless the jets somehow escape detection.


\subsection{Gluinos and squarks}

In mSUGRA the gluino mass is determined mainly by $m_{1/2}$ \cite{slm}:
\begin{equation}
M_{\tilde{g}} \approx 2.5  m_{1/2}
\end{equation}
The five squark flavors ($\tilde{d}, \ \tilde{u}, \ \tilde{s},
\ \tilde{c}, \ \tilde{b}$) with their left and right chiral states
are assumed to be mass degenerate, giving altogether ten degenerate 
squark flavors with:
\begin{equation}
M_{\tilde{q}} \approx \sqrt{m_{0}^2 + 6m_{1/2}^2}
\end{equation}
The $\tilde{t}_L$ and $\tilde{t}_R$ are however treated differently.
Through their mixing they give rise to two mass eigenstates 
$\tilde{t}_1$ and $\tilde{t}_2$ with $\tilde{t}_1$ most often
significantly lighter than all other $\tilde{q}$ states, which requires a
separate discussion in terms of both, production and decay.
Mixing also introduces a small non-degeneracy among the
$\tilde{b}_1$ and $\tilde{b}_2$ states, but this is usually
neglected compared to behavior of the stop system.

At LHC energies, the total SUSY particle production cross-section
is dominated by strongly interacting gluinos
and squarks, which through their cascade decays, (Figs.~2.13 and 2.14)
can produce many jets and leptons, with missing energy due to at least 2
escaping LSPs and possibly  neutrinos.
Figure 2.15 shows the total squark/gluino production cross-section
contours versus $m_0$ and $m_{1/2}$. Figure 2.16 shows the
($m_{0}$,$m_{1/2}$) parameter space subdivided into several domains
corresponding to the characteristic kinematic configurations of
gluino and squark events and the predominant sources of
lepton production. Isomass contours for squarks, gluinos,
light and pseudoscalar Higgses are also shown in Fig.~2.16 for
tan$\beta = 2, \  A_{0} = 0, \ \mu < 0$, with h masses calculated at 
the two-loop radiative correction level.

The first domain is characterised by  $ m_{\tilde{g}} >
m_{\tilde{q}}$ and  $ m_{\tilde{\chi}_{2}^{0}} > m_{\tilde{l}}$.
This means that gluinos decay to
quark-squark pairs 
 but the contribution from gluino production
to the total squark/gluino cross-section
is small compared  to the squarks.
 Both  the \chnb \  and  \chha \
 in this domain  decay via sleptons (see Figs.~2.2 and 2.3) giving rise to
a significant number of leptons in the final
 state. This region extends to  higher values of $m_{0}$
at fixed  $m_{1/2}$ than just
two-body decays of \chnb \ and \chha \ to sleptons,
 since the three-body  decays (e.g. $\tilde{\chi}_{2}^{0}$
$\rightarrow$ $\tilde{\chi}_{1}^{0}$ $l^{+}l^{-}$)
 still remain even when two-body
modes $\tilde{\chi}_{2}^{0} \rightarrow \tilde{\chi}_{1}^{0}$h and
$\tilde{\chi}_{1}^{\pm} \rightarrow \tilde{\chi}_{1}^{0}$W$^\pm$ are    
open.

In the second domain of the ($m_{0}$,$m_{1/2}$) plane,
the gluino mass is still somewhat larger than the
squark masses, so that gluinos still decay into quark-squark pairs. The
yield of stop-top final states is however higher than in the first domain
since the contribution of the $ \tilde{g}\tilde{g}$ process in the total
$\tilde{g}/\tilde{q}$ production cross-section is larger and
the $ \tilde{g} \rightarrow  \tilde{t} t$
branching ratio is also higher. This results in an increase of the 
jet multiplicity and in a higher fraction of $b$-jets in the final
state.
The $ \tilde{\chi}_{2}^{0}$ and $\tilde{\chi}_{1}^{\pm}$ are now
lighter than
the sleptons, and they decay to LSP + h$/$W$^{\pm}$, respectively,
which means a reduced yield of isolated leptons when compared to the first
domain.

In the third characteristic region of the  ($m_{0}$,$m_{1/2}$) plane,  
the gluino mass is smaller
than the mass of squarks,
except for $ \tilde{t}_{1}$ and $\tilde{b}_{L}$, and hence
the stop-top final state dominates in overall production.
 The cross section for
  $\tilde{q}\tilde{q} $
 is significantly smaller than that for $ \tilde{g}\tilde{g} $. All this
gives rise to a significant jet multiplicity in the event and some
increase
of the  yield of isolated leptons,
as a result  of $t \rightarrow$ W$b \rightarrow l \nu b $ decays,
and an enrichment in the fraction of $b$-jets.
Compared with the previous domain,
nothing changes from the point of view
of the \chnb \ and \chha \ decays.
 
    The  fourth domain  is characterised
 by very massive  squarks compared
to  gluinos, thus  $ \tilde{g}\tilde{g}$ production dominates and the
gluino decays predominantly via three-body final states (mainly LSP + $ t
\bar{t}$). This increases further
 the jet (and $b$-jet) multiplicity of the event. Again the decays of the
$ \tilde{\chi}_{2}^{0}$  and   $\tilde{\chi}_{1}^{\pm}$ are the same as in
domains 2 and 3.
  
   There is a
fifth domain which has approximately the shape of a  band  along the 
 $m_{0}$ axis, under domains 2-4,  for  $m_{1/2}$ $\leq$ 180 - 190 GeV.
 In this region the masses of $\tilde{\chi}_{1}^{\pm}$ and
$\tilde{\chi}_{2}^{0}$
are smaller than $\tilde{\chi}_{1}^{0}$ + W and
$\tilde{\chi}_{1}^{0}$ + h, respectively, and smaller than the  slepton
and sneutrino masses, thus 3-body decays of $\tilde{\chi}_{2}^{0}$
($\rightarrow \tilde{\chi}_{1}^{0} \tilde{q} \tilde{\bar{q}} /
\tilde{\chi}_{1}^{0} l^{+}l^{-}$)
and $\tilde{\chi}_{1}^{\pm}$
($\rightarrow \tilde{\chi}_{1}^{0} \tilde{q} \tilde{\bar{q'}} /
\tilde{\chi}_{1}^{0} l^{\pm}\nu$) take place. However, the yield of
leptons does not differ significantly from that of domain 1.

\newpage

\ \ \ 

\vspace{10mm}

\ \ \ 

\begin{figure}
\vspace{-25mm}
        \rotatebox{0}{
        \includegraphics[ bb=103 77 519 750
                ,height=19.cm,width=13.cm,clip=true,draft=false]
                {D_Denegri_1018n.ill}}
\end{figure}

Figure  2.1: Isomass contours for: \ a) light (\chha, \chna,
 \chnb) \ and \ b) heavy (\chhb, \chnc, \chnd) charginos/neutralinos.

\begin{figure}[hbtp] 
\vspace{50mm}
\hspace*{-20mm}
    {\includegraphics{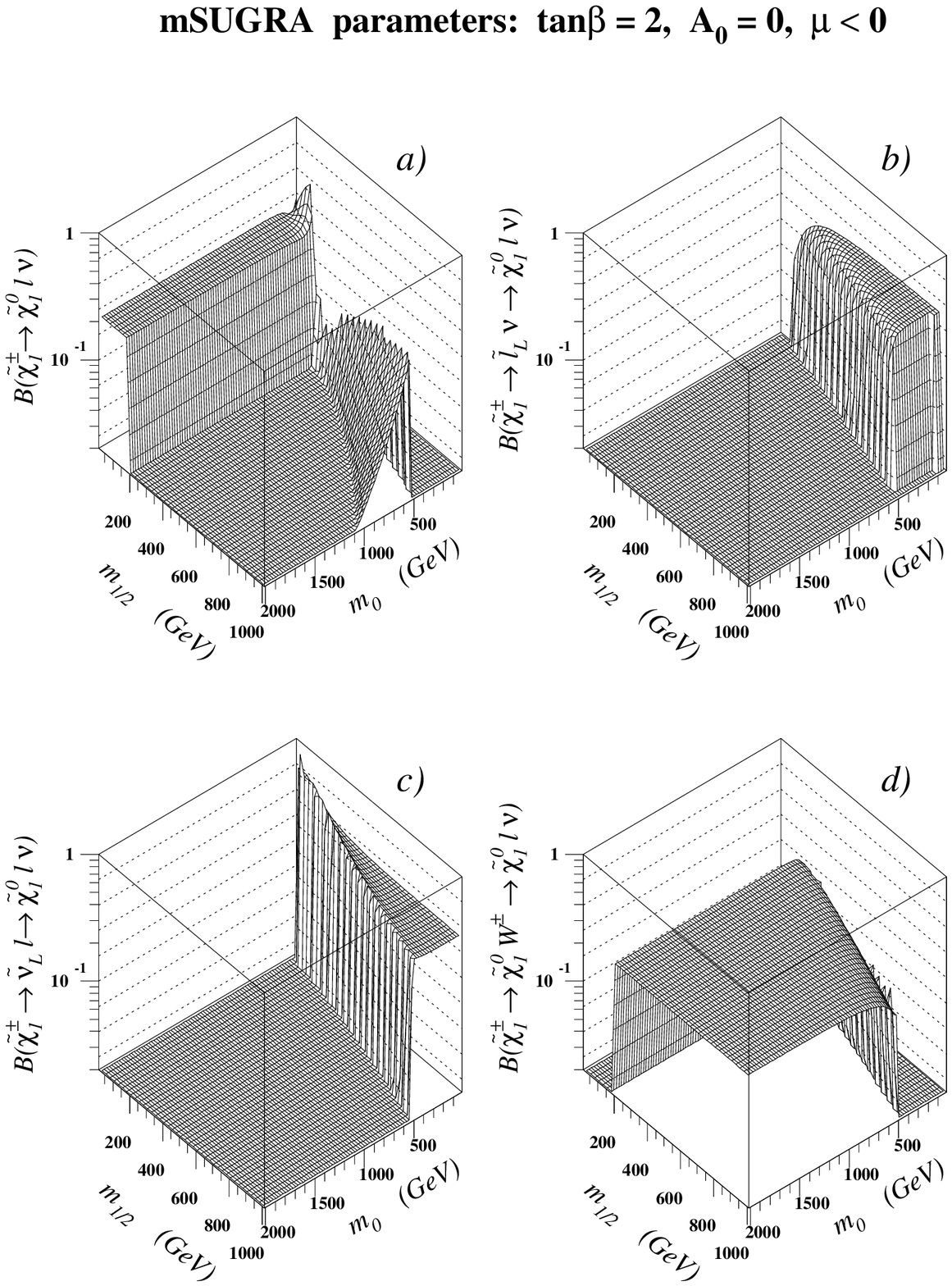}}

 \vspace{-30mm}

Figure  2.2: Chargino decay branching ratio versus ($m_0, m_{1/2}$): \
a) $\tilde{\chi}_1^{\pm}$ $\rightarrow$$\tilde{\chi}_1^0$
$l^{\pm}$ $\nu$, \ 
b) $\tilde{\chi}_1^{\pm}$ $\rightarrow$
$\tilde{l}_{L}^{\pm}$ $\nu$ $\rightarrow$
$\tilde{\chi}_1^{0}$ $l^\pm$ $\nu$, \
c) $\tilde{\chi}_1^{\pm}$ $\rightarrow$
$\tilde{l}_{R}^{\pm}$ $\nu$ $\rightarrow$
$\tilde{\chi}_1^{0}$ $l^\pm$ $\nu$ \ and \
d) $\tilde{\chi}_1^{\pm}$ $\rightarrow$
$\tilde{\chi}_{1}^{0}$  $W^\pm$ $\rightarrow$
$\tilde{\chi}_1^{0}$
$l^\pm$ $\nu$.
\end{figure}

\begin{figure}[hbtp]
\vspace{-15mm}
\hspace*{0mm}
    {\includegraphics{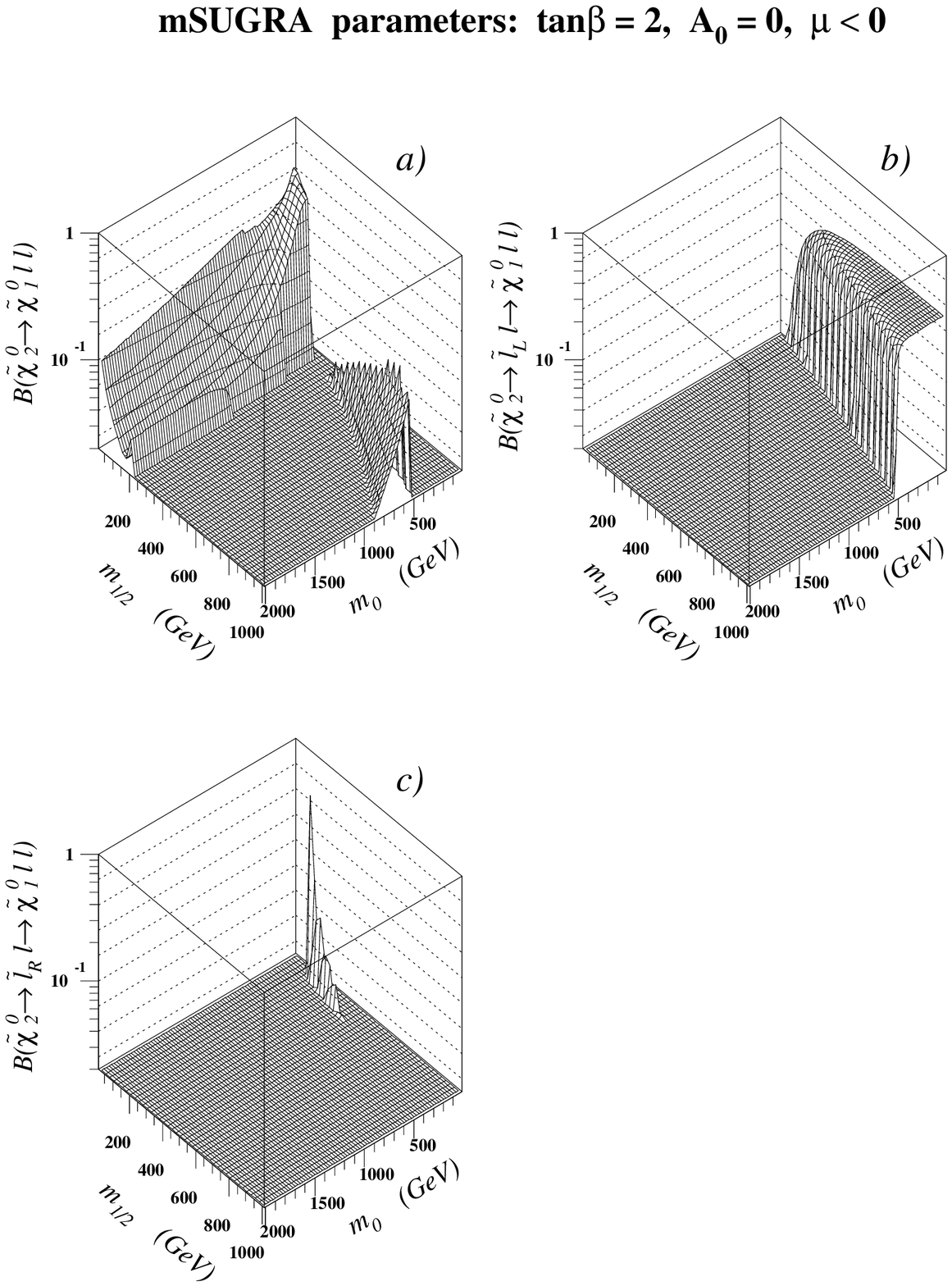}}

 \vspace{-0mm}

Figure 2.3: Neutralino decay branching ratio versus ($m_0, m_{1/2}$): \
a) $\tilde{\chi}_2^{0}$ $\rightarrow$  $\tilde{\chi}_1^0$
$l^+$ $l^-$, \
b) $\tilde{\chi}_2^{0}$ $\rightarrow$ $\tilde{l}_{L}^{\pm}$
$l^{\mp}$ $\rightarrow$  $\tilde{\chi}_1^0$ $l^+$ $l^-$ \ 
and \ c) $\tilde{\chi}_2^{0}$ $\rightarrow$ $\tilde{l}_{R}^{\pm}$
$l^{\mp}$ $\rightarrow$  $\tilde{\chi}_1^0$ $l^+$ $l^-$.
\end{figure}

\begin{figure}[hbtp]
\vspace{50mm}
\hspace*{-20mm}
    {\includegraphics{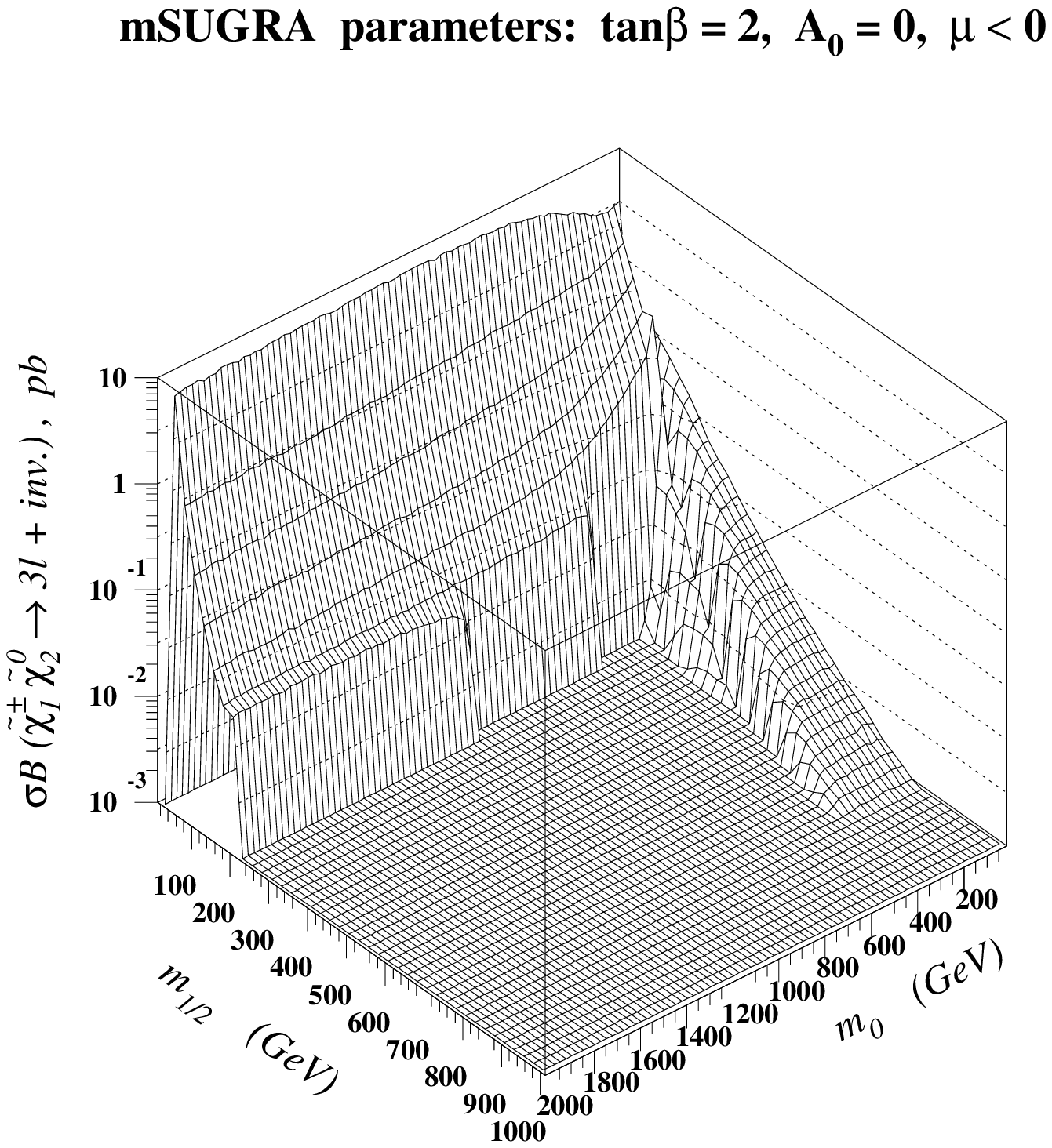}}

\vspace{-50mm}

Figure 2.4: Cross-section times branching ratio versus ($m_0, m_{1/2}$)
for \chha\chnb \ direct production followed by leptonic decays.
\end{figure}

\vspace{-50mm}

\begin{figure}[hbtp]
\vspace{-35mm}
\hspace*{-20mm}
    {\includegraphics{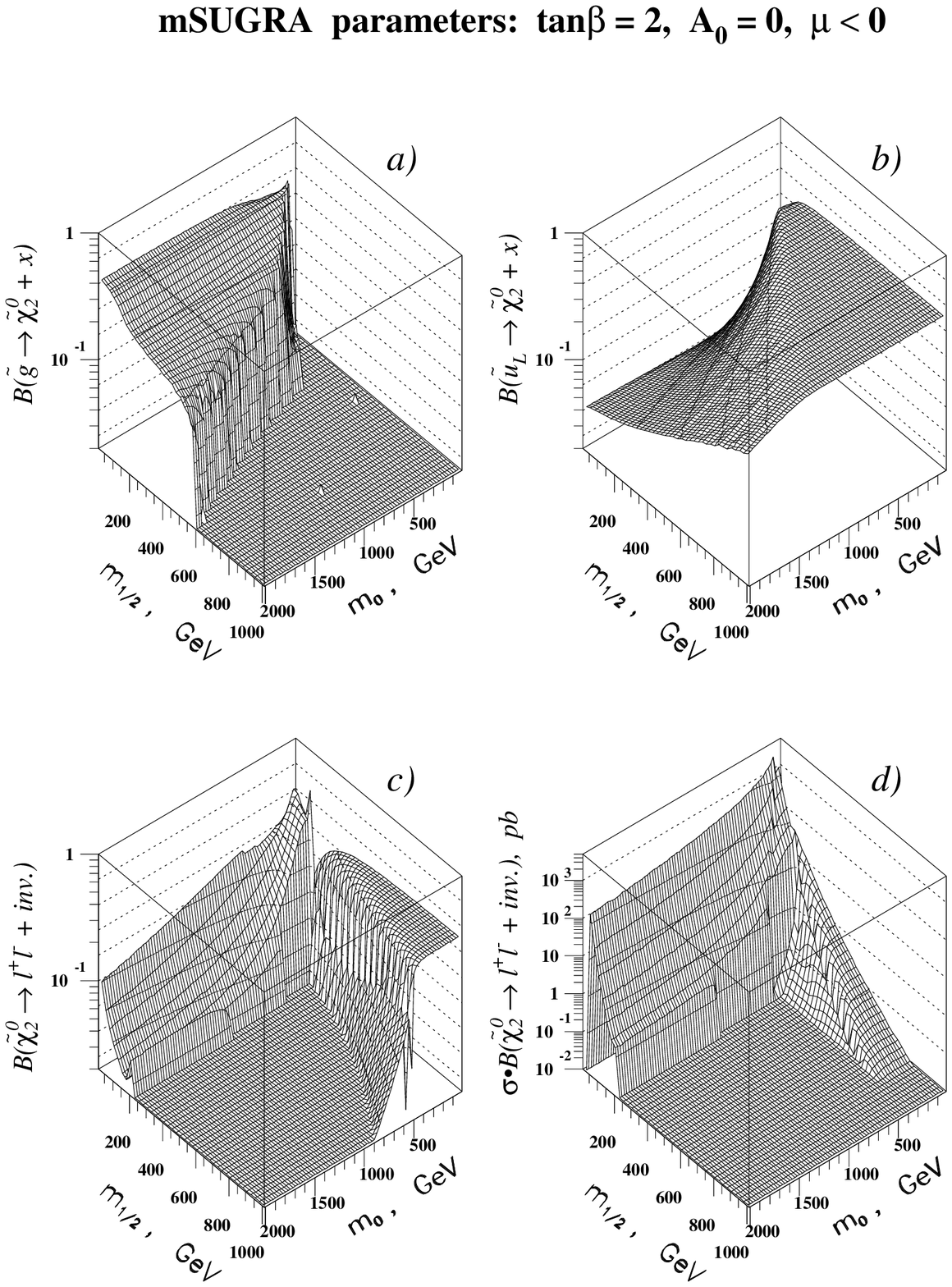}}

\vspace{10mm}

Figure 2.5: $\tilde{\chi}^{0}_{2}$ inclusive production
and decay: \ a) \ branching ratio
B($\tilde{g} \rightarrow \tilde{\chi}^{0}_{2} +$ x), \
b) \ B($\tilde{u}_{L} \rightarrow \tilde{\chi}^{0}_{2} +$ x), \ 
c) \ B($\tilde{\chi}^{0}_{2} \rightarrow l^+l^- + invisible$) \
and \ d) \ $\tilde{\chi}^{0}_{2}$ inclusive production cross-section
times branching ratio into $l^+l^-$.
\end{figure}

\vspace{-50mm}

\begin{figure}[hbtp]
\vspace{-35mm}
\hspace*{-20mm}
    {\includegraphics{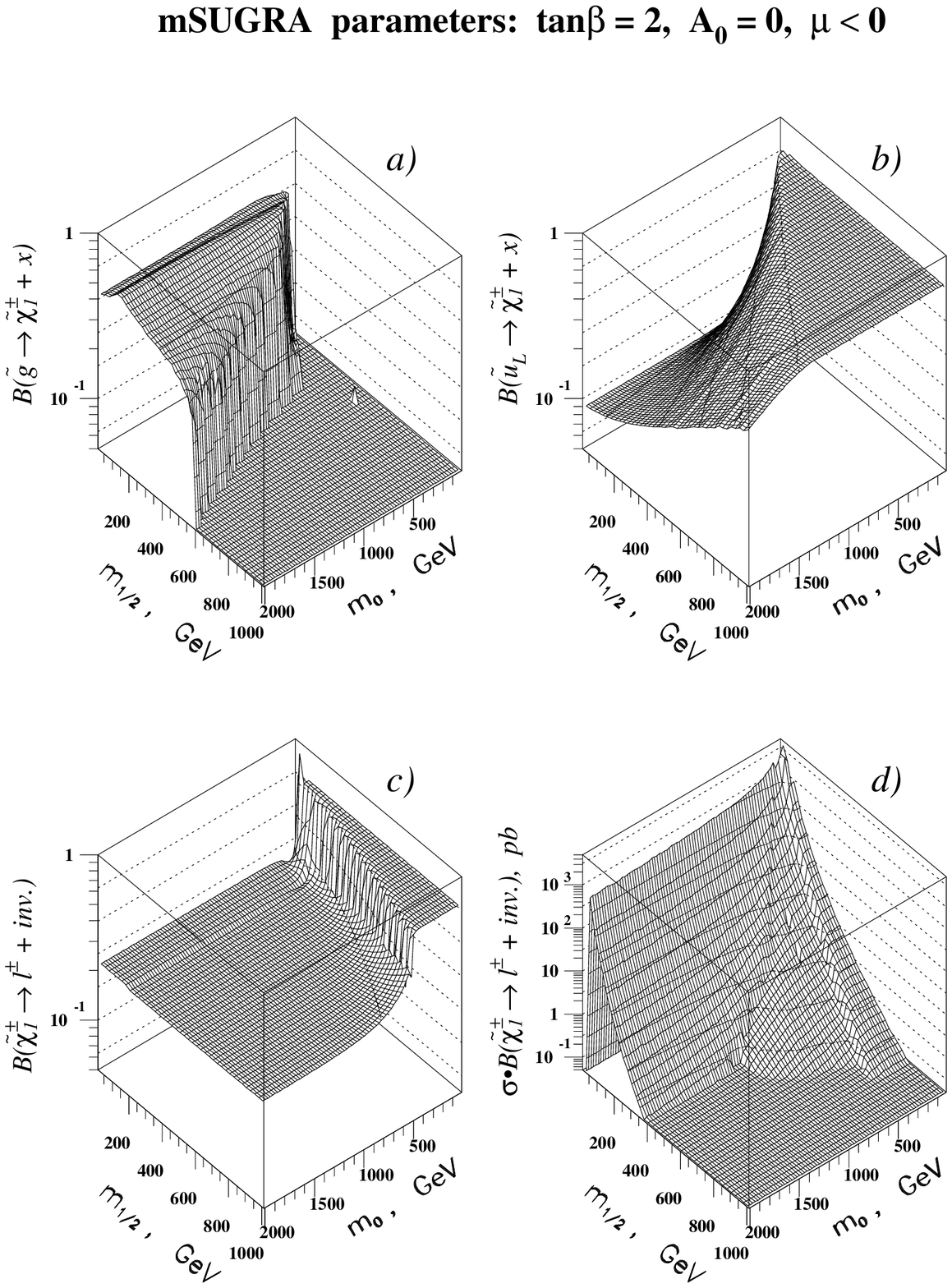}}

 \vspace{10mm}

Figure 2.6: \ $\tilde{\chi}^{\pm}_{1}$ inclusive production
and decay: \ a) \ branching ratio
B($\tilde{g} \rightarrow \tilde{\chi}^{\pm}_{1} +$ x), \   
b) \ B($\tilde{u}_{L} \rightarrow \tilde{\chi}^{\pm}_{1} +$ x), \ 
c) \ B($\tilde{\chi}^{\pm}_{1} \rightarrow l^{\pm} + invisible$) \
and \ d) \ $\tilde{\chi}^{\pm}_{1}$ inclusive production cross-section
times branching ratio into $l^{\pm}$.
\end{figure}

\newpage
 
\begin{figure}[hbtp]
\hspace*{5mm}
{\includegraphics[height=17.cm,width=13.cm,clip=true,draft=false]{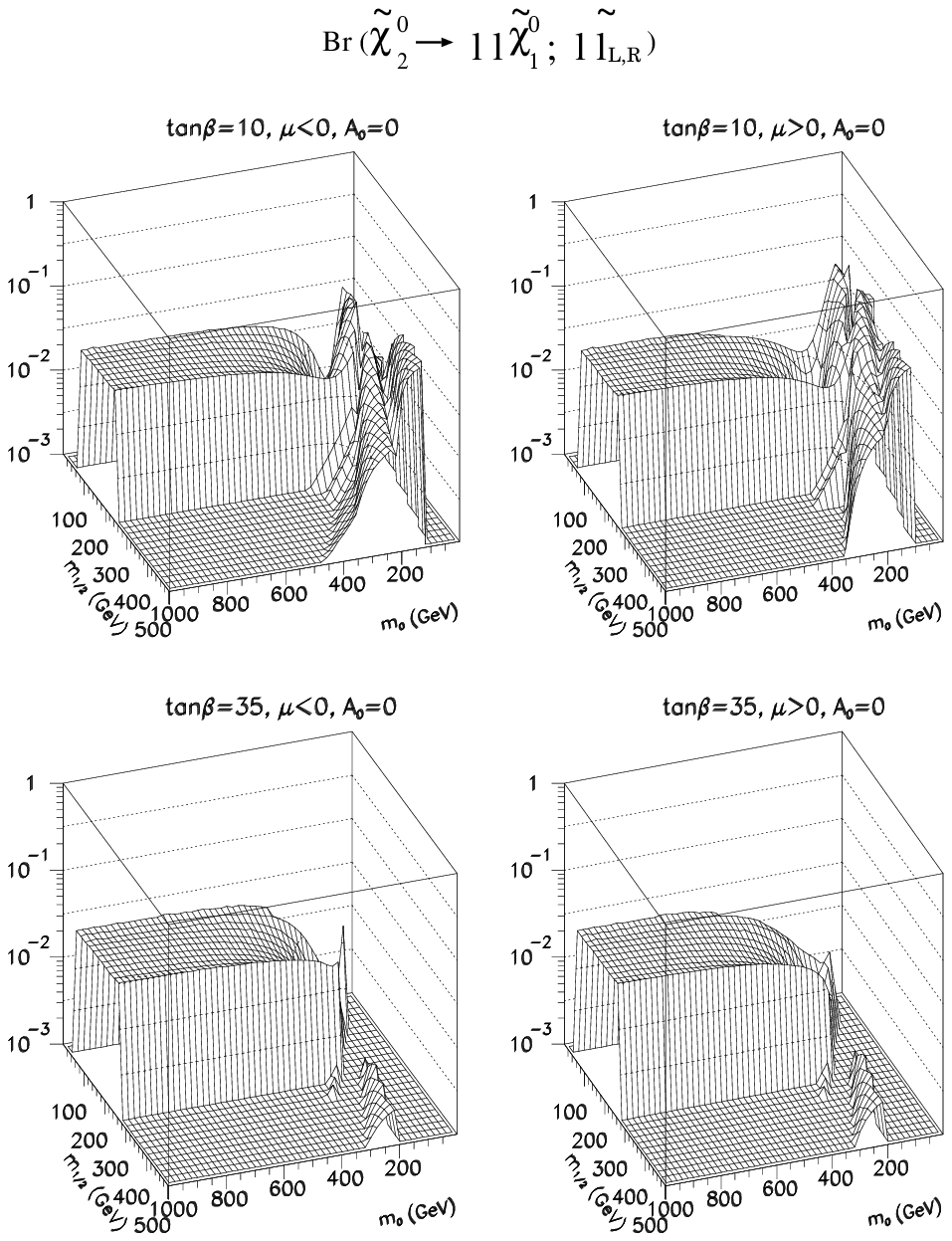}}
\end{figure}

\vspace{10mm}
 
Figure 2.7: The sum of
$\tilde{\chi}_2^{0} \to \tilde{\chi}_1^0 l^+ l^-$ and
$\tilde{\chi}_2^{0} \to \tilde{l} l \to \tilde{\chi}_1^0 l^+ l^-$
branching ratios for various values of tan$\beta = 10$ and 35 for
both signs of $\mu$.

\newpage

\ \ \\

\vspace{10mm} 

\begin{figure}[hbtp]
\vspace{-30mm}  
\hspace*{-20mm}
    {\includegraphics{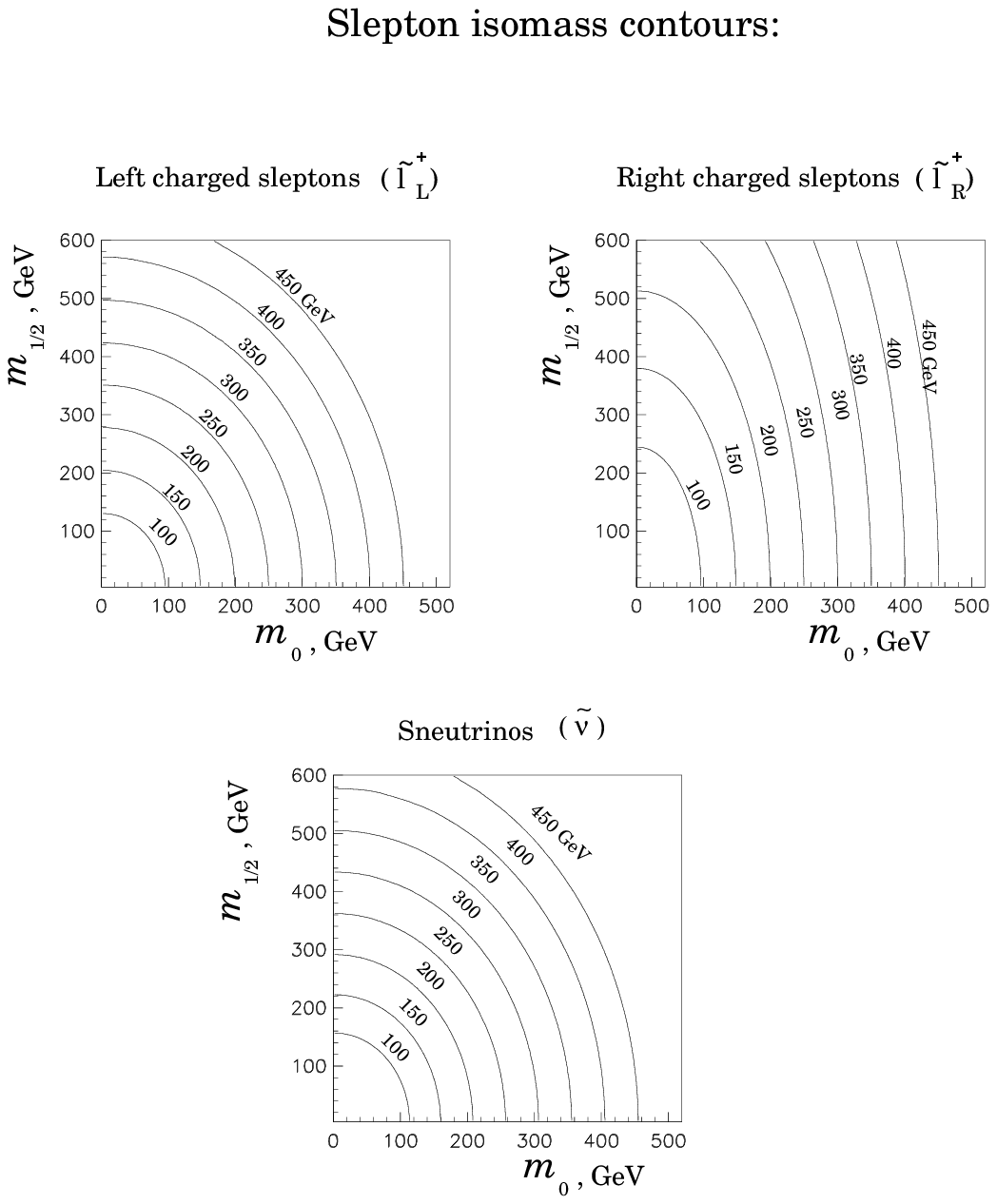}}
\end{figure}

 \vspace{5mm}

Figure 2.8a: Isomass contours for a) left charged sleptons, b) right
charged sleptons and c) sneutrinos in mSUGRA parameter space
($m_0,m_{1/2}$) for $tan\beta =2, \, A_0=0, \mu<0$.

\newpage

\begin{figure}[hbtp]
\vspace{10mm}
\hspace*{-10mm}
\resizebox{17.cm}{!}{\rotatebox{0}{\includegraphics{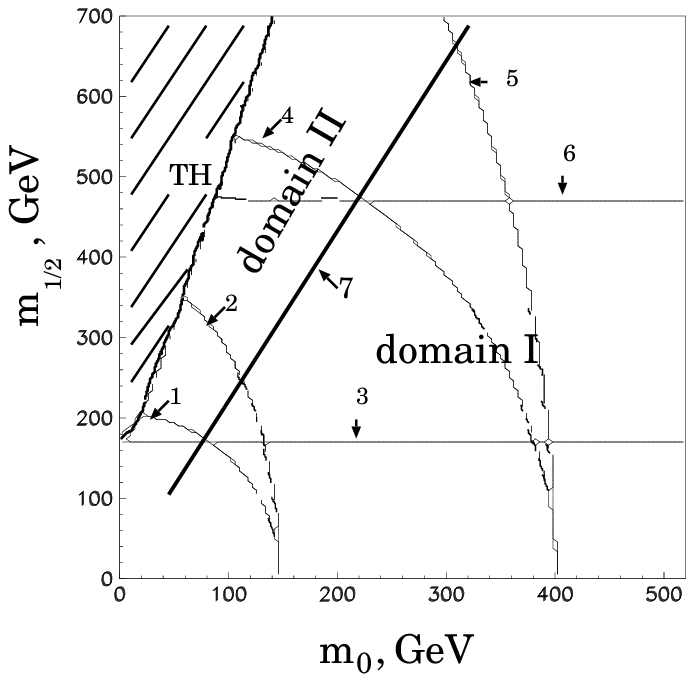}}}
\end{figure}

 \vspace{10mm}

Figure 2.8b: Slepton isomass contours for
$m_{\tilde{l}_L}=m_{\tilde{l}_R}=m_{\tilde{\chi}_2^0}$=150 GeV
corresponding
to lines 1-3 respectively and
$m_{\tilde{l}_L}=m_{\tilde{l}_R}=m_{\tilde{\chi}_2^0}$=400 GeV, lines 4-6
respectively. Line 7 separates domains I and  II of ($m_0, m_{1/2}$)
parameter space; in domain II charginos, neutralinos can decay to left
sleptons.

\newpage

\begin{figure}[hbtp]
\vspace{0mm}
\hspace*{-0mm}
\resizebox{16.5cm}{!}{\rotatebox{0}{\includegraphics{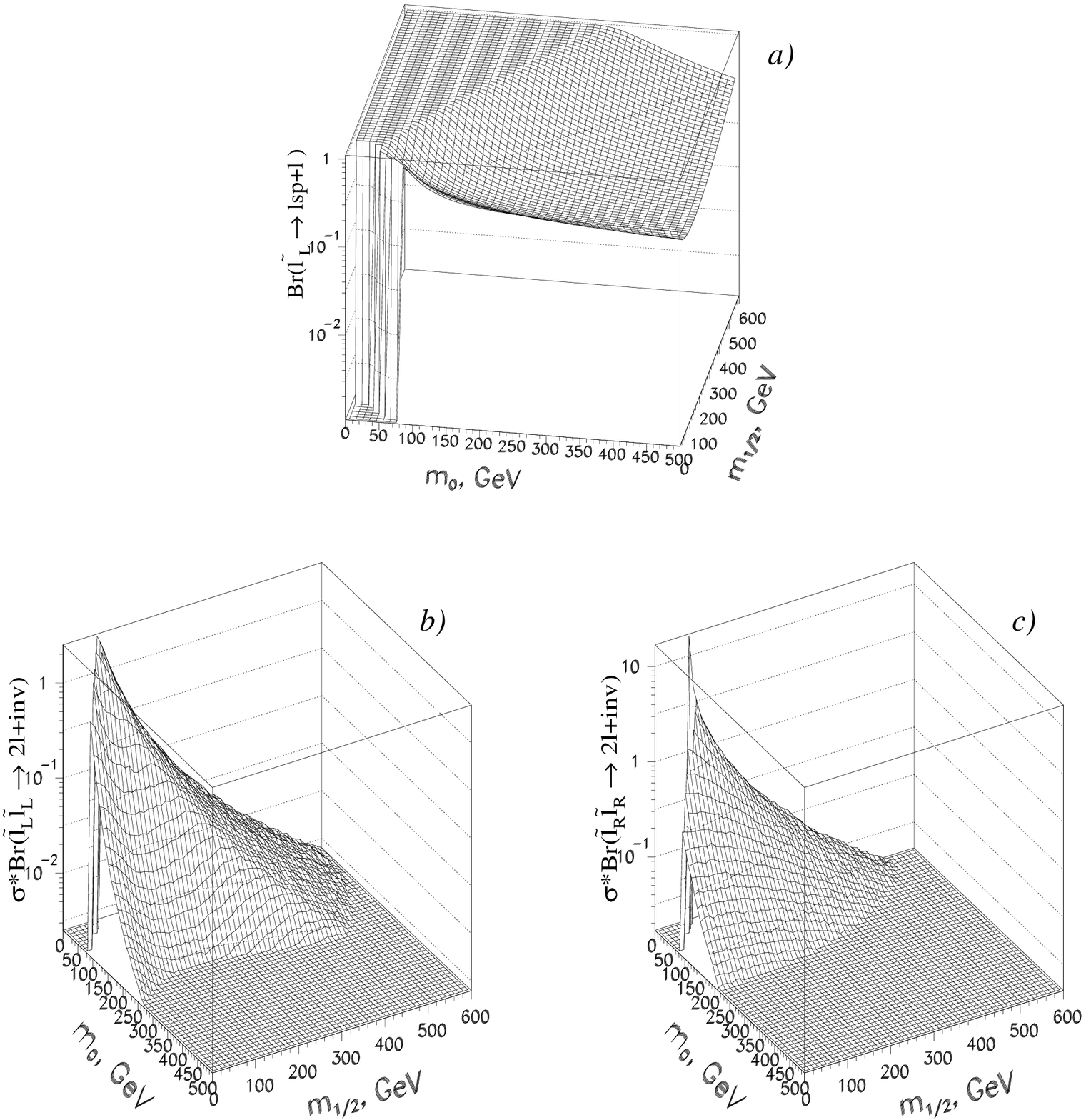}}}
\end{figure}

\vspace{20mm}

Figure 2.9: a)  
Branching ratio of $\tilde{l}_L \rightarrow$ \chna \ + $l$; b)
cross-section times branching ratio for $2\, leptons\, +
\,E_T^{miss}+\,no\, jets$
events  from direct left slepton pair production;
c) cross-section times branching ratio for
$2\, leptons\, + \,E_T^{miss}\,+ \,no\, jets$ 
events from direct right slepton pair production.

\newpage

\begin{figure}[hbtp]
\vspace{10mm}
\hspace*{10mm}
\resizebox{15.cm}{!}{\rotatebox{0}{\includegraphics{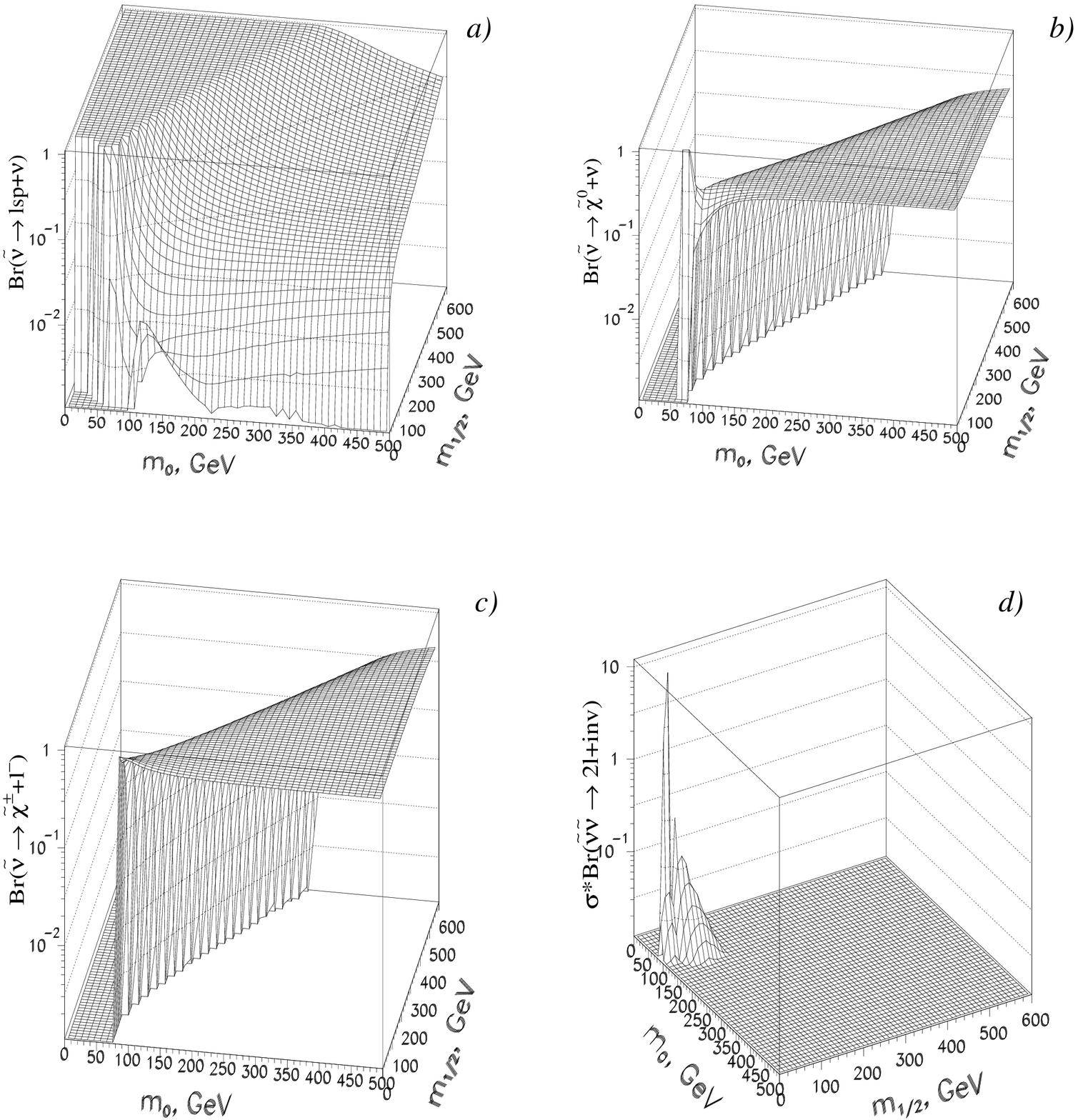}}}
\end{figure}

 \vspace{30mm}
Figure 2.10: 
Decay branching ratios of sneutrinos:
a) $\tilde{\nu} \rightarrow \chi_1^0 + \nu$,
b) $\tilde{\nu} \rightarrow \tilde{\chi}_2^0 + \nu$ and
c) $\tilde{\nu} \rightarrow \tilde{\chi}_1^{\pm} + l^{\mp}$ 
as a function of ($m_0, \, m_{1/2}$);
d) cross-section times branching ration
for $2\, leptons\, + \,E_T^{miss}\,+\,no\, jets$
events topology  from sneutrino pair production.

\newpage

\begin{figure}[hbtp]
\vspace{25mm}
\hspace*{5mm}
\resizebox{14.cm}{!}{\rotatebox{0}{\includegraphics{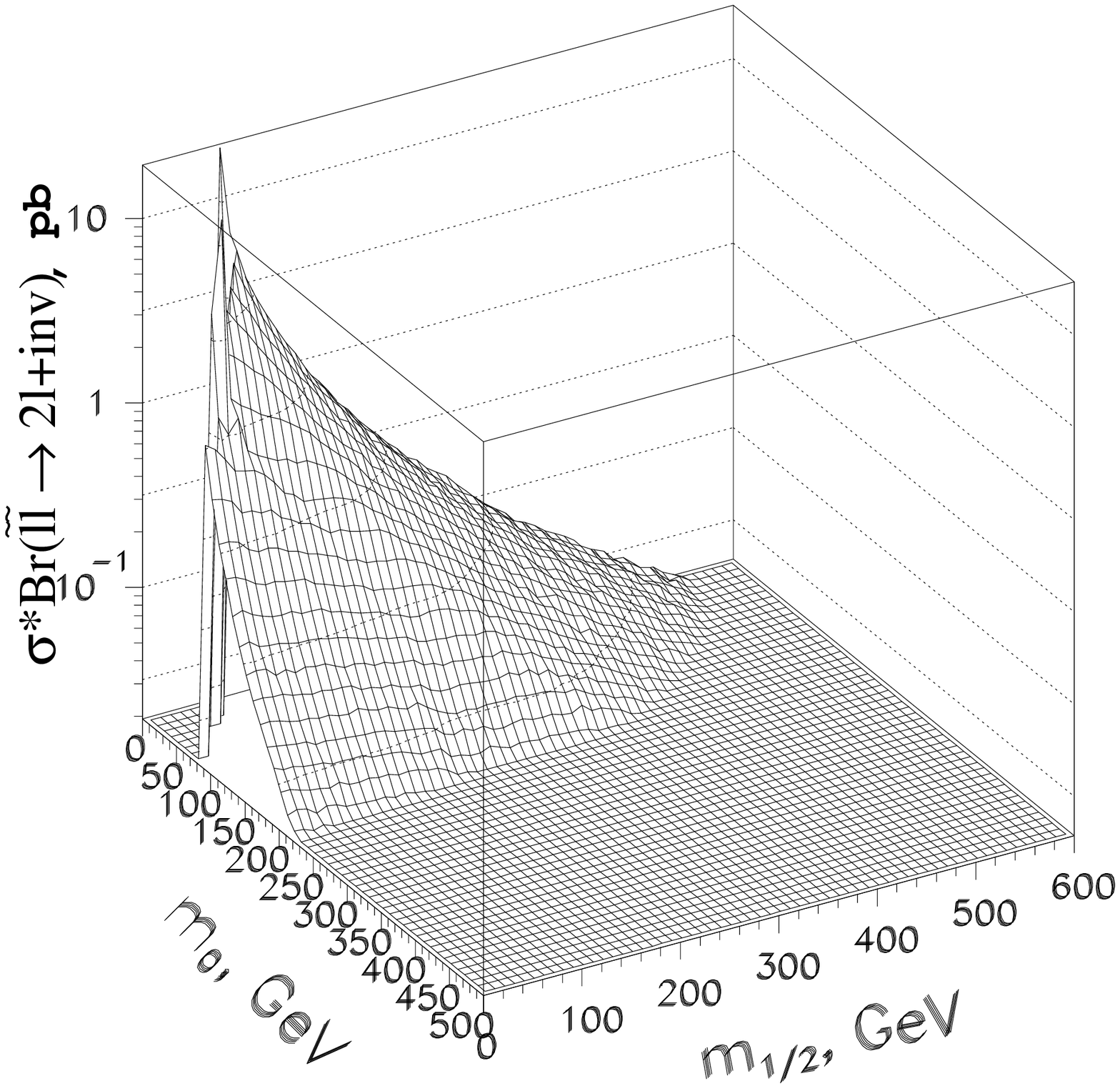}}}
\end{figure}

 \vspace{30mm}

Figure 2.11:
Total cross-section times branching ratio for 2
$leptons + E_T^{miss}$ events arising from all possible combinations of
directly produced slepton pairs, namely $\tilde{l}_L\tilde{l}_L$,
$\tilde{l}_R\tilde{l}_R$ and $\tilde{\nu}\tilde{\nu}$.

\newpage
 
\begin{figure}[hbtp]
\vspace{0mm}
\hspace*{0mm}
\resizebox{16.cm}{!}{\rotatebox{0}{\includegraphics{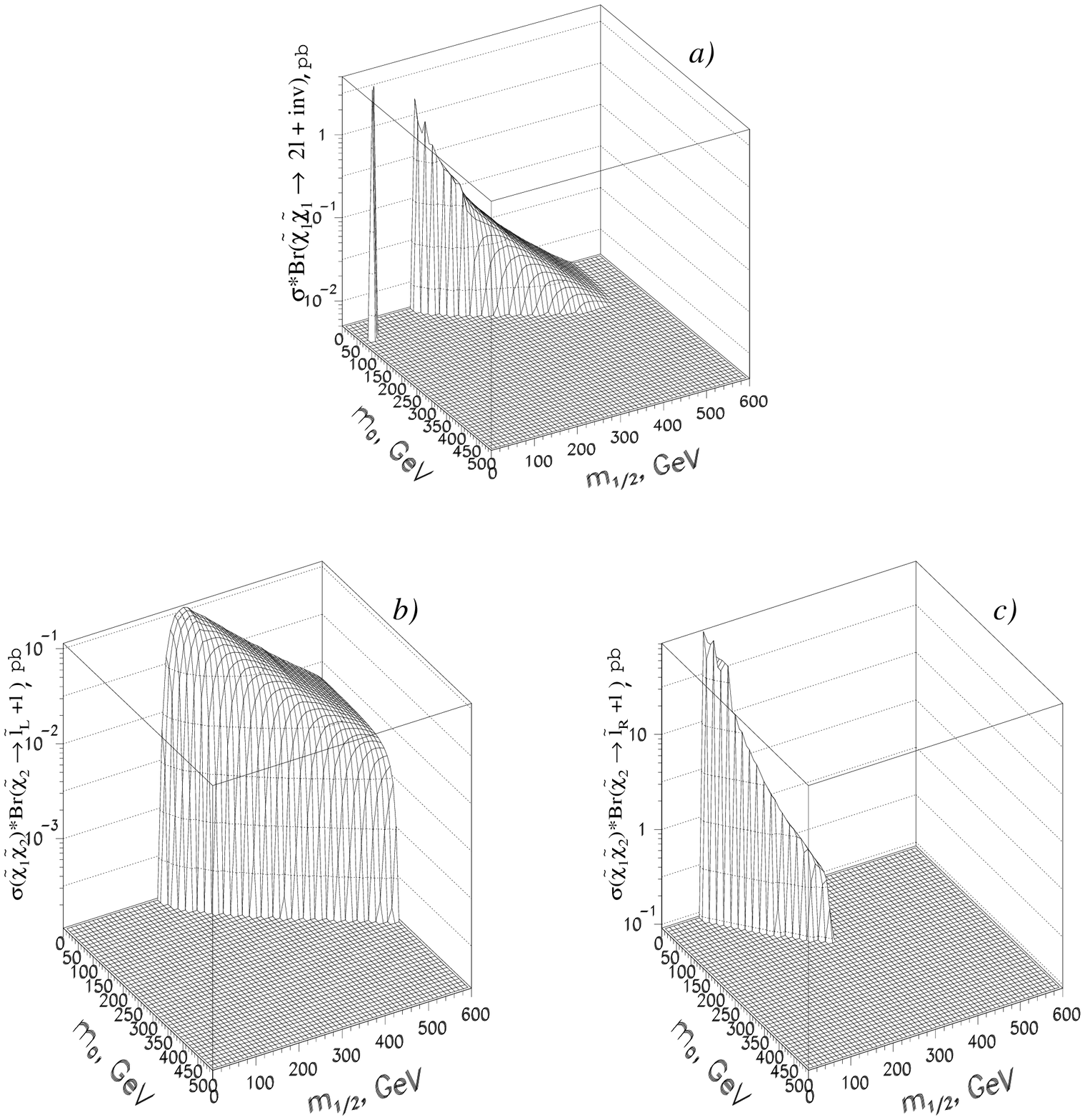}}}
\end{figure}

 \vspace{30mm}

Figure 2.12:
Cross-section times branching ratio for a)
$\tilde{\chi}_1^{\pm}\tilde{\chi}_1^{\pm}$ direct production followed
by leptonic decays of at least one of the charginos;
b) single sleptons production
from directly produced $\tilde{\chi}_1^{\pm}\tilde{\chi}_2^0$  with
$\chi_2^0$ decays to left sleptons $\tilde{\chi}_2^0 \rightarrow
\tilde{l}_L l$ (b);
c)single sleptons production from directly produced
$\tilde{\chi}_1^{\pm}\tilde{\chi}_2^0$  with $\chi_2^0$ decays to
right sleptons
$\tilde{\chi}_2^0 \rightarrow \tilde{l}_R  l$ (c).

\newpage
 
\ \ \\
 
\vspace{5mm}
 
\begin{figure}[hbtp]
\vspace{-30mm}
\hspace*{0mm}
\resizebox{13.cm}{!}{\rotatebox{0}{\includegraphics{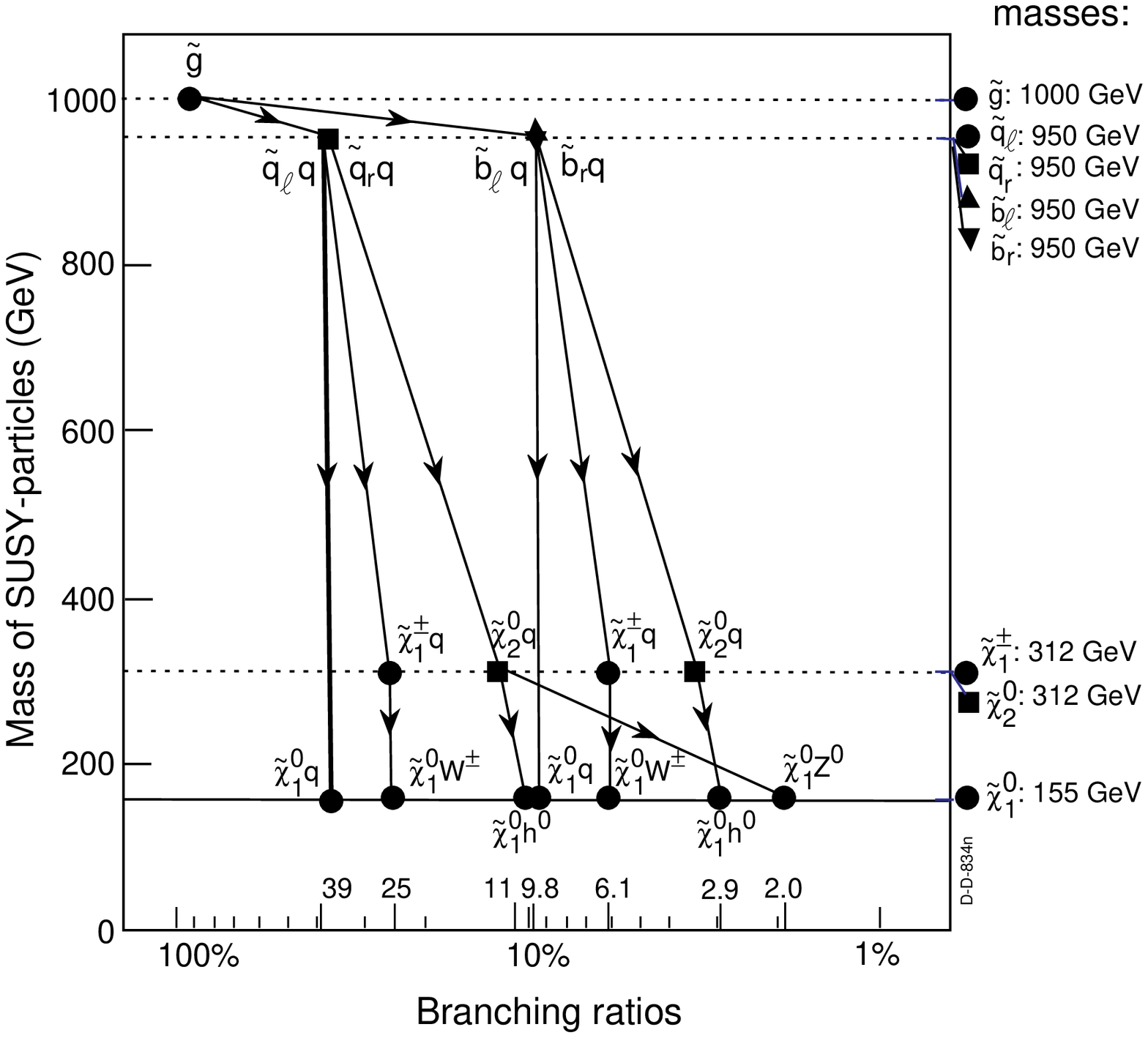}}}
\end{figure}
 
 \vspace{0mm}
Figure 2.13: Typical decay modes for the massive (1 TeV)
gluino illustrating the variety of possible cascade decays.
 
\vspace{40mm}
 
\begin{figure}[hbtp]
\vspace{-30mm}
\hspace*{0mm}
\resizebox{16.cm}{!}{\rotatebox{0}{\includegraphics{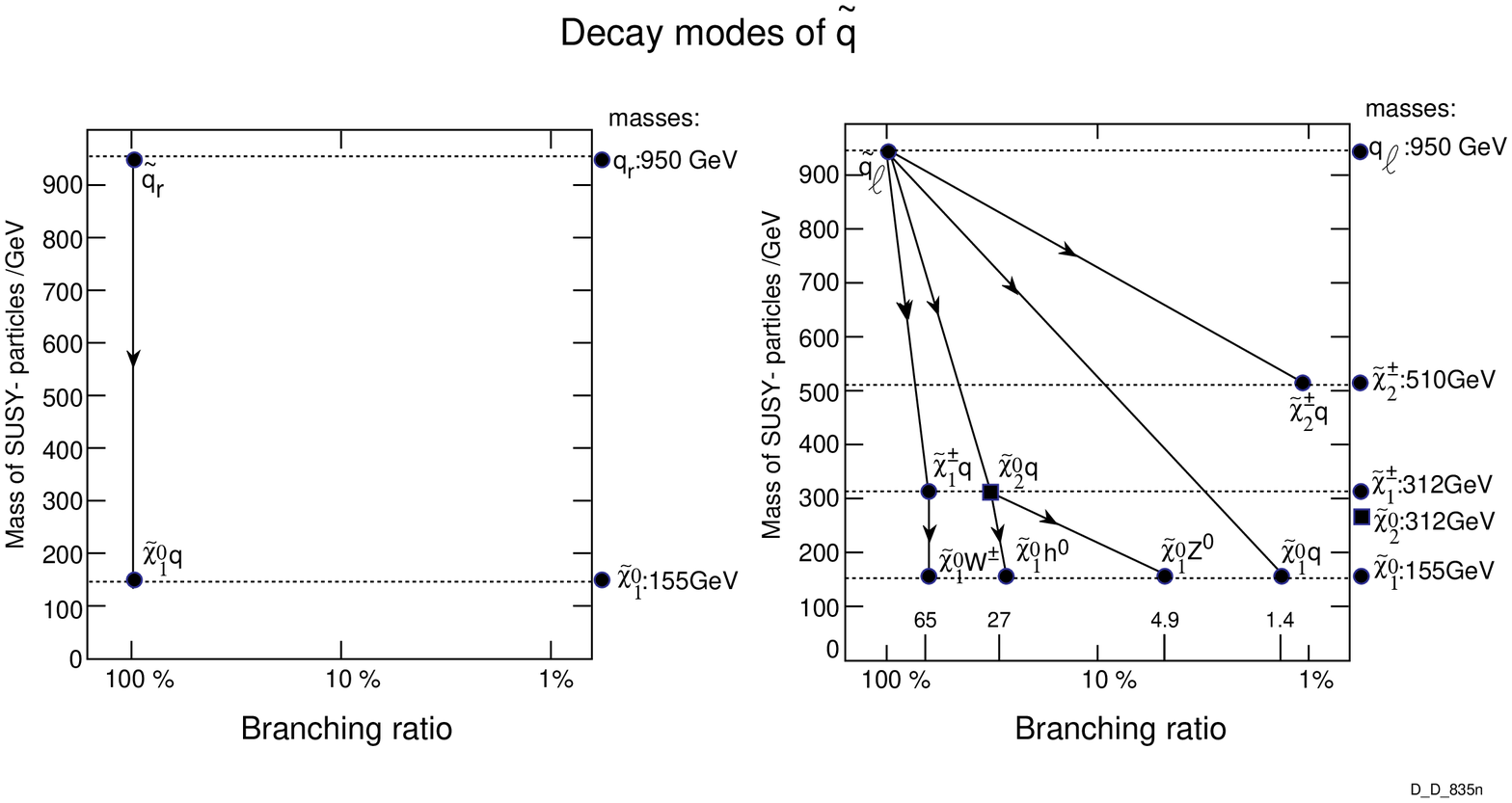}}}
\end{figure}
 
 \vspace{0mm}
Figure 2.14: Typical decay modes of the massive (0.95 TeV) squark.

\newpage  

\ \ \\

\vspace{15mm}

\begin{figure}[hbtp]
\vspace{-30mm}
\hspace*{0mm}
\resizebox{16.cm}{!}{\rotatebox{0}{\includegraphics{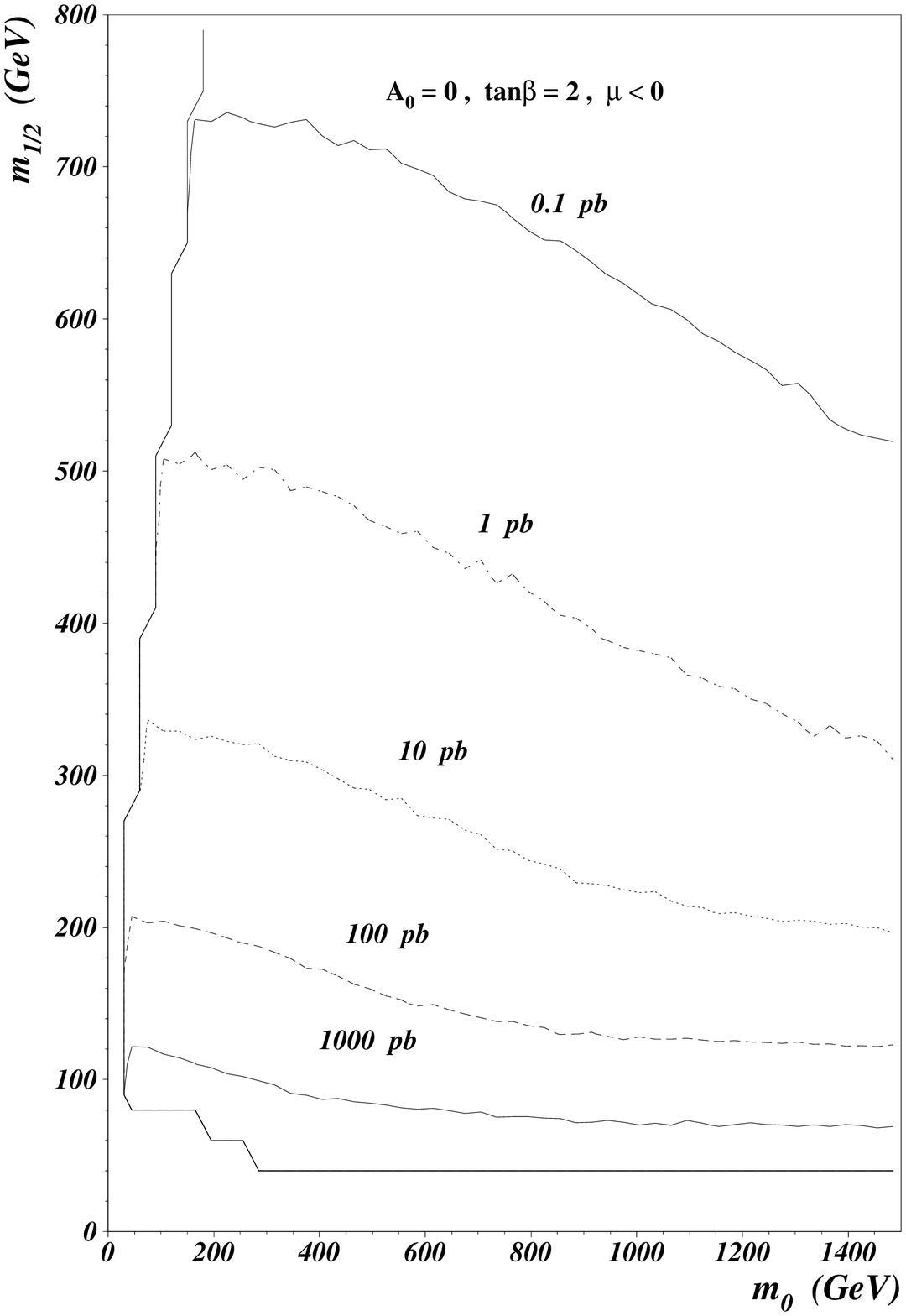}}}
\end{figure}

 \vspace{-5mm}

\hspace*{15mm}Figure 2.15 Gluino/squark production cross-section versus 
($m_0, m_{1/2}$).

\newpage

\ \ \\

\vspace{15mm}

\begin{figure}[hbtp]
\vspace{-30mm}
\hspace*{0mm}
\resizebox{16.cm}{!}{\rotatebox{0}{\includegraphics{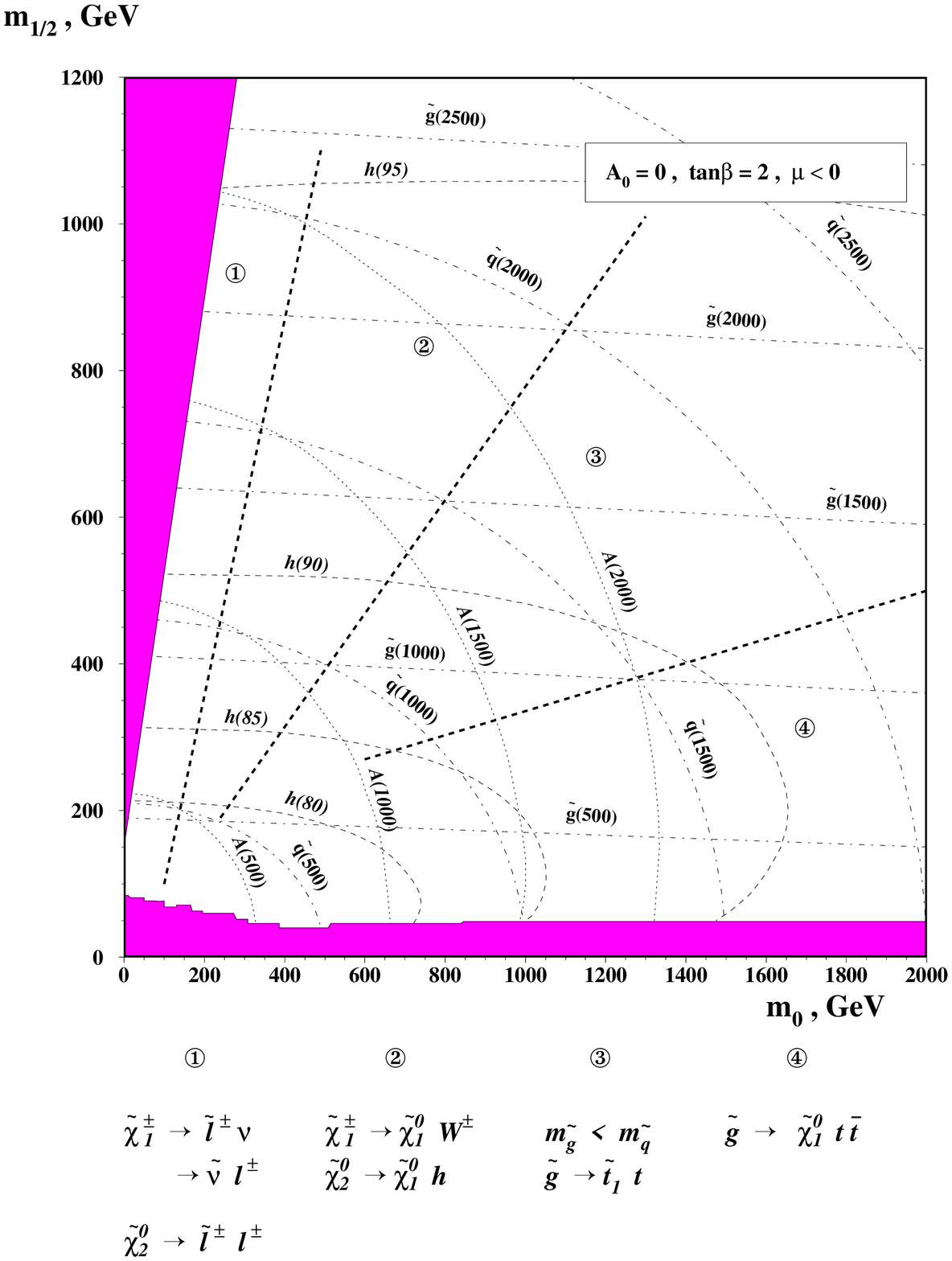}}}
\end{figure}

 \vspace{-5mm}

Figure 2.16: Domains of the ($m_0, m_{1/2}$) parameter space
with charachteristic predominant decay modes.
Isomass contours for squarks, gluinos,
light and pseudoscalar higgses are also shown.
The shaded region near the $m_{1/2}$ axis
shows the theoretically forbidden region of parameter space,
and a similar
region along the $m_{0}$ axis corresponds to both, theoretically
and experimentally excluded portions of parameter space.

\newpage

\ \ \\

\vspace{15mm}

\begin{figure}[hbtp]
\vspace{-30mm}
\hspace*{0mm}
\resizebox{16.cm}{!}{\rotatebox{0}{\includegraphics{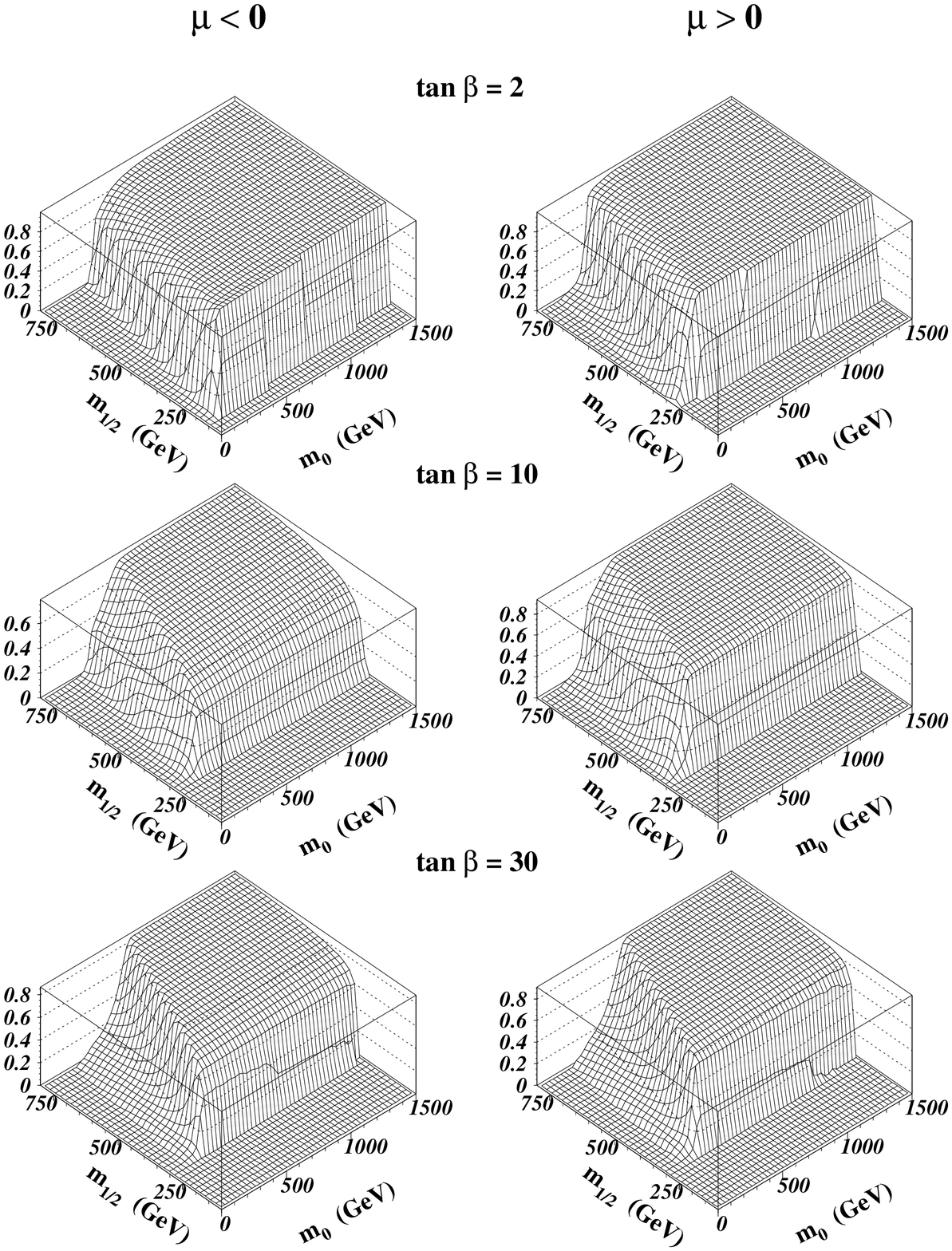}}}
\end{figure}

 \vspace{-5mm}

Figure 2.17: $\tilde{\chi}_{2}^{0} \rightarrow \tilde{\chi}_{1}^{0}$h
branching ratio versus ($m_0, m_{1/2}$) for fixed $A_0=0$ and
different values of tan$\beta$ and $\mu$.

\newpage


\section{Experimental signatures considered}

As discribed earlier, the highest cross-section for 
high mass $R$-parity conserving
SUSY at hadron colliders is due to
squarks and/or gluinos which decay
through a number of steps to quarks, gluons, charginos, neutralinos,
sleptons, W, Z, Higgses and ultimately  a stable LSP (lightest 
neutralino  in mSUGRA),
which is weakly interacting and escapes detection.
The  final state has missing energy (2 LSP's + neutrinos),
a number of jets, 
and a variable number of leptons, depending
on the decay chain. Due to the escaping LSP's, which appear
at the end of each sparticle decay chain, the masses of
sparticles can not generally be reconstructed explicitely. 
However, the sparticle production and
decay characteristics, discussed in the previous section,
lead to a number of specific event topologies,
which should allow the discovery of SUSY in general, and 
specifically the separation of certain SUSY sparticles and
processes from SM and  other SUSY processes.
The LHC must therefore not only be able to discover SUSY, if it
is realized at electroweak scale, and if it has not
yet been discovered at the Tevatron (which covers a very
promising energy/mass range), but the LHC experiments sholud also 
be able to disantagle the various SUSY production mechanisms
and reconstruct most of the sparticle spectrum, determine or
limit the models/scenarios, constrain model parameters, etc.
Our present understanding shows that all this is possible
at the LHC, but a lot of methodical studies are still required.
Usually, one characterizes the general SUSY signal significance by an
excess over the expected SM event rates, while for specific SUSY
processes the background includes besides SM processes
other SUSY reactions contributing to the same final state 
topology.

The following final states have been
investigated in substantial detail in CMS:

1) $leptons + jets$ + \etm \ final states with a variable
number of leptons and jets and possible requirements on
$b$-jets. These channels
provide the maximum mass and parameter
reach of SUSY through the production of strongly interacting
particles, which in their cascade decays can
give rise to many leptons and jets.

2) inclusive $2~leptons$ + \etm \ + $(jets)$ and
$3~leptons$ + (\etm) final states.
In these channels SUSY
reveals itself through \chnb \ inclusive production with subsequent
decay -- directly or via a slepton --
into two leptons and \chna. For some regions
of parameter space, the two lepton
invariant mass with its characteristic shape, a sharp edge at the end
point of the spectrum, allows the determination of the sparticle
masses, including the slepton mass, and model parameters.
These channels could well be the first
indication/signature for SUSY production at the LHC.

3) exclusive $2~leptons$ + $no \ jets$ + \etm \ final state,
which is enriched in direct (Drell-Yan)
slepton pair production
as discussed in section 2.2. 
The main issue here is to understand and
keep under control both the SM background (with many processes to be
considered) and internal SUSY backgrounds.
This channel could allow discovery and study
of slepton production.

4) exclusive $3~leptons$ + $no \ jets$ + \etm \ final state, which
 is the
signature for direct \chha \chnb \
production in a Drell-Yan process.
It is therefore theoretically
most reliable (along with slepton production).
For this and the slepton study, to suppress the SM and internal SUSY
backgrounds  
a good response to jets down to $E_T^{jet} \sim 30$ GeV,
with a calorimetric coverage up to $|\eta| \sim 4.5$, is required.
This exclusive channel could also play a central role in a precise
determination of the \chna \ -- the SUSY dark matter candidate.

For the present study, production of
SUSY processes has been simulated using the ISAJET 7.14 Monte Carlo event
generator \cite{isajet} and the SM backgrounds by PYTHIA 5.7
\cite{pythia}, both with the CTEQ2L structure functions \cite{cteq2l}.
Before presenting the physics expectations, we discuss
the CMS detector optimisation studies done specifically with
SUSY searches in mind.


\section{Detector issues}

\subsection{CMS detector optimisation for SUSY studies}

The design goals of CMS are to measure muons, electrons and photons
with a resolution of $\lappeq$ 1$\%$ over a large momentum range 
($\le $100 GeV), to measure jets with a resolution of 10$\%$ at 
$E_T$ = 100 GeV and to be highly hermetic,  
with a missing $E_T$ performance as required for SUSY searches.  
        The central element of CMS is a 13 m long, 6 m diameter solenoid 
generating a uniform magnetic field of 4 T \cite{tp, MTDR}, Fig.~4.1.  The
magnetic flux is 
returned through a 1.8 m thick saturated iron yoke instrumented 
with muon chambers.  Muons are precisely measured in the
 inner tracker and are identified and measured in four muon stations 
inserted in the return yoke. Precision tracking in the muon
stations is carried out
with drift tube planes in the barrel and with cathode strip chambers 
in the endcaps \cite{mudet}. The goal  is to achieve a spatial resolution
of $\sim$100 $\mu$m and an angular accuracy on a local muon track 
segment of $\sim$1 
mrad per station. Excellent time resolution is needed to identify 
the bunch crossing (with a periodicity of 25 nsec). Muon stations
therefore 
also include resistive plate chamber triggering planes with time resolution
$\sim$2 nsec \cite{mudet}. 

    The inner tracking system of CMS is designed to reconstruct
 high-$p_T$  muons, isolated electrons and hadrons over $|\eta|<$ 2.5 with
a momentum resolution of $\Delta p_T/p_T \simeq 0.15 p_T \oplus 0.5\%$
($p_T$ in TeV). 
 Silicon and gas microstrip detectors are used to provide
 the required precision and granularity.
In the present design there are about 9$\times 10^6$ MSGC detector
channels, about 4$\times 10^6$ Si microstrip channels and
about 8$\times 10^7$ Si pixel channels \cite{tracker}.
For high momentum muons the
combination of tracker and muon chamber measurements greatly
improves
the resolution: $\Delta p_T/p_T \simeq$ 0.06 for a $p \simeq 1$ TeV muon
in $|\eta| <$ 1.6 \cite{tp}.  

        The calorimeter system of CMS is made of a high resolution 
lead-tungstate (PbWO$_4$) crystal electromagnetic calorimeter and a
copper-scintillator
hadron calorimeter behind it. The primary function of the electromagnetic 
calorimeter is to precisely
measure electrons and photons, and in 
conjunction with the hadron calorimeter to measure also jets. The 
arrangement of crystals is shown in Fig.~4.1.  In the barrel the crystals
are 25 radiation lengths ($X_0$) deep and the lateral granularity is
$\simeq 2$ cm $\times$
2 cm corresponding to
$\Delta \eta \times \Delta \phi = 0.014 \times 0.014$.  
They are read out with Si avalanche 
photodiodes \cite{ecal}. The electromagnetic calorimetry extends over
$|\eta| < 3$. The total number of crystals is $\sim 1.1 \times 10^5$.
With a prototype PbWO$_4$ calorimeter in a test beam, an energy resolution 
of $\sigma_E/E \simeq 0.6\%$  has been obtained for electrons of $E =
120$ GeV \cite{ecalbt}. 
        Hadron calorimetry with large geometrical coverage for 
measurement of multi-jet final states and missing transverse energy (\etm)
is essential in all sparticle searches, as it is the \etm \ which
provides evidence for the escaping LSP's (lightest
neutralinos). The hadron 
calorimeter of CMS is made of copper absorber plates interleaved with
scintillator tiles read out with embedded wavelength shifting fibers
\cite{HTDR}. The readout in the 4 Tesla field is done with hybrid
photodetectors \cite{pmt}. The tiles are organized in towers (Fig.~4.1)
giving a lateral
segmentation of $\Delta \eta \times \Delta \phi \simeq 0.09 \times 0.09$.
The hadronic resolution obtained 
in a test beam is $\sigma_E/E \simeq 100\%/\sqrt{E} \oplus 5\%$ 
for the combined PbWO$_4$ and hadronic
 calorimeter system \cite{calbt}. This central hadron calorimetry extends
up to $|\eta| = 3.0$.  It is complemented in the forward region 
$3.0 <|\eta|< 5.0$ by
quartz-fiber ``very forward calorimeters'' (Fig.~4.1) \cite{HTDR, vf}.
Their function is to ensure detector hermeticity for good missing
transverse energy resolution, and to extend the forward jet detection and
jet vetoing capability of CMS which is essential in slepton, chargino,
neutralino searches as discussed in the following.  Detector hermeticity
is particularly important for processes where the physical (real) missing 
$E_T$ is on the order of few tens of GeV  as is the case in 
h, H, A $\rightarrow \tau \tau$, W $\rightarrow l\nu, \ t \rightarrow l\nu
b, \ t \rightarrow$ H$^{\pm} b \rightarrow \tau \nu b$, etc. and in 
particular in slepton, chargino and neutralino searches connecting the 
LEP2/Fermilab and the LHC search ranges.  

\subsection{HCAL optimisation and tail catcher}

        A number of steps have been undertaken in the design of CMS to 
optimise its \etm \ response in view of SUSY searches. These include
the inclusion
of very forward calorimetry, the addition of a tail catcher behind 
the coil \cite{tailc} and optimisation of the crack for the passage
of services
between the barrel and endcap calorimeters. We now briefly discuss these
issues. 
        The central ($|\eta| < $3.0) calorimetry of CMS is complemented
by very
 forward quartz-fiber calorimeters \cite{vf} covering the rapidity range
 3 $< |\eta| <$ 5, Fig.~4.1. Figure 4.2 illustrates the effect of the
forward
 energy flow containment and its effect on missing $E_T$ measurements.
  It shows the expected instrumental (fake) missing $E_T$ distributions 
for QCD di-jet events as a function of calorimetric coverage at large 
rapidities. Calorimetry extending up to $|\eta| \simeq$ 5 reduces the fake 
(instrumental) \etm \ by an order of magnitude in the 20 - 120 GeV range.
Furthermore, the missing $E_T$ resolution,
$\sigma (E_T^{miss})/\Sigma E_T$ where $\Sigma E_T$ is the calorimetric
transverse energy sum, is also much reduced in the presence of very
forward calorimetry, see Fig.~4.3.
        There is a second requirement on  forward calorimetry which
 turns out to be essential if we hope to eventually extract slepton and 
chargino/neutralino signals.  In the search for direct DY slepton pair
 production $\tilde{l}\tilde{l} \rightarrow l^+l^-$\chna \chna , or
 for the associated direct (DY) chargino/neutralino
\chha \chnb $\rightarrow l^{\pm}\nu$\chna $l^+l^-$\chna \ production, 
leading to final states with two  or three isolated leptons, no jets, 
and missing  $E_T$, it is essential to have the capability to recognise
and veto on forward jets. This is needed to suppress the large
backgrounds due to $t\bar{t}$, $\tilde{q}\tilde{q}$,
$\tilde{g}\tilde{q}$, $\tilde{g}\tilde{g}$ and the associated production 
modes $\tilde{q}\tilde{\chi}$, $\tilde{g}\tilde{\chi}$ which would
otherwise overwhelm the signals. 
The extension into the forward direction of the central jet vetoing 
capability provides the needed additional rejection factors to keep 
these backgrounds under control. 

Figure 4.4a shows for the case of direct \chha \chnb \ production,
(discussed in more detail in section 8), the expected rejection factors
against the main backgrounds as a function of the rapidity range and
detection threshold over which the jets can be recognized.
Figure 4.4b shows the rejection factors
against internal SUSY and the SM $t\bar{t}$ backgrounds.
Here the jet veto
is applied after lepton isolation and an \etm \ cut, thus the value of the 
rejection factor is much reduced. In both cases,
the jet coverage needed to obtain sufficient background
rejection is $|\eta|$ up to $\simeq 4$ and the loss of signal acceptance
due to the jet veto is typically $\sim$10$\%$. 

The locations of the tail-catcher scintillator layers, two in 
the barrel region behind the coil and one in the endcap, are shown on the
longitudinal cut through the CMS detector in Fig.~4.1. Figure 4.5 shows
the depth in interaction lengths $\lambda$ of calorimetry in CMS,  and
the total 
sampled depth including the tail catcher layers, as a function of rapidity
 \cite{tailc1}. Whilst the total calorimetric absorber thickness
within the coil
 is somewhat marginal at rapidity $\eta \sim$ 0 for full hadronic shower
containment,
inclusion of tail-catcher layers allows hadron energy measurements with at
least 10.5 interaction lengths everywhere.

Figure 4.6 shows the effects on hadron energy measurements of the
inclusion of the tail-catcher layers, specifically on the reduction 
of the low-energy tail in the response to (200 GeV) hadrons \cite{tailcs}.
What is shown is 
the ratio of the (GEANT) simulated response to the incident energy for
single pions in the rapidity range 0.33 to 0.86 with
no tail catcher, with one, and with two tail catcher layers included in
the readout.  Mismeasurements of  hadronic energies, and in particular the
presence of  low-energy tails in hadronic (jet) energy  measurements is
one of the main sources of fake instrumental missing $E_T$. When due to
insufficient  shower containment in depth, inclusion of the tail-catcher
layers cures this problem to a large extent.

There are two other sources of hadronic
mismeasurements and low-energy tails: dead areas and volumes due to
detector cracks and matching the hadronic response of ECAL to
HCAL.

Special attention has been paid in CMS to the effects on hadronic
energy measurements -- and thus on missing $E_T$ -- of the 
main crack for services
between the barrel and endcap calorimetry at rapidity $\sim$ 1.2 - 1.5
(Fig.~4.1).
  The various configurations which have been studied  for this crack are
shown in Fig.~4.7 \cite{calconf}. The aim was to minimize the degradation
of the hadronic energy resolution and the level of the low energy tail for 
hadrons and jets straddling the region of the gap.  The optimal choice, 
from the point
 of view of hadron response, weight of the cantilevered endcap hadron
 calorimeter, manoeuvrability of endcaps and margin of freedom to close the
 detector, is a conical endcap shape with a non-pointing crack at
52 degrees from the beam line labeled TP7 in Fig.~4.7. Figure 4.8 shows
the reconstructed energy response (GEANT) to single pions of 100 GeV when
 incident on and off the crack region, for three crack widths of 11, 18
 and 25cm. With the present estimate of the volume of services a gap of
 12 cm is needed, plus 2 cm for clearance.

        An essential ingredient to obtain optimal \etm \ response of a 
detector is the linearity of its response as a function of incident hadron
(jet) energy. The non-compensating mixed
calorimeter of CMS -- with a PbWO$_4$ crystal ECAL
compartment followed by a Cu/scintillator sampling HCAL -- has 
significantly
different electromagnetic to hadronic (e/h)  responses in the two parts, 
e/h $\sim$ 1.8 in the ECAL vs. 1.2 in the HCAL. Linearity is more
difficult to achieve in such a mixed calorimeter than in a more
homogeneous system. To restore linearity as much as possible,
and to compensate for the effects of dead material in the space between
 the ECAL and HCAL, weighting techniques for responses must be used
between ECAL and HCAL, and within HCAL compartments and tail catcher
layers. An essential limiting factor is the number of readout channels
we can afford for the longitudinal HCAL tower segmentation.
The optimal set of
weighting factors must be determined on basis of test beam data and 
detailed simulations of responses to hadrons and jets,
taking into account financial limitations. In the present
HCAL design \cite{HTDR} a good compromise between expected performance
and cost is found by having a very shallow (1 scintillator plane) front
HCAL tower readout and second deep ($\simeq$ 5 $\lambda$) HCAL segment 
readout. The first shallow HCAL segment just behind the ECAL
detects hadron interactions in the ECAL
-- where their energy deposition is underestimated (e/h $\sim$ 2)
-- and compensates by overweighting this layer.
Present investigations show that with this technique the
expected non-linearity should not exceed 4$\%$ for pions
between 20 and 300 GeV \cite{HTDR}.

\subsection{The role of the tracker in SUSY searches}

        The inner tracking system of CMS is primarily
designed to reconstruct high-$p_T$ muons, isolated electrons and hadrons
over $|\eta|<$ 2.5 with a momentum 
resolution of $\Delta p_T/p_T \sim 0.15p_T \oplus 0.5\%$ ($p_T$ in TeV).
Hadrons must be reconstructed down to $\sim$ 1 - 2 GeV, as lepton and
photon isolation is 
a very important selection criterion for a number of physics signals, and 
in particular in the SUSY searches for sleptons, charginos and
neutralinos. Another important task of the tracker is to measure
track impact parameters allowing the detection and measurement of
long-lived particles and the tagging of $b$-jets.
The main problem in tracking is that of pattern recognition. At a luminosity
 of $10^{34}$ cm$^{-2}$s$^{-1}$, interesting events will be superimposed
on a background of about 500 soft charged tracks within the rapidity 
range considered, coming from $\sim$ 15
minimum bias events which occur in the same bunch crossing. To solve the
pattern recognition problem, detectors with small cell sizes are required. 
Silicon and gas microstrip detectors provide the required precision and 
granularity to maintain cell occupancies below 1$\%$, but the number of
detector channels is large ($\sim 1.5 \times 10^7$).  The design of the
CMS tracker is shown in Fig.~4.9. 

A track in the barrel part of the tracker, in its present design
(summer 1997), first encounters two layers
of pixel detectors, then four layers of microstrip Si detectors of 
67 and 100
$\mu$m pitch followed by seven layers of 200 $\mu$m pitch microstrip gas
chambers
(MSGC; $\sim$50 $\mu$m precision) \cite{tracker}. Alternate MSGC
and Si microstrip layers are
double-sided to allow determination of the z coordinate of a track by a
small-angle stereo measurement. The inner cylindrical volume ($r \lappeq$
50 cm) with Si detectors will be kept at $\sim -5^0$ C temperature, whilst
the outer (MSGC) part at
$\sim 18^0$ C. Figure 4.10 shows the expected track  momentum
resolutions
in the CMS tracker alone. As already mentioned,
for high momentum muons the combination of
tracker and muon
chamber measurements greatly improves the resolution: 
$\Delta p_T/p_T \simeq$ 0.06 for a
$p \simeq 1$ TeV muon in $|\eta|<$ 1.6 \cite{tp}.  This requires that the
relative alignment of
 the inner and outer systems be known to within $\sim$100 $\mu$m
\cite{mudet}. The pixel
layers, at radii of 7.7 and 11 cm  
in the high luminosity pixel detector option,
ensure precise impact
 parameter measurements, with an asymptotic (high momentum) accuracy of 
$\sigma_{IP}$ = 23 $\mu$m in the transverse plane (Fig.~4.10), and of
$\simeq
90$ $\mu$m along the z axis \cite{tp}. For the initial ``low luminosity'' 
($\lappeq 10^{33}$ cm$^{-2}$s$^{-1}$) running, the impact
 parameter performance is improved by having the pixel layers at  4 and 7.7 
cm radii.

 \subsection{Lepton isolation}

        In the search for direct DY slepton pair production
$\tilde{l}\tilde{l} \rightarrow l^+l^-$\chna \chna \ or
for the associated direct (DY)
chargino/neutralino
\chha \chnb $\rightarrow  l\nu$\chna$l^+l^-$\chna \ production
leading to a final state
 with two or three isolated leptons, no jets and missing $E_T$, there are
two essential instrumental requirements allowing separation of the
signal. The first is lepton isolation, to suppress the 
copious backgrounds from processes such as
$t\bar{t}$,
Z$b\bar{b}$, $b\bar{b}$.
The rejection factors expected on basis of tracker isolation criteria as a
function of the $p_T$ cut on the accompanying tracks are shown in
Fig.~4.11
for low and high luminosity running conditions. The second requirement, as 
discussed previously, is on the detector capability to veto on jets.

\subsection{Tagging of $b$-jets in CMS}

        As discussed in Section 2.3, an important ingredient for SUSY
physics studies is the capability of the detector to tag $b$-jets. 
Figures 2.13, 2.14 and 2.15 illustrate the many ways $b$-jets
can serve as final state signatures of $\tilde{b}$,
$\tilde{t}$, $t$ production in the $\tilde{g}/\tilde{q}$ cascades.
Particularly important is the possibility to detect h $\rightarrow b
\bar{b}$ in $\tilde{g}/\tilde{q} \rightarrow$ \chnb $\rightarrow$ \chna h
decay channel with $S/B \sim 1$ and potentially with much lower
integrated luminosity than required to observe h $\rightarrow \gamma
\gamma$. Techniques for $b$-tagging have
been studied in detail in CMS. The inner part of the CMS tracker, in 
particular the two pixel layers, has been optimised in detector
positioning
and required resolution with $b$-tagging performance  as the main
criterion
\cite{microv}. The expected $b$-tagging efficiency and sample purity has
been studied with several tagging algorithms based on track 
impact parameter measurements. Figure 4.12 shows, for example, the
expected
tagging efficiency as a function of jet $E_T$ averaged over the pixel
detector acceptance. For $b$-jets in $t\bar{t}$ events it is $\sim$
30$\%$,
 whilst the expected fake rate probability is $\sim$ 1$\%$.  The tagging
algorithm used
 for Fig.~4.12 requires $\ge$ 3 tracks of $p_T \ge$ 2 GeV with $\ge$ 2
$\sigma$ significance for
 impact parameter measurements in the transverse plane. 

How much the expected tagging efficiency is dependent on the tagging
algorithm and what is the possible trade-off between efficiency and sample
purity is illustrated by Fig.~4.13 where several tagging algorithms have
been exercised. Figure 4.14 shows how the tagging efficiency and purity
depends
on the radius of the first barrel pixel layer, changing it from 4 cm as 
possible for the low luminosity running to 7.7 cm as required to sustain 
the radiation damage at $10^{34}$ cm$^{-2}s^{-1}$. 
There would be an important gain in tagging efficiency
if the first pixel layer could be always kept at 4 cm radius, at a cost
of replacing it (almost) every year. This is particularly important for
channels like $h \rightarrow b \bar{b}$ which go with the square
of the tagging efficiency. Whilst the tagging efficiency is a simple
function of the pixel point resolution and the radial position of the
barrel pixel layers, the mistagging probability is dependent on the
non-Gaussian
part of the measured impact parameter distribution and is
more difficult to control as it depends on the overall pattern recognition
performance and multiple scattering of the tracker.
 All the $b$-tagging efficiencies discussed
here rely on impact parameter measurements; further improvements in
$b$-tagging efficiency are possible, including explicit reconstruction
of secondary vertices, lepton tags, etc.


\subsection{CMSJET -- an approximate description of detector response}

Understanding of the detector response with an adequate simulation  
is a very important part of correctly estimating the
SUSY discovery potential of the CMS experiment.   
We have used the CMSJET program, which
is a fast, non-GEANT simulation package for the
CMS detector \cite{hcalp, cmsjet}. CMSJET provides a satisfactory
approximate description of detector
components and of the response to
hadrons, leptons and gammas.
It is a good compromise between
performance speed and precision.
The following aspects of CMSJET are
relevant to this study:
 
$\bullet$ Charged particles are tracked in a 4 T magnetic field.
A reconstruction efficiency of 90$\%$ per charged track
with $p_T > 1$ GeV and $|\eta| < 2.5$ is assumed.
 
$\bullet$ The geometrical acceptances for $\mu$ and $e$ are
$|\eta|<2.4$ and 2.5, respectively. The
lepton momentum is smeared according to parameterizations   
obtained from full GEANT simulations. For a 10 GeV lepton
the momentum resolution
$\Delta p_T / p_T$ is better than 1$\%$ over the full $\eta$ coverage.
For a 100 GeV lepton the resolution becomes $\sim 1 \div 5 \%$ depending
on $\eta$. We have assumed a 90$\%$
triggering plus reconstruction efficiency
per lepton within the geometrical acceptance of the CMS detector.
This value is probably pessimistic for muon tracks, but more
realistic for electrons.
 
$\bullet$ The electromagnetic calorimeter of CMS extends
up to $|\eta| = 2.61$. There is a pointing crack in the
ECAL barrel/endcap transition region between $|\eta|=1.479-1.566$
(6 ECAL crystals) with significantly degraded resolution.
The hadronic calorimeter covers
$|\eta| < 3$. The Very Forward calorimeter extends
from $|\eta| = 3$ to $|\eta| = 5$.
The full granularity of the calorimetric system is implemented in CMSJET.
The granularity, assumed
energy resolutions and electronic noise of calorimeters
are listed in Table 4.1.
Noise terms have been simulated with Gaussian distributions
and zero suppression cuts have been applied.

{\small

\begin{table}[htb]
Table 4.1: CMS calorimeter description
\label{tab1}
\begin{center}
\begin{tabular}{|c|c|c|c|c|}
\hline
& $\eta$  region & $\Delta \eta \times \Delta \phi$ & $\Delta E / E$ &
Noise (MeV)  \\
\hline\hline
ECAL &  $|\eta|<1.479$ & 0.0145$\times$0.0145
& $5\%/\sqrt{E} \oplus 0.5\%$ & 25, $3\sigma$ zero sup. \\
 & 1.566$<$$|\eta|$$<$2.00 & 0.0217$\times$0.0218 &  &   \\
     & 2.00$<$$|\eta|$$<$2.35 &
0.0292$\times$0.0291 &
                     &                           \\
 & 2.35$<$$|\eta|$$<$2.61 & 0.0433$\times$0.0436 &  &  \\
\hline
Crack &
 1.479$<$$|\eta|$$<$1.566 & 0.0870$\times$6.283
& $50\%/\sqrt{E} \oplus 2\%$ & -- \\
\hline
HCAL & $|\eta|<2.262$ & 0.087$\times$0.087 &  $\eta$
dependent& 250, $2\sigma$ zero sup. \\
     & 2.262$<$$|\eta|$$<$2.61 & 0.174$\times$0.175 &
parameterization; &
                       \\
  & 2.61$<$$|\eta|$$<$3.0 & 0.195$\times$0.349 & 
$82\%/\sqrt{E} \oplus 6.5 \%$ at $\eta$=0
& \\
\hline
VF  & $3<|\eta|<5$ &
0.17$\times$0.17 $\div$  &
$172\%/\sqrt{E} \oplus 9 \%$ & 250, $2\sigma$ zero sup. \\
  &  &
 0.175$\times$0.175 & for hadrons &  \\
\hline
\end{tabular}
\end{center}
\end{table}
}

$\bullet$ e/$\gamma$ and hadron shower developments are taken into
account by parameterization of the lateral and
longitudinal profiles of the showers.
The starting point of a shower
is fluctuated according to an exponential low.
 
$\bullet$ Jets are reconstructed using a cone
algorithm, with a cone radius
$R = 0.4 \div 0.9$ and variable transverse energy threshold
on jets depending on event kinematics.
 
$\bullet$ For the high luminosity ($L=10^{34}$ cm$^{-2}$s$^{-1}$)
study, event pile-up is taken into
account by superposition of ``hard'' pile-up events
on top of signal and background.
These are represented by PYTHIA QCD jet events with $\hat{p}_T>5$ GeV.
The number of superimposed pile-up events
has been fluctuated with a Poisson distribution having a
mean value of 15.

The CMSJET model of the CMS detector is depicted in Fig.~4.15.

\newpage     

\ \\

\vspace{10mm}

\begin{figure}[h]
\vspace{-30mm}
\hspace*{5mm}
\resizebox{14.cm}{!}{\rotatebox{90}{\includegraphics{D_Denegri_1041n1.ill}}}
\end{figure}

\vspace{0mm}

Figure 4.1:  Longitudinal cut through the CMS detector. The tower   
structure of the hadron calorimeter  and the arrangement of the
electromagnetic calorimeter crystals is indicated,
as well as the locations of the tail catcher layers.

\newpage
 
\ \ \\

\vspace{10mm}

\begin{figure}[hbtp]
\vspace{-10mm}
\hspace*{0mm}
\resizebox{15.cm}{!}{\rotatebox{0}{\includegraphics{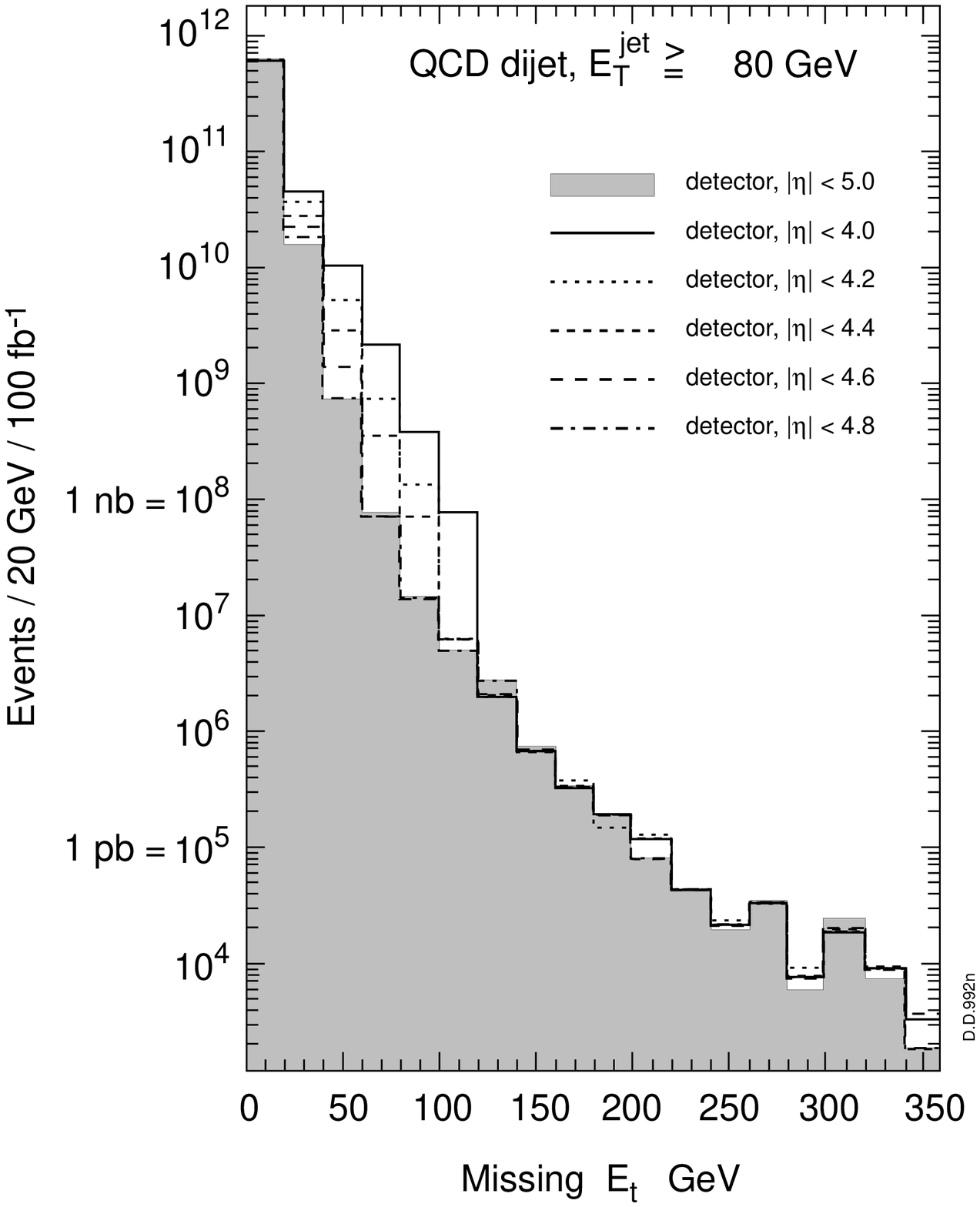}}}
\end{figure}
 
\vspace{10mm}

Figure 4.2: Missing $E_T$ distributions for QCD di-jet events as a
function of calorimetric coverage at large rapidities.

\newpage  

\begin{figure}
\vspace{-0mm}  
\begin{center}
\resizebox{120mm}{150mm}
{\includegraphics{D_Denegri_1053n.ill}}
\vspace{0.5cm}
\end{center}  
\end{figure}

Figure 4.3: Improvement on missing $E_T$ resolution in the presence
of very forward calorimetry.

\newpage

\ \ \\  
 
\vspace{10mm}
 
\begin{figure}[hbtp]
\vspace{-40mm}
\hspace*{0mm}
\resizebox{13.cm}{!}{\rotatebox{0}{\includegraphics{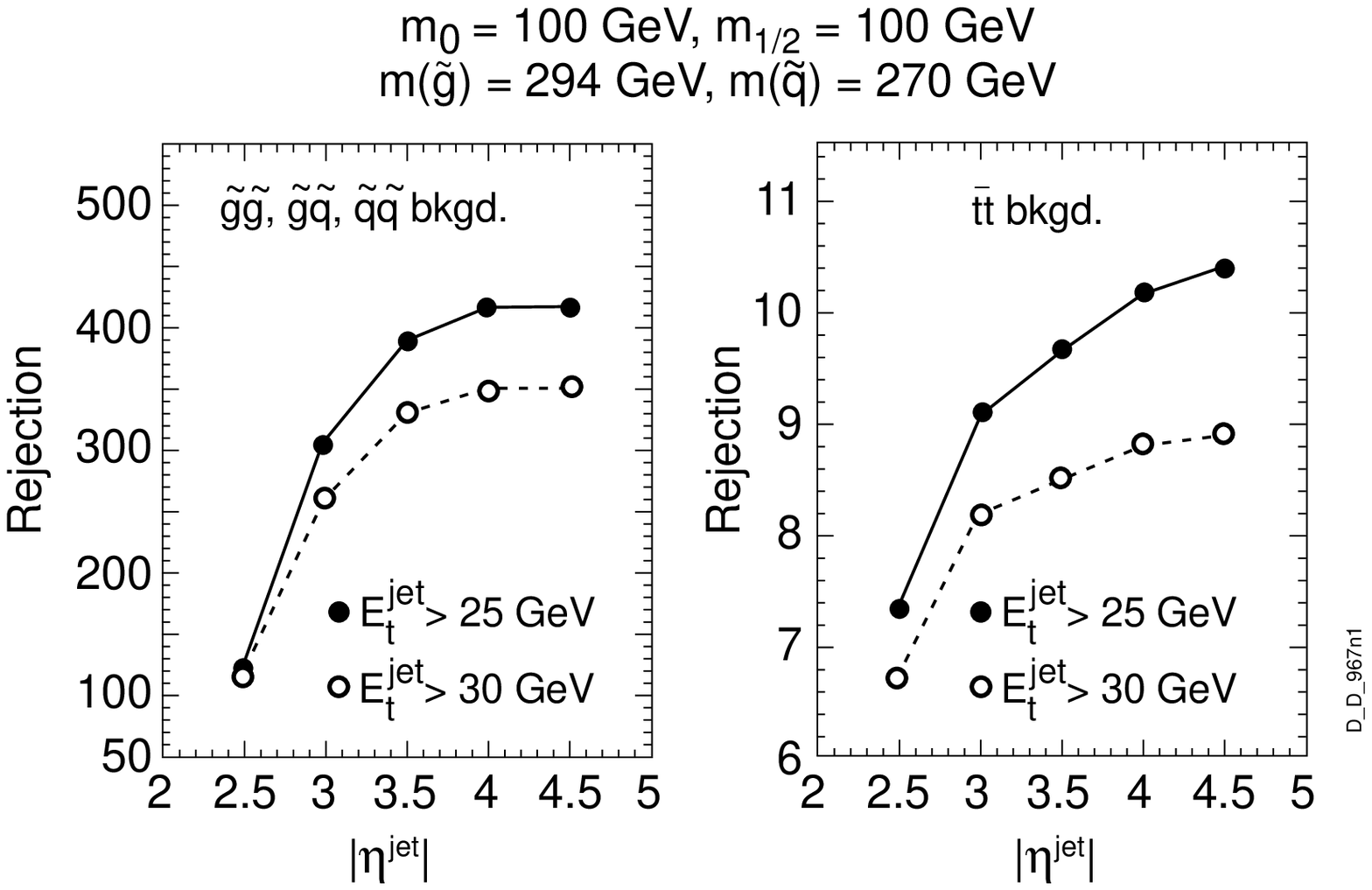}}}
\end{figure}
 
 \vspace{2mm}
 
Figure 4.4a: Rejection factors which can be achieved in
chargino/neutralino search using a jet veto, as a function
of jet acceptance and jet detection threshold.

\begin{figure}[hbtp]
\vspace{10mm}
\hspace*{0mm} 
\resizebox{14.cm}{!}{\rotatebox{0}{\includegraphics{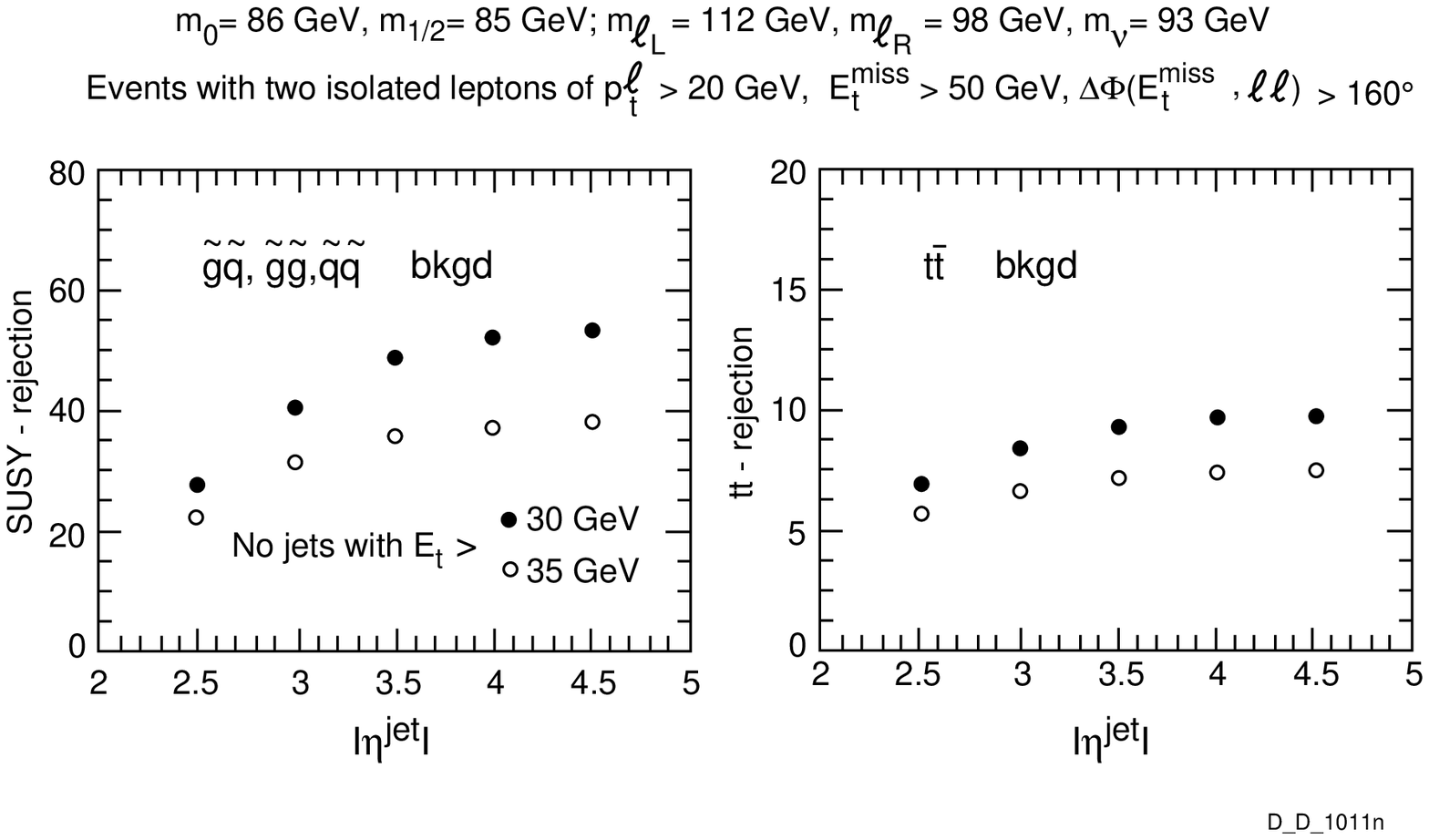}}}
\end{figure}
 
 \vspace{2mm}

Figure 4.4b: Jet veto rejection factors against squark/gluino
and  $t\bar{t}$ backgrounds in slepton searches in two-lepton final   
states.

\newpage

\ \ \\

\vspace{10mm}

\begin{figure}[hbtp]
\vspace{-10mm}
\hspace*{0mm}
\resizebox{15.cm}{!}{\rotatebox{0}{\includegraphics{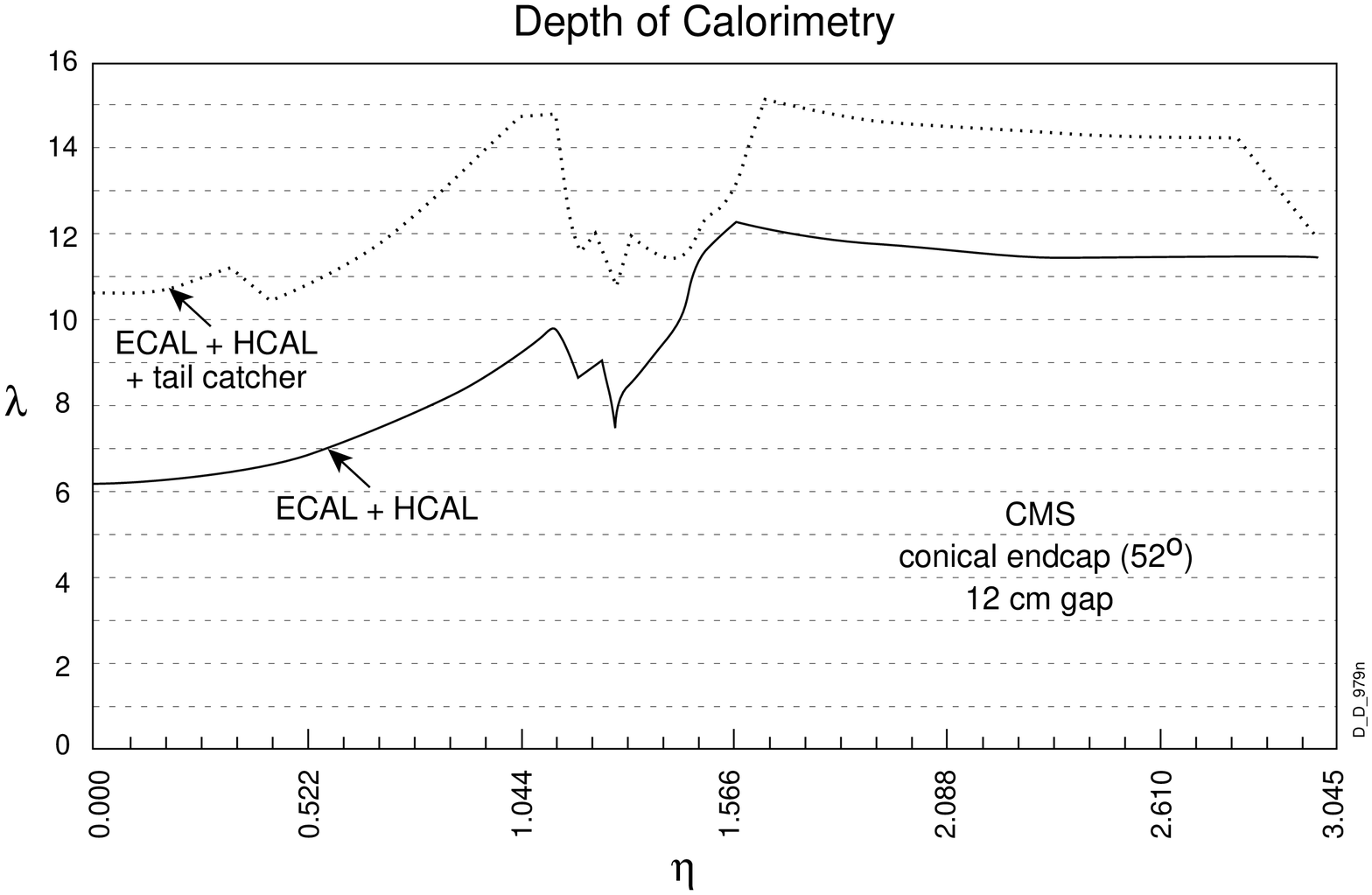}}}
\end{figure}

\vspace{20mm}

Figure 4.5: Depth of calorimetry  and total sampled depth including the
tail
 catcher layers as a function of rapidity in CMS.

\newpage

\ \\

\vspace{10mm}

\begin{figure}[hbtp]
\vspace{-20mm}
\hspace*{10mm}
\resizebox{13.cm}{!}{\rotatebox{0}{\includegraphics{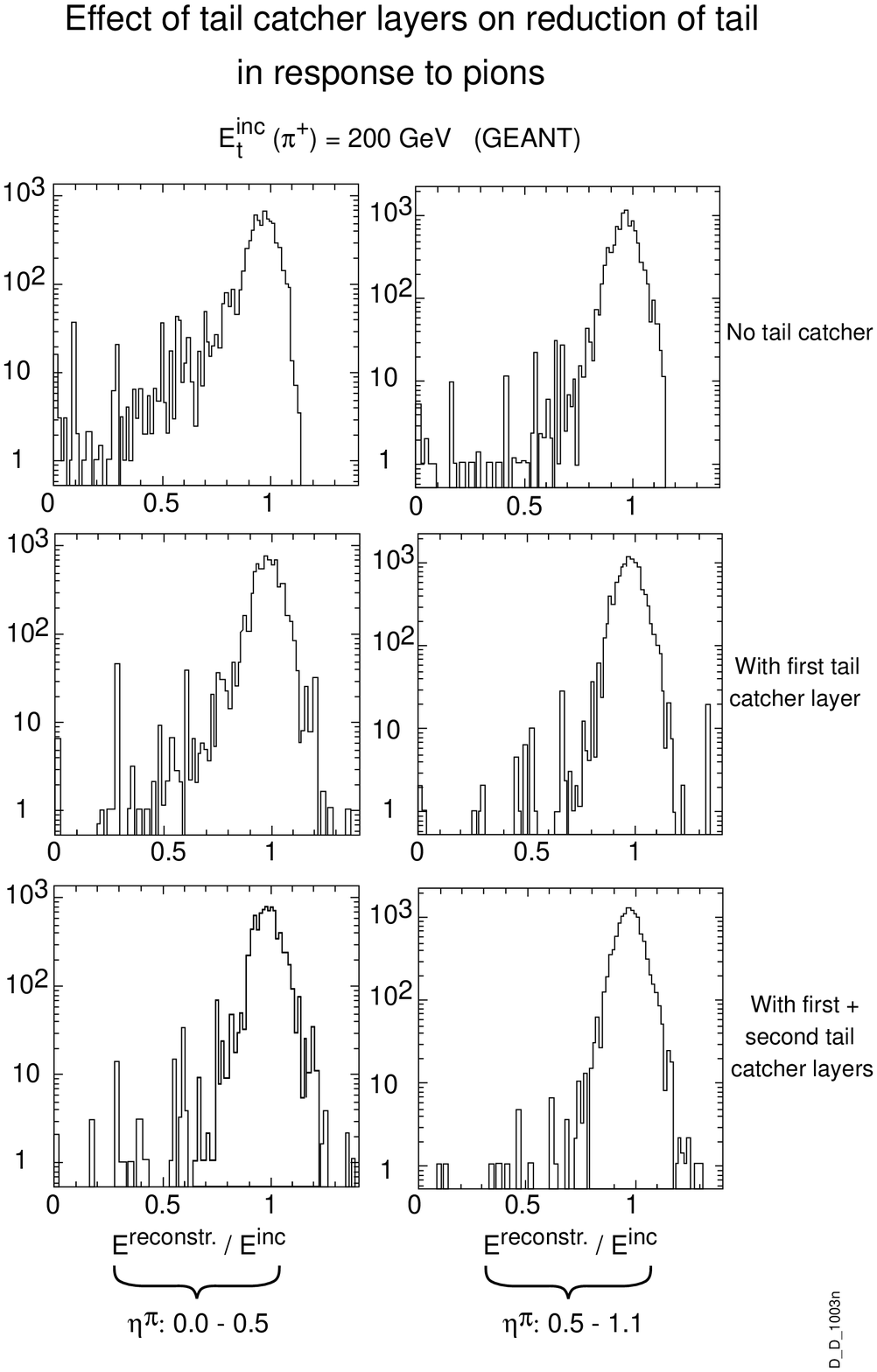}}}
\end{figure}

\vspace{10mm}

Figure 4.6: Effects on hadron energy measurements of tail catcher layers
in the
 central and next-to-central CMS wheels i.e. in two adjacent rapidity
 ranges.

\newpage

\ \\
 
\vspace{10mm}
 
\begin{figure}[hbtp]
\vspace{-30mm}
\hspace*{10mm}
\resizebox{12.5cm}{!}{\rotatebox{0}{\includegraphics{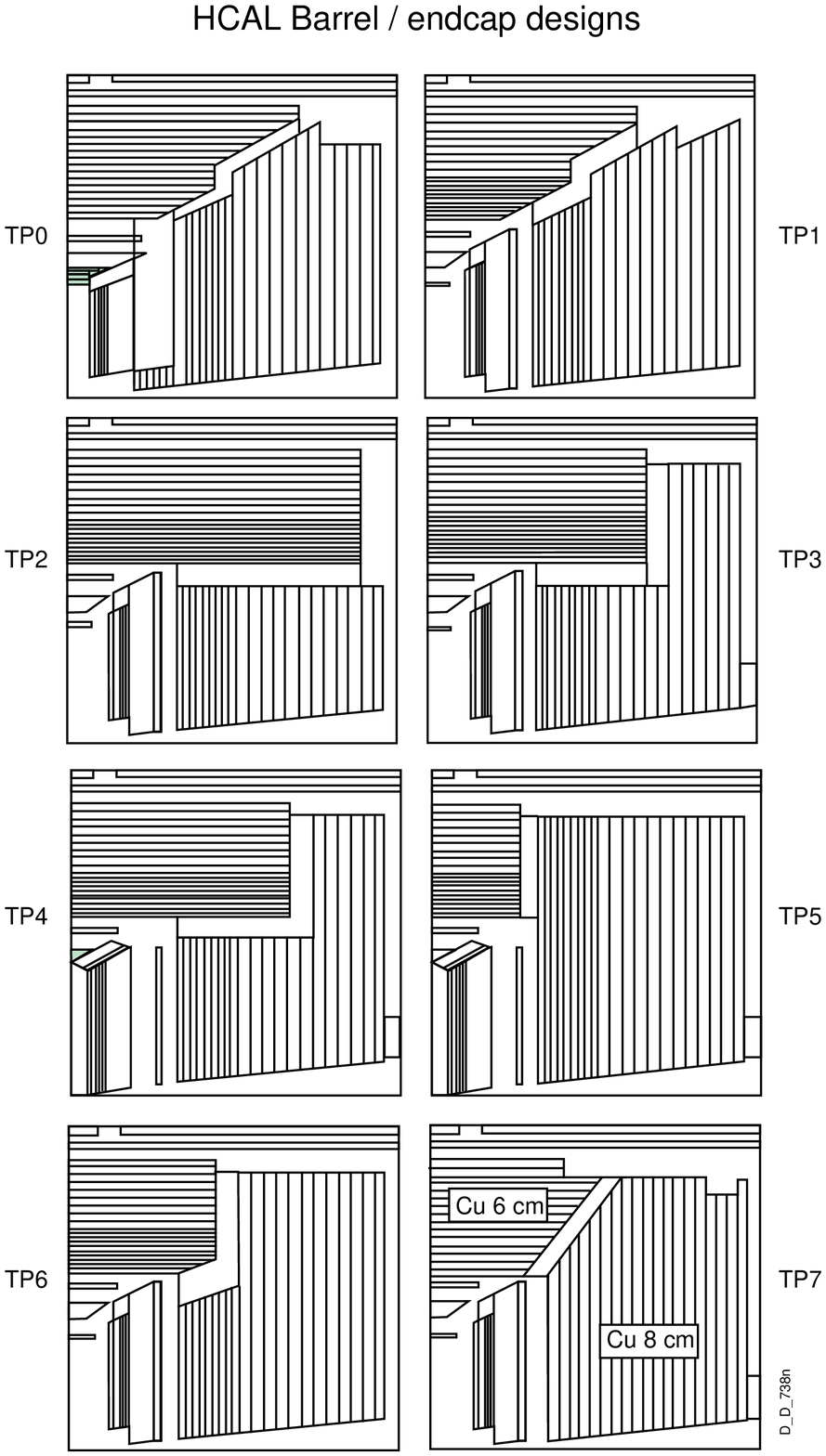}}}
\end{figure}

\vspace{10mm}

 Figure 4.7: Various configurations of the barrel/endcap  calorimetry
transition
 region considered.

\newpage

\ \\
 
\vspace{10mm}
 
\begin{figure}[hbtp]
\vspace{-30mm}
\hspace*{10mm}
\resizebox{12.5cm}{!}{\rotatebox{0}{\includegraphics{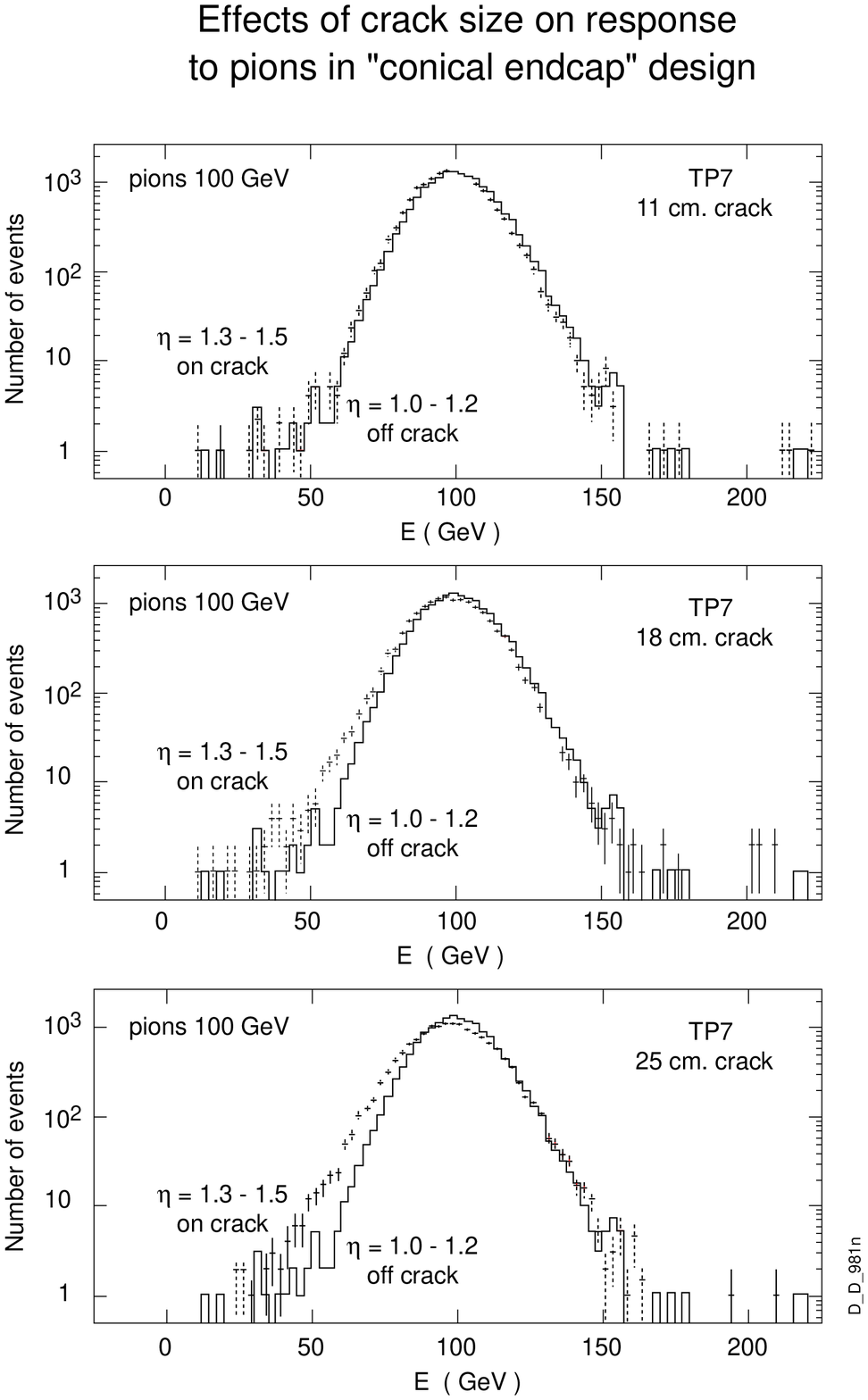}}}
\end{figure}
 
\vspace{10mm}

Figure 4.8:  Effects on hadron energy measurements of the crack for
services
 between the barrel and the endcap calorimetry for three crack sizes.
 
\newpage

\begin{figure}[hbtp]
\vspace{-15mm}
\hspace*{10mm}
\resizebox{55.cm}{!}{\rotatebox{0}{\includegraphics{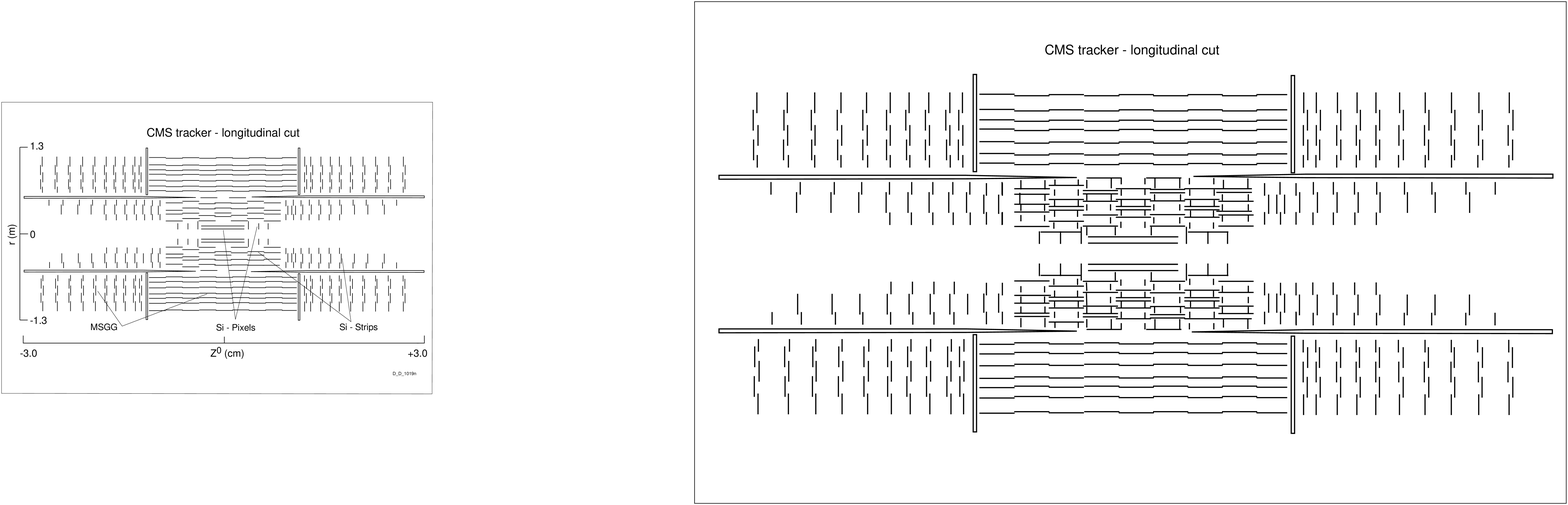}}}
\end{figure}

\hspace*{30mm} Figure 4.9: Layout of the CMS  tracker.

\newpage

\ \\

\vspace{10mm} 

 \begin{figure}[hbtp]
\vspace{-15mm}
\hspace*{2mm}
\resizebox{13.5cm}{!}{\rotatebox{0}{\includegraphics{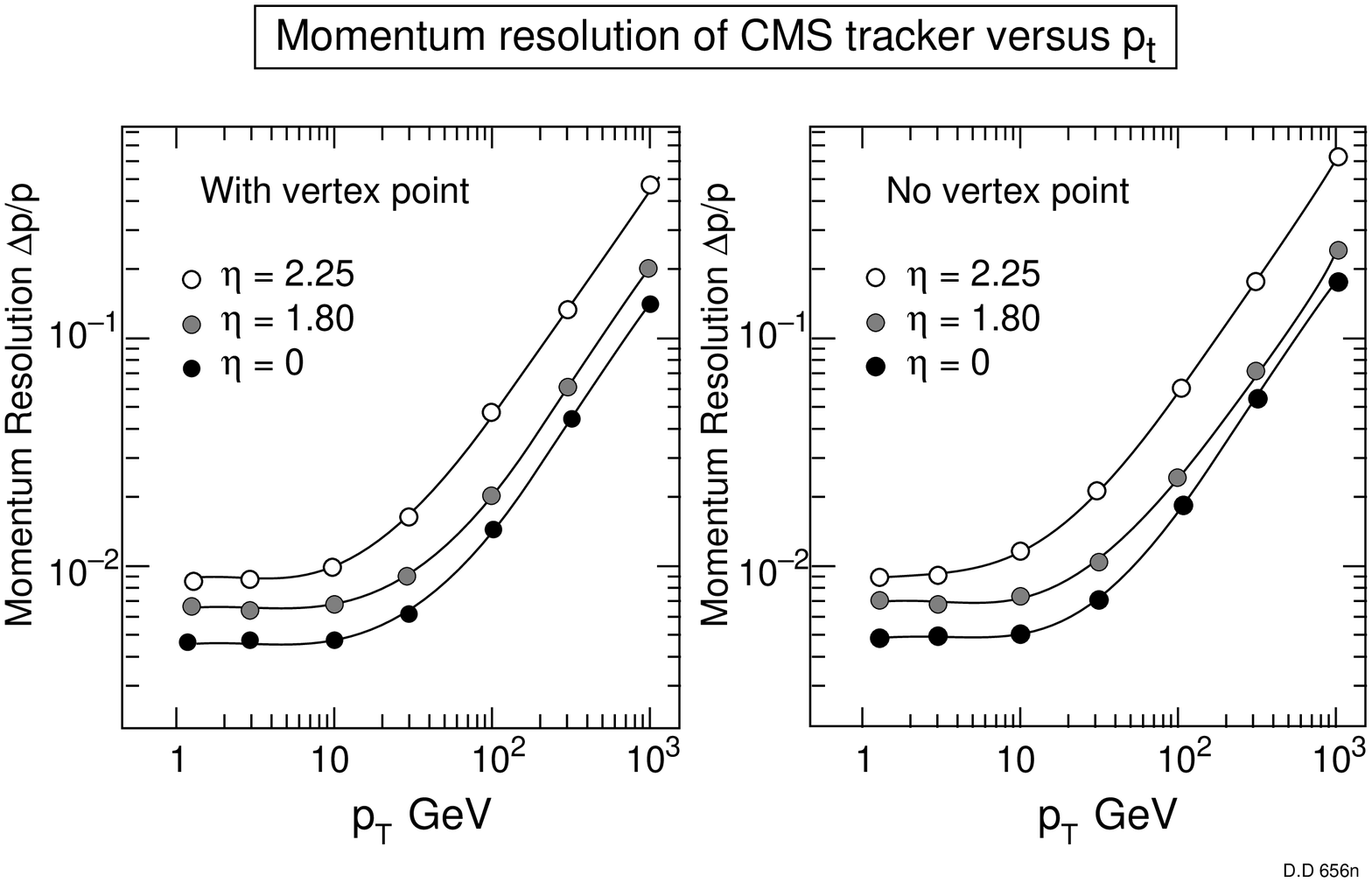}}}
\end{figure}
 
\vspace{2mm}

\begin{figure}[hbtp]
\vspace{-10mm}
\hspace*{20mm}
\resizebox{10.cm}{!}{\rotatebox{0}{\includegraphics{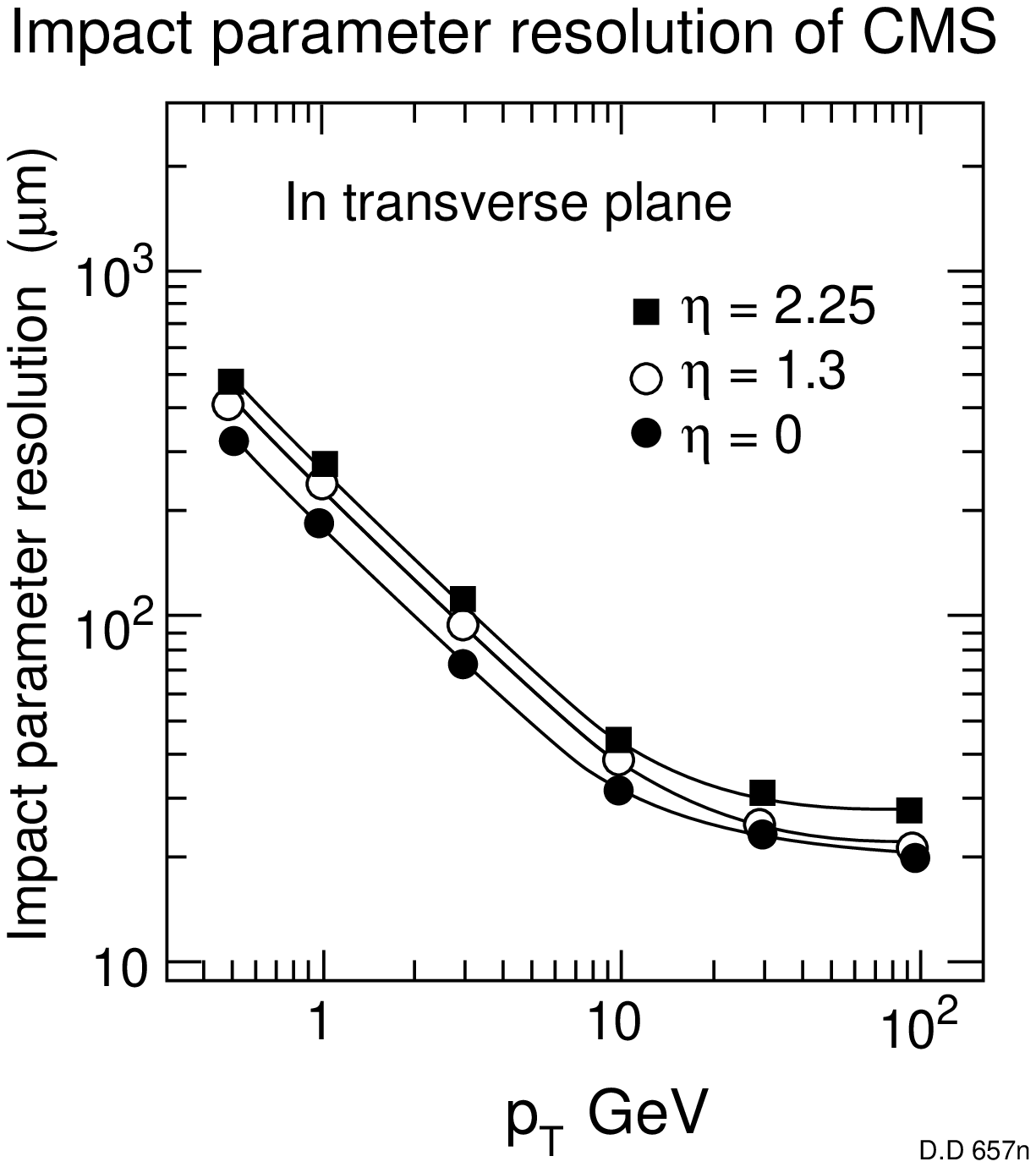}}}
\end{figure}
 
\vspace{2mm}

Figure 4.10: Momentum and impact parameter resolutions as a function of
$p_T$ at various rapidities.
 
\newpage

\ \\

\vspace{40mm}
 
\begin{figure}[h]
\vspace{-38mm}
\hspace*{10mm}
{\rotatebox{0}{\includegraphics[bb=100 97 499 606,
   width=13.cm,height=16.5cm, clip=true,draft=false]
{D_Denegri_0961n2.ill}}}
\end{figure}
 
\vspace{5mm}
 
Figure 4.11: The rejection factors against $t\bar{t}$,
expected on, the basis of lepton tracker
isolation criteria as a function of the $p_T$ cut on the accompanying tracks
in a cone of $R = 0.3$.

\newpage

\ \\
 
\vspace{20mm}
 
\begin{figure}[h]
\vspace{-38mm}
\hspace*{20mm}
{\rotatebox{0}{\includegraphics[bb=78 264 525 685,
   width=9.5cm,height=8.cm, clip=true,draft=false]
{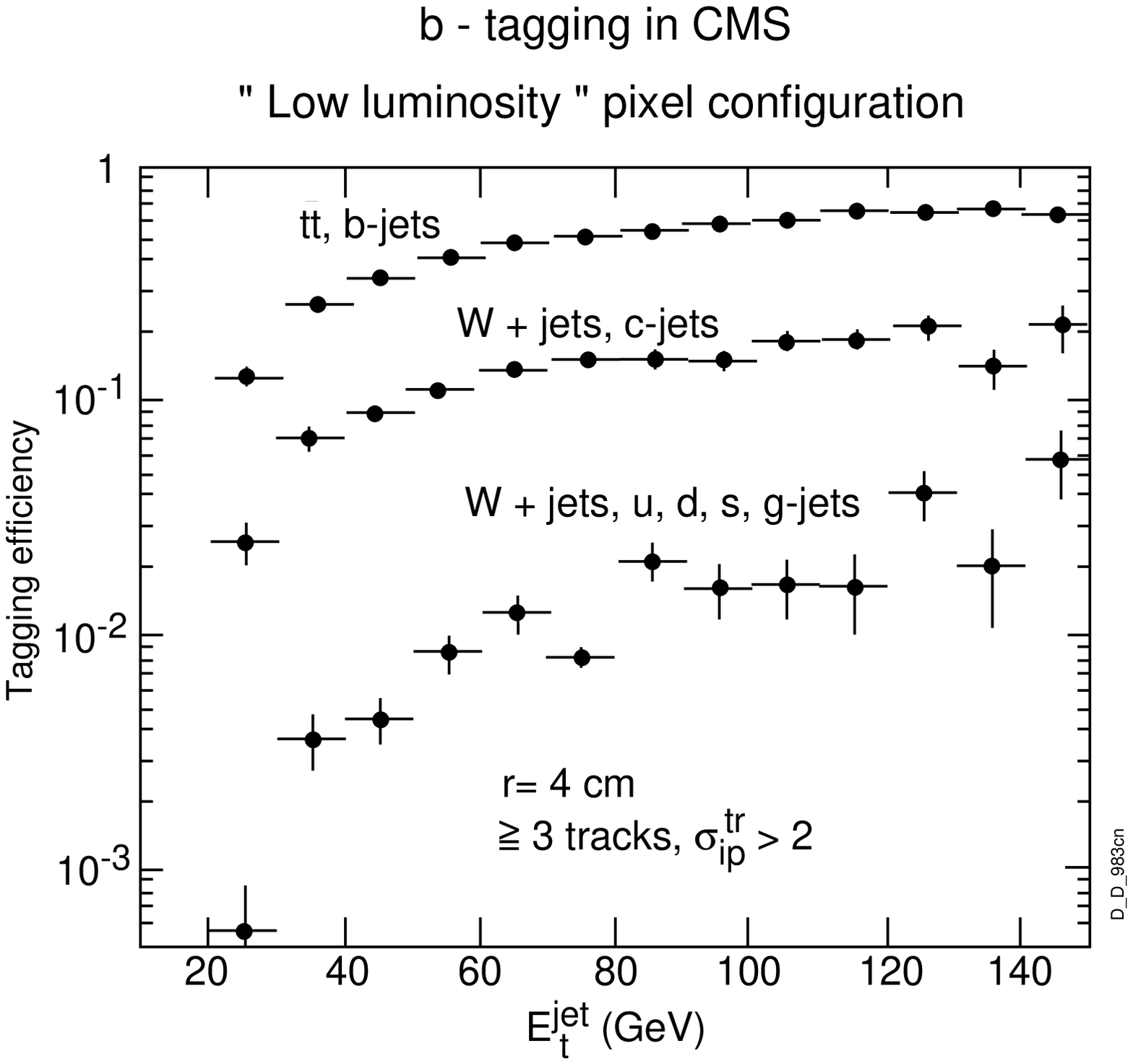}}}
\end{figure}
   
\vspace{5mm}

Figure 4.12: Tagging efficiency for $b$- and $c$-jets and mistagging
probability for
light quark and gluon jets. The tagging algorithm requires $\ge$ 3 tracks
of $p_T \ge$  2 GeV with $\ge$ 2$\sigma$ significance for impact
parameter measurements in the transverse plane.

\begin{figure}[hbtp]
\vspace{5mm}
\hspace*{20mm}
{\rotatebox{0}{\includegraphics[bb=67 51 533 731,
   width=9.5cm,height=10.cm, clip=true,draft=false]
{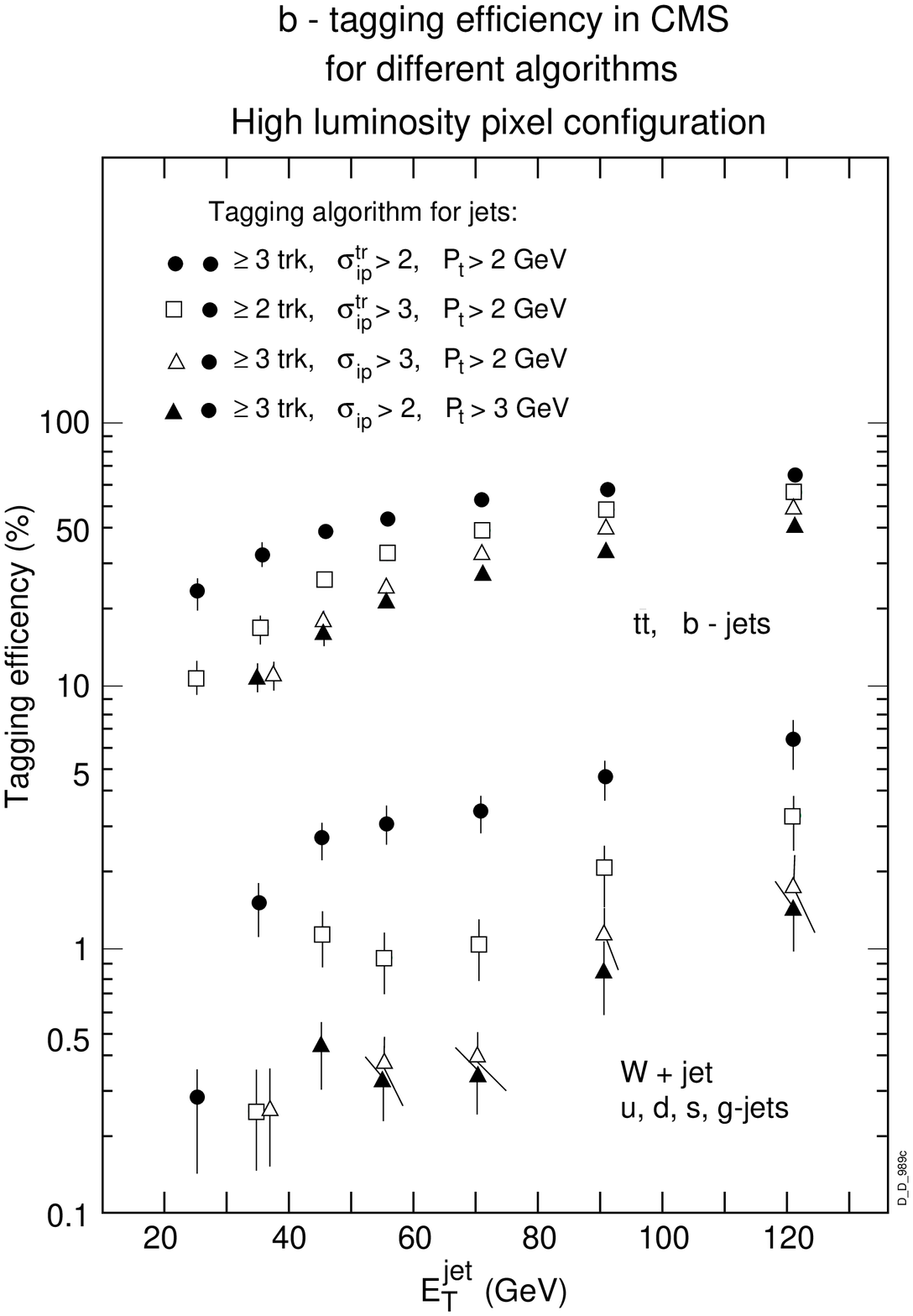}}}
\end{figure}
 
\vspace{2mm}
 
Figure 4.13: Tagging efficiency for $b$-jets and mistagging probability
for light quark and gluon jets for various tagging algorithms.

\newpage

\begin{figure}

\vspace*{10mm}

\resizebox{160mm}{120mm}
{\includegraphics{D_Denegri_0986n.ill}}
\end{figure}

\vspace*{10mm}

Figure 4.14: Tagging efficiency for $b$-jets and mistagging probability as
a function of the inner barrel pixel layer radius.

\newpage
 
\begin{figure}[hbtp]
\vspace{-10mm}
\hspace*{0mm}
\resizebox{14.cm}{!}{\rotatebox{0}{\includegraphics{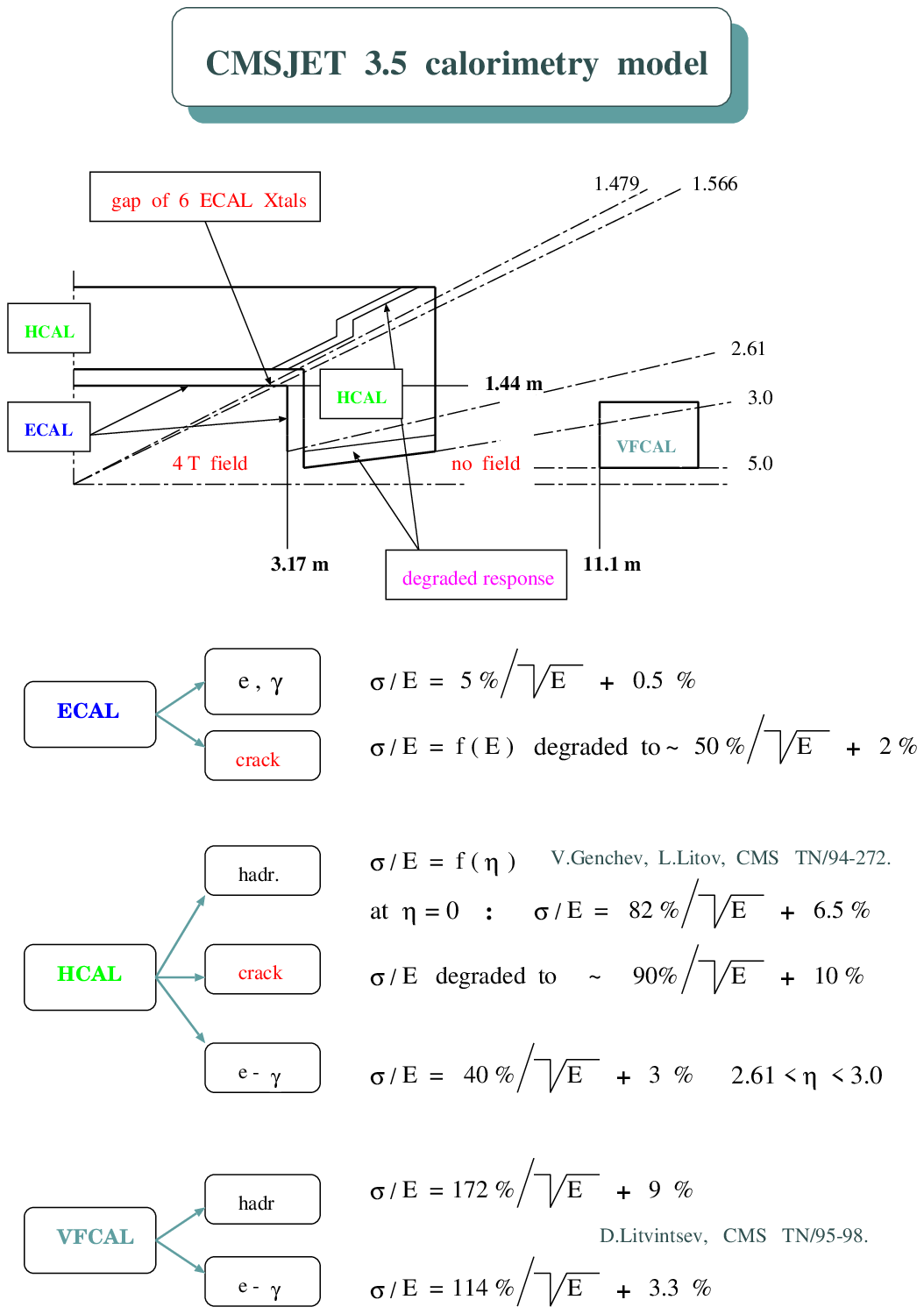}}}
\end{figure}
 
 \vspace{30mm}
 
Figure 4.15: The CMS calorimetry model in CMSJET 3.5

\newpage


\section{$Leptons + jets +$ \etm \ channel -- \\
search for gluinos and squarks}

     To investigate the maximum reach in parameter space for SUSY
via the production of strongly interacting sparticles 
we study $multilepton + multijet$ + \etm \ final states,
resulting from  the cascade decays of these sparticles \cite{gq, gqmp}.
As mentioned previously, a complete mass reconstruction of gluinos 
and squarks is impossible due to the presence of escaping particles,
two LSP's plus possibly neutrinos. So SUSY signal observability is based
on an excess of events of a given  topology  over known (expected)
backgrounds. Among the leptonic final states  studied, including always at
least $2~jets$ + \etm \ in the final state, are one lepton ($1l$), two
leptons of the same sign ($2l$ SS), two leptons of opposite
sign ($2l$ OS), and 3,4,5 leptons regardless the sign ($3l$,
$4l$, $5l$).
As discussed in the following, these topologies have been investigated
both with and without explicit requirements on muon isolation. 
The $jets$ + \etm \ topology, with no isolated leptons, is not discussed,
since it is the most sensitive one to detailed detector simulation,
and preliminary studies show that the mass reach is inferior to
one of the channels discussed here \cite{baer}.

 In order to  establish the limits of the SUSY reach in the
 $(m_{0},  m_{1/2})$  parameter space for various integrated luminosities,
 the signal was generated at more than 100 points.
At each point the signal is generated as a mixture of 
 all possible combinations of the strongly interacting sparticle
 production processes
($\bf \tilde{g}\tilde{g}, \tilde{g}\tilde{q}, \tilde{q}\tilde{q} $) \rm
 with masses and decay branching ratios corresponding to this point.
 The set of  $m_{0},  m_{1/2}$  points where the signal samples were
 generated for  tan$\beta$ = 2, A$_{0}$ = 0, $\mu$ $<$ 0  is  shown in
 Fig.~5.1. Figures 5.2a-d give, at four representative points in the
parameter space,
the main sparticle masses and decay chains with corresponding branching ratios.

The backgrounds considered in this study  are:
$t \bar{t}$, W + jets, Z + jets, WW, ZZ, ZW, Z$b\bar{b}$ and
QCD (2 $\rightarrow$ 2 processes including $b\bar{b})$.
There  could be some double counting between Z + jets and Z$b\bar{b}$
processes, but in practice this means more conservative background
estimates.
  
No pseudorapidity cuts are applied during background generation to avoid 
distortion of the event kinematics.  All processes were generated in the
following $\hat{p}_{T}$  intervals: 40 - 100, 100 - 200,  200 - 400, 400 -
800, $\geq$ 800 GeV. This allows the accumulation of significant
statistics in the  high-$p_{T}$ range in a more economical way than
without subdivision of the generation $\hat{p}_{T}$ range \cite{gq}.

The kinematics of signal events is usually  harder than for the SM
backgrounds, in particular for the interesting regions corresponding
to the maximum reach in terms of squark/gluino masses.    
The cross-sections of the background processes can however be higher
by orders of magnitude, and the high-$p_{T}$ tails of different backgrounds 
can have a kinematics similar to that of the signal.
Figures 5.3-5.5 illustrate kinematical distributions for the signal 
at three different points in parameter space and compare it to background.
More specifically in Fig.~5.3 one can see the comparison between
kinematical distributions for the signal point ($m_{0}=200$ GeV,
$m_{1/2}=760$ GeV) and the distributions for the sum of backgrounds in
$2l$ SS, $2l$ OS and $3l$ final states. Both signal and
background histograms contain only events satisfying first level selection 
criteria described below. Only the hardest jet and lepton in the event are
shown in distributions. The last histogram in each Figure shows the
distribution  of $E_{T}^{sum}$, the scalar sum of all relevant transverse
energies for leptons, jets and \etm. It reflects the overall  transverse
energy flow in the event, the event ``hardness''. This variable is also
rather insensitive to  the transverse energy flow induced by event pile-up
since it accounts only for those calorimeter cells which are attached to a
jet, and is thus a useful quantity to cut on in the high luminosity
regime.

    The topology of signal and background events is already rather  
 similar after the first level selection cuts,  in terms of
 angular distributions and circularity, as can be seen in Fig.~5.3a, thus  
it is not very useful to apply explicit cuts on these variables.
The difference between signal and background in the lepton $p_{T}$
distributions is also not as pronounced as might have been expected,
as signal leptons are produced in cascade decays, thus losing  ``memory''
of the hardness of the original process. For extremely high masses of
squarks or gluinos in the $\sim$ 2 TeV mass range, the difference in the
angular and $p_T^{l}$ distributions between signal and the total SM
background is still significant and helps in obtaining an effective data
selection. Ultimately, all the cuts have to be justified from the point of
view of the best observability of the signal over expected background.
 
Since the kinematics of the signal vary significantly with the variation 
of  $m_{0}$ and $m_{1/2}$ parameters, Figs.~5.3-5.5,  the set of different
cuts is applied and optimised independently in each point investigated.
These cuts are listed in Table 5.1. The optimisation procedure takes into
account  the number of jets per event,  and parameters of lepton
isolation, as discussed in details in the next subsection. The \etm \ cut
can be incremented beyond the values listed in Table 5.1, keeping all
other cuts untouched, to take advantage of the large values this variable
can take when the signal is close to the boundaries of explorable  
parameter space where  masses of squarks and gluino are in the 1-2 TeV range.

We find  that requiring more than 2 jets (up to $\geq$ 4-5 for 1, 10 
fb$^{-1}$, up to $\geq$ 6 for 100 fb$^{-1}$)  in the event is useful in
many cases, especially in domains 3 and 4 as defined in Fig.~2.16. The
exception  is  domain 1, where  requiring  $\geq$ 2 jets (as well as
requiring muon isolation for all cases except for 4 and 5 lepton final
states) is sufficient to get the best signal observability.

\vspace*{5mm}

Table 5.1: The various cuts applied to optimise the reach.

\begin{center} 
\begin{tabular}{|c|c|c|c|c|c|} \hline 
 & & & & & \\ 
 CUT & p$_{T}^{e}$ (GeV)
& p$_{T}^{\mu}$ (GeV) &   p$_{T}^{jet 1,2,3}$ (GeV)& \etm (GeV)&
 $E_{T}^{sum}$ (GeV) \\ 
          &    &    &    &     &    \\ \hline  \hline
 {\bf C1} & 20 & 10 & 40 & 100 & -  \\ \hline
 {\bf C2} & 20 & 10 & 150,80,40 & 150 & 500 \\\hline
 {\bf C3} & \multicolumn{2}{c}{40,20,20}  \vline
 & 150,80,40 & 150 & 500 \\\hline
 {\bf C4} & \multicolumn{2}{c}{40,20,20}  \vline
 & 200,120,40 & 200 & 700 \\\hline
 {\bf C5} & \multicolumn{2}{c}{50,30,20}  \vline
 & 250,150,40 & 200 & 1000 \\\hline
 {\bf C6} & \multicolumn{2}{c}{80,50,20}  \vline
 & 300,200,40 & 250 & 1200 \\\hline
 {\bf C7} & \multicolumn{2}{c}{100,70,20} \vline
& 300,200,40 & 250 & 1200 \\\hline
\end{tabular}
\end{center}

It is also found that some moderate cuts:
\begin{itemize} 
 \item $ Circularity$  $>$  0.1
 \item $ \delta \varphi$ (hardest lepton, \etm) $>$ 10-15 deg
\end{itemize} 
are useful to improve signal observability everywhere in $1l$ final
states, and in $2l$ OS final states, primarily in the region $m_{1/2}$
$>$ 600 GeV for an integrated luminosity $\geq$ 10 fb$^{-1}$.

From the chosen multiparametric space of cuts and constraints, we look
for those which provide the highest value of $m_{1/2}$ with a 5$\sigma$
significance excess for signal at each probed $m_0$ point and for each
lepton multiplicity final state investigated \cite{gq}.

An important issue in this study is the lepton isolation criterion.
The term ``isolated lepton''  means satisfying simultaneously the 
following two  requirements:
\begin{itemize}
\item 
no charged particle with $p_{T} >$ 2 GeV in a cone with 
defined radius in $\eta$ - $\varphi$ space around the lepton, and    
\item  
$\Sigma E_{T}^{calo}$ in a cone ring $0.1 < R < 0.3$ around the 
lepton impact point has to be less than a specified percentage of the
lepton transverse energy.
\end{itemize}

Figure 5.6 shows the acceptance for the signal in two points
of parameter space,  ``{\bf SIGNAL 1}'' at $m_{0}$ = 200 GeV,  $m_{1/2}$
= 700 GeV chosen in the  first characteristic domain in Fig.~2.16; ``{\bf
SIGNAL 2}'' at  $m_{0}$ = 1000 GeV,  $m_{1/2}$ = 600 GeV in the third domain, 
and the acceptance for background. The comparison is done  for the three
final state topologies, namely, two lepton of same sign ($2l$ SS), two
leptons of opposite sign ($2l$ OS) and three leptons ($3l$). This is
done as a function of the tracker isolation cone size  
$R=\sqrt{\delta\eta^2~+~\delta\varphi^2}$ and for the set of cuts denoted
as C2 in Table 5.1. Both electrons and muons are required to be isolated.
Each curve corresponds to a specified calorimetric cut:  no cut
(circles), $E_{T}^{calo}$ in the cone ring  $0.1 < R < 0.3$ not exceeding
10$\%$ (squares) and 5$\%$ (triangles) of the lepton $E_{T}$.

 It is rather difficult to deduce from these curves what is in general
 the most advantageous isolation criterion. Tight isolation is not very
useful in case of ``{\bf SIGNAL 2}'' in $2l$ OS and $3l$ final
states, where it leads to a significant loss of signal. At the same  time,
the background is not suppressed sufficiently, especially in case of
$2l$ OS, where it is mainly due to $t\bar{t}$ production giving well 
isolated leptons through the  decay chain $t \rightarrow$ W$b \rightarrow
l \nu b$.  So, the isolation criterion could also in principle be
optimised at every $m_{0},  m_{1/2}$ point  and for each leptonic
topology so as to obtain the best signal observability.  In practice,
since the identification  of non-isolated electrons is experimentally
difficult, a minimal isolation is always required, e.g. charged particle
isolation in a cone  of $R = 0.1$ and a requirement of 
$E_{T}^{calo}$ in $R = 0.3$ not exceeding 10$\%$ of the electron $E_{T}$.
We find that a still tighter isolation  does not further improve the
signal observability at all the $m_{0}, m_{1/2}$ points investigated.
As for the muons, we find that  a  gain of $\sim$ 50-150 GeV
in reach in  $m_{1/2}$ is possible if no isolation is applied at all on
them. This is valid for all lepton topologies and almost everywhere in
the  $m_{0}, m_{1/2}$ plane, except in  the first domain in Fig.~2.16.
This is due to the copious production of $b$-quarks in the signal,
especially in domains 2-4, giving rise to a significant yield of non-isolated
leptons from $B$-decays. So except when explicitly stated,
we do not apply in the following any isolation requirement on muons.

    The case of high instantaneous luminosity running requires a 
separate discussion  of lepton isolation due to event pile-up
deteriorating lepton (electron) isolation. However, as we use only a
loose isolation criterion to begin with, the criterion can  easily be
changed with only minor loss of efficiency. Furthermore,  with a good
tracker it should be possible to separate most of the tracks originating
from vertices other than the leptonic one.
Even in case of confusion with tracks from event pile-up vertices,
the probability of the signal lepton to be ``moderately'' isolated
(as we have defined above) in spite of pile-up 
is still 97 - 98$\%$ when averaged over the entire ECAL, as we
calculated  using PYTHIA pile-up (MSEL = 2) with
an average number of pile-up events per bunch-crossing of
$\bar{n}$ = 25. So event pile-up at high instantaneous luminosity
does not affect significantly the results of our study.

    Figures 5.7-5.9 show the final results of our study of
squark and gluino observability and mass search in $m_{0}$, $m_{1/2}$
parameter  space in the various $multilepton + multijet$ + \etm \ final
states for our standard set of parameters tan$\beta$ = 2, A$_{0}$ = 0,
$\mu$ $<$ 0. The results are shown for increasing  values
of integrated luminosity from 1 to 100 fb$^{-1}$. All the 5$\sigma$
boundaries shown are obtained relaxing muon isolation whenever possible,
but keeping always the minimal electron isolation criteria needed for
unambiguous instrumental electron identification.
The maximal squark and gluino mass reach is obtained in 1 lepton final states.
With 10$^{5}$ pb$^{-1}$ the maximal gluino mass reach varies from $\approx$
2.5 to 2.0 TeV, depending on $m_{0}$, whilst the squark mass reach varies 
between 2.1 - 2.5 TeV. What is important is that already with 
10$^{3}$ pb$^{-1}$ the $\lsim$ 1 TeV domain can be thoroughly explored,
and according to the ``no-fine tuning'' arguments
this mass domain is the most plausible one.     

In Fig.~5.8 one can see that the cosmologically interesting region
\cite{cosmo} is entirely contained within the domain explorable already
with 10 fb$^{-1}$ in all topologies considered.
The Figure shows the relic neutralino dark matter density
contours corresponding to $\Omega h^{2}$ values of 0.25, 0.5 and 1,
where $\Omega$ is the neutralino density relative to the critical density
of the Universe and $h$ is the Hubble constant scaling factor
($h \sim 0.7$). The prefered mixed dark matter scenario corresponds
to values of $\Omega h^{2}$ between 0.15 and 0.4.

Figure 5.10 illustrates
the  gain in mass reach which can be obtained when non-isolated muons are
taken into account in the analysis. Figure 5.11 shows the expected
signal-to-background ratio at the 5$\sigma$ boundaries in each topology.
It is always bigger than 1, thus we may expect that the maximum reach will
not be very sensitive to a possible underestimation of the background
cross-sections. To estimate the influence of systematic uncertainties on
the background cross-sections on the $\tilde{g}$, $\tilde{q}$ observability
contours, we increased all the background cross-sections 
by a factor of 2, and the W/Z + jets background by a factor
of 5. Figure 5.12  shows how the 5$\sigma$ significance contours are
modified under these conditions for an integrated luminosity of 10 fb$^{-1}$.
The loss of reach is small in all cases.

We also checked the sensitivity of our results to changes in some of our
basic parameters. The change of the sign of $\mu$, keeping $tan\beta = 2,
A_0 = 0$, affects primarily charginos and neutralinos.
It changes their masses, decay modes and branching ratios,
and these alter the final state kinematic and topology.
In the squark sector, a positive sign of $\mu$
decreases the $\tilde{t}_1$ mass by $\approx$ 30 -- 50 GeV
compared to the $\mu < 0$ case. Consequently, final state leptons
coming from stops and top quarks will be slightly ``softer'', thus requiring
lower threshold cuts and causing modified background levels compared to
the $\mu < 0$ case.
Figure 5.13 gives the maximum expected reach of the CMS detector
for squark and gluino searches in different multilepton final states
for 10$^{4}$ pb$^{-1}$ with $\mu > 0$ \cite{gqmp}.
The reach is again greatest in the single lepton channel, and extends
up to gluino and squark masses of $\sim 2$ TeV.
Relaxation of the muon isolation requirements increases the reach
by $\approx$~50~GeV in the single lepton case, and by $\approx 200$ GeV in
the trilepton case.
The $4l$ channel with 10$^{4}$ pb$^{-1}$ has an interesting explorable domain 
only if muons are not isolated \cite{4lmp}.
Figure 5.14 gives the same results for $L_{int}=10^{5}$ pb$^{-1}$
\cite{gqmp}. With this luminosity
and with full muon isolation (not shown), 
the $4l$ maximal reach region in fact becomes
``a band of observability'' approximately between
$m_{1/2}$~$\approx$~200~GeV and $m_{1/2} \approx 350$ GeV \cite{4lmp}.
The lower boundary is mainly caused by a decrease of the branching ratio
for
$\tilde{g} \rightarrow \tilde{\chi}^0_1t\bar{t}$.
For $m_0 < 200$~GeV the region of observability extends up to 
$m_{1/2}\approx 650$~GeV, as here there is a high yield of    
leptons from chargino and neutralino decays.

The main conclusions of this particular study 
are the following: within the  mSUGRA model, the SUSY
signal due to production of  $\tilde{q}$ and $\tilde{g}$ would be detectable
as an excess of events over SM expectations up to masses $m_{\tilde{q}}$
$\sim$ $m_{\tilde{g}}$ $\sim$ 1 TeV with only $10^{3}$ pb$^{-1}$. The
ultimate reach, for $10^{5}$ pb$^{-1}$, would extend up to $m_{\tilde{q}}$
$\sim$ 2500 GeV for small $m_{0}$ ($\lsim$ 400 GeV) and up to $\sim$ 2000
GeV for any reasonable value of $m_{0}$ ($\lsim$ 2000 GeV);
squark masses can be probed for values in excess 
of 2000 GeV. This means that the entire plausible domain of EW-SUSY 
parameter space (for tan$\beta$ = 2)
 can be probed with these $lepton(s) + jets$ + \etm \
final states. Furthermore, the $S/B$ ratios are $>$ 1 everywhere in
parameter space (with the appropriate cuts) thus allowing a study of the
kinematics of $\tilde{q}$, $\tilde{g}$ production and obtaining information  
on their masses \cite{paige2}. The cosmologically interesting region
$\Omega h^{2} \leq$ 1 can be entirely probed
with an integrated luminosity not in excess of $\sim 10^{3}$ pb$^{-1}$,
thus with a very large safety margin.

\vspace{20mm}
\begin{figure}[hbtp]
\begin{center}
\resizebox{8cm}{!}{\includegraphics{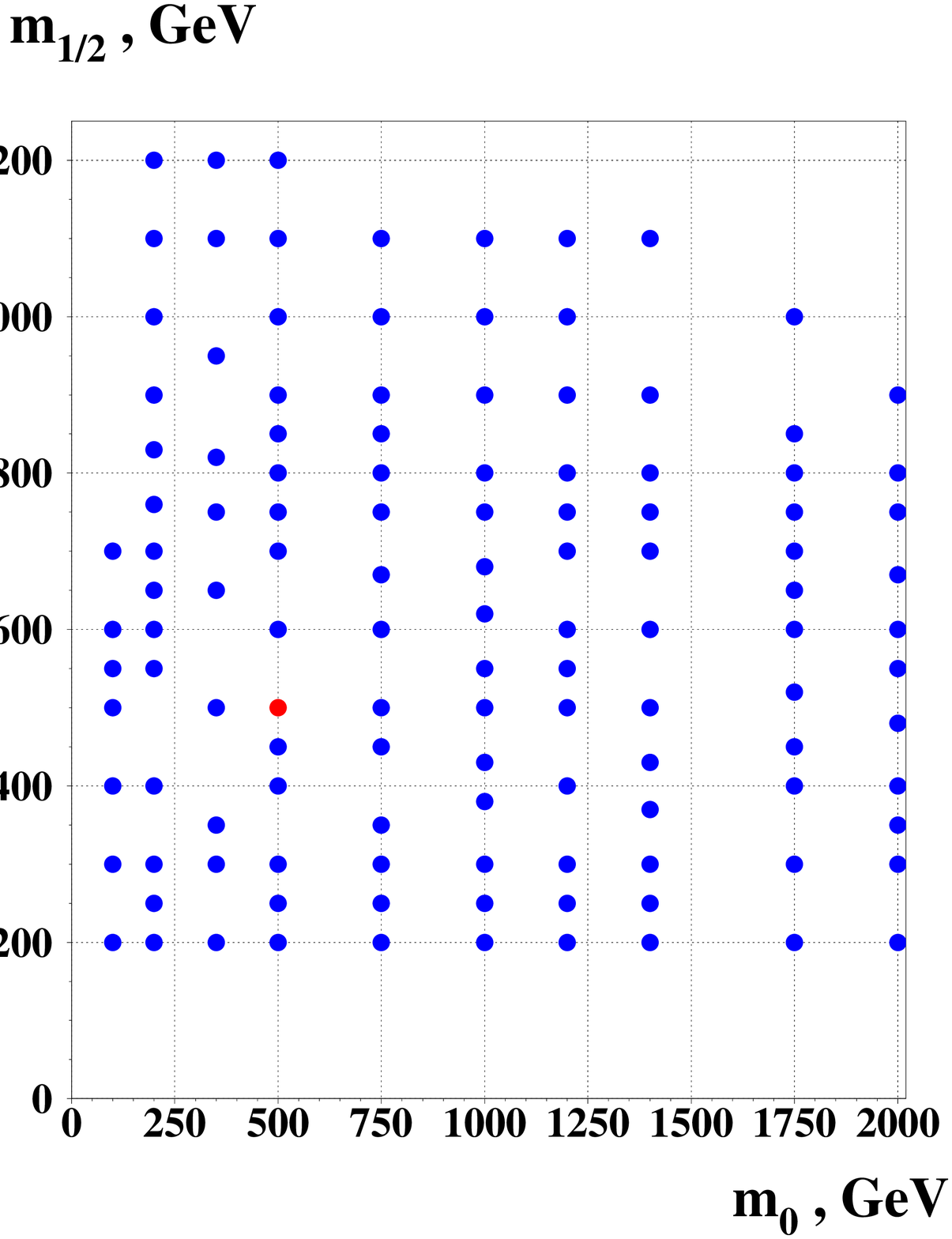}}
\end{center}
\end{figure}
\vspace{5mm}

Figure 5.1: The set of generated data points in the ($m_{0},
m_{1/2}$) mSUGRA plane for tan$\beta$ = 2, A$_{0}$ = 0, $\mu$ $<$ 0. 

\begin{figure}[hbtp]
\vspace{-30mm}
\hspace*{5mm}
\resizebox{15cm}{!}{\includegraphics{fig9.2.eps}}
\hspace*{30mm}Figure 5.2: Some typical points in parameter space.
\end{figure}

\newpage
 
\ \\

\begin{figure}[hbtp]
\vspace*{-35mm}
\begin{center}
\resizebox{15.5cm}{!}{\includegraphics{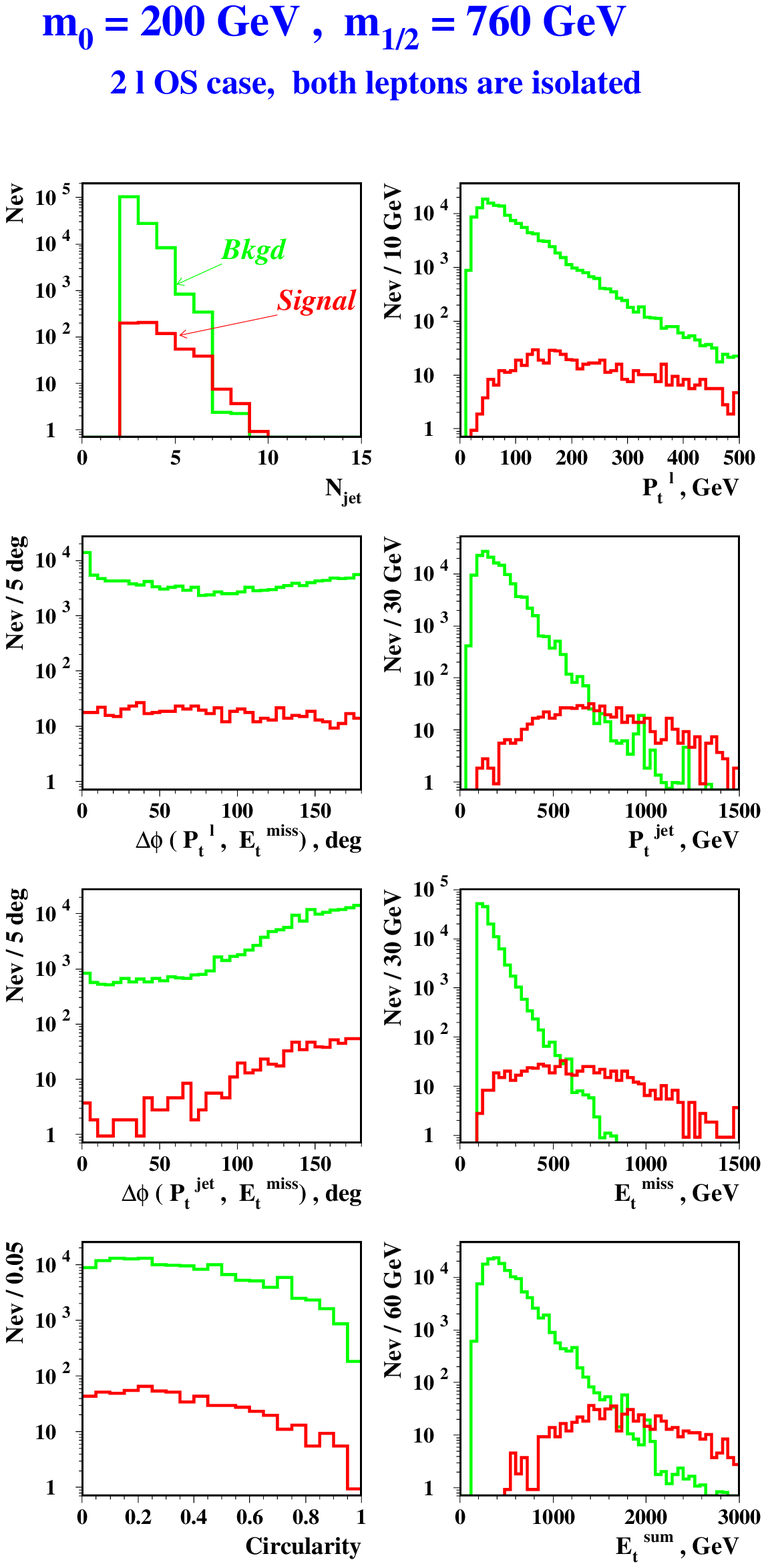}}
\end{center}
\end{figure}

\vspace*{-5mm}

Figure 5.3a:
Comparison of the signal and background kinematics at one
point in mSUGRA parameter space, after requiring two leptons
of opposite sign (plus jet and \etm).

\newpage
 
\ \\

\vspace*{10mm}

\begin{figure}[h]
\vspace{-40mm}
\hspace*{2mm}
\resizebox{15cm}{!}{\rotatebox{0}{\includegraphics{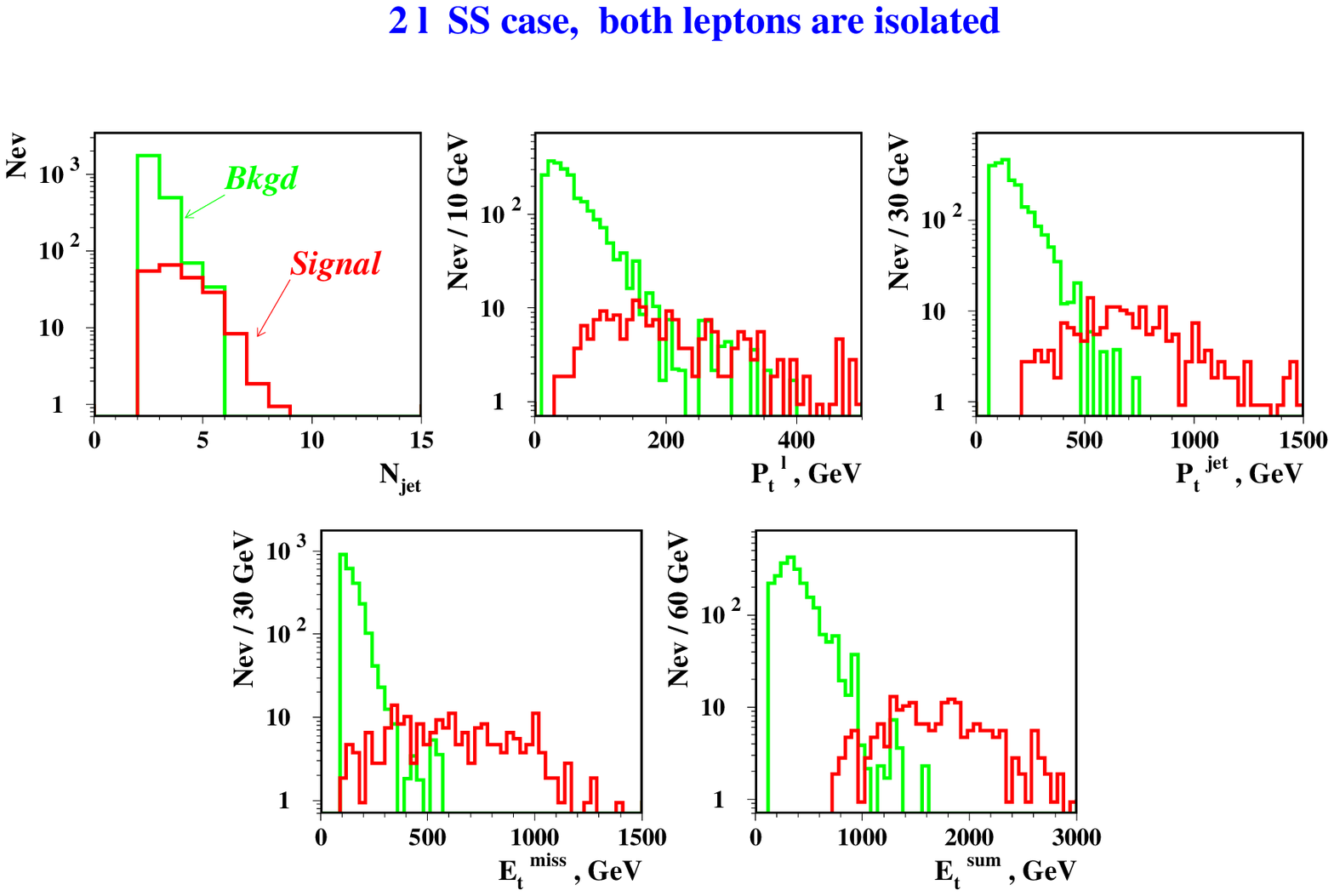}}}
~\\
\hspace*{0mm}Figure 5.3b:
Comparison of the signal and background kinematics in the same point as in
in Fig.~5.3a, but for $2l$ SS final state topology.
\end{figure}

\vspace*{10mm}

\begin{figure}[h]
\vspace{-30mm}
\hspace*{2mm}
\resizebox{15cm}{!}{\rotatebox{0}{\includegraphics{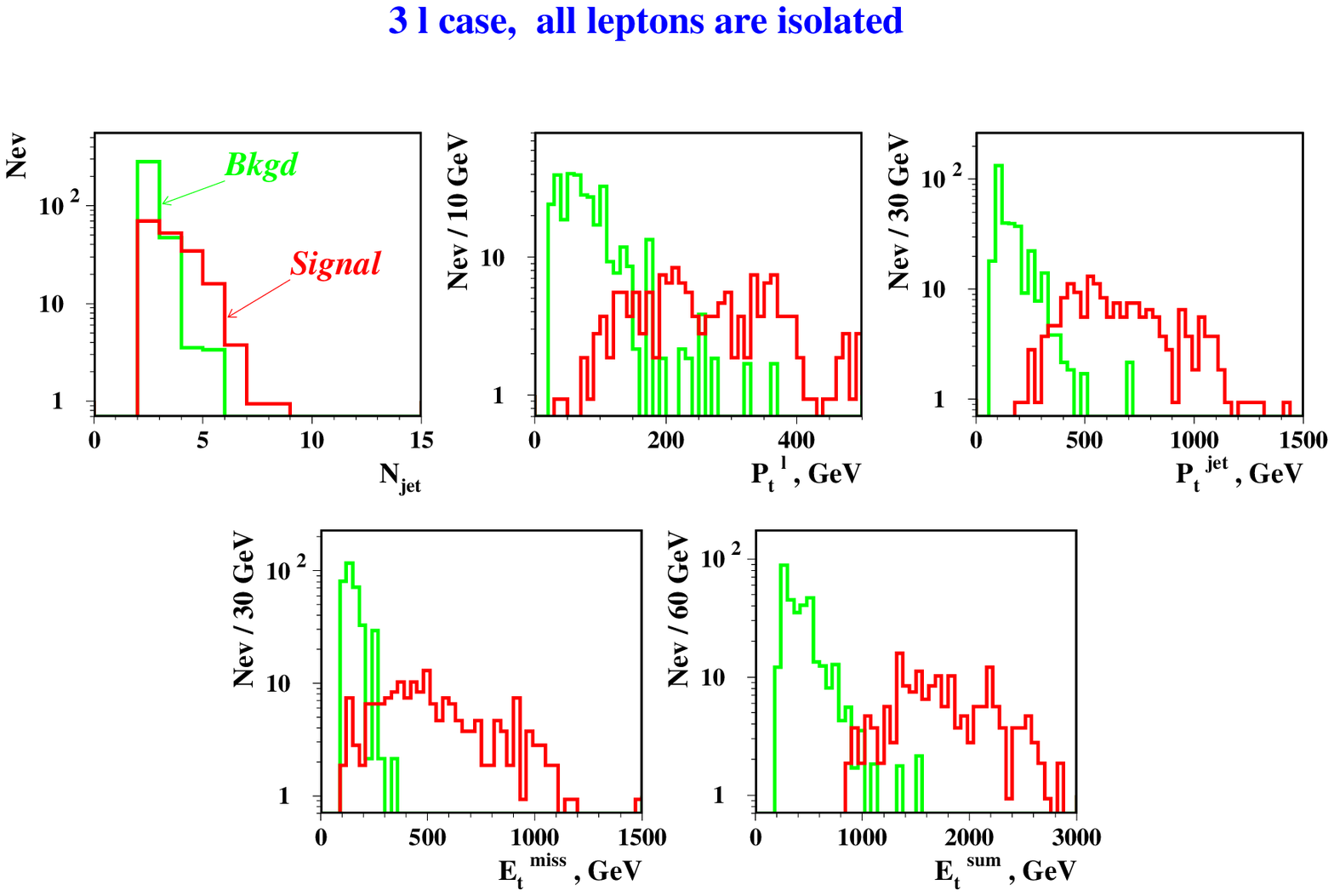}}}
~\\
\hspace*{0mm}Figure 5.3c:
Comparison of the signal and background kinematics in the same point as in
in Fig.~5.3a, but for $3l$ final state topology.
\end{figure}

\newpage
 
\ \\

\vspace*{10mm}

\begin{figure}[h]
\vspace{-40mm}
\hspace*{2mm}
\resizebox{15cm}{!}{\rotatebox{0}{\includegraphics{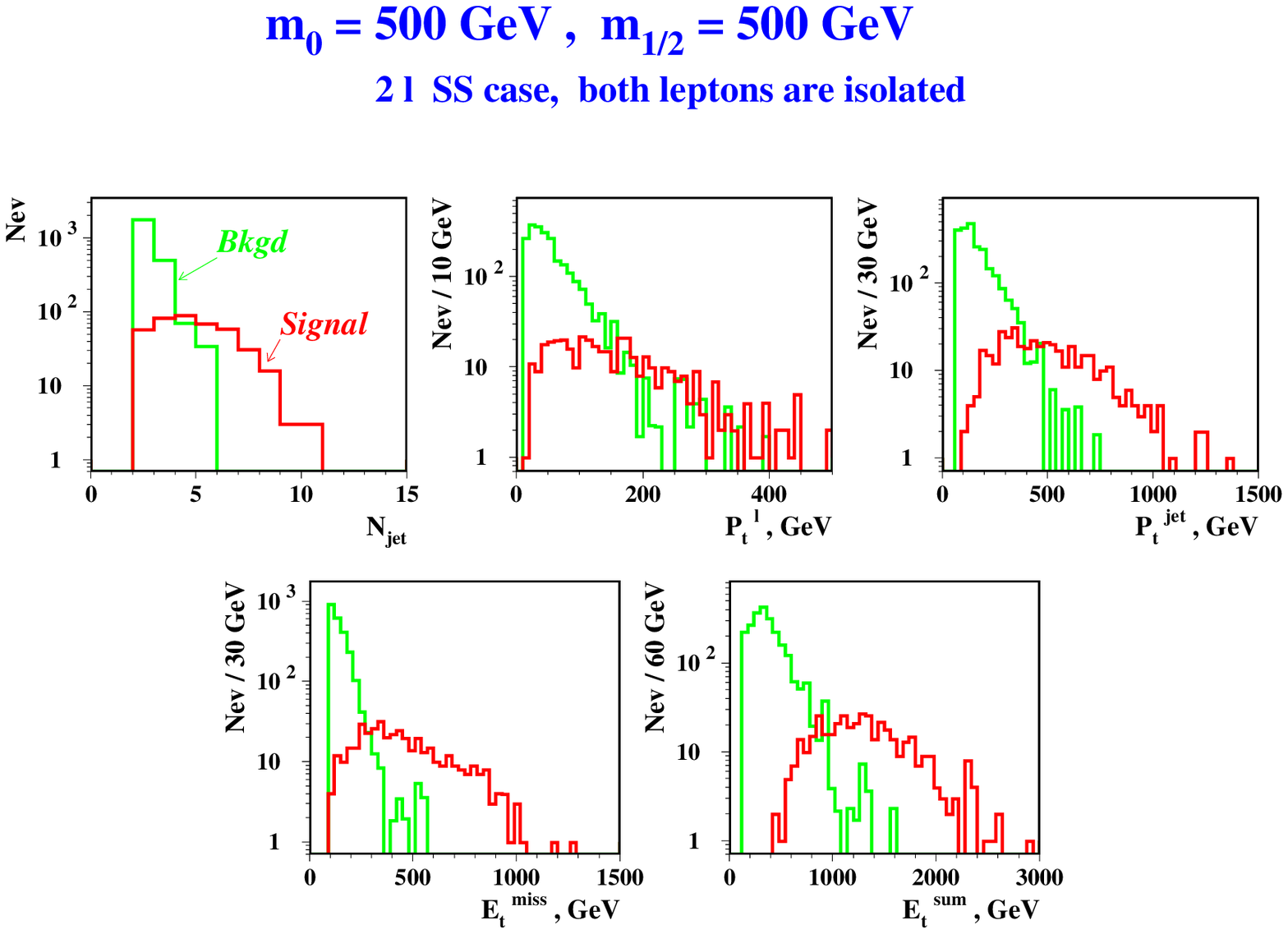}}}
~\\
\hspace*{15mm}Figure 5.4:
Comparison of the signal and background kinematics
for a second point in parameter space.
\end{figure}

\vspace*{20mm}

\begin{figure}[h]
\vspace{-30mm}
\hspace*{2mm}
\resizebox{15cm}{!}{\rotatebox{0}{\includegraphics{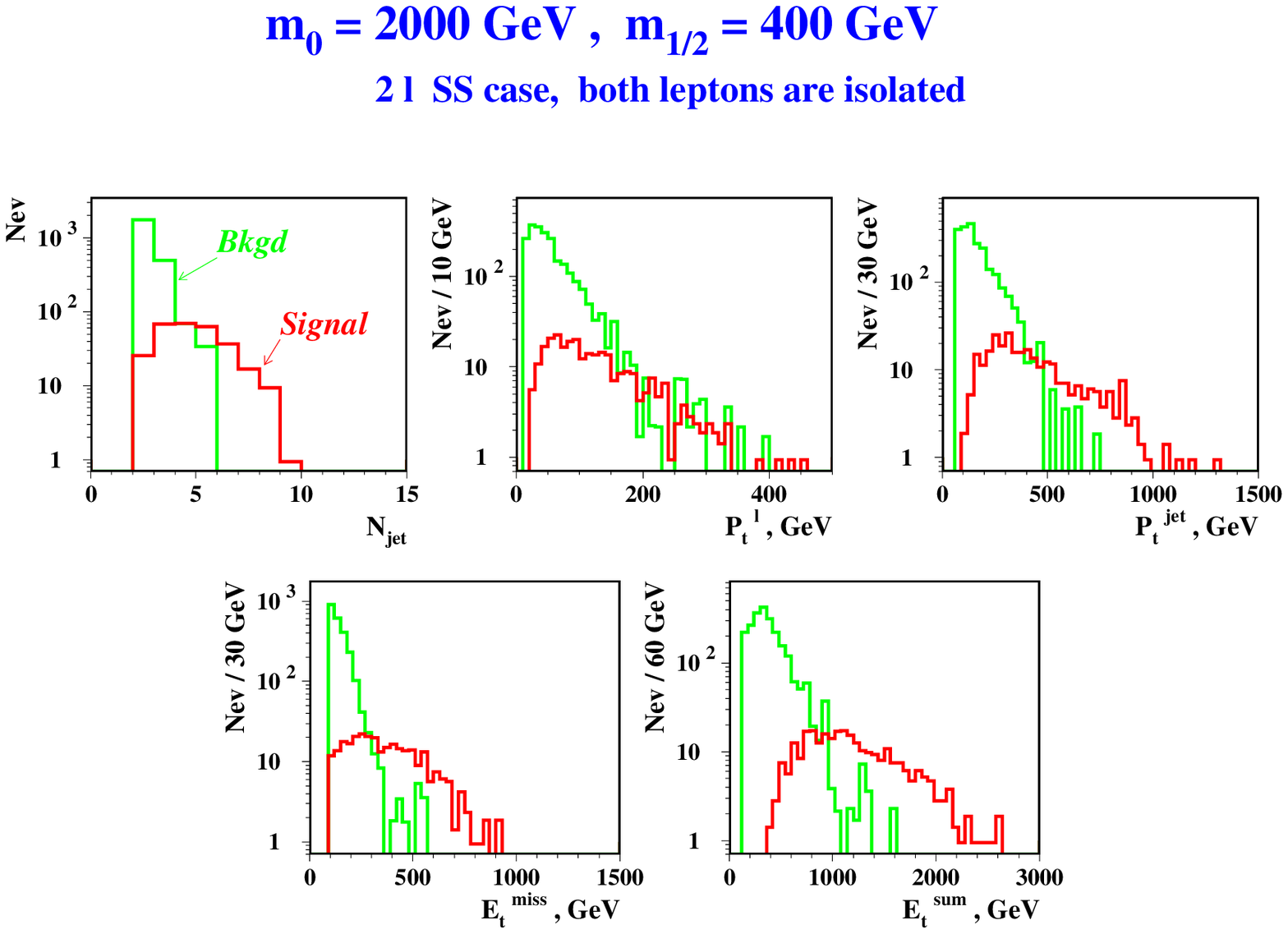}}} 
~\\
\hspace*{15mm}Figure 5.5:
Comparison of the signal and background kinematics
for a third point in parameter space.

\end{figure}

\newpage
 
\ \\

\vspace{-20mm}

\begin{figure}[hbtp]
  \begin{center}
\resizebox{12cm}{!}{\includegraphics{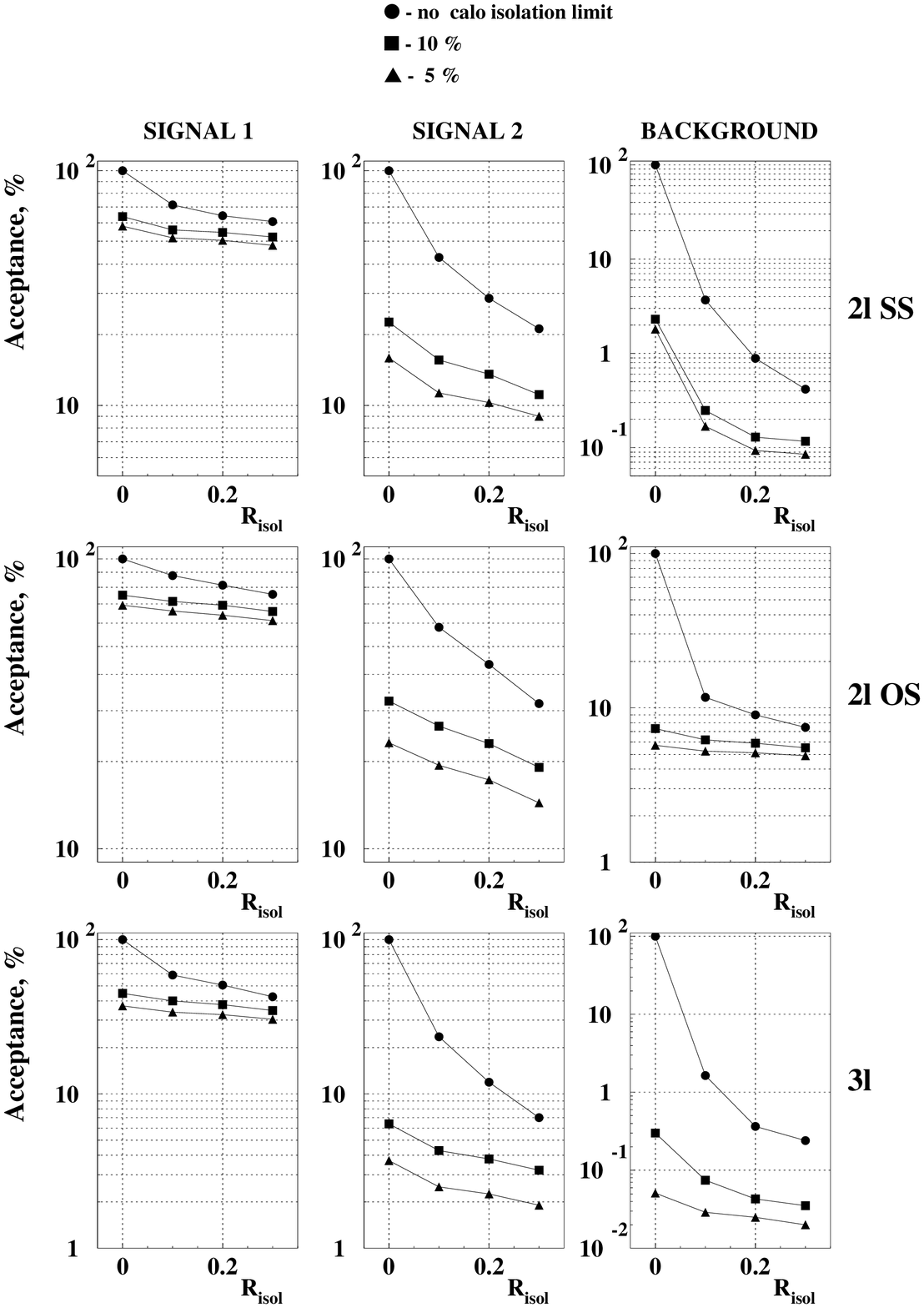}}
  \end{center}
\vspace{-5mm}
\end{figure}

Figure 5.6:
Effect of the lepton isolation criterion on signal and background,
at two different points in parameter space: ``{\bf SIGNAL 1}'' is
 $m_{0}$ = 200 GeV,  $m_{1/2}$ = 700 GeV, ``{\bf SIGNAL 2}'' is
 $m_{0}$ = 1000 GeV,  $m_{1/2}$ = 600 GeV, as a function of $R_{isol}$,
the tracker isolation cone size $R_{isol}$ =
$\sqrt{\delta\eta^2~+~\delta\varphi^2}$. The results are shown for three
choices of calorimetric isolation parameter (sum of calorimeter energy
deposit in a cone ring $0.1 < R < 0.3$ around the lepton impact point) and
for three event topologies.

\newpage
 
\ \\

\vspace{-20mm}

\begin{figure}[hbtp]
  \begin{center}
\resizebox{12cm}{!}{\includegraphics{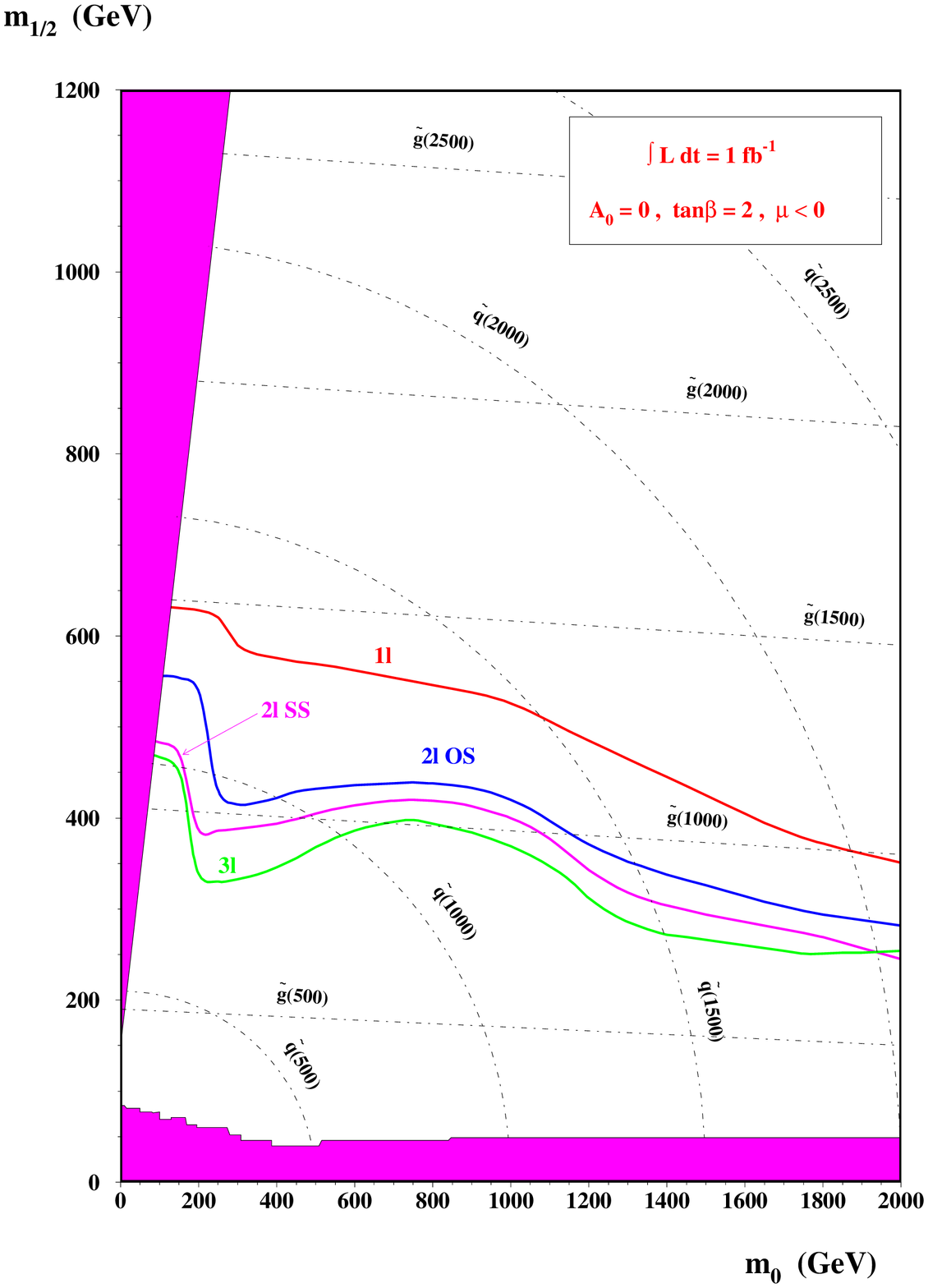}}
  \end{center}
\vspace{-5mm}
\end{figure}

Figure 5.7:
5$\sigma$ gluino/squark contours in various lepton multiplicity final
states
for 1 fb$^{-1}$ and our standard set of model parameter values tan$\beta$
= 2,
A$_{0}$ = 0, $\mu$ $<$ 0. Squark  and gluino isomass curves shown by 
dash-dotted lines.

\newpage
 
\ \\

\vspace{-20mm}
 
\begin{figure}[hbtp]
  \begin{center}
\resizebox{12cm}{!}{\includegraphics{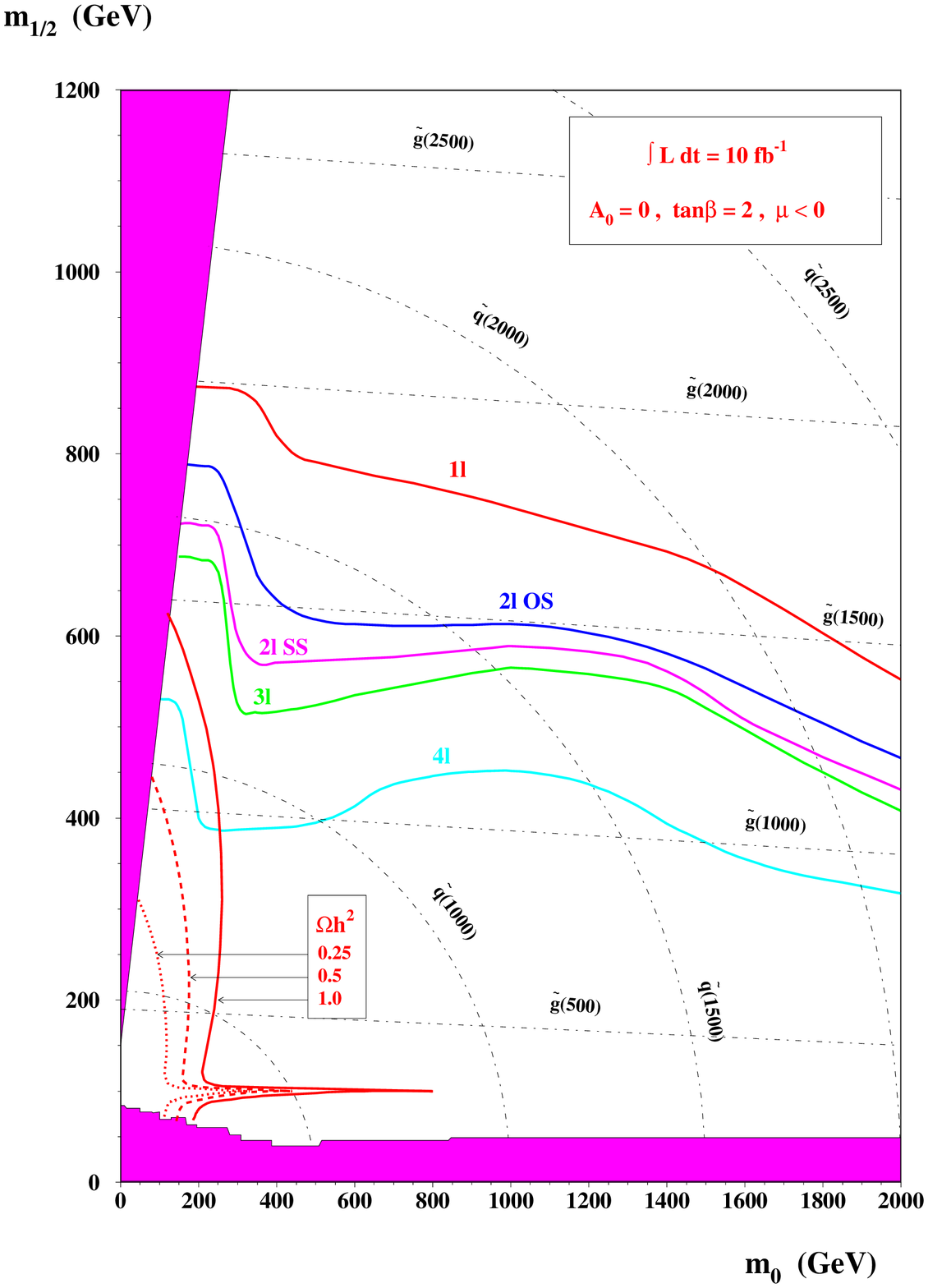}}
  \end{center}
\vspace{-5mm}
\end{figure}

Figure 5.8:
5$\sigma$ contours as in Fig.~5.7, but for 10 fb$^{-1}$.
The relic neutralino dark matter density contours of $\Omega h^{2}$ =
0.25, 0.5 and 1.0 are also shown. 
The region of $\Omega h^{2}$ between 0.15 and 0.4 is prefered by mixed
dark matter models \cite{cosmo}; the $\Omega h^{2}$ = 1 contour is
the upper limit.

\newpage
 
\ \\

\vspace{-20mm}

\begin{figure}[hbtp]
  \begin{center}
\resizebox{12cm}{!}{\includegraphics{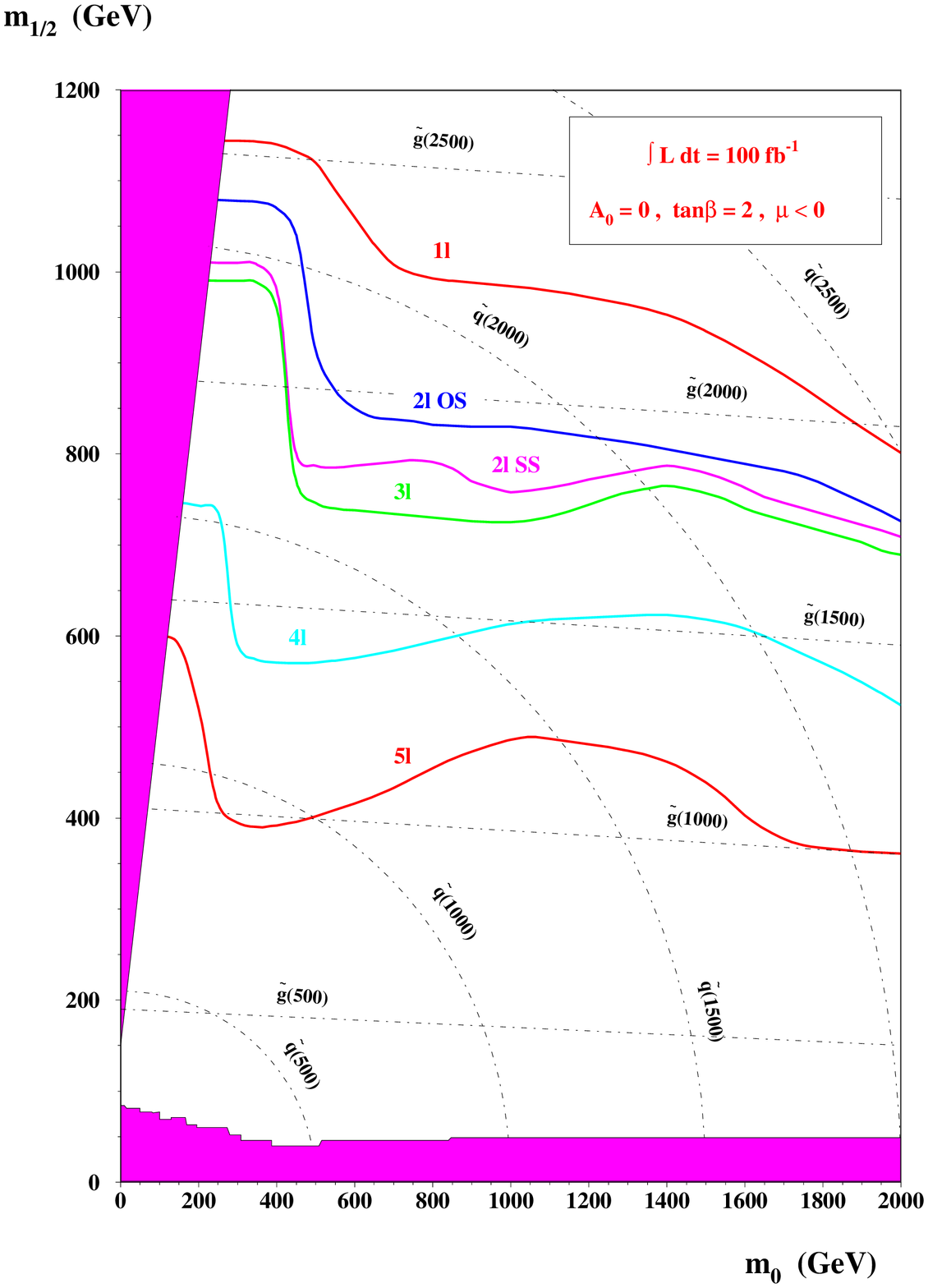}}
  \end{center}
\vspace{-5mm}
\end{figure}

Figure 5.9:
Ultimate squark and gluino mass reach, in terms of a $\geq 5 \sigma$
observation contour for  100 fb$^{-1}$,
in various lepton multiplicity final states .

\newpage
 
\ \\

\vspace{-20mm}

\begin{figure}[hbtp]
  \begin{center}
\resizebox{12cm}{!}{\includegraphics{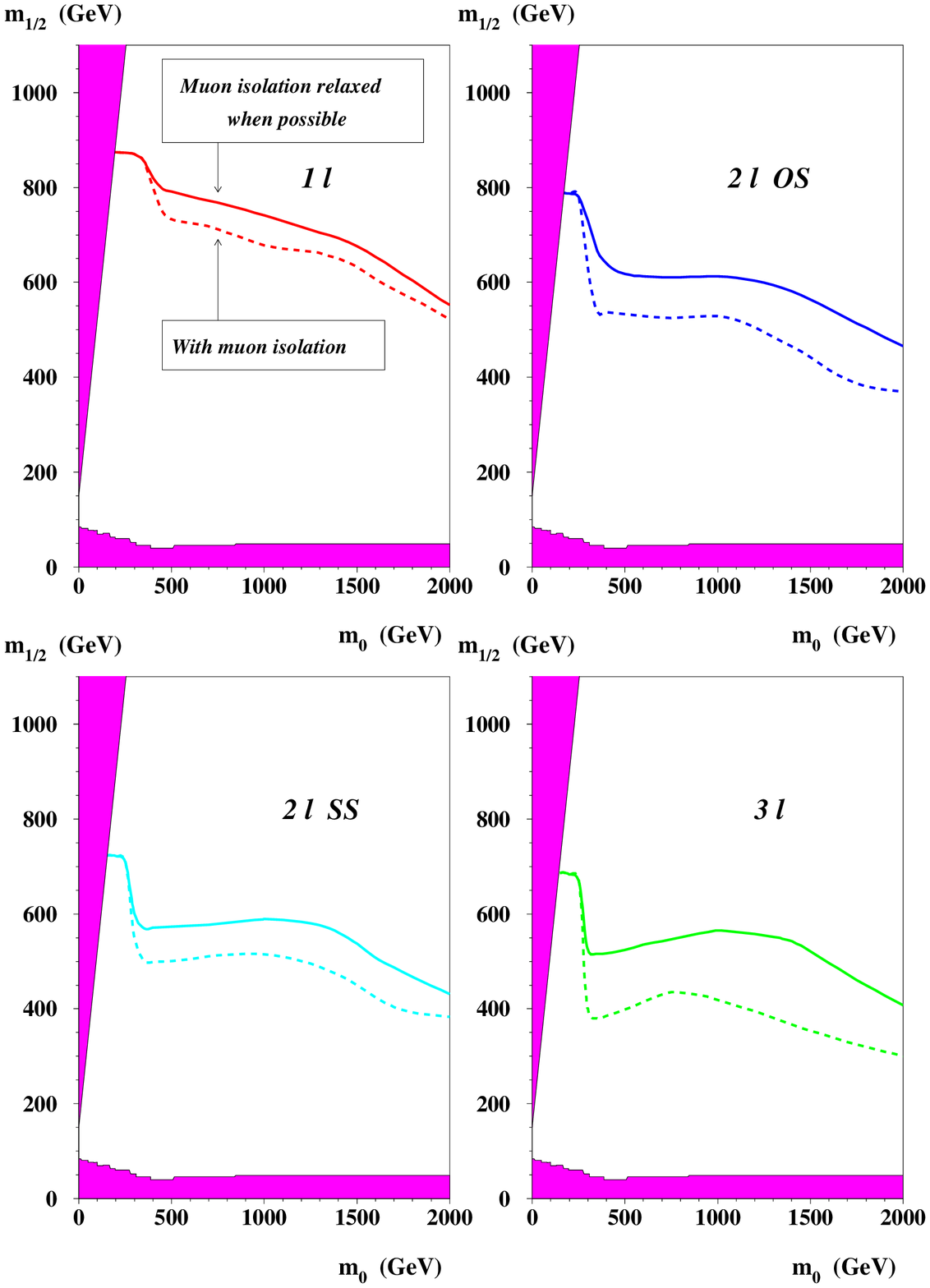}}
  \end{center}
\vspace{-5mm}
\end{figure}

Figure 5.10:
Influence of muon isolation on the 5$\sigma$ contours in various
topologies for 10 fb$^{-1}$.

\newpage
 
\ \\

\vspace{-20mm}

\begin{figure}[hbtp]
  \begin{center}
\resizebox{12cm}{!}{\includegraphics{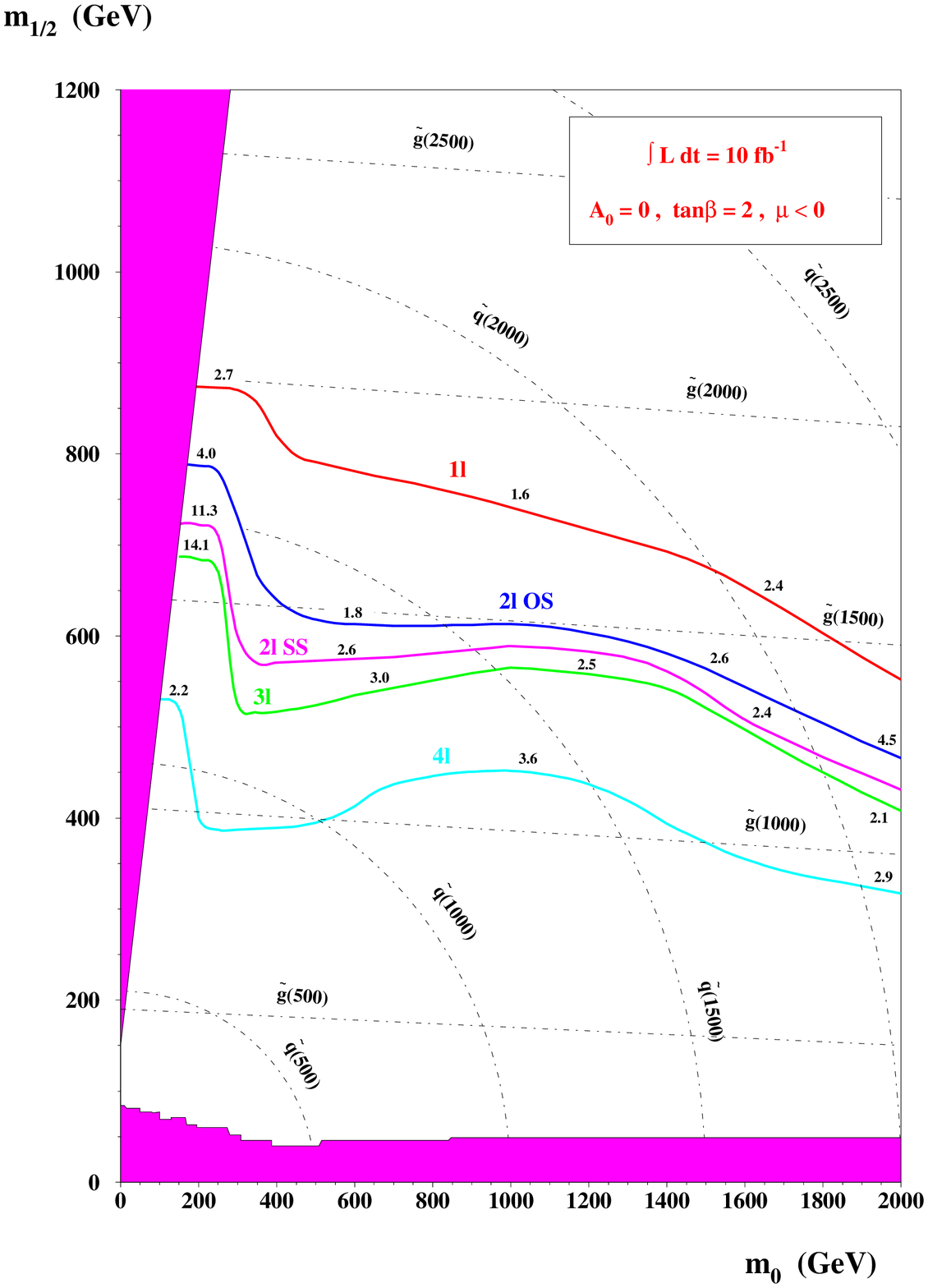}}
  \end{center}
\vspace{-5mm}
\end{figure}

Figure 5.11:
Signal to SM background ratio on the 5$\sigma$ reach boundaries
in in various lepton multiplicity final state topologies, for 10
fb$^{-1}$.
 
\newpage
 
\ \\

\vspace{-20mm}

\begin{figure}[hbtp]
  \begin{center}
\resizebox{12cm}{!}{\includegraphics{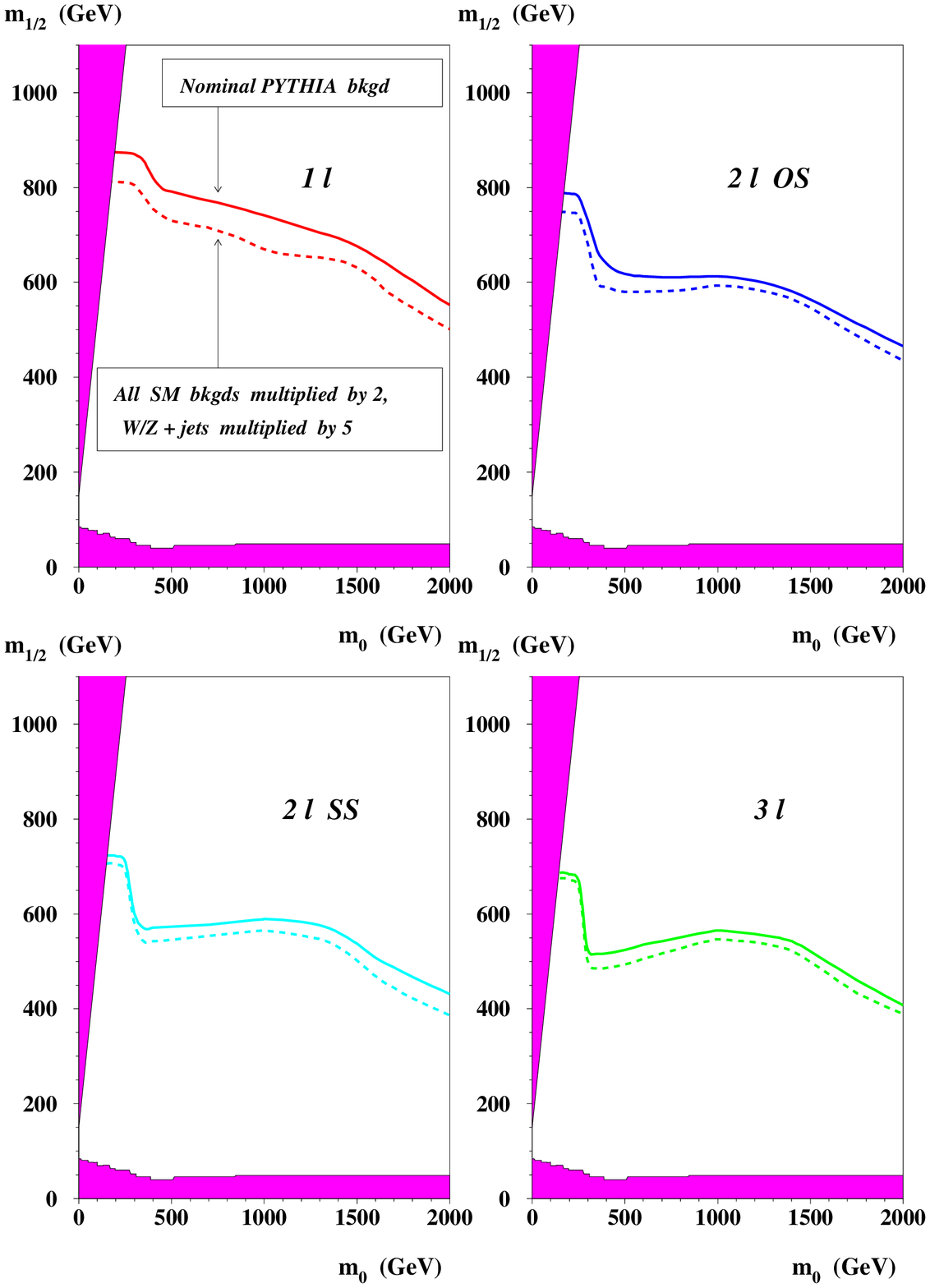}}
  \end{center}
\vspace{-5mm}
\end{figure}

Figure 5.12:
Sensitivity in the systematic uncertainties in background cross section on
the 5$\sigma$ contours in various lepton multiplicity final state
topologies, for 10 fb$^{-1}$.

\newpage
 
\ \\

\vspace{-20mm}

\begin{figure}[hbtp]
  \begin{center}
\resizebox{12cm}{!}{\includegraphics{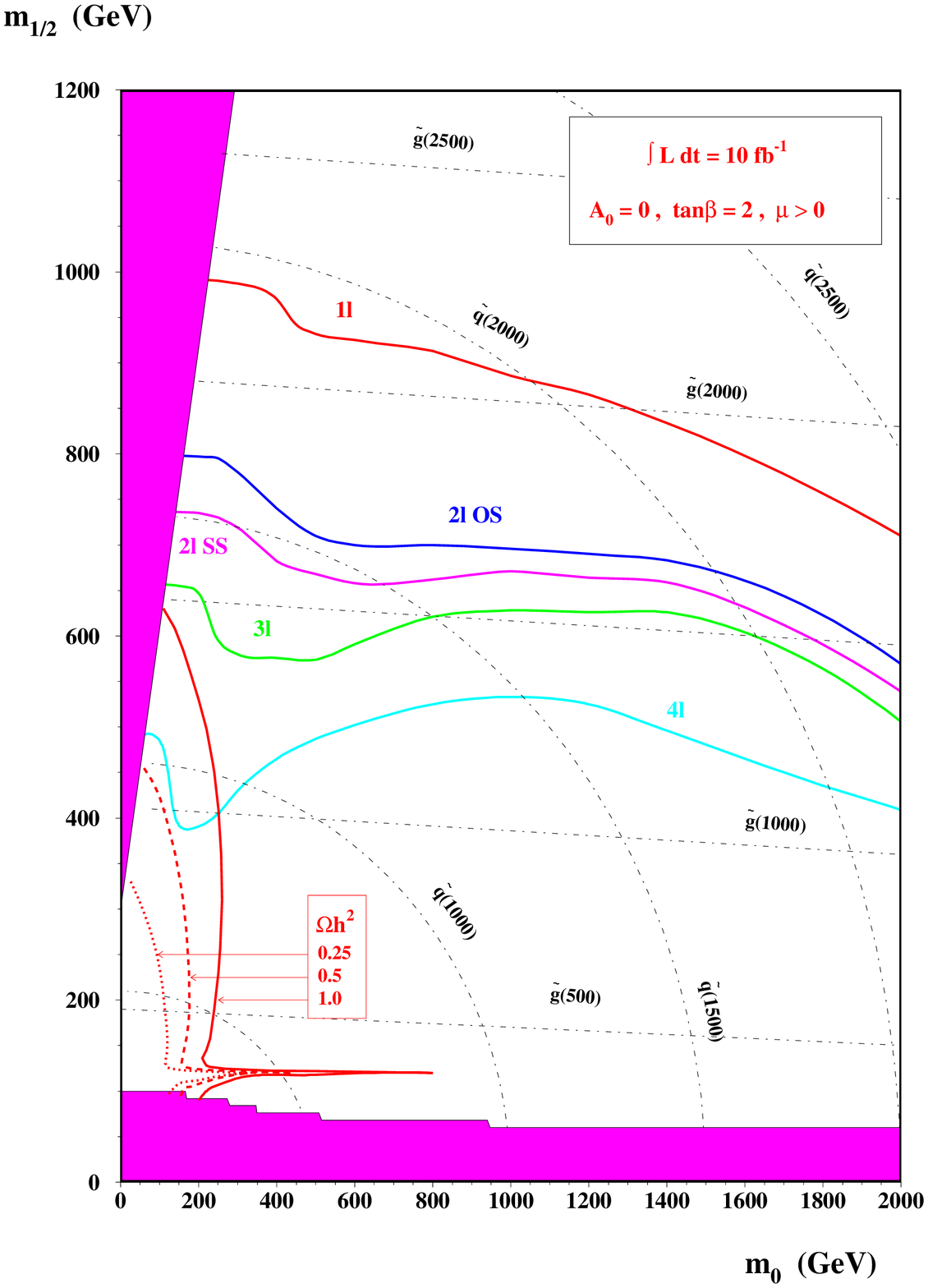}}
  \end{center}
\vspace{-5mm}
\end{figure}

Figure 5.13: 
Maximal squark/gluino reach for $\mu$ $>$ 0 for 10 fb$^{-1}$ in different final
state topologies.  The relic neutralino dark matter density contours 
of $\Omega h^{2}$ = 0.25, 0.5 and 1.0 are also shown for this set of model
parameters.

\newpage
 
\ \\

\vspace{-20mm}

\begin{figure}[hbtp]
  \begin{center}
\resizebox{12cm}{!}{\includegraphics{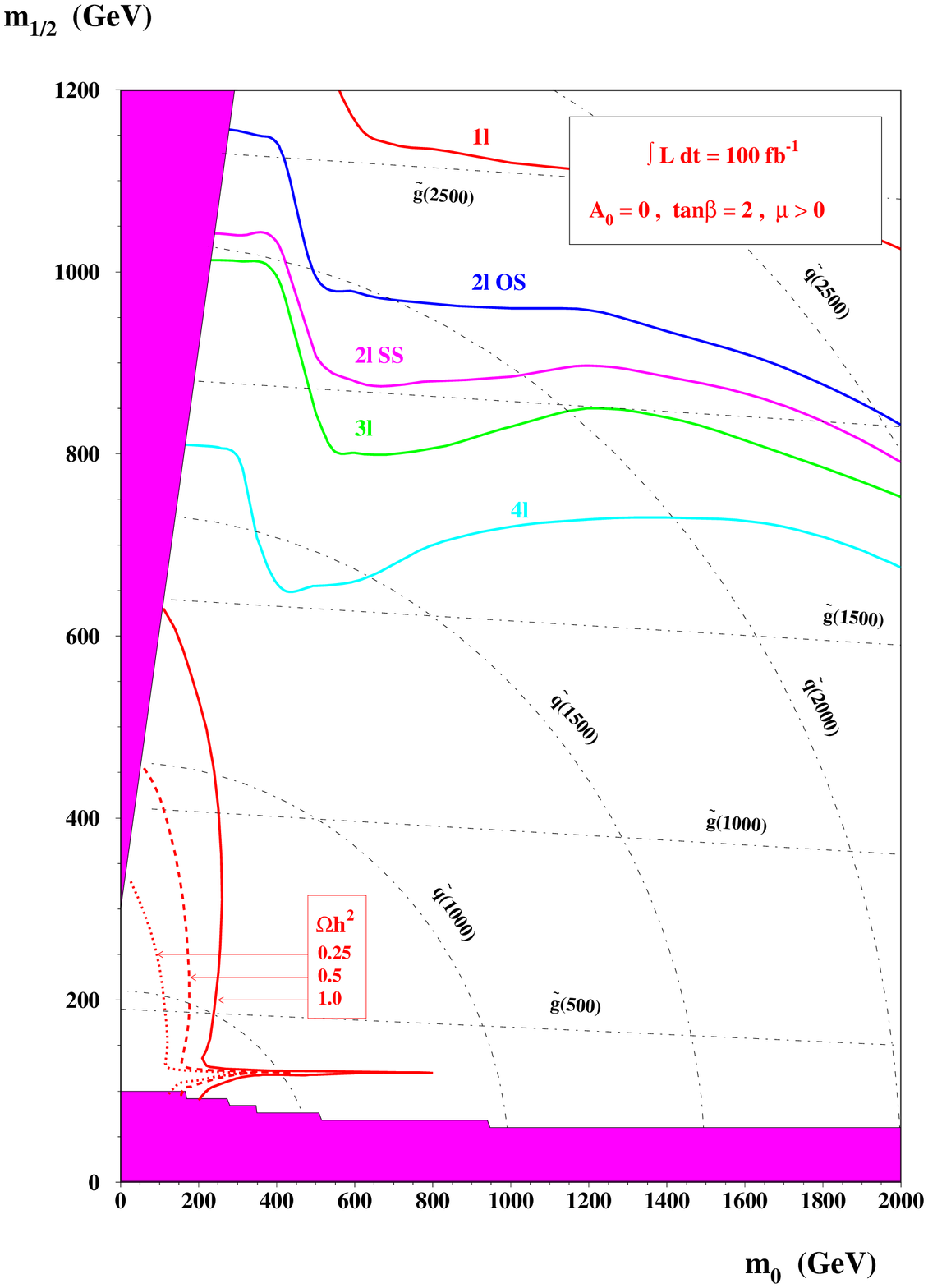}}
  \end{center}
\end{figure}

Figure 5.14: 
Same as in Fig.~5.13, but for 100 fb$^{-1}$.

\newpage


\section{Search for next-to-lightest neutralino}

Leptonic decays of the second-lightest neutralino,
\chnb \ have a useful kinematical feature: 
the dilepton invariant mass spectrum has a sharp edge near the
kinematical upper limit with
$M^{max}_{l^{+}l^{-}} = M_{\tilde{\chi}^{0}_{2}} -
M_{\tilde{\chi}^{0}_{1}}$ in the case of direct three-body decays
\chnb $\rightarrow l^+ l^-$\chna \ and
$M^{max}_{l^{+}l^{-}} = \sqrt{
(M^{2}_{\tilde{\chi}^{0}_{2}} - M^{2}_{\tilde{l}})
(M^{2}_{\tilde{l}} - M^{2}_{\tilde{\chi}^{0}_{1}}) } /
M_{\tilde{l}}$ in the case of
two-body cascade decays
\chnb $\rightarrow l^{\pm} \tilde{l}_{L,R}^{\mp}  
\rightarrow l^+ l^-$\chna.
This feature was first discussed in Ref. \cite{edge1} for
\chha \chnb \ direct (EW) production. On the other hand,
at LHC, the \chnb \ is
abundantly produced from gluinos and
squarks over whole ($m_0, m_{1/2}$) plane, see e.g. Figs.~2.5a-b, and
this source of \chnb \ is much more prolific than direct EW production.
Therefore, we propose to perform an inclusive search for \chnb \
and to use the specific shape of the dilepton mass spectrum
as evidence for SUSY \cite{incl_lhcc, inclchi2}.
For the SM background suppression,
besides {\it two same-flavor opposite-sign leptons},
one can ask for an additional signature characterizing SUSY.
This can be: missing energy taken away by escaping
LSPs, or additional jets coming from gluinos and squarks cascade decays,
or an extra high-$p_T$ lepton from charginos, neutralinos,
sleptons, vector bosons, or $b$-quarks copiously produced in SUSY.
Depending on the sparticle masses and predominant decay modes one of these
extra requirements can turn out to be more advantageous than the others,
or to be complementary. In the following we discuss
the inclusive $2 \,\,leptons\, + \, E_T^{miss} \, + \,(jets)$
and $3 \ leptons + \, (E_T^{miss})$ channels and determine
the region of ($m_0, m_{1/2}$) parameter space
where the dilepton invariant mass edge is visible in these final
states.

\subsection{Inclusive $2 \,\,leptons\, + \, E_T^{miss} \, + \,(jets)$
channel}

The most straightforward  signature for selecting the $\tilde{\chi}^0_2$
decays is provided by the topology with
two same-flavor opposite-sign leptons accompanied by large missing
transverse energy and usually accompanied by a high multiplicity of jets.
Here we concentrate on the
$2 \ leptons + E_T^{miss} + (jets)$
channel, where the final state leptons are electrons and muons
\cite{ll_incl, ll_sem}.

The largest SM background is due to $t\bar{t}$ production,
with both $W$'s decaying into leptons, or one of the leptons 
coming from a $W$ decay and the other from the $b$-decay of the same
$t$-quark.  We also considered other SM backgrounds: $W+jets$, $WW$, $WZ$, 
$Wtb$, $b\bar b$ and  $\tau \tau$-pair production, with decays into
electrons and muons. Chargino pair production 
$\tilde{\chi}_1^{\pm}\tilde{\chi}_1^{\mp}$ is the largest SUSY background
but gives a small contribution compared to the signal.

To observe an
edge in the $M_{l^+l^-}$ distributions with the statistics provided
by an integrated luminosity  $L_{int}=10^3 \,pb^{-1}$ in a significant
part
of the ($m_0,m_{1/2}$) parameter plane, it is enough to require two
hard isolated leptons ($p_T^{l_{1,2}}>15$ GeV) accompanied by large
missing
energy, $E_T^{miss}>100$ GeV. Our criterion for observing an edge in the
$M_{l^+l^-}$ distribution contains two requirements:
$(N_{EV}-N_B)/\sqrt{N_{EV}}\ge 5$ and $(N_{EV}-N_B)/N_B \ge 1.3$, where
$N_{EV}$ is the number of events with $M_{l^+l^-} \le M_{l^+l^-}^{max}$,
and $N_B$ is number of the expected background events.
Fig.~6.1a shows the invariant mass spectra of the two leptons at various
($m_0,m_{1/2}$) points for tan$\beta = 2$, $\mu < 0$.
The observability  of the ``edge''
varies from $77 \sigma$ and signal to background ratio
31 at point (200,160) to $27  \sigma$ and a signal to background ratio 2.3
at
point (60,230).
The appearance of the edges in the distributions is sufficiently
pronounced
already with $L_{int}=10^3$ pb$^{-1}$ in a significant part of
$(m_0,m_{1/2})$ parameter plane.
The edge position can be measured with a precision of $\sim 0.5$ GeV.

With increasing  $m_0$ and $m_{1/2}$ the cross-sections  are
decreasing and therefore higher luminosity and harder cuts
are needed. To achieve maximum reach in $m_{1/2}$ with
$L_{int}=10^4 \,pb^{-1}$ for points 
where \chnb \ has two-body decays via $\tilde{l}_L$ ($m_0 \lappeq 0.45\cdot m_{1/2}$),
a cut up to
$E_T^{miss}>300$ GeV is necessary to sufficiently
suppress the background.
For points with large $m_0$, 
in the region where \chnb \ has three-body decay
($m_0 \gappeq 0.5\cdot
m_{1/2},\,\, m_{1/2} \lappeq 200\,{\rm GeV})$
 the transverse momentum $p_T$ of the
leptons and $E_T^{miss}$ are not very large, but there are more
hard jets due to gluino and squark decays.
Thus for these points we keep the same cuts for leptons and missing
energy as before ($p_T^{l_{1,2}}>15$ GeV, $E_T^{miss}>100$ GeV)
and require in addition a jet multiplicity $N_{jet}\ge 3$,
with energy $E_T^{jet}>100$ GeV, in the rapidity range
$\mid\eta_{jet}\mid<3.5$. To optimise the edge visibility
we also apply an azimuthal angle cut,
$\Delta \phi(l^+l^-) < 120^0$. 
This jet multiplicity requirement is also helpful
for the first domain, where \chnb \ decays
to leptons through $\tilde{l}_L$ ($m_0 \lappeq 0.45\cdot m_{1/2}$),
Right sleptons are too light to
provide large lepton $p_T$ and $E_T^{miss}$, and to use
cuts on $p_T^l$ and $E_T^{miss}$ alone is not very advantageous.
With $L_{int}=10^5 \,pb^{-1}$, to suppress the background  at larger
accessible $m_0,\,m_{1/2}$ values,
we have to require at least 2 or 3 jets, depending on the $m_0,\,m_{1/2}$
region to be explored.
Fig.~6.1b shows invariant dilepton mass distributions at some
($m_0,m_{1/2}$) points close to maximum reach with
$L_{int}=10^4 \,pb^{-1}$ and $L_{int}=10^5 \,pb^{-1}$ respectively.

Figure 6.2 shows the inclusive dilepton spectrum for larger values of
tan$\beta = 10$ and 35. Despite decreasing branching ratios as visible in
Fig.~2.7, the dilepton (e$^+$e$^-$, $\mu^+\mu^-$) edge is visible for
low $m_{1/2}$ values already with $10^3 \,pb^{-1}$ \cite{ll_sem}.

The regions of the ($m_0,m_{1/2}$) parameter plane where an edge in the
$M_{l^+l^-}$ spectra for tan$\beta = 2$, $\mu < 0$
can be observed at different luminosities are shown
in Fig.~6.3. In Fig.~6.4 we show
separately the three domains where an edge due to $\tilde{\chi}_2^0
\rightarrow l l \tilde{\chi}_1^0$, $\tilde{l}_R l$ and $\tilde{l}_L l$
decays can be observed at $L_{int}=10^3 \,pb^{-1}$.
One can notice a small overlapping region, where we expect to observe
two edges, due to $\tilde{\chi}_2^0 \rightarrow l^+ l^- \tilde{\chi}_1^0$
and to $\tilde{\chi}_2^0 \rightarrow \tilde{l}_R^{\pm} l^{\mp}
\rightarrow l^+ l^- \tilde{\chi}_1^0$
decays ({\it case 1}). With increasing luminosity
and correspondingly higher statistics, this overlapping region
increases, see Figs.~6.5 and 6.6. These plots
show the same as Fig.~6.4, but for $L_{int}=10^4 \,pb^{-1}$ and
$L_{int}=10^5 \,pb^{-1}$, respectively.

\subsection{Inclusive $3 \ leptons$ channel}

In this channel an extra high-$p_T$ lepton is required to suppress
the SM backgrounds \cite{incl_lhcc, inclchi2}. This can be either an
isolated lepton, e.g. from the $\tilde{\chi}^{\pm}_{1}$ decay which
has a similar behavior to $\tilde{\chi}^{0}_{2}$ (Fig.~2.6),
or a non-isolated lepton from the abundant production of high
transverse momentum $b$-jets in SUSY events.
Thus in the inclusive 3~$lepton$ channel we require:

$\bullet$
two opposite-sign, same-flavor leptons ($\mu$ or e) with
$p_T > 10$ GeV;

$\bullet$
a third ``tagging'' lepton with $p_T > 15$ GeV;

$\bullet$
lepton isolation: if there is no track with $p_T > 2 \ (1.5)$
GeV within $\Delta R = \sqrt{\Delta \eta^2 + \Delta \phi^2} < 0.3$
about the lepton direction, it is considered as isolated.

The lepton $p_T$ thresholds can be varied depending on the region
of ($m_0, m_{1/2}$) studied and in some cases leptons are not
required to be isolated. In the selected 3 $lepton$ events we reconstruct
the invariant mass $M_{l+l-}$ of a lepton pair with the same flavor and
opposite sign. When several $l^{+}l^{-}$ combinations per event are
possible, the one with minimal separation $\Delta R_{l^{+}l^{-}}$ is
chosen. Any significant deviation from the expected SM $l^{+}l^{-}$
invariant mass spectrum then provides evidence for SUSY.
 
We stress that no other requirements, such as missing transverse
energy, jet activity, etc., are imposed in the inclusive $3 \ leptons$
channel. The price for this simplicity is a decrease of signal acceptance
due to the requirement on the third lepton, and somewhat higher $p_T$ of
leptons needed to sufficiently suppress the SM backgrounds.
Thus there is complementarity between this search which relies on lepton
(e, $\mu$) measurements and the tracker for lepton isolation and momentum
measurement, and the search, in inclusive $2 \ leptons$ + \etm \
channel which relies on lepton detection and  overall calorimetry, HCAL in
particular, for \etm \ measurement.

The main SM  sources of three leptons considered are production of
WZ, $t \bar{t}$, ZZ and Z$b \bar{b}$. The expected SM background to
$3~leptons$ final states at an integrated luminosity of $L_{int}=10^4$
pb$^{-1}$ is shown in Fig.~6.7.
Near the Z mass the main contribution comes from the WZ production,
while $t \bar{t}$ production dominates at lower invariant masses.

Dilepton spectra for the mSUGRA ``Point 4'' (see Table 1.1), superimposed on
the SM background, are shown in Fig.~6.8. The number of events corresponds
to an integrated luminosity of $L_{int}=20$ pb$^{-1}$, i.e. just a few
hours of initial LHC running. The $p_T$ threshold on all three leptons is
15 GeV. In Fig.~6.8a no isolation requirement is applied,
whereas in Fig.~6.8b the two leptons entering the dilepton mass
distribution are isolated. The contribution from the SM background is
negligible in the latter case. In both cases the specific shape of the
$l^+l^-$ distribution, with its sharp edge, reveals \chnb \ production.
At this mSUGRA point the \chnb \ has a three-body decay mode with a
branching ratio B$(\tilde{\chi}^{0}_{2} \rightarrow l^{+}l^{-} +
\tilde{\chi}^{0}_{1})=0.32$ ($l=$ e, $\mu$) and the masses are
$M_{\tilde{\chi}^{0}_{2}} = 97$ GeV, $M_{\tilde{\chi}^{0}_{1}} = 45$ GeV.
The sharp edge is situated at the expected value of
$M^{max}_{l^{+}l^{-}} = 52$ GeV.

Figure 6.9 shows the $M_{l+l-}$ distribution for
mSUGRA ``Point 1'' (see Table 1.1) with an integrated luminosity
of $L_{int}=10^4$ pb$^{-1}$. The $p_T$ thresholds
for leptons are 15, 15, 30 GeV and all three
leptons are isolated. At this mSUGRA point the \chnb \
possesses a two-body decay mode with the branching ratio 
B$(\tilde{\chi}^{0}_{2} \rightarrow \tilde{l}_{R} + l) = 0.24$.
The sparticle masses are
$M_{\tilde{\chi}^{0}_{2}} = 231$ GeV,
$M_{\tilde{\chi}^{0}_{1}} = 122$ GeV,
$M_{\tilde{l}_{R}} = 157$ GeV.
The ``edge'' of the distribution is situated near
108 GeV, as expected. We define a signal significance as
$S = N_S / \sqrt{N_S + N_B}$,
where $N_S$ and $N_B$ are the numbers of signal and background events,
respectively, in a mass window below the edge. In this
particular case signal and background events are calculated in a mass
window of 100 GeV$<M_{l^{+}l^{-}}<110$ GeV,
$N_S=140$ and $N_B=39$, resulting in signal significance of
$S=10.5$.
$N_B$ consists of 18, 8, 11 and 2 events coming from
WZ, $t \bar{t}$, ZZ and Z$b \bar{b}$ production, respectively.

An example for the mSUGRA ``Point 5'' is shown in Fig.~6.10. The
cuts are the same as in the previous case.
At this point \chnb \ has three-body
decay modes with B$(\tilde{\chi}^{0}_{2} \rightarrow l^{+}l^{-}
+ \tilde{\chi}^{0}_{1})=0.06$ and the masses are
$M_{\tilde{\chi}^{0}_{2}} = 124$ GeV,
$M_{\tilde{\chi}^{0}_{1}} = 73$ GeV.
The observed value of $M^{max}_{l^{+}l^{-}}$ is close to the  
expected value of 51 GeV. In a mass
window of 20 GeV$<M_{l^{+}l^{-}}<52$ GeV the numbers of signal
and background events are
$N_S=588$ and $N_B=402$ giving a signal significance of $S=19$.
$N_B$ consists of 113, 139, 65 and 85 events coming from
WZ, $t \bar{t}$, ZZ and Z$b \bar{b}$ production, respectively.

By the same procedure
the mSUGRA parameter space $(m_{0}, m_{1/2})$ was scanned for fixed
tan$\beta$ = 2, $A_0 = 0$ and $\mu < 0$ \cite{inclchi2}.
Figures 6.11a-c show the $M_{l+l-}$ distributions for
three different mSUGRA points with the optimal
thresholds on lepton $p_T$ and isolation  
requirement in inclusive 3 $lepton$ events. Generally,
with increasing $m_{1/2}$ the observation 
of the edge becomes more difficult due to the rapidly
decreasing gluino and squark production cross-sections.
Nonetheless, this spectacular structure in the dilepton
mass spectrum reflecting \chnb \ decays, may be among
the first ways through which presence of SUSY may reveal
itself. It may appear with very modest statistics,
in a non-sophisticated analysis and with only modest
demands on detector performance. The precision
for ``edge'' measurement in most of the region,
where it is detectable is expected to be $\lsim 0.5$ GeV.
Note, that the Z peak seen in Figs.~6.8-6.11 serves as an overall
calibration signal; it allows to control the mass scale as well as the
production cross-section. 

Asking for a statistical significance $S > 7$ with at least 40 signal
events calculated in a mass interval $>10$ GeV below the edge
the domain explorable with an integrated luminosity of
$L_{int} = 10^{4}$ pb$^{-1}$ is shown in Fig.~6.12. It entirely
covers the cosmologically prefered region
$0.15 < \Omega h^2 < 0.4$ \cite{cosmo}; here $\Omega$ is the ratio
of the relic particle density to critical density, $h$ is
the Hubble constant scaling factor and  bounds are obtained
assuming $\tilde{\chi}^{0}_{1}$ is the
cold dark matter particle in the mixed dark matter
scenario for the Universe, varying $h$ over the allowed range
$0.5 < h < 0.8$.

For some regions of mSUGRA parameter space $M^{max}_{l^{+}l^{-}}$ is
close to, or even hidden by the Z signal. At these points
the accuracy of an edge measurement is $\sim$10 GeV.
Applying a cut on missing transverse energy (and/or jets)  
suppresses the contribution from the SM Z production,
improving signal visibility.

\subsection{Inclusive $3 \ leptons$ + \etm \ channel}

With increasing $m_{1/2}$, gluino and squark masses
increase and the missing transverse energy becomes larger. Requiring
\etm $>$ 200 (300) GeV rejects most of the SM background
leaving a big fraction of signal events.
Figures 6.11d-e show the dilepton spectra for various mSUGRA
points in the 3 $lepton$ + \etm \ final states.
In the region of high $m_{1/2}$, dileptons are
mainly produced in the cascade decays of \chnb. The third,
``tagging'' lepton predominantly comes from \chha \ which
also has cascade leptonic decays.
Thus in this region of parameter space a large fraction of
$3~lepton$ + \etm \ events are in fact inclusively produced
\chha \chnb \ pairs. Therefore
all three leptons are required to be isolated.
The lepton $p_T$ thresholds
are chosen to be quite asymmetric: 50, 25, 10 GeV, to account for
the cascade nature of the \chnb \ and \chha \ decays.
The region of mSUGRA parameter space, where the \chnb \
is observable from the dilepton mass distribution shape
in the $3~lepton$ + \etm \  events at $L_{int} =  10^5$ pb$^{-1}$
luminosity is shown in Fig.~6.12 \cite{inclchi2}.
The reach extends up to $m_{1/2} \sim 900 $ GeV
and a detectable edge is seen as
long as $\sigma \cdot $B $\gsim 10^{-2}$ pb (see Fig.~2.5).

\vspace{5mm}

We conclude that

$\bullet$   
Observation of an edge in the dilepton invariant mass spectrum
reflects the production of $\tilde{\chi}^{0}_{2}$ and hence
establishes the existence of SUSY; this observation
in some cases is possible with very small statistics and could be
the first evidence for SUSY at LHC.

$\bullet$
With no great demand on detector performance
a large portion of mSUGRA parameter space
including the cosmologically
prefered domain can be explored with an
integrated luminosity as low as $L_{int} = 10^{4}$ pb$^{-1}$.

\newpage

\begin{figure}[hp]

\resizebox{140mm}{140mm}
{\includegraphics{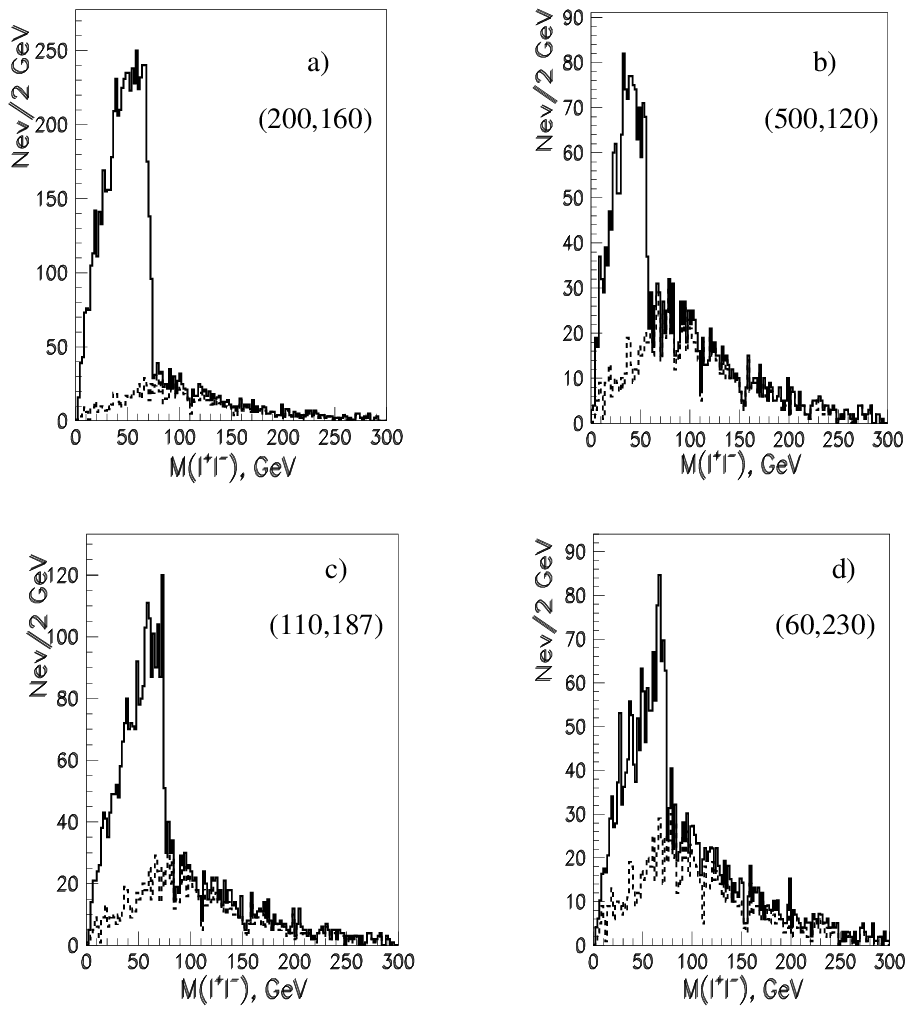}}

\vspace{10mm}

Figure 6.1a:
Invariant mass distribution of two same-flavor, opposite-sign
leptons at various $(m_0,m_{1/2})$ points for tan$\beta = 2$, $\mu <0$
with $L_{int}=10^3$ pb$^{-1}$. SM background is also shown (dashed line).
\end{figure}

\newpage

\ \ \\

\begin{figure}[hp]

\vspace*{-5mm}

\resizebox{145mm}{160mm}
{\includegraphics{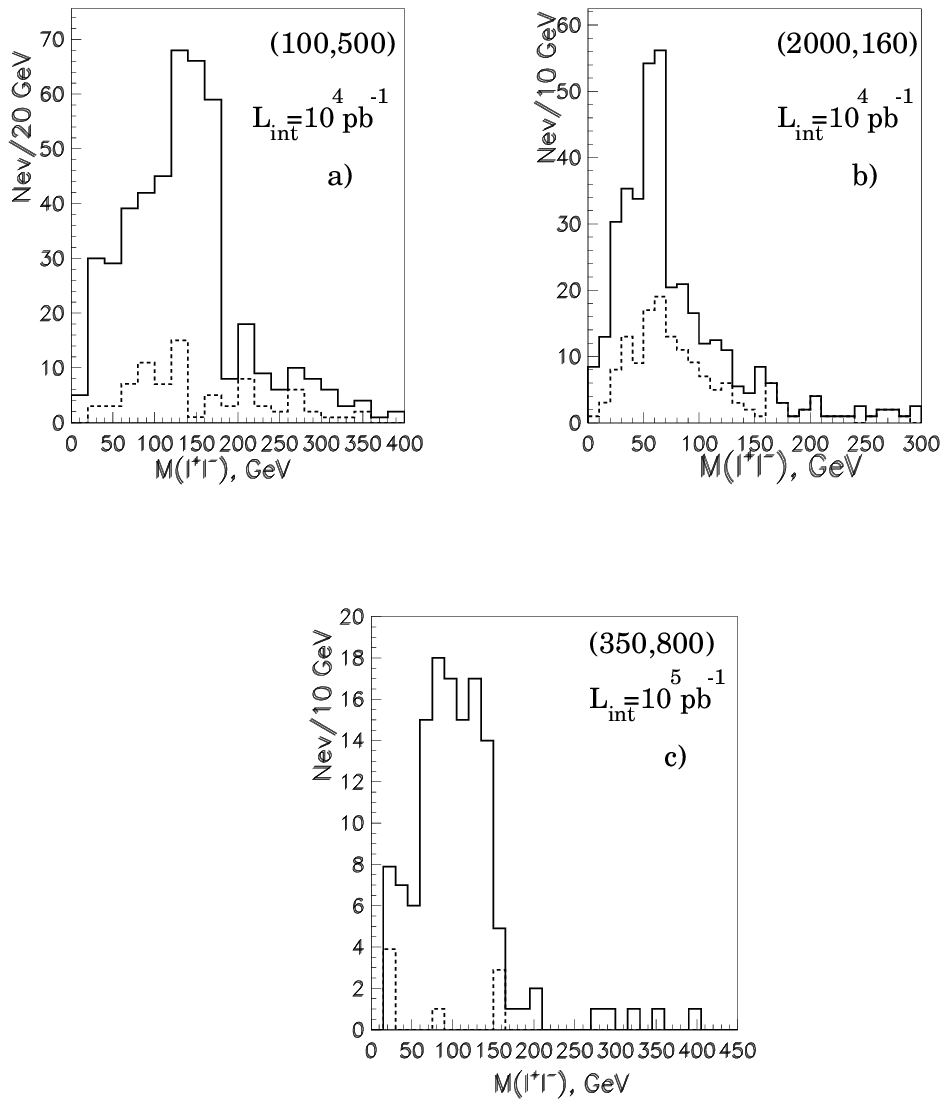}}

\vspace{1.5cm}

Figure 6.1b: Invariant mass distribution of two same-flavor,
opposite-sign leptons at $(m_0,m_{1/2})$ points close to the experimental
reach at corresponding luminosities $L_{int}=10^4$ pb$^{-1}$ and $10^5$
pb$^{-1}$ for tan$\beta = 2$, $\mu <0$.
SM background is also shown  (dashed line).
\end{figure}

\newpage
 
\begin{figure}[hp]
 
\resizebox{140mm}{150mm}
{\includegraphics{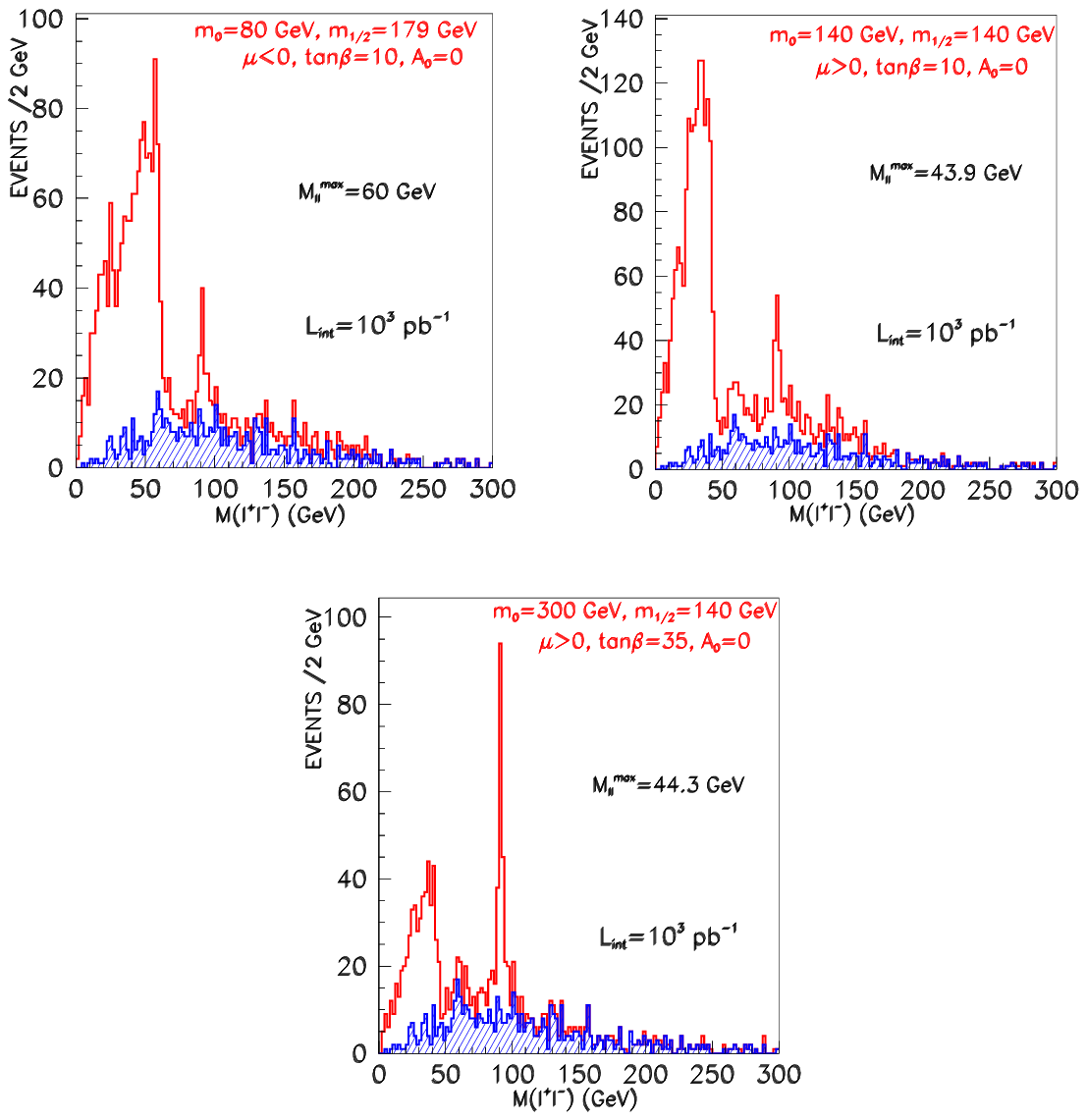}}
 
\vspace{1cm}
 
Figure 6.2:
Invariant mass distribution of two same-flavor, opposite-sign
leptons at various $m_0, m_{1/2}$, tan$\beta$ and $\mu$ points for
$L_{int}=10^3$ pb$^{-1}$. SM background is also shown.
\end{figure}

\newpage

\begin{figure}[hp]

\resizebox{140mm}{160mm}
{\includegraphics{fig5-3.ill}}
\vspace{0.5cm}

Figure 6.3: Observability of edges in invariant dilepton mass
distribution with luminosities $10^3$ pb$^{-1}$ (dashed line),
$10^4$ pb$^{-1}$ (solid line) and $10^5$ pb$^{-1}$ (dashed-dotted line).
Also shown are the explorable
domain in sparticle searches at LEP2 (300 pb$^{-1}$) and the Tevatron
(1 fb$^{-1}$), theoretically and experimentally excluded regions.
\end{figure}

\begin{figure}[hp]

\resizebox{140mm}{160mm}
{\includegraphics{fig5-4.ill}}
\vspace{0.5cm}

Figure 6.4:  Domains where the observed  edge in the
$M_{l^+l^-}$ distribution is due to the decays
$\tilde{\chi}_2^0 \rightarrow \tilde{l}_L^{\pm} l^{\mp} \rightarrow
\tilde{\chi}_1^0 l^+ l^-$  (solid line),
$\tilde{\chi}_2^0 \rightarrow \tilde{l}_R^{\pm} l^{\mp} \rightarrow
\tilde{\chi}_1^0 l^+ l^-$ (dashed-dotted line),
$\tilde{\chi}_2^0 \rightarrow \tilde{\chi}_1^0 l^+ l^-$ (dashed line),
$L_{int}=10^3$ pb$^{-1}$.
\end{figure}

\newpage

\begin{figure}[hp]

\resizebox{140mm}{160mm}
{\includegraphics{fig5-5.ill}}
\vspace{0.5cm}

Figure 6.5:
Domains where the observed edge in the
$M_{l^+l^-}$ distribution is due to the decays
$\tilde{\chi}_2^0 \rightarrow \tilde{l}_L^{\pm} l^{\mp} \rightarrow
\tilde{\chi}_1^0 l^+ l^-$ (solid line),
$\tilde{\chi}_2^0 \rightarrow \tilde{l}_R^{\pm} l^{\mp} \rightarrow
\tilde{\chi}_1^0 l^+ l^-$ (dashed-dotted line),
$\tilde{\chi}_2^0 \rightarrow \tilde{\chi}_1^0 l^+ l^-$ (dashed line),
$L_{int}=10^4$ pb$^{-1}$.
\end{figure}

\newpage

\begin{figure}[hp]

\resizebox{140mm}{160mm}
{\includegraphics{fig5-6.ill}}

\vspace{0.5cm}

Figure 6.6:
Domains where the observed edge in the
$M_{l^+l^-}$ distribution is due to the decays
$\tilde{\chi}_2^0 \rightarrow \tilde{l}_L^{\pm} l^{\mp} \rightarrow
\tilde{\chi}_1^0 l^+ l^-$ (solid line),
$\tilde{\chi}_2^0 \rightarrow \tilde{l}_R^{\pm} l^{\mp} \rightarrow
\tilde{\chi}_1^0 l^+ l^-$ (dashed-dotted line),
$\tilde{\chi}_2^0 \rightarrow \tilde{\chi}_1^0 l^+ l^-$ (dashed line),
$L_{int}=10^5$ pb$^{-1}$.
\end{figure}

\newpage

\ \ \\  

\begin{figure}[hp]
\vspace*{-30mm}
\hspace*{10mm}
\resizebox{16.cm}{!}{\rotatebox{0}{\includegraphics{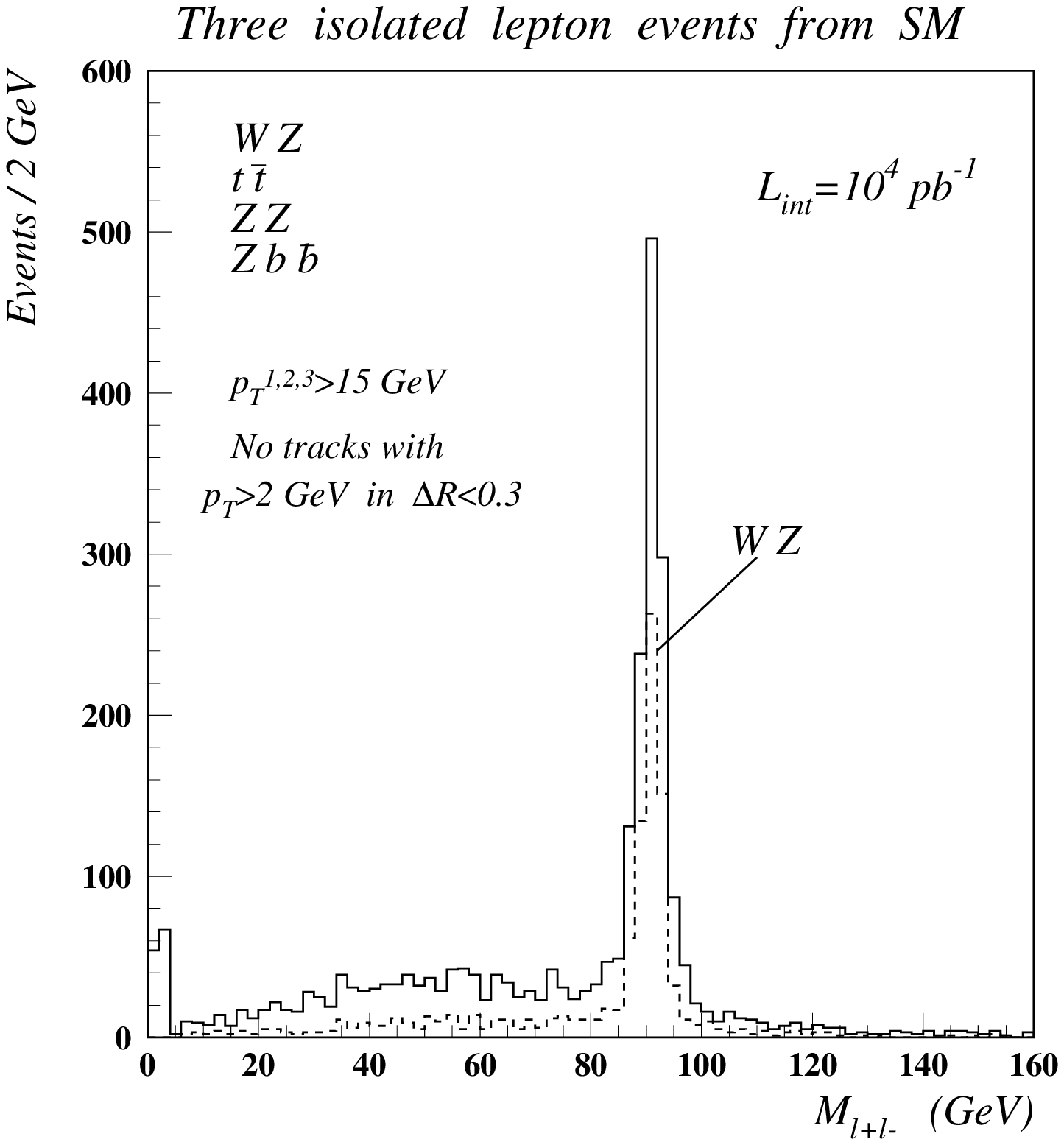}}}

Figure 6.7: Expected Standard Model background in inclusive
3 $lelpton$ final state.

\end{figure}

\newpage

\ \ \\

\begin{figure}[hp]
\vspace{-35mm}
\hspace*{10mm} 
\resizebox{18.cm}{!}{\rotatebox{0}{\includegraphics{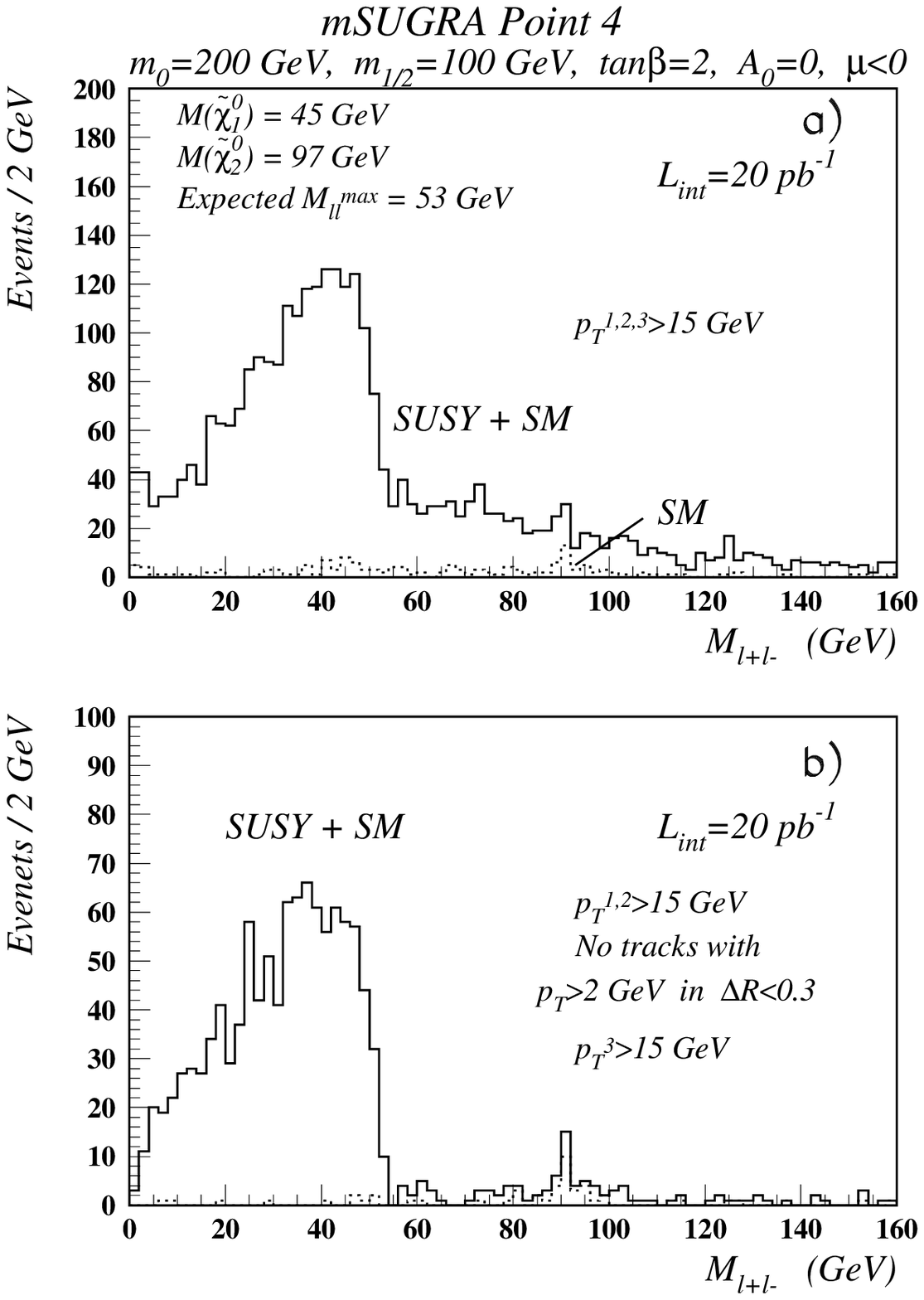}}}

Figure 6.8: Expected $l^+l^-$ mass spectrum for mSUGRA Point 4:
$m_0 =200$ GeV, $m_{1/2} =100$ GeV,
tan$\beta$=2, $A_{0} =0$ and $\mu < 0$.
a) Leptons with $p_T>15$ GeV in $|\eta|<2.4$ are considered;
b) as in previous case, but two leptons which enter invariant
mass spectrum are required to be isolated.
The dotted histogram corresponds to the SM background.

\end{figure}

\newpage

\ \\

\vspace{0mm}

\begin{figure}[hp]
\vspace{-30mm}
\hspace*{30mm}
\resizebox{11.3cm}{!}{\rotatebox{0}{\includegraphics{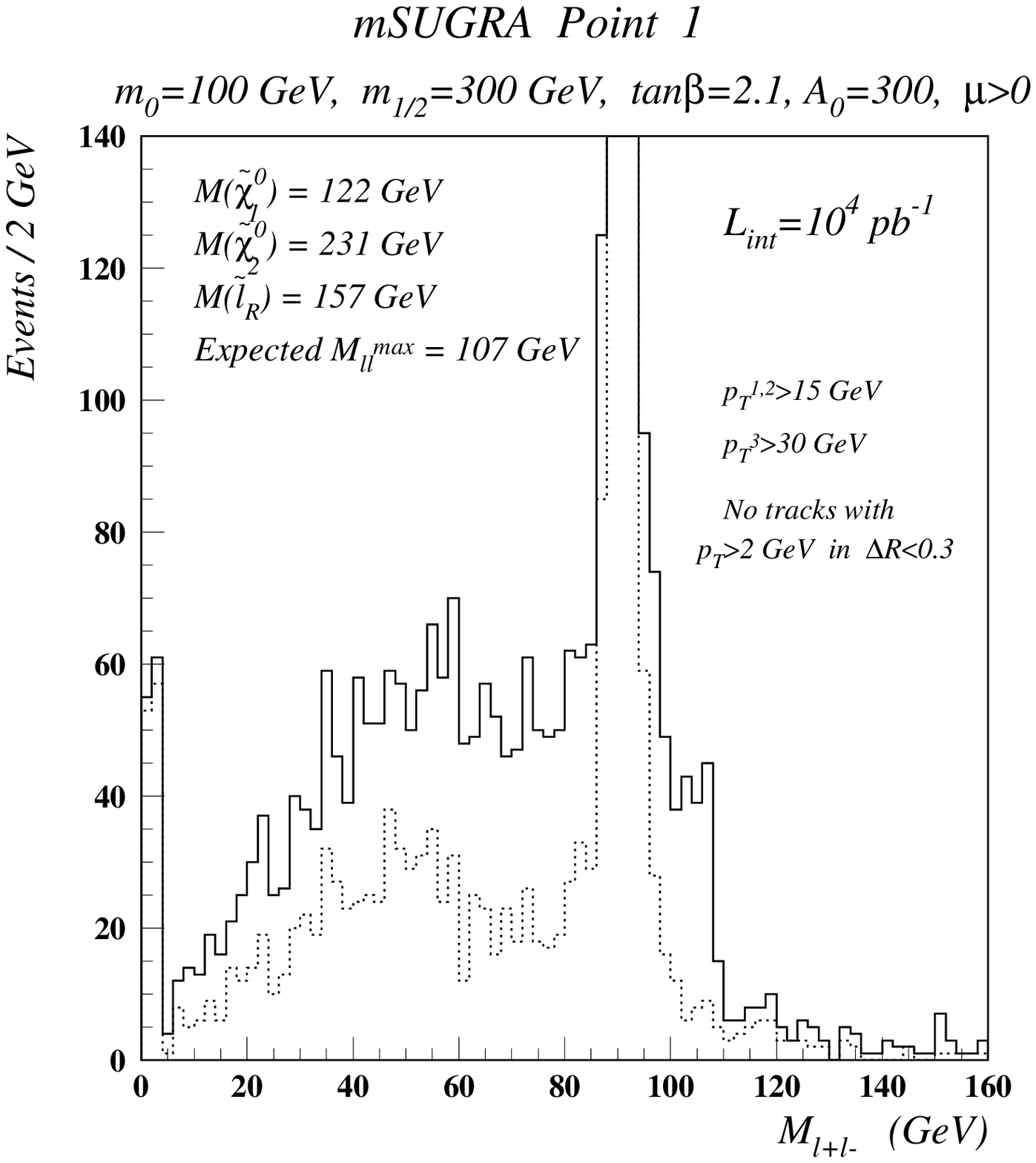}}}
\end{figure}

\vspace{-10mm}

Figure 6.9: Expected $l^+l^-$ mass spectrum for mSUGRA Point 1:
$m_0 =100$ GeV, $m_{1/2} =300$ GeV,
tan$\beta$=2.1, $A_{0} =300$ GeV and $\mu > 0$.
The dotted histogram corresponds to the SM background.

\vspace{15mm}

\begin{figure}[hp]
\vspace{-20mm}
\hspace*{30mm}
\resizebox{9.3cm}{!}{\rotatebox{0}{\includegraphics{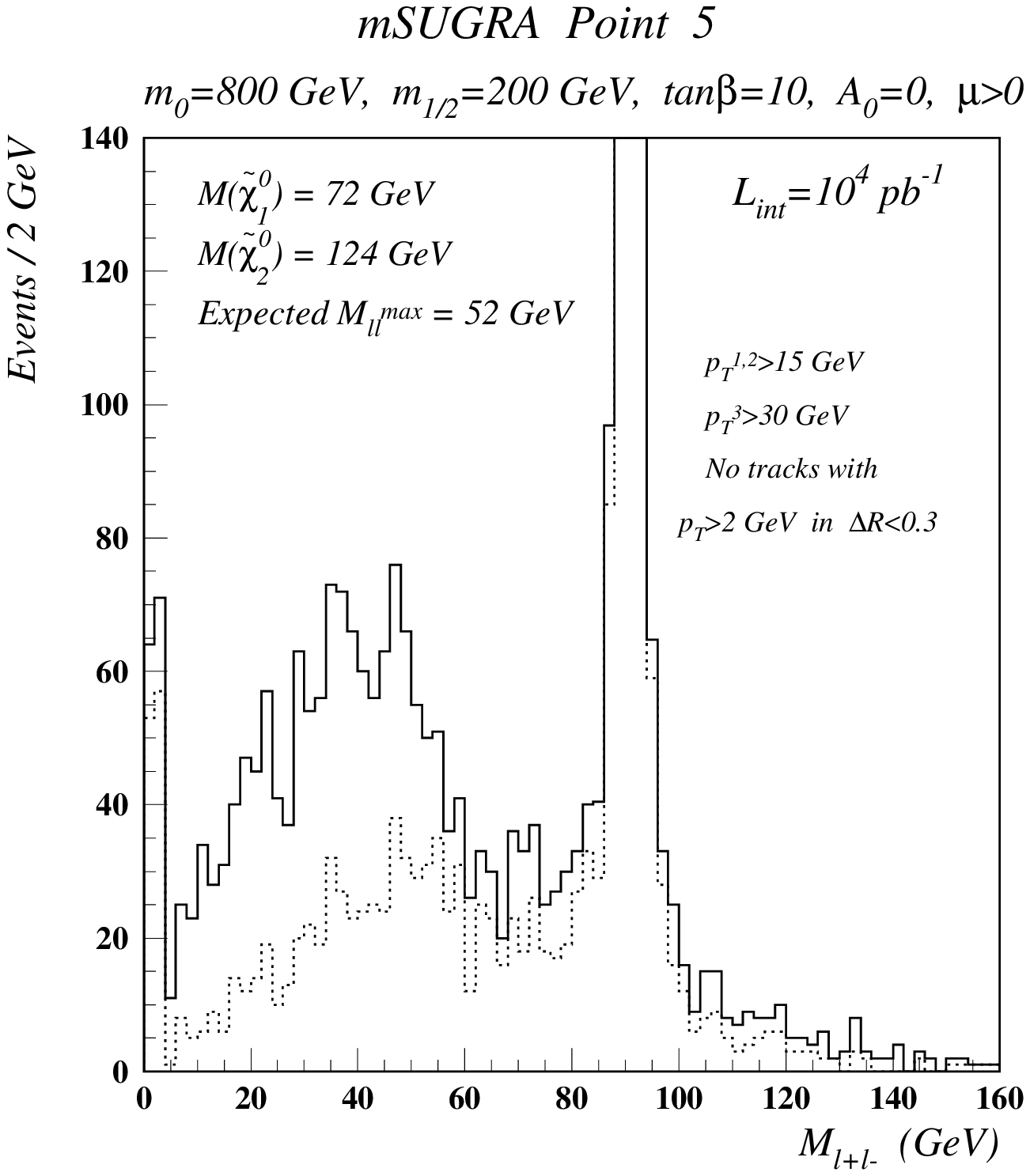}}}
\end{figure}

\vspace{-10mm}

Figure 6.10:
Expected $l^+l^-$ mass spectrum for mSUGRA Point 5:
$m_0 =800$ GeV, $m_{1/2} =200$ GeV,
tan$\beta$=10, $A_{0} =0$ and $\mu > 0$.
The dotted histogram corresponds to the SM background.

\newpage

\ \\

\vspace{10mm}

\begin{figure}[hp]
\vspace{-30mm}
\hspace*{-0mm}
\resizebox{20.cm}{!}{\rotatebox{0}{\includegraphics{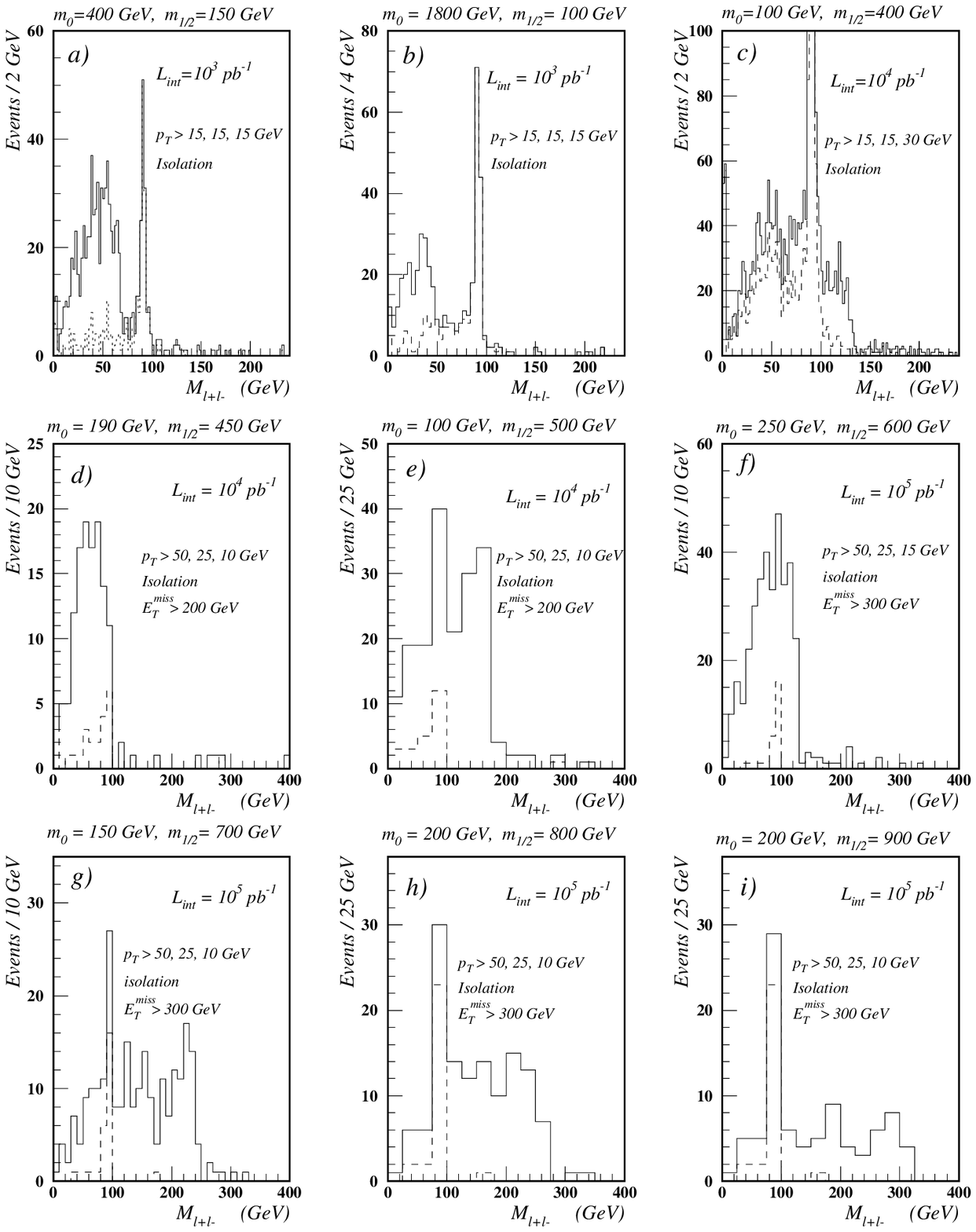}}}
\end{figure}

\vspace{10mm}

Figure 6.11: Dilepton mass spectrum for mSUGRA Point from
different ($m_0, m_{1/2}$) regions in inclusive
3 $leptons$ and 3 $leptons$ + \etm \ channels. 
Contribution from SM background is shown with dashed
histogram.
The other mSUGRA parameters are:
tan$\beta$=2, $A_{0} = 0$ and $\mu < 0$.

\newpage

\ \\

\vspace{10mm}

\begin{figure}[hp]
\vspace{-30mm}
\resizebox{18.cm}{!}{\rotatebox{0}{\includegraphics{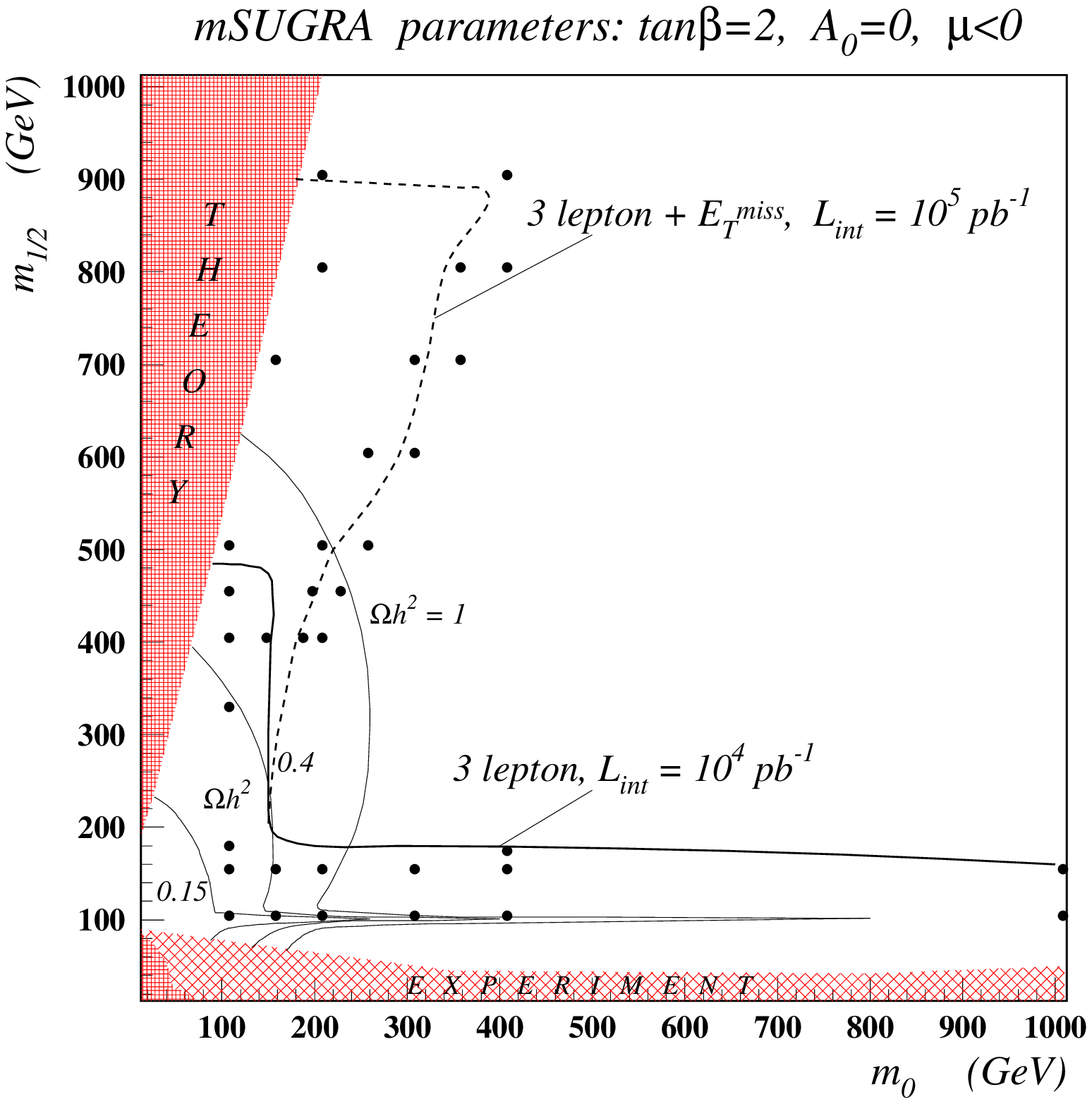}}}
\end{figure}

\vspace{0mm}

Figure 6.12:
Explorable region of mSUGRA in inclusive
3 $leptons$ and 3 $leptons$ + \etm \ channels
at $L_{int}= 10^4$ pb$^{-1}$ and $L_{int}= 10^5$ pb$^{-1}$,
respectively.
Shaded regions are excluded by theory ($m_0 \lsim 80$ GeV)
and experiment ($m_{1/2} \lsim 90$ GeV). The cosmologically preferable
region 0.15 $< \Omega h^2<0.4$ is also given.
Simulated mSUGRA Points are shown as well.   

\newpage


\section{Exclusive $2 \ leptons + no \ jets$ + \etm \ channel -- \\
search for sleptons}

The previous two sections described inclusive searches suitable
for the discovery of SUSY. We now focus on more exclusive final states
where individual SUSY particles may be targeted.
 
The slepton production and decay features described in section 2.2 
give rise to several interesting experimental signatures. 
In this study we concentrate on the possibilities to detect a slepton
signal in the simplest event topology $two\,\,leptons\,+$ \etm +
$no\,\,jets$  taking into account all possible SUSY and SM contributions
to this topology \cite{ll_excl}. The observability of the slepton signal
is evaluated summing all flavors and all production mechanisms which
contribute to the {\it two same-flavor opposite-sign leptons + \etm + no
jets} final state. Evidence for a $slepton \ signal$ will be judged from
an excess over full $SUSY$ and $SM \,\, backgrounds$, but presence of a
$SUSY signal$ will be estimated from  an excess  over only the expected 
$SM\,\, backgrounds$.

The main expected Standard Model background is   $t\bar{t}$ 
production, with both $W$'s decaying to leptons, or one of the leptons 
coming from a $W$ decay and the other from the $b$-decay of the same
$t$-quark.
The other SM backgrounds considered are $Wt\bar b$, WW, WZ, $b \bar b$ and
$\tau \tau$-pair production, with decays to electrons and muons. 
The assumed cross-sections are:
$\sigma_{t\bar t}$=660 pb ($M_{t}=175$ GeV), $\sigma_{Wt\bar b}$=160 pb,
$\sigma_{WW}$=70 pb, $\sigma_{WZ}$=26 pb, $\sigma_{\tau \tau }$=7.5 pb,
$\sigma_{b \bar b}$=5000 pb ($p_T^b \geq 100$ GeV).
No forcing of decay modes is implemented.

As internal SUSY backgrounds we consider processes $\tilde{q} \tilde{q}$, 
$\tilde{g} \tilde{q}$, $\tilde{g} \tilde{g}$ which, through cascade decays 
with jets outside the acceptance, below the detection threshold, or lost
in cracks,
can also lead to $two \,leptons $ + \etm + $\,\, no\,\, jets$ final states.
The $\tilde{\chi}_1^{\pm} \tilde{\chi}_2^0$ process for example leads
mainly to three-lepton final states. However, it can also contribute 
to two-lepton  final states if
one of the leptons is outside the acceptance or is misidentified,
or if only the \chnb \  decays to leptons, and 
jets from the \chha \  decay are not reconstructed.

We perform a mapping of the mSUGRA parameter space to determine 
the optimal set of signal selection cuts in the various regions 
of ($m_0, m_{1/2}$) plane, taking into account event kinematical 
properties, 
and evaluate the observability of the signal. 
An important issue is to find out if there is a gap in slepton
detectability, especially in the low mass range at the junction between
the observability domains at LEP2, Tevatron and the LHC.
The mSUGRA-parameter space  points ($m_0,m_{1/2}$) probed are shown
in Fig.~7.1. This figure also shows the masses $m_{\tilde{l}_L}$ and 
$m_{\tilde{\chi}_1^0}$ at each point, the
theoretically and experimentally excluded regions and 
the explorable domains in sparticle  searches at LEP2 (300 pb$^{-1}$)
and the Tevatron (1 fb$^{-1}$).

Taking into account the signal and background event topologies the
following kinematical variables are found useful for slepton signal
extraction:

\vspace{0.2cm}

i) for leptons:
 
 \hspace{0.5cm} $\bullet p_T^l$-cut on leptons and lepton isolation 
   ($Isol$), which is defined here as the calorimetric energy flow around the
 lepton in a cone of $\Delta R<0.5$ divided by the lepton energy;

\hspace{0.5cm} $\bullet$ effective mass of two same-flavor opposite-sign 
leptons, to suppress WZ and potential ZZ backgrounds by rejecting 
events in a $m_Z \pm \delta M_Z$ band;

 \hspace{0.5cm} $\bullet$ $\Delta \phi_{l^+l^-}$, the relative azimuthal
angle between two same-flavor opposite-sign leptons;

\vspace{0.2cm}

ii) for \etm: 

      \hspace{0.5cm} $\bullet $ \etm-cut,

 \hspace{0.5cm} $\bullet  \Delta \phi (E_T^{miss},l^+l^-)$,
the  relative azimuthal angle between \etm \
and vector sum of the transverse momenta of the two
leptons.

\vspace{0.2cm}

iii) for jets:

\hspace{0.5cm} $\bullet$  ``jet veto''-cut: no jet
with 
  $E_T^{jet}$ lower than certain threshold, in some rapidity interval,
typically $\mid \eta_{jet}\mid<4.5$.

\vspace{0.2cm}

The data selection cuts based on these variables are separately optimised
in various parts of the explored parameter space \cite{ll_excl}.
For illustration we show some typical distributions.
Figure 7.2 shows the \etm \ distributions for signal and 
various SUSY backgrounds at two $(m_0,m_{1/2}$) 
points,  one with low and one with high slepton masses. The hardest
spectra in  both figures  correspond to $\tilde q \tilde q$ production
and the softest ones to chargino-neutralino production. This figure also
shows the main SM backgrounds. For low slepton mass there is no big difference 
in shape between signal and  SM-background spectra, thus it is not very
advantageous to apply a hard  \etm \ cut.

Figure 7.3 shows the  $\Delta \phi (E_T^{miss},l^+l^-)$ distribution
at the point from the low slepton mass region,  $m_0=86$ GeV, $m_{1/2}=85$
GeV, after $p_T^{l_{1,2}}$, cuts on lepton isolation
and \etm . The internal SUSY and SM backgrounds are still much larger 
in magnitude than the signal. Keeping only events with large relative
azimuthal angles significantly improves the signal to background ratio. 

Figure 7.4 illustrates the expected  rejection factors which can be
obtained by a ``central'' jet veto for two  $E_T^{jet}$ thresholds  
as a function of jet acceptance at point $m_0=86$ GeV, $m_{1/2}=85$ GeV. 
Signal acceptance decreases slightly with increasing
$|\eta_{jet}|$ of the jet veto region, whilst for sufficient
rejection of SUSY and SM backgrounds  ($t\bar t$) it is important
to have calorimetric coverage up to $\mid\eta\mid=3.5-4.0$.
The rejection power against $\tilde g, \tilde q$ and $t\bar t$ 
improves by a factor of 1.5-1.8 with increasing coverage from
$\mid\eta\mid<2.5$ to $\mid\eta\mid<3.5$. 
To achieve significant background rejection reliable
jet detection down to $E_T \approx 30$ GeV is also essential.

Figure 7.5 shows  the $\Delta \phi_{l^+l^-}$ distribution for the main 
SM and SUSY backgrounds and events with direct slepton production at the same 
point $m_0=86$ GeV,  $m_{1/2}=85$ GeV.
For the signal, small $\Delta \phi_{l^+l^-}$ events come from  $\tilde{\nu}_l
\bar{\tilde{\nu}_l}$  production whilst events with large $\Delta
\phi_{l^+l^-}$ are due to  charged slepton production. The main contribution
to the slepton signal from $\tilde{\chi}_1^{\pm} \tilde{\chi}_2^0$
production in the two-lepton final state 
comes from decays of  $\tilde{\chi}_2^0$, thus the
distribution of these events also peaks at low relative azimuthal angles.
All processes which go through  $\tilde{\chi}_2^0$ decays to leptons,
including $\tilde{g},\tilde{q}$ contributions have almost the same shape
$\Delta \phi$-distribution.  In some parts of parameter space
it may be more advantageous to keep only events with large 
$\Delta \phi_{l^+l^-}$, which rejects most of the internal SUSY
backgrounds. However, this cut also eliminates  direct $\tilde{\nu}$ 
production events and the indirect contribution from \chnb  \,\,
decays. Only $\tilde{\chi}_1^{\pm}\tilde{\chi}_1^{\pm}$ 
events with leptons coming from different parent charginos
survive as a significant SUSY background. 

To extract the slepton signal at points from domain I in Fig.~2.8b
($m_0 \gappeq 0.5 m_{1/2}$) with $m_{\tilde{l}_L} \sim 100$ GeV, 
just beyond of the LEP2 and Tevatron sparticle reaches, 
where the SM backgrounds are dominant, the 
selection {\bf Set 0} is appropriate at low luminosity. The {\bf Set 0$'$}
is better adapted for the points from domain II ($m_0 \lappeq 0.5
m_{1/2}$), see Table 7.1. 

The expected number of signal and background events in this particular
difficult region, after all cuts, is shown in Table 7.2. 
We also give the signal to background ratio
and significance for slepton production and the same quantities for
overall SUSY production.

\begin{table}[t]
\begin{center}
Table 7.1.  Low mass slepton selection criteria

\vspace*{3mm}

\begin{tabular}{|c|c|c|} 
\hline
&{\bf Set 0}&{\bf Set 0$'$}\\   
\hline
$Isol \, <$  & 0.1 & 0.1  \\
\hline
$p_T^{l_{1,2}}>$ (GeV) & 20 & 20  \\
\hline
\etm$>$ (GeV) & 50 & 50 \\
\hline
$\Delta \phi (E_T^{miss},l^+l^-)>$ ($^0$)& 160 & 160 \\
\hline
No jets in $|\eta_{jet}|<4.5$ & &\\
 with $E_T^{jet}>$ (GeV)&  30 & 30 \\
\hline
No $M_{ll}  \in M_Z \pm \delta m$ (GeV)& $91 \pm 5$ & $91 \pm 5$ \\  
\hline
$\Delta \phi_{l^+l^-}>$ ($^0$)& 130 & no cut \\
\hline
\end{tabular}
\end{center}
\end{table} 

It is clear that in this domain of parameter space, the slepton
($m_{\tilde{l}} \sim 100$ GeV) contribution to the finale state
topology is only a modest one. The slepton signal can be detected
from an excess of events over background expectations, but not
from any particular kinematical feature. The $S/B$ is about 0.3.
Of the SM backgrounds, WW is the largest and essentially irreducible,
$t\bar t$ and W$t\bar b$ are comparable and at the same level as the
signal after hard jet veto cuts. The main SUSY background is due to
$\tilde{\chi}_1^{\pm}\tilde{\chi}_1^{\mp}$.
For the point $m_0=$85 GeV, $m_{1/2}=$86 GeV, from domain I,
the total number of signal events $N_S^{Sl-tot}$ is just
due to direct slepton production, i.e. there are no indirect
slepton contributions, but there are additional SUSY contributions to
this topology.

For points with $m_0 \lappeq 0.5 m_{1/2}$ (domain II),
no $\Delta \phi_{l^+l^-}$ cut  is applied so as
to take advantage of
all SUSY processes contributing to slepton production.
$N_S^{Sl-tot}$ in this case has contributions
from both direct and indirect slepton production.
The contribution from indirect slepton production is comparable to the
direct one
and helps to reach a $5\sigma$ significance level needed
for slepton signal visibility.
In this region the only backgrounds are SM contributions.
The irreducible WW is still the dominant, and
$Wt\bar b$ increases significantly.

To achieve the  highest mass reach  with  $10^4\, pb^{-1}$  
a somewhat harder set of cuts is appropriate,
 ({\bf Set 1} in Tab. 7.3). The reach in parameter space
with $10^4\,pb^{-1}$ is shown in  Fig.~7.6; it corresponds to left-slepton
masses up to  $\simeq 160-180$ GeV,
i.e. well above the LEP2 reach ($m_{\tilde{l}} \lappeq 95$ GeV).
In the region $m_0 \lappeq 0.5 m_{1/2}$
(domain II) the total
slepton signal coincides with the overall
SUSY signal, but for points from region $m_0 \gappeq 0.5 m_{1/2}$
(domain I) the additional SUSY
processes contribute to the overall SUSY signal, 
but not to the slepton signal.

\begin{table}[t]
\begin{center}
Table 7.2: Expected number of direct ($N_S^{Sl-dir}$) and
indirect ($N_S^{Sl-indir}$) slepton signal events,
and SM ($N_B^{SM}$) and SUSY background ($N_{B}^{SUSY}$) events,
after cuts,  at $10^4$ pb$^{-1}$. The
signal to background ratio $N_S^{Sl-tot}/N_B^{tot}$  and significance of
slepton signal  $N_S^{Sl-tot}/\sqrt{N_S^{Sl-tot}+N_B^{tot}}$
are also given; the same
for overall SUSY signal  $N_S^{SUSY}/N_B^{SM}$,
$N_S^{SUSY}/\sqrt{N_S^{SUSY}+N_B^{SM}}$.

\vspace*{3mm}

\begin{tabular}{|c|c|c|c|c|}
\hline
($m_0,m_{1/2}) \rightarrow$ & (86,85)  & (4,146)  & (20,160) & (20,190)   \\
\hline
\hline
&{\bf Set 0} & {\bf Set 0$'$}& {\bf Set 0$'$} & {\bf Set 0$'$} \\
\hline
\hline
$\tilde{l}\tilde{l}$ & 323 & 315 &319& 214  \\
\hline
\hline
$\tilde{\chi}_1^{\pm}\tilde{\chi}_2^0$ &14 &45  & 48 & 40  \\
\hline
$\tilde{\chi}_1^{\pm}\tilde{\chi}_1^{\mp}$ &46 & 253 & 217 & 168  \\
\hline
$\tilde g \tilde q$&13 & 8 &$-$ &$-$ \\
\hline
$\tilde q \tilde q$&35 & 6 &$-$ &$-$ \\
\hline
$\tilde g \tilde g$&0  & 2 &$-$ &$-$  \\
\hline
\hline
WW            &454  & 1212 &1212 & 1212 \\
\hline
Wt$\bar b$    &163  & 577 & 577 & 577  \\
\hline
$t \bar t$    &345 & 574 &574 & 574 \\
\hline
WZ            &15& 43  & 43 & 43 \\
\hline
$\tau \tau$   &15 & 15&15 &15   \\
\hline
\hline
$N_S^{Sl-dir}$ &323& 315 & 319  & 214   \\
\hline
$N_S^{Sl-indir}$ &$-$& 314 & 265  & 208   \\
\hline
$N_S^{SUSY}$ &431& 629  &584  &422   \\
\hline
$N_B^{SUSY}$  & 108 & $-$  & $-$  & $-$ \\
\hline
$N_B^{SM}$  & 992 & 2421  &2421  &2421 \\
\hline
\hline
$N_S^{Sl}/N_B^{tot}$  & 0.3 &  0.26  & 0.24 &  0.17  \\
\hline
$N_S^{Sl-tot}/\sqrt{N_S^{Sl-tot}+N_B^{tot}}$& 8.6 & 11.4&10.7&7.9  \\
\hline
\hline
$N_S^{SUSY}/N_B^{SM}$  & 0.4 & 0.26 & 0.24 & 0.17  \\
\hline
$N_S^{SUSY}/\sqrt{N_S^{SUSY}+N_B^{SM}}$& 11.4 & 11.4 & 10.7 & 7.9 \\
\hline
\end{tabular}
\end{center}
\end{table}

With $10^5\, pb^{-1}$,  a significantly
higher slepton mass range becomes accessible.
The average lepton transverse momentum and
missing energy  increasing with $m_{\tilde{l}}$ become
significantly harder than for the main backgrounds.
Harder kinematical cuts are therefore preferable.
Slepton production becomes the main and ultimately the dominant
contribution to this topology. We use four main different
sets of cuts (Table 7.3)
optimised for different regions of $(m_{0},m_{1/2})$
parameters space. These sets are chosen taking into account the mass
relations, production mechanisms and decay patterns discussed in section
2.2.

{\bf ``Set 2''} is used to explore the $m_{\tilde{l}_L} \sim 200$ GeV
domain. The $t\bar t$, $Wt\bar b$ backgrounds
are suppressed with hard cuts on $p_T$ of the   
leptons, \etm \ and the ``jet veto''. This last cut eliminates entirely
the backgrounds from strongly interacting sparticles. We keep events with
small relative azimuthal angle between the
leptons $\Delta \phi_{l^+l^-}$ mainly to improve the rejection of SM
backgrounds, and to eliminate
$\tilde{\chi}_1^{\pm}\tilde{\chi}_1^{\mp}$ production when it is necessary.
The slepton S/B ratio is about 1.

{\bf ``Set 3''} is used for $m_{\tilde{l}_L} \sim 300$ GeV at points
with $m_0 \lappeq 0.5 m_{1/2}$, whilst
for points with $m_0 \lappeq 0.5 m_{1/2}$
{\bf ``Set 4''} is more appropriate, as the cross-sections are smaller.
Furthermore, there is no source of indirect slepton 
production. {\bf ``Set 4''} is also used to explore the slepton
mass range 300 GeV $\lappeq m_{\tilde{l}_L} \lappeq$ 350 GeV.
{\bf ``Set $4'$''} is used for masses 350 GeV $\lappeq m_{\tilde{l}_L}
\lappeq$ 440 GeV, and
{\bf ``Set $4''$''} for $ m_{\tilde{l}_L} \gappeq 440$ GeV.
Details of the signal and background contribution at each simulated point
can be found in \cite{ll_excl}.

{\small

\begin{table}[t]
\begin{center}
Table 7.3. High mass slepton selection criteria

\vspace*{0.3cm}

\begin{tabular}{|c|c|c|c|c|c|c|}
\hline
&{\bf Set 1}&{\bf Set 2}&{\bf Set 3}&{\bf Set 4}&{\bf Set 4}$'$&
{\bf Set 4}$''$\\
\hline
$Isol\, <$ & 0.1 & 0.1 & 0.1 & 0.1 & 0.1 & 0.03 \\
\hline
$p_T^{l_{1,2}}>$ (GeV) & 20 & 50  & 50  & 60  & 60  & 60  \\
\hline
\etm $>$ (GeV)&100  & 100  & 120  & 150  & 150  & 150 \\
\hline
$\Delta \phi (E_T^{miss},l^+l^-)>$ ($^0$) & 150 & 150 & 150 & 150
& 150 & 150 \\
\hline
No jets in $|\eta_{jet}|<4.5$ & & & & & &\\
with $E_T^{jet}>$ (GeV) & 30 & 30 & 30 & 45 & 45 & 45 \\
\hline
No $M_{ll} \in M_Z \pm \delta m$ (GeV) & $91 \pm 5$
&$91 \pm 5$ & $91 \pm 5$  & $91 \pm 5$  & $91 \pm 5$ &$91 \pm 5$ \\
\hline
$\Delta \phi_{l^+l^-}<$ ($^0$)& 130 & 130 & 130 & 130 & 140 & 130 \\
\hline
\end{tabular}
\end{center}
\end{table}
}

Figure 7.6 summarizes the results of this study.
At each point investigated the expected slepton signal significance,
$n\sigma$, and the  signal to background ratio (in parentheses) is given,
for an integrated luminosity of $10^5$ $pb^{-1}$. The $5\sigma$ reach is
shown as well. Recall that the selection procedure has been
optimised in terms of signal significance $N_S/\sqrt{N_S+N_B}$ to
investigate the maximum slepton mass reach, and not in terms
of sample purity (i.e. S/B ratio). The reason is that if SUSY were
found at Fermilab where $\tilde{g}/\tilde{q}$ masses up to
$\lappeq 400-450$ GeV can be probed, sleptons can be probed
at the Tevatron and LEP2 only up to a mass of $m_{\tilde{l}} \lappeq$
100 GeV, thus the main task of LHC would be to explore slepton
spectroscopy, with masses up to $m_{\tilde{q}/\tilde{g}} \lappeq$   
400 GeV.
Figure 7.7 shows the same information as Fig.~7.6, but  in terms of left
slepton and lightest neutralino LSP masses. The domain that can be
explored extends  up to
 $m_{{\tilde l}_L} \sim 340$ GeV for all allowed $m_{LSP} \, (<200$ GeV),
and up to $m_{{\tilde l}_L} \sim 340-440$ GeV only if $m_{LSP} \gappeq
(0.45-0.6) \cdot m_{{\tilde l}_L}$ for a given $m_{{\tilde l}_L}$.
The mass reach for right sleptons is up to 340 GeV. As the main SM  
background is WW production, to study the stability of our results we
assumed a $50 \%$ uncertainty in the WW cross-section.
Figure 7.8 shows how much the 5$\sigma$ reach boundary would be
modified if the WW cross-section were twice as large as predicted
by PYTHIA.

A very interesting aspect of slepton searches at LHC from the cosmological
point of view is that, in this topologically
simplest channel,  the explorable domain for $10^5$ $pb^{-1}$ almost 
coincides with the regions of parameter space where the LSP would be
cosmologically relevant (Figs.~7.6 and 7.7). The boundaries for the 
domains we can probe in slepton searches
and the neutralino dark matter
$\Omega h^2$ density contours have shapes similar to
slepton isomass curves as both are largely determined
 by slepton masses. In particular,
neutralino annihilation is governed by slepton
exchange with $\sigma_{ann}\sim 1/m_{\tilde{l}}^4$. Too large a slepton
mass would induce too small an annihilation cross-section and thus too
large a relic density which would overclose the Universe.
The $\Omega h^2$ range prefered by mixed dark matter scenarios
is for  0.15 $\lappeq \Omega h^2 \lappeq$ 0.4 (with the Hubble
scaling constant 0.5 $\lappeq h^2 \lappeq$ 0.8) which corresponds
to a slepton mass range 100 GeV $\lappeq m_{\tilde{l}} \lappeq$  
250 GeV, (see \cite{daniel1}, \cite{cosmo}.)

We conclude that
to search for direct production of sleptons the most appropriate channel is
$2\,\,leptons +$ \etm $\,\, and \, \,no \,\,jets$.
In regions of low $m_0, m_{1/2}$, just beyond the sparticle
reaches at LEP2 and the Tevatron, the slepton signal
can be detected, but the search will be difficult as
$\tilde{l}$ is not the dominant contribution to this final state.
In significant parts of the explorable ($m_0,\,m_{1/2}$) space,
however, direct slepton pair
production is the main contribution to this final state, with
non-negligible contributions from indirect production too, mostly through
$\tilde{\chi}_1^{\pm},\tilde{\chi}_2^0$. The inclusion of this 
indirect slepton
production  improves the mass reach.
With $10^4\,pb^{-1}$ luminosity, CMS is sensitive up to  
$m_{{\tilde l}_L} \lappeq 160$ GeV.
With $10^5\,pb^{-1}$ luminosity the reach extends up to
$m_{{\tilde l}_L} \sim 340$ GeV for all allowed $m_{LSP} \, (<200$ GeV),
and up to $ m_{{\tilde l}_L} \sim 340-440$ GeV if
$m_{LSP} \sim (0.45-0.6)\cdot m_{{\tilde l}_L}$ for a given
$m_{{\tilde l}_L}$. Throughout the domain where the slepton
signal should be detectable, it is always contaminated by backgrounds.
The dominant background is always WW production. In the $m_{\tilde{l}_L}   
\sim 100$ GeV domain the WW background exceeds the number of
signal events, whilst for $m_{\tilde{l}_L}\gappeq 200$ GeV 
the number of background events is comparable or smaller than
the expected number of signal events.
The ``jet veto'' cut effectively eliminates the $\tilde{g},\tilde{q}$
backgrounds, but also the potentially important indirect  slepton
production from $\tilde{g}$ and $\tilde{q}$. This issue
requires a separate study.

In the region  $m_0 \lappeq 0.45\cdot m_{1/2}$ sleptons decay directly
into the  LSP and sneutrinos
thus decay invisibly, $\tilde{\nu} \rightarrow \tilde{\chi}_1^0\nu$.
In  this portion of parameter space only charged  sleptons can be
detected. As right sleptons are also significantly lighter than left
ones they  are mostly eliminated by the hard selection cuts needed to   
suppress
the backgrounds, so mainly left sleptons are present in the final event
sample. Left sleptons will thus be detectable  in the entire region where
a slepton signal
is seen, whilst right sleptons will be detectable mainly in domain
$m_0 \gappeq 0.45\cdot m_{1/2}$.
Sneutrinos will contribute to this final state only in a very
limited area with $m_{1/2} \lappeq$ 160 GeV and $m_0 \lappeq$ 150
GeV.

The search for sleptons is important and significant, as  LEP2 and
the Tevatron  can only explore the slepton mass  range up to $m_{\tilde
{l}_L} \simeq 100$ GeV.   
The domain of ($m_0,\,m_{1/2}$) parameter space explorable through
this simplest experimental channel, with two {\it same-flavor,
opposite-sign leptons + \etm  +  no jets}, covers almost the entire
domain of parameter
space where SUSY is plausible in general terms (minimal fine-tuning)
and covers all of the parameter space where  SUSY would be of cosmological
significance (at least for $tan \beta =2, A_0=0, \mu<0$),
with $>10^4 pb^{-1}$ of integrated luminosity.

\newpage
\begin{figure}[hbtp]
\vspace{0mm}
\hspace*{0mm}
\resizebox{14.cm}{!}{\rotatebox{0}{\includegraphics{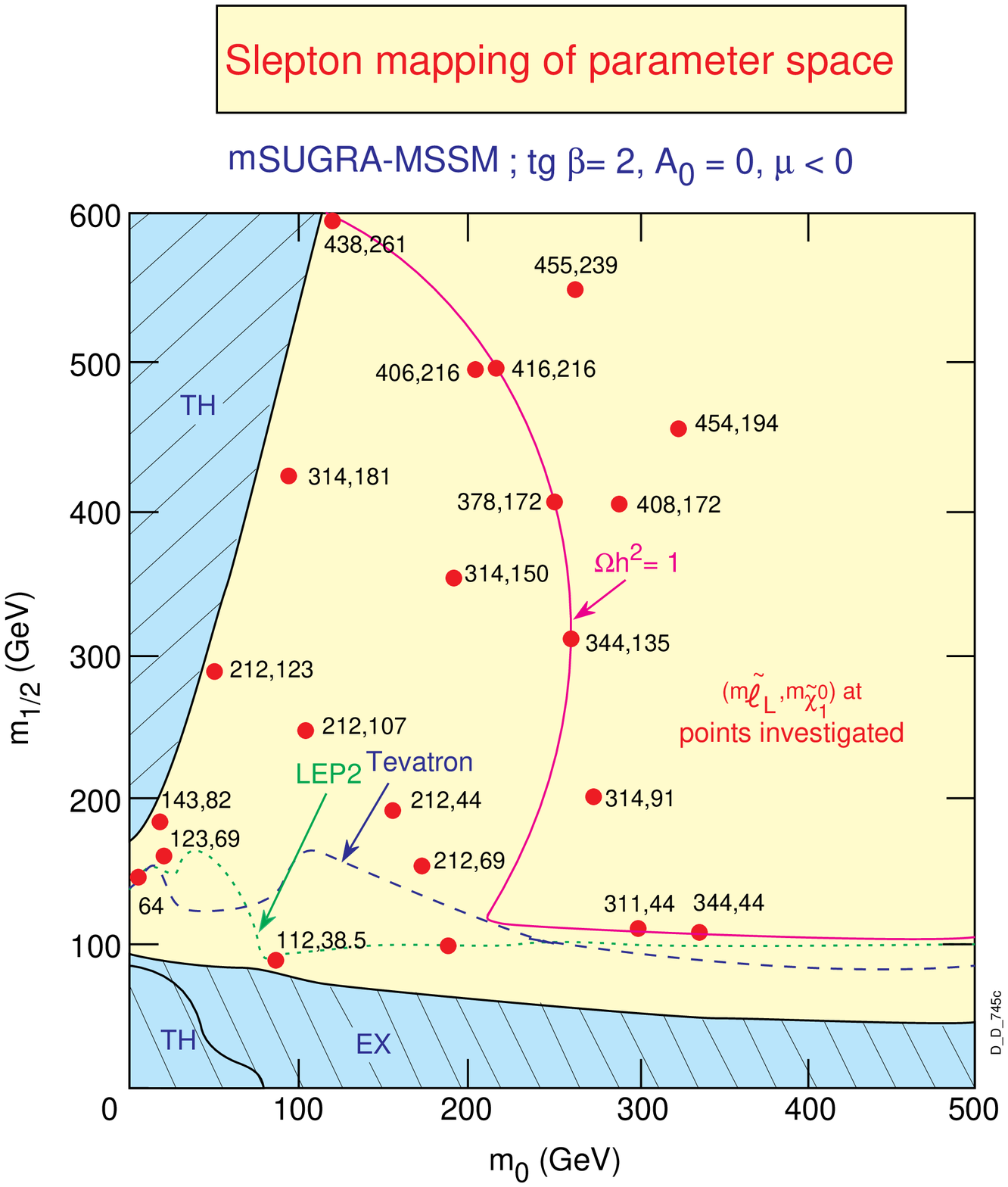}}}   
\end{figure}

 \vspace{30mm}
Figure 7.1: Points in mSUGRA parameter space where slepton
production has been studied. The numbers at each point give
corresponding values of $m_{\tilde{l}_L}$ and $m_{\chi_1^0}$.
The shaded areas represent excluded regions on the bases of
LEP and Tevatron data. The dashed lines represent the expected
reach with sparticle searches of LEP2 and Tevatron with 1 fb$^{-1}$.
The $\Omega h^2=1$ absolute upper limit contour for neutralino dark matter
is also shown.

\newpage
\begin{figure}[hbtp]
\vspace{0mm}
\hspace*{0mm}
\resizebox{14.cm}{!}{\rotatebox{0}{\includegraphics{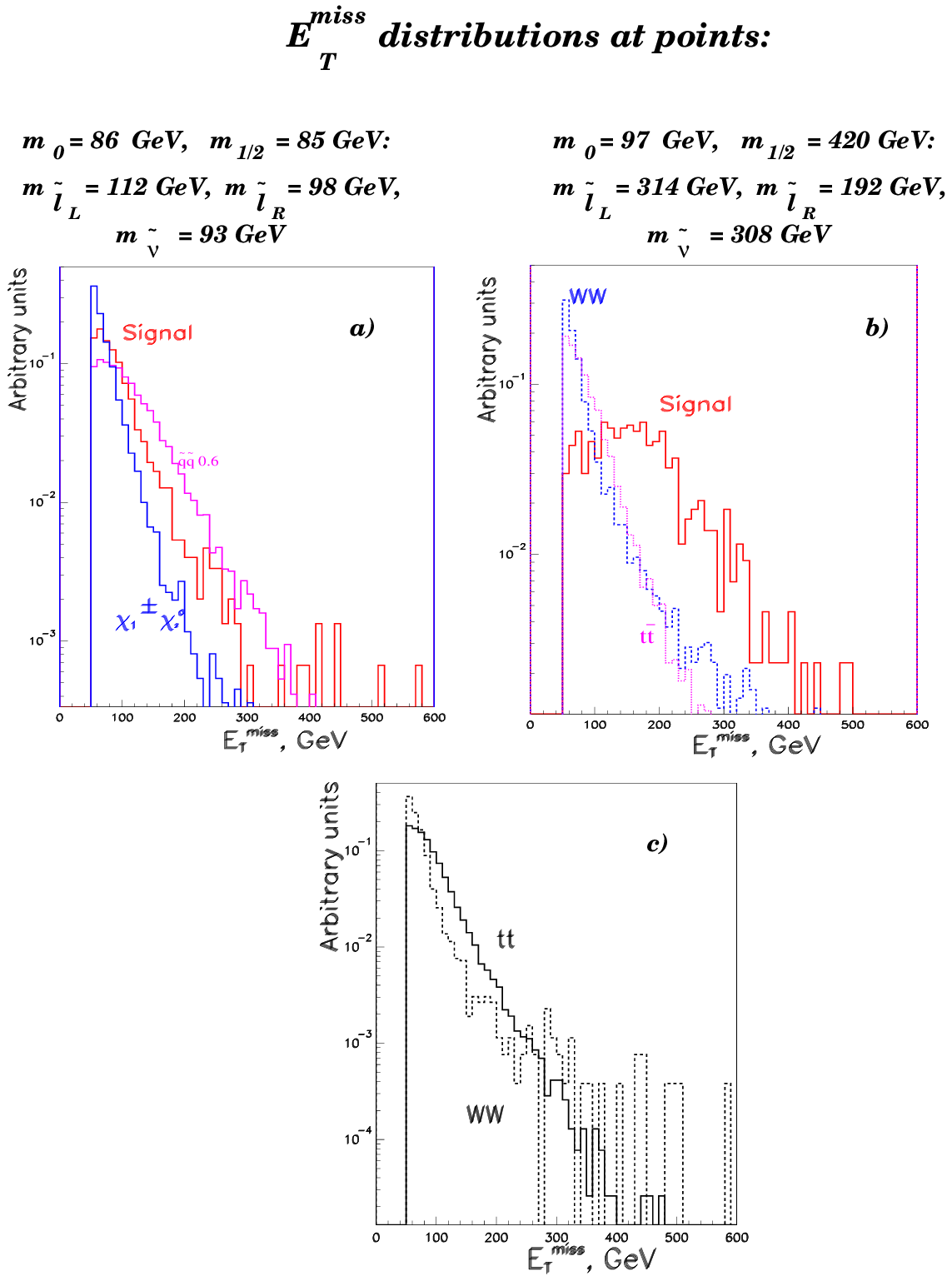}}}
\end{figure}

 \vspace{30mm}
Figure 7.2: 
\etm \ distribution for slepton signal (DY production), SUSY backgrounds
($\tilde{g},\,\tilde{q},\,\tilde{\chi}_1^{\pm}\tilde{\chi}_2^0$)
at two parameter space points:
a) $m_0=86$ GeV, $m_{1/2}=85$ GeV,
b) $m_0=97$ GeV, $m_{1/2}=420$ GeV
and c) the main SM backgrounds ($WW,\,t\bar t$).

\newpage

\ \ \\

\vspace{40mm}

\begin{figure}[hbtp]
\vspace{-40mm}
\hspace*{20mm}
\resizebox{12.cm}{!}{\rotatebox{0}{\includegraphics{D_Denegri_1095n.ill}}}
\end{figure}

\vspace{20mm}

Figure 7.3: Relative azimuthal angle between  \etm \ and the resulting
dilepton momentum for the slepton signal and main SM and SUSY backgrounds
in 2$leptons$ + \etm \ + $no jets$ final states.

\newpage

\begin{figure}[hbtp]
\vspace{-0mm}
\hspace*{10mm}
\resizebox{14.cm}{!}{\rotatebox{0}{\includegraphics{D_Denegri_1011n1.ill}}}
\end{figure}

\vspace{20mm} 

Figure 7.4: Rejection factors a) for internal SUSY backgrounds,
b) for SM  backgrounds
by  central ``jet veto'', and c) slepton
signal acceptance, as a function of jet
rapidity
acceptance
$\mid\eta_{jet}\mid$ and for two jet detection thresholds.

\newpage

\begin{figure}[hbtp]
\vspace{-0mm}
\hspace*{10mm}
\resizebox{12.cm}{!}{\rotatebox{0}{\includegraphics{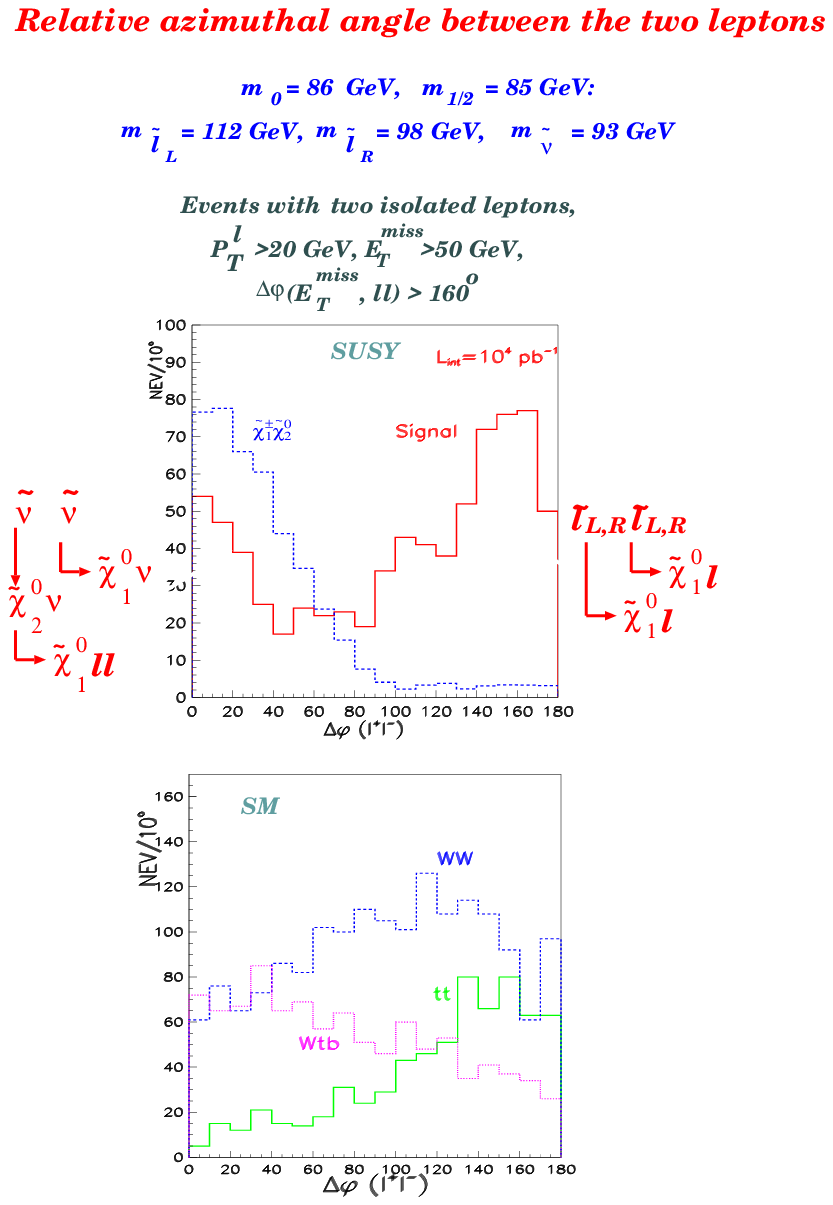}}}
\end{figure}

\vspace{30mm}

Figure 7.5: Relative azimuthal angle between two same-flavor
opposite-sign
leptons at one point on parameter space,
a) for the various contributions to the SUSY slepton
signal and for a SUSY
background ($\tilde{\chi}_1^{\pm}\tilde{\chi}_2^0$), and b) for the main
SM backgrounds, in 2$leptons$ + \etm \ + $no \ jets$ final states.

\newpage

\ \\

\vspace{10mm}

\begin{figure}[hbtp]
\vspace{-25mm}
\hspace*{-5mm}
\resizebox{16.cm}{!}{\rotatebox{0}{\includegraphics{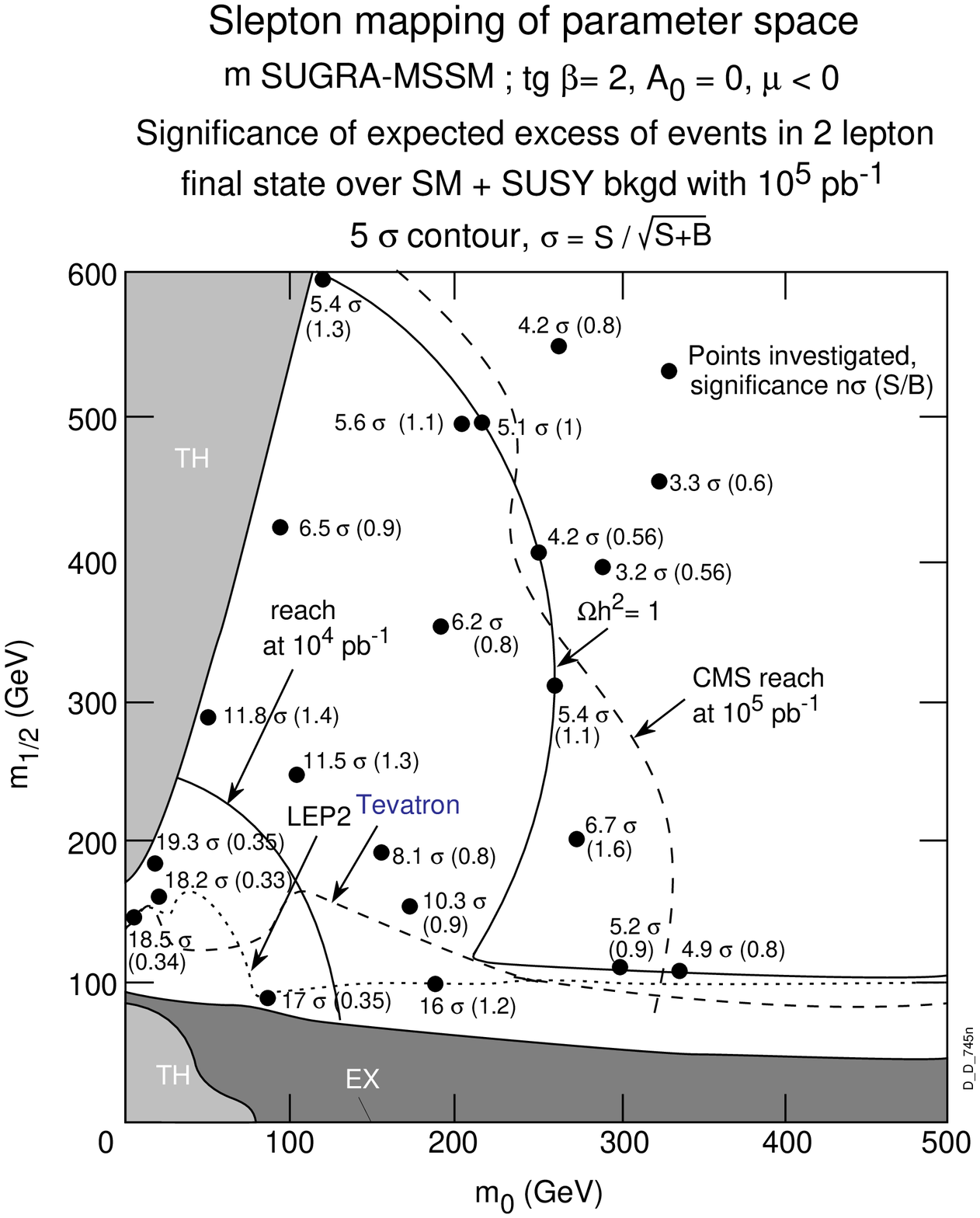}}}
\end{figure}

\vspace{5mm}

Figure 7.6: Expected reach in parameter space for slepton searches in the
2$leptons+$ \etm $+\,no\,\,jets$ final states with $10^4\,pb^{-1}$ and  
$10^5\,pb^{-1}$ in CMS.
The reach is given as a 5$\sigma$ contour. At each point
investigated the slepton signal significance is indicated as $n\sigma$,
and the S/B ratio is given in parenthesis, where B is the full SM + SUSY
background
to the slepton signal. The cosmological absolute upper bound
($\Omega_{\tilde{\chi}_1^0}$=1, $h$=1) density contour for
neutralino dark matter $\Omega h^2=1$ is indicated.

\newpage

\begin{figure}[hbtp]
\vspace{10mm}
\hspace*{10mm}
\resizebox{13.cm}{!}{\rotatebox{0}{\includegraphics{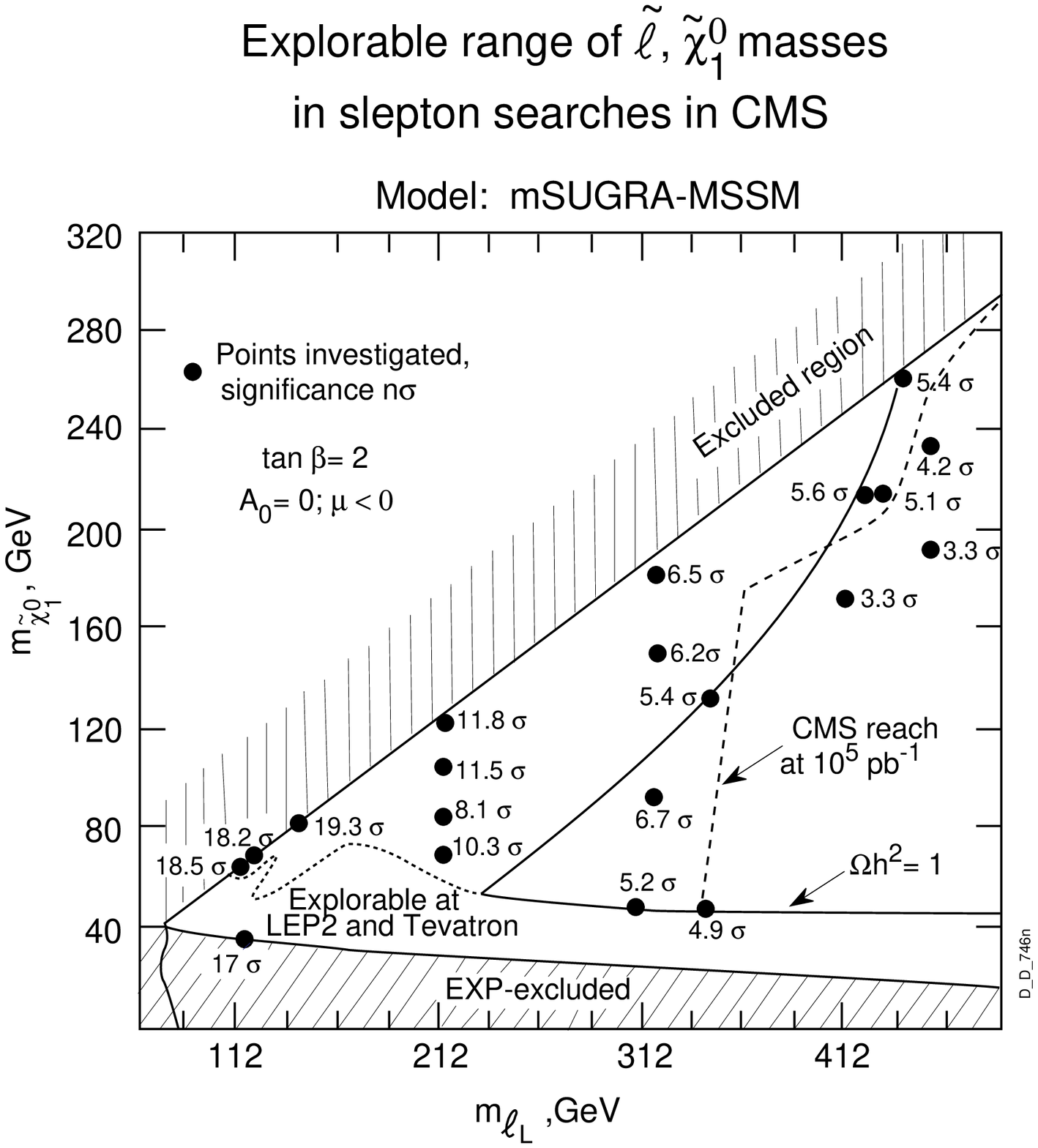}}}
\end{figure}

\vspace{30mm}

Figure 7.7: Same mapping as in  Fig.~7.6, but in terms of masses of the
 left slepton and the LSP.

\newpage  

\begin{figure}[hbtp]
\vspace{10mm}
\hspace*{10mm}
\resizebox{13.cm}{!}{\rotatebox{0}{\includegraphics{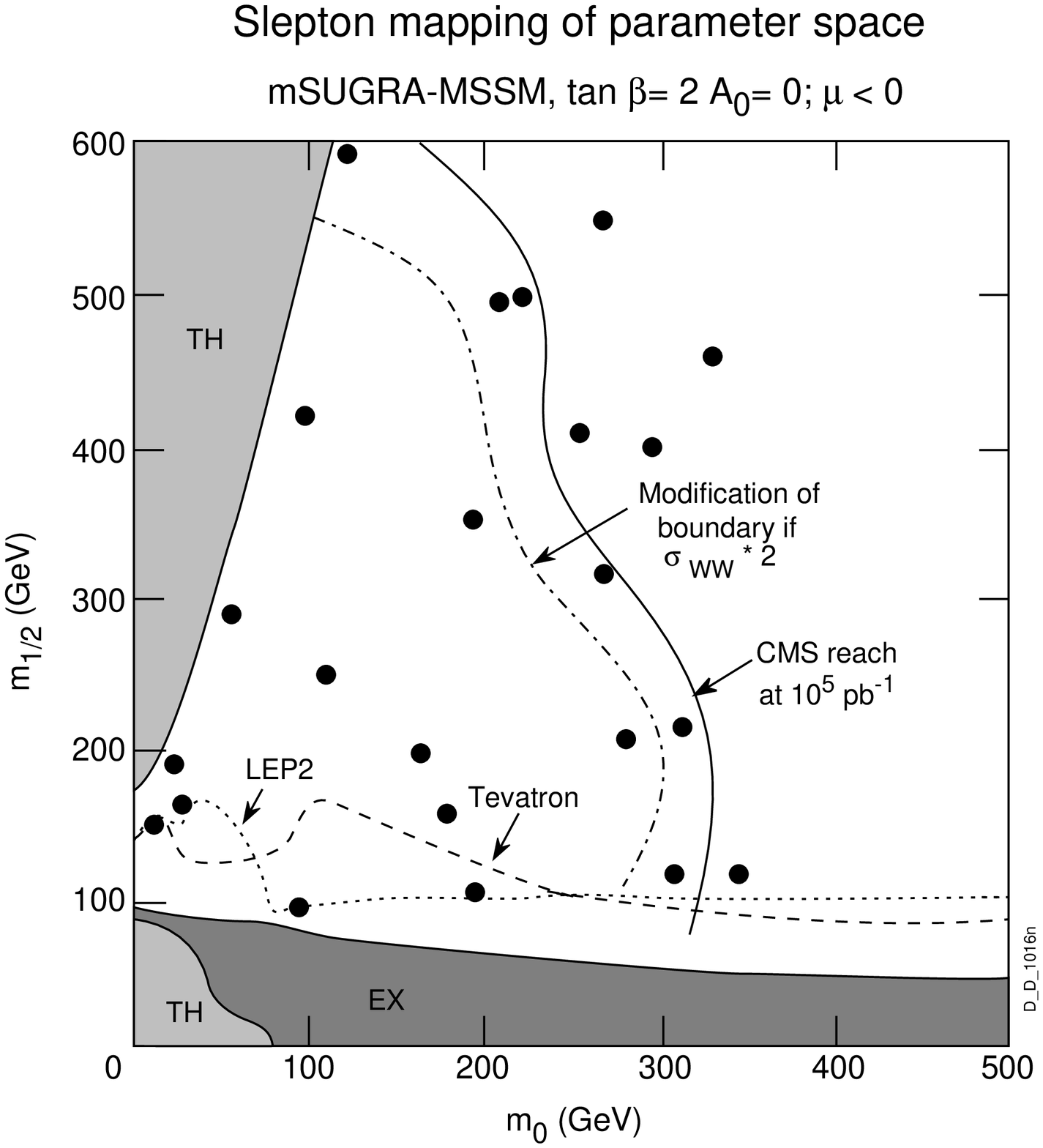}}}
\end{figure}

\vspace{30mm}
Figure 7.8: Modification of the $5\sigma$ reach in slepton searches if
the
production cross-section for the main SM background of WW pair production
is increased by a factor of 2.

\newpage


\section{Exclusive $3 \ leptons + no \ jets$ + \etm \  channel -- \\
search for chargino/neutralino pair production}

Direct production of \chha\chnb \ followed by leptonic decays
of both sparticles (see section 2.1) gives three high-$p_T$ isolated
leptons
accompanied by missing energy.
These events are free from jet activity except for those
coming from initial-state QCD radiation. Therefore we search  
for \chha\chnb \ pair production in 3 $leptons\ + no \ jets +$ \etm \
final states \cite{wz}.

We have considered the following SM processes contributing to the
background: WZ, ZZ, $t\bar{t}$, W$tb$, Z$b\bar{b}$, $b\bar{b}$
with the cross-sections $\sigma_{WZ}$  =  26 pb,  $\sigma_{ZZ}$  =  15   
pb,  $\sigma_{t\bar{t}}$  =  670 pb for a top mass of 170 GeV,
$\sigma_{Wtb}$  =  160 pb, $\sigma_{Zb\bar{b}}$  = 600 pb,
$\sigma_{b\bar{b}}$  = 500 $\mu$b. Three leptons in the final state
are obtained by folding the appropriate leptonic decay modes.
We have forced Z~$\rightarrow l^{+}l^{-}$, W$^{\pm} \rightarrow l^{\pm}
\nu$ and $b \rightarrow cl\nu$ decays, where $l=\mu, e, \tau$.
In case of $t\bar{t}$ and W$tb$ only one W has been forced.
These SM backgrounds, except ZZ production, can be
suppressed by requiring that all three leptons be isolated, and
vetoing central jets. ZZ events contain four isolated leptons and
asking for exactly three isolated leptons
in the event leaves only a small fraction of the ZZ continuum,
when one of the leptons escapes detector acceptance. The
WZ, ZZ and Z$b\bar{b}$ backgrounds can be suppressed
by removing events with a dilepton invariant mass close to the
Z mass.
At each investigated point in the ($m_0, m_{1/2}$) plane
we have studied all SUSY production processes which
can lead ultimately to three leptons in the final state.
These are the strong interaction production of
\ $\tilde{g} \tilde{g}$, \ $\tilde{g} \tilde{q}$, \
$\tilde{q} \tilde{q}$ pairs; associated production of
 $\tilde{g}\tilde{\chi}, \  \tilde{q}\tilde{\chi}$ pairs;
chargino-neutralino $\tilde{\chi}\tilde{\chi}$ pair
production, and slepton pair production,
$\tilde{l}\tilde{l}, \ \tilde{l}\tilde{\nu}, \ \tilde{\nu}\tilde{\nu}$.
The SUSY event sample with three high-$p_T$ isolated leptons
is dominated by strong production; but, of course,
the jet veto requirement is very efficient in reducing
\g \ and \q \ events which in their cascade decays
produce many central high-$E_T$ jets.
It has to be pointed out that SUSY events other than
the directly  produced \chha\chnb and which survive all 
selection criteria, may still contain some \chha\chnb's produced in
cascade decays of heavier sparticles. These
events can be considered as an ``indirect'' \chha\chnb \ signal,
a situation similar to the one encountered in discussing
slepton searches.

In Fig.~8.1 we show the fully simulated SUSY points in the
($m_0, m_{1/2}$) plane, as representatives of various regions
according to the predominant or characteristic decay modes.
In order to illustrate event kinematics
we choose three points from different ($m_0, m_{1/2}$)
regions among those shown in Fig.~8.1:  

 S1 \ \ \ ($m_0, m_{1/2}$) = (100 GeV, 100 GeV) \ with

 \ \ \ \ \ \ \ \
M(\chha) $\simeq$ M(\chnb) = 98 GeV, \ \ M(\chna) = 45 GeV, \ \

 \ \ \ \ \ \ \ \
M(\g) = 294 GeV, \ \ M(\q) = 270 GeV

 S2 \ \ \ ($m_0, m_{1/2}$) = (2000 GeV, 175 GeV) \ with

 \ \ \ \ \ \ \ \
M(\chha) $\simeq$ M(\chnb) = 154 GeV, \ \ M(\chna) = 76 GeV, \ \

 \ \ \ \ \ \ \ \
M(\g) = 551 GeV, \ \ M(\q) = 2032 GeV

 S3 \ \ \ ($m_0, m_{1/2}$) = (100 GeV, 400 GeV) \ with

 \ \ \ \ \ \ \ \
M(\chha) $\simeq$ M(\chnb) = 341 GeV, \ \ M(\chna) = 172 GeV, \ \
 
 \ \ \ \ \ \ \ \
M(\g) = 983 GeV, \ \ M(\q) = 861 GeV

Figure 8.2 shows the $d\sigma/dp_T^l$ distribution in the
three-lepton event sample with $p_T^l >$ 15 GeV.
Distributions are shown for signal and SM background.
At this stage the dominant SM backgrounds are $t\bar{t}$ and
$Zb\bar{b}$. For high $m_{1/2}$,
e.g. point S3, the lepton $p_T$ cut can be set at 20 GeV.
The loss in signal acceptance is $\sim$12$\%$ compared to a
15 GeV cut, whilst $t\bar{t}$ and $Zb\bar{b}$ are reduced
by an additional factor of $\sim$2.

To suppress background events with leptons originating from
semileptonic decays of $b$-quarks, we apply tracker isolation,
requiring all three leptons to be isolated. If there is no
charged track with $p_T$ greater than a certain threshold
$p_T^{cut}$ in a cone of $R=\sqrt{\Delta \eta^2 + \Delta \phi^2} 
=0.3$ about the lepton direction, it is considered as isolated.
Figure 8.3 a shows the lepton isolation rejection  factor against
$t\bar{t}$ background versus the $p_T^{cut}$ on charged tracks. The two
curves correspond to two different instantaneous luminosity regimes,
$L = 10^{33}$ cm$^{-2}$s$^{-1}$ and $L = 10^{34}$ cm$^{-2}$s$^{-1}$.
In the latter case we have superimposed 15 ``hard'' pile-up events. Figure
8.3 b shows signal acceptance versus $p_T^{cut}$. We choose $p_T^{cut}$
= 1.5 GeV and 2 GeV for low and high luminosity running, respectively.
The corresponding rejection factors against $t\bar{t}$ are about 60 and 50,
whereas signal acceptance decreases from about 90$\%$ to 70$\%$.
The next requirement is the central jet veto, which allows us reject
the internal SUSY background coming from \g \ and \q \ cascade decays
overwhelming  \chha\chnb \ direct production. This cut also reduces
some SM backgrounds. In Fig.~8.4 we show jet multiplicity and
jet $E_T$ distributions for signal, SUSY \g \g, \g \q, \q \q \
background events at point S1, and for the $t\bar{t}$ background. Since
at this ($m_0, m_{1/2}$) point the gluino is heavier than the squark, it
decays to a squark-quark pair. Therefore gluino events are richer in
jets compared to squark events. Figure 8.5 shows jet veto rejection
factors against \g \g, \g \q, \q \q \ and $t\bar{t}$ backgrounds and the
signal acceptance as a function of jet pseudorapidity, $|\eta^{jet}|$,
for two jet detection thresholds: $E_T^{jet} > 25$ GeV and $E_T^{jet} >
30$ GeV. As the jet veto region is decreased down to $|\eta^{jet}|  
\simeq$ 3.5, the  rejections/acceptance are not
too sensitive to jet rapidity coverage, but below this value of
$|\eta^{jet}|$, rejection factors decrease rapidly, degrading
significantly the separation of direct \chha\chnb \ production from   
internal SUSY and SM backgrounds. The reconstruction of jets at
$|\eta^{jet}| \sim$ 3.5 implies calorimetric coverage up to $|\eta|
\sim$ 4.3 at least, as the jets are collected in a cone of $\Delta R =
0.8$.With increasing $m_0$ and/or $m_{1/2}$ squarks and/or gluinos  
become heavier. Thus jets are harder and SUSY events are easier to   
suppress by a jet veto with relaxed requirement on $|\eta^{jet}|$ and
$E_T^{jet}$. Furthermore, for the signal events the QCD radiation
becomes harder (\chha, \chnb \ are more massive) and the low  
production cross-section requires the highest LHC luminosity,
with an event pile-up contaminated environment.
For this region, the jet veto threshold can be set with
$E_T^{jet}$ as high as 50 GeV and is justified by the need to
suppress the SM (mainly $t\bar{t}$) background.

Despite the fact that missing energy is always present
in SUSY signals (at least in $R$-parity conserving models),
for some regions of mSUGRA parameter space, it is advantageous not
to apply any specific cut on it. Figure 8.6 shows the distribution of
\etm \ in events with three isolated high-$p_T$ leptons and no jets
for the signal and the two main SM backgrounds: $t\bar{t}$ and $WZ$.  
For point S1 the shape of the distribution is very similar to that of the
background and therefore no cut on \etm \ is
applied. For point S3 requiring \etm $ > 80$ GeV
significantly reduces $t\bar{t}$ and $WZ$ backgrounds whilst
keeping more than 75$\%$ of signal events \cite{wz}.

For WZ background suppression, which after all the previous 
requirements remains the dominant SM background, an $M_Z$ cut is
necessary. In the signal events, the shape of the (same-flavor
opposite-sign) dilepton invariant mass $M_{l^{+}l^{-}}$ distribution
is determined by the \chnb \ decay kinematics and has a very sharp edge.
Figure 8.7 shows the $M_{l^{+}l^{-}}$ distribution for the
signal at three parameter space points S1, S2, S3 and for WZ
events. In events having all leptons of the same flavor, the
combinatorial   background obscures the sharp mass edge in the signal
events. A rejection factor of $\sim$ 17 for WZ is obtained by   
removing events with $M_{l^{+}l^{-}}> M_Z - 10 $ GeV, whilst the
small loss of signal events, $\lappeq 10\%$ for point S1,
comes from events when all three leptons have
the same flavor. However, for higher values of $m_{1/2}$,
e.g. point S3, the endpoint of the $M_{l^{+}l^{-}}$ distribution
is above the $Z$ peak; in this case we cut
only events falling within the $Z$
mass window: $M_Z - 10$ GeV $ < M_{l^{+}l^{-}} < M_Z + 10 $ GeV.

On the basis of the above considerations we adopt
four sets of selection criteria, appropriate for different    
($m_0, m_{1/2}$) regions, given in Table 8.1.

\begin{table}[t]
\begin{center}
Table 8.1: Selection criteria

\vspace{3mm}

\begin{tabular}{|l|l|l|}
\hline
Set 1 & $p_T$ cut       &
$3 l$ with $p_T^l>15$ GeV in $|\eta^l|<2.4(2.5)$
\\ \cline{2-3}
appropriate for&Isolation       &
No track with  $p_T>1.5$ GeV in  $R=0.3$
\\ \cline{2-3}
$m_{1/2} \lappeq $ 140 GeV&Jet veto        &
No jet with $E_T^{jet}>$  25 GeV in $|\eta^{jet}|<3.5$
\\ \cline{2-3}
& $Z$-mass cut \ \   &
$M_{l^{+}l^{-}} < 81$ GeV
\\ \hline \hline
Set 2 &$p_T$ cut       &
$3 l$ with $p_T^l>15$ GeV in $|\eta^l|<2.4(2.5)$
\\ \cline{2-3}
appropriate for &Isolation       &
No track with  $p_T>1.5$ GeV in  $R=0.3$
\\  \cline{2-3}
140 GeV $\lappeq$ $m_{1/2}$&Jet veto        &
No jet with $E_T^{jet}>$  30 GeV in $|\eta^{jet}|<3$
\\ \cline{2-3}
$\lappeq $ 180 GeV&$Z$-mass cut  \ \   &
$M_{l^{+}l^{-}} < 81$ GeV
\\ \hline \hline
Set 3 &$p_T$ cut       &
$3 l$ with $p_T^l>15$ GeV in $|\eta^l|<2.4(2.5)$
\\ \cline{2-3}
appropriate for &Isolation       &
No track with  $p_T>2$ GeV in $R=0.3$
\\ \cline{2-3}
180 GeV $ \lappeq $ $m_{1/2}$ &Jet veto        &
No jet with $E_T^{jet}>$  40 GeV in $|\eta^{jet}|<3$
\\ \cline{2-3}
$\lappeq$ 300 GeV& missing $E_T$       &
\etm $>50$ GeV
\\ \cline{2-3}
&  $Z$-mass cut \ \   & $M_{Z} \pm 10$ GeV
\\ \hline \hline
Set 4 & $p_T$ cut       &
$3 l$ with $p_T^l>20$ GeV in $|\eta^l|<2.4(2.5)$
\\ \cline{2-3}
appropriate for & Isolation       &
No track with  $p_T>2$ GeV in  $R=0.3$
\\ \cline{2-3}  
$m_{1/2} \gappeq $ 300 GeV& Jet veto        &
No jet with $E_T^{jet}>$  50 GeV in $|\eta^{jet}|<3$
\\ \cline{2-3}
& missing $E_T$ & \etm $>80$ GeV  
\\ \cline{2-3}
& $Z$-mass cut  \ \   & $M_{Z} \pm 10$ GeV
\\ \hline
\end{tabular}
\end{center}
\end{table}

The $m_{1/2}$ boundaries are approximate and, for optimal cuts,
they depend also on $m_{0}$.

Except for Set 4, which we apply for rather massive \chha\chnb,
the $p_T$ cuts on leptons are the same, i.e. $p_T^l>$ 15 GeV.
Set 1 and Set 2 are appropriate for the regions where the
signal is accessible already at low luminosity.
In the regions where we apply the cuts of Set 3 or Set 4
a high luminosity is required, as they correspond to smallest
cross-sections; the luminosity determines the isolation
requirement. With increasing $m_{1/2}$ the jet veto is relaxed,
varying the $E_T^{jet}$ threshold from 25 to 50 GeV. A cut on \etm \ is
applied only for Set 3 and Set 4; and finally, the $M_Z$ cut is 
looser for higher $m_{1/2}$.
 
Table 8.2 gives the SM background cross-sections for these
selection criteria after each cut.
The two main SM backgrounds surviving these cuts are
WZ and $t\bar{t}$. With Set 1 and Set 2 the dominant background
is WZ and with  Set 3 and  Set 4 it is $t\bar{t}$.

Details on each simulated point can be found in \cite{wz}.
The signal significance is defined as the
number of signal events divided by the square root of
all events (SUSY + SM) passing our selection criteria.
With this definition we give: i) evidence for
existence of Supersymmetry in
$3l + no \ jets +$ \etm \ channel; ii) overall
evidence for \chha\chnb \ production and iii) evidence for direct
production of \chha\chnb. The ratio of direct \chha\chnb \ events over
other events surviving the cuts, \nchd /(\nall - \nchd ), is also given.
All investigated points are analyzed in a similar
manner and 5$\sigma$ significance contours are obtained
in the ($m_0, m_{1/2}$) plane, as shown in Fig.~8.8.

In the regions of small $m_0$ and $m_{1/2}$ ($\lappeq 200$ GeV and
$\lappeq 150$ GeV, respectively) indirectly produced \chha\chnb \ 
events contribute significantly in the $3l + no \ jets +$ \etm \
channel. However, this contribution is expected to be less than
$\sim$15$\%$ of the overall \chha\chnb \ production. Therefore, we
can give 5$\sigma$ significance contours for direct \chha\chnb \
production.

\begin{table}[t]

\begin{center}
Table 8.2: SM background cross-sections (in fb)
at each step of applied cuts.

\vspace{3mm}

\begin{tabular}{|c||l||c|c|c|c|c|c|}  \hline
Set of cuts   & Cuts      & WZ & ZZ & \bt & W$tb$ &\zbb  & \bb
 \\ \hline\hline
Set 1 & $p_T^l$  & 111 & 21.7  & 2544 & 139.1 & 731 & 9300
\\ \cline{2-8}
&Isolation  & 102  & 19.7 & 39.8 & 4.09  & 28.8  & $< 0.30$
 \\ \cline{2-8}
&Jet veto & 73.8 & 11.6 & 4.12  & 1.71 & 14.6 & $< 0.15$
\\ \cline{2-8}
&$M_Z$  & 4.19 & 0.89 & 2.18 & 0.86 & 0.76 & $< 0.15$
  \\ \hline \hline
Set 2 & $p_T^l$   & 111 & 21.7  & 2544 & 139.1 & 731 & 9300
\\ \cline{2-8}
&Isolation  & 102  & 19.7 & 39.8 & 4.09  & 28.8  & $< 0.30$
 \\ \cline{2-8}
&Jet veto & 76.1 & 12.7 & 4.84  & 1.90 & 16.1 & $< 0.15$
\\ \cline{2-8}
&$M_Z$  & 4.49 & 0.99 & 2.48 & 1.05 & 0.91 & $< 0.15$   
  \\ \hline \hline
Set 3 &$p_T^l$        & 111   & 21.7 & 2544 & 139.1 & 731 & 9300
\\ \cline{2-8}
& Isolation       & 75.4  & 14.7 & 46.3 & 4.82  & 32.9  & $< 0.45$
 \\ \cline{2-8}
& Jet veto        & 51.3  & 8.69 & 4.77  & 1.91 & 15.7 & $< 0.40$
\\ \cline{2-8}
& \etm \ & 22.0  & 2.61 & 3.01  & 0.48 & 1.42 & $< 0.01$
\\ \cline{2-8}
&$M_Z$      & 1.93  & 0.35 & 2.63 & 0.40 &  0.06 & $< 0.01$
  \\ \hline\hline
Set 4 &$p_T^l$       & 88.4  & 16.6 &  1330  & 65.2 & 378  & 2800
\\ \cline{2-8} 
&Isolation       & 60.5  & 11.4 &  19.0  & 1.33  & 10.3  & $< 0.04$
 \\ \cline{2-8}
&Jet veto        & 46.4  & 7.92 &  3.25  & 0.53 & 4.78   & $< 0.04$
\\ \cline{2-8}
&\etm \  & 5.77  & 0.68 &  0.73  & 0.05 & 0.30   & $< 0.001$
\\ \cline{2-8}
&$M_Z$       & 0.45  & 0.09 &  0.60  & $<0.05$ & $<0.04$  & $< 0.001$
 \\ \hline
\end{tabular}
\end{center}
\end{table} 
 
From Fig.~8.8 one can see that at a low integrated luminosity of
$L_{int}$ = $10^4$ pb$^{-1}$ the direct production of \chha\chnb \
can be distinguished from background up to $m_{1/2}$ $\lappeq$ 150 GeV
for all $m_0$. A further increase of luminosity up to $L_{int}$ = $10^5$
pb$^{-1}$ extends the explorable region only by about $10 \div 20$ GeV
for $m_0 \gappeq 120$ GeV. However,for $m_0 \lappeq 120$ GeV the
gain in parameter space is much larger -- up to $m_{1/2}$ $\lappeq$ 420
GeV, reflecting the more advantageous behavior of the production
cross-section times branching ratio shown in Fig.~2.4.

It should be pointed out that the internal SUSY backgrounds are
particularly dangerous for small $m_0$, $m_{1/2}$, in the region $m_0
\lappeq 200$ GeV,  $m_{1/2} \lappeq 150$ GeV, generating a background
higher or comparable to the SM contribution because of the high
production cross-sections of the relatively light gluinos and squarks.
Softness of these gluino/squark events makes it difficult to suppress
them below the SM level, even with a very strict jet veto requirement
(selection criteria Set 1). For some regions of the ($m_0, m_{1/2}$) 
parameter plane, chargino/neutralino and slepton pair productions are
comparable to gluino/squark contributions. As $m_0$ and $m_{1/2}$
increase, more massive gluinos and squarks result in harder jets,
and decreased cross-sections, and hence the SUSY background becomes
negligible. At the 5$\sigma$ boundary with an integrated luminosity of
$10^5$ pb$^{-1}$, background events dominantly
consist of SM processes. At this boundary the signal to background  
ratio is expected to be $\gappeq$ 0.20 with about 150 signal
events ($L_{int}=10^5$ pb$^{-1}$)
for $m_{0} \gappeq$ 120 GeV. In this region
the dominant SM background is WZ production which cannot
be suppressed by a stronger jet veto or lepton isolation requirements.
For $m_{0} \lappeq$ 120 GeV the signal to background ratio is $\gappeq$
0.60 with about 70 signal events. Here the dominant SM background is
$t\bar{t}$ production. The lower number of signal events
needed to attain a 5$\sigma$ significance and the higher
signal to background ratio in this region is the result
of the harder event kinematics of \chha\chnb,
allowing for more stringent selection criteria (Set 4). As a
consequence the SM contamination is smaller \cite{wz}.

At small $m_0$, the \chha\chnb \ reach in parameter space
is very luminosity-dependent. This region of small $m_0$
is of particular importance, since it is the prefered
one from the cosmological point of view,
where \chna \ is the candidate for the cold dark matter in the Universe.
With the present lifetime of the Universe ($\gappeq 10^{10}$
years) and the estimate of the
relic neutralino abundance at freeze-out time
as a function of ($m_0, \ m_{1/2}$),
one can determine regions of parameter space
where $\Omega h^2 < 1$;
here $\Omega$ is the ratio of the relic neutralino density to
the critical
density which closes the Universe and $h$ is the Hubble constant
scaling factor, $0.5 \lappeq h \lappeq 0.8$.
In Fig.~8.8, the  $\Omega h^2$ = 0.15 and $\Omega h^2$ = 0.4
contours
mark the most probable region of the parameter space,
corresponding to a mixed dark matter scenario
with 60$\%$ of neutralino dark matter.
One can see that \chha\chnb \ direct production
covers a significant part of this prefered region of
parameter space.
 
We can conclude, that:

$\bullet$ by measuring the excess of events
in the $3l + no \ jets +$ \etm channel
over SM and internal SUSY backgrounds,
\chha\chnb \ direct production
can be investigated in mSUGRA
parameter space up to $m_{1/2}$ $\lappeq$ 170 GeV (150 GeV) for all
$m_0$ at a luminosity $L_{int}$ = $10^5$ pb$^{-1}$  ($10^4$ pb$^{-1})$;
 
$\bullet$ with a high integrated luminosity of $L_{int} = 10^5$
pb$^{-1}$ and for $m_0 \lappeq $ 120 GeV,
the discovery region for direct  production
of \chha\chnb \ extends up to $m_{1/2}$ $\lappeq$ 420 GeV.

\newpage  
 
\begin{figure}[hbtp]
\vspace{25mm}  
\hspace*{-10mm}
    \resizebox{17cm}{!}{\rotatebox{0}{\includegraphics{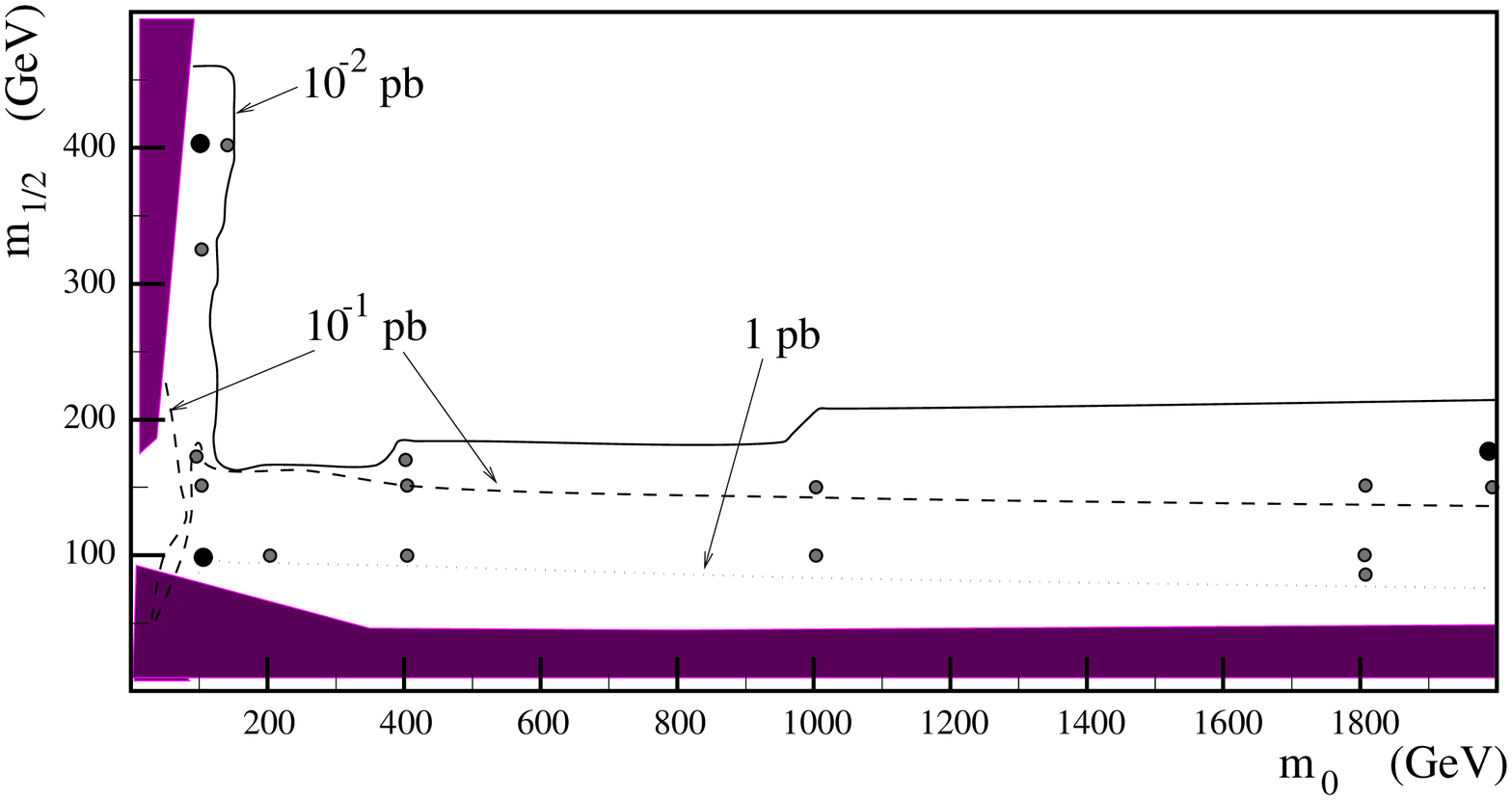}}}
\end{figure}
 
 \vspace{20mm}

Figure 8.1: Simulated SUSY points for chargino/neutralino pair
production. Sigma times branching ratio contours are also shown.
Shaded regions are excluded by theory ($m_0 \lappeq 80$ GeV)
and experiment ($m_{1/2} \lappeq 90$ GeV).

\newpage

\ \ \\

\begin{figure}[hbtp]
\vspace{-30mm}
\begin{center}
    \resizebox{15cm}{!}{\rotatebox{0}{\includegraphics{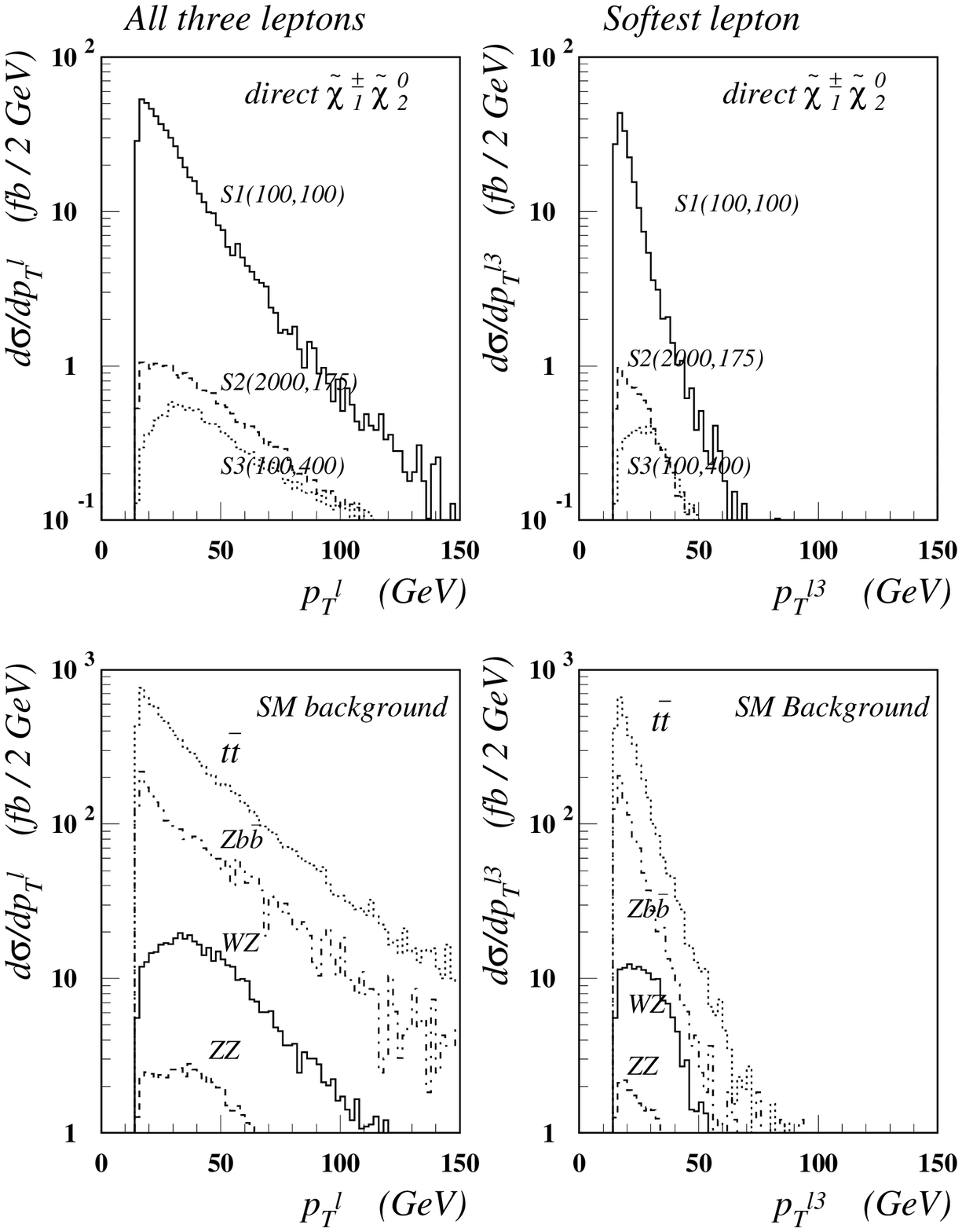}}}
\end{center}
\end{figure}

Figure 8.2: $d\sigma / dp_T^l$ distributions for
signal and SM background. Left plots show distributions for
all three leptons and the right ones  for softest lepton
in the event.

\newpage

\ \ \\

\begin{figure}[hbtp]
\vspace{-40mm}
\begin{center}
  \resizebox{14.cm}{!}{\includegraphics{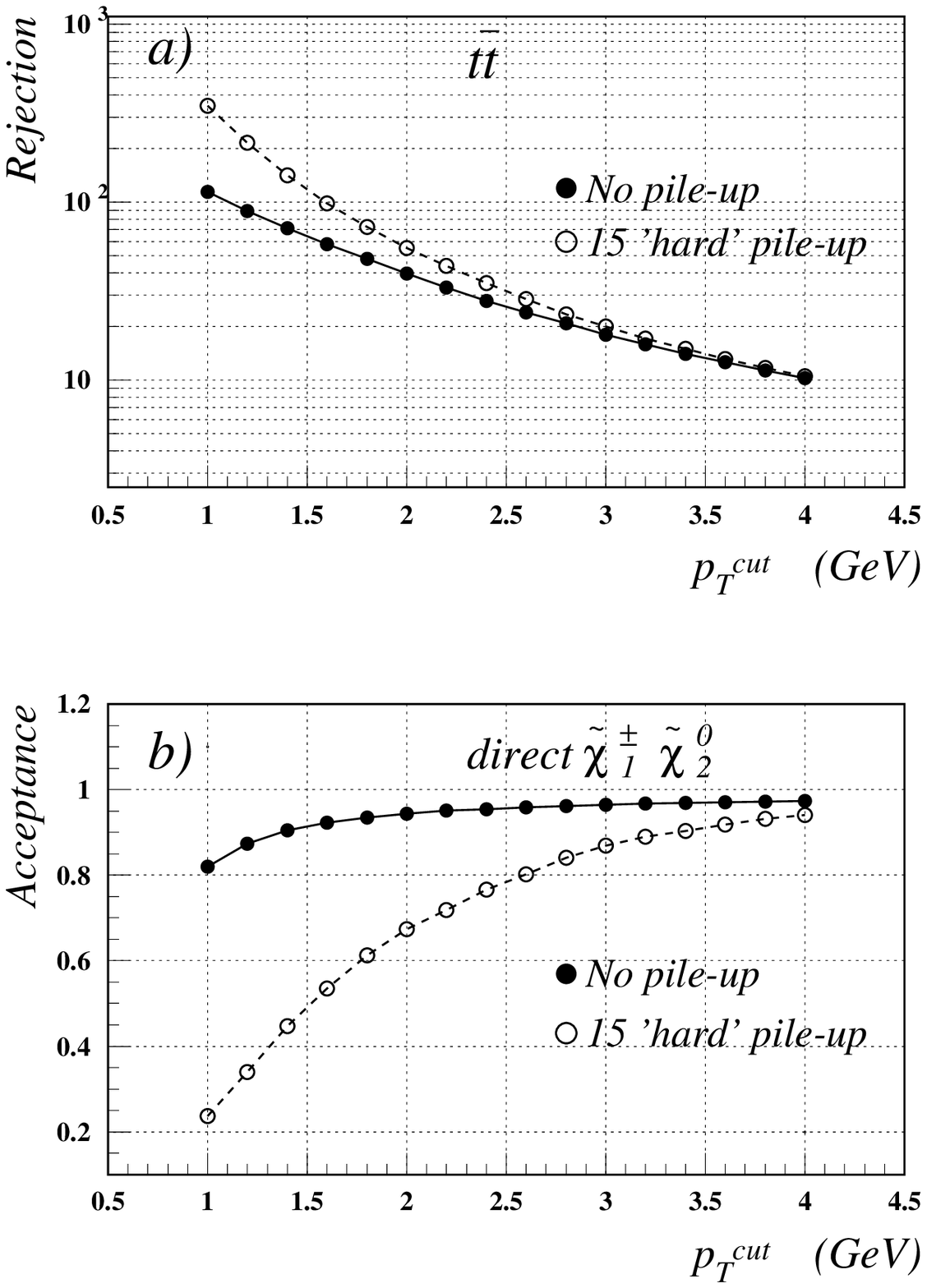}}
\end{center}
\end{figure}

Figure 8.3: Tracker isolation: a) $t\bar{t}$ background rejection
versus $p_T^{cut}$ on charged track, b) signal acceptance versus
$p_T^{cut}$ on charged track. Events are selected with three
leptons of $p_T > 15$ GeV and all of them are required to be isolated.
Full and open circles correspond to low, $L = 10^{33}$
cm$^{-2}$ s$^{-1}$, and high, $L = 10^{34}$ cm$^{-2}$ s$^{-1}$,
luminosity regimes, respectively.

\newpage

\ \ \ \\

\begin{figure}[hbtp]
\vspace{-25mm}  
\begin{center}
    \resizebox{13cm}{!}{\includegraphics{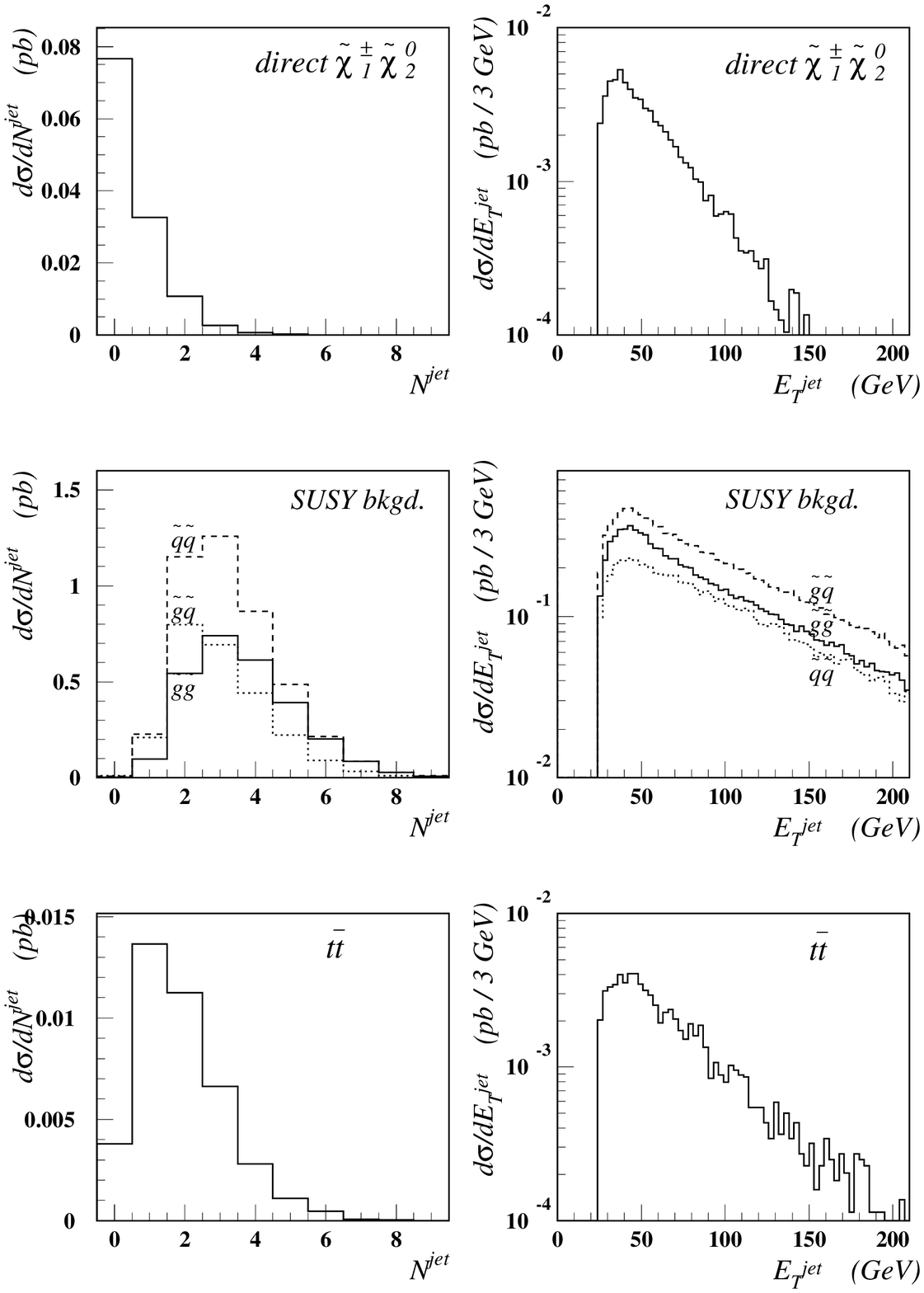}}
\end{center}
\end{figure}

\vspace{-10mm}

Figure 8.4: Jet multiplicity and transverse
energy distributions for the signal and internal SUSY background at mSUGRA
point S1(100 GeV, 100 GeV) and the $t\bar{t}$ background. Events are
selected with three isolated leptons of $p_T > 15$ GeV.

\newpage

\ \\

\vspace{50mm}  

\begin{figure}[hbtp]
\vspace{-30mm}
\begin{center}
    \resizebox{17cm}{!}{\includegraphics{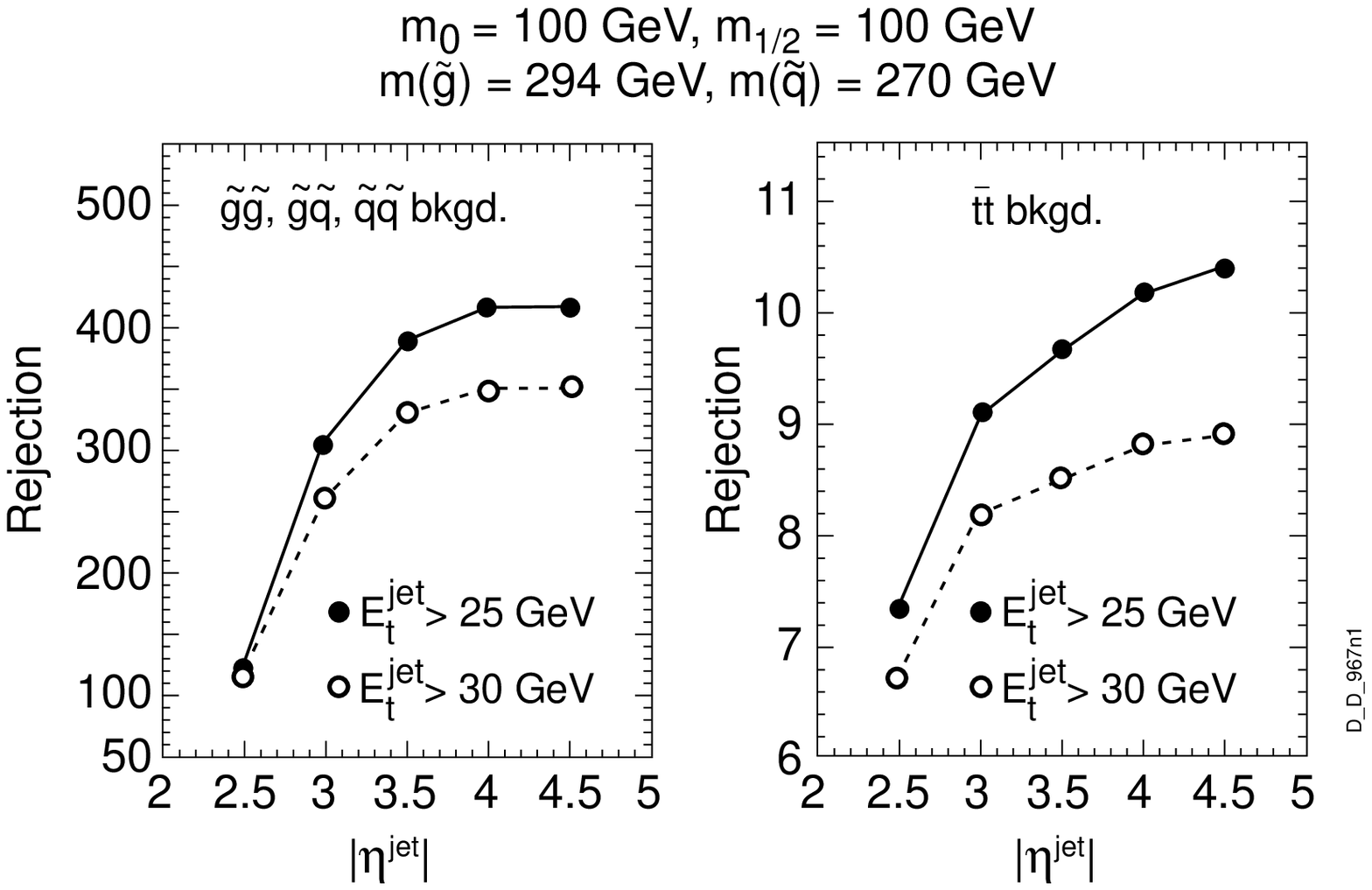}}
\end{center}
\end{figure}   

 \vspace{10mm}

Figure 8.5: Jet veto rejection/acceptance versus $|\eta^{jet}|$
in \chha \chnb \ searches.
Full circles correspond to $E_T^{jet} > 25$ GeV and the open
ones to $E_T^{jet} > 30$ GeV.

\newpage
 
\ \\
 
\begin{figure}[hbtp]
\vspace{-30mm}
\begin{center}
    \resizebox{14cm}{!}{\includegraphics{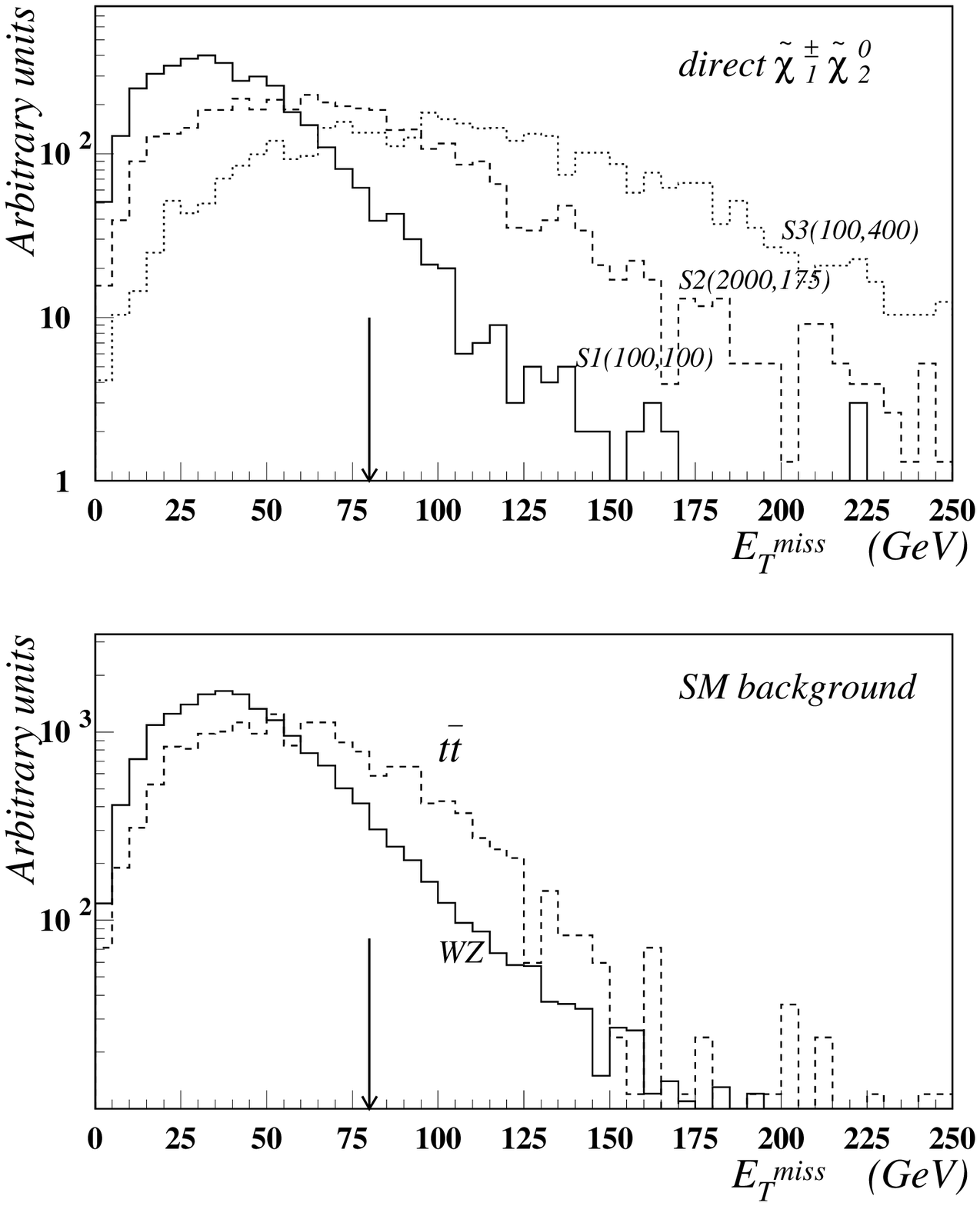}}
\end{center}
\end{figure}
 
Figure 8.6: Missing transverse energy distributions in signal and
background events with three isolated leptons of $p_T > 15$ GeV
and no jets.

\newpage     
 
\ \\
 
\begin{figure}[hbtp]
\vspace{-30mm}
\begin{center}
    \resizebox{15cm}{!}{\includegraphics{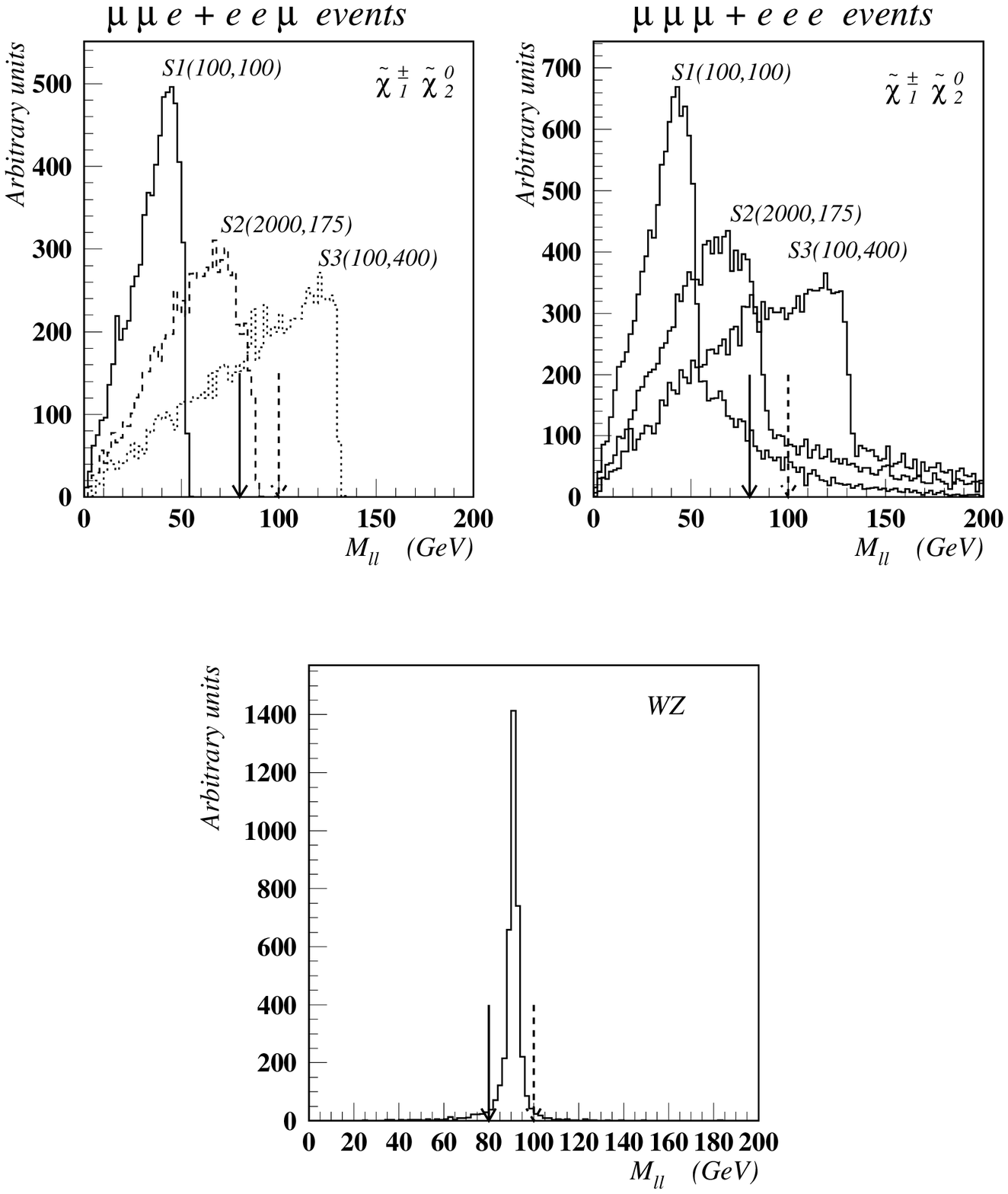}}
\end{center}
\end{figure}

Figure 8.7: Dilepton invariant mass distributions
in signal and WZ background
events with three isolated leptons of $p_T > 15$ GeV
and no jets.

\newpage
 
\ \\
 
\vspace{15mm}

\begin{figure}[hbtp]
\vspace{-30mm}
\hspace*{-5mm}
\resizebox{15.cm}{!}{\rotatebox{90}{\includegraphics{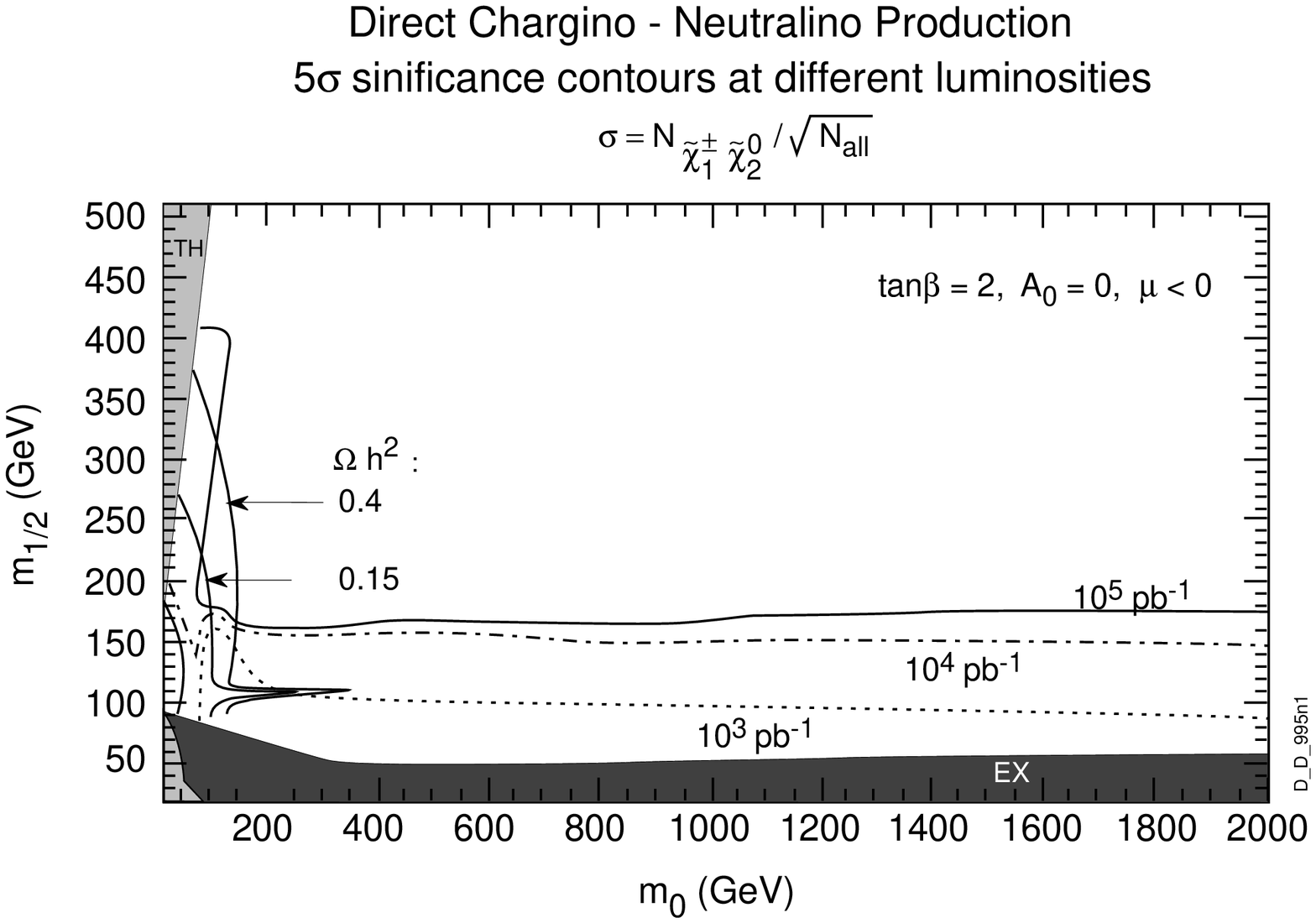}}}
\end{figure}

\vspace{0mm}  

\hspace*{15mm}Figure 8.8: $5\sigma$ significance contours at different
luminosities.

\newpage


\section{Possibility of observing h $\rightarrow b \bar{b}$
in squark and gluino decays}

    The usual way to search for the lightest CP-even Higgs (h)
of the MSSM is through its direct production by gluon fusion or in
association with W, $b\bar{b}$ or $t\bar{t}$ and its decay into
$\gamma\gamma$, ZZ$^{*} \rightarrow 4l$, $\mu\mu$ or $b\bar{b}$
\cite{higgs, tp, seez}. All these production mechanisms are Electroweak.
Each of these channels has some limitations, such as high tan$\beta$ for
the $\mu\mu$ mode,  or high mass, $m_{h}$ $\gsim$ 130 GeV and large stop
mixing for the ZZ$^{*} \rightarrow 4l$ mode, for example. The most
general way is to search for inclusive h~$\rightarrow$ $\gamma\gamma$,
which, with $L_{int}=10^{5}$ pb$^{-1}$, would allow us to explore a
domain approximately given by $m_A \gsim 250$ GeV and tan$\beta \gsim 3$
\cite{higgs}. With the h~$\rightarrow$ $\gamma\gamma$ mode we expect 
a signal on top of a large irreducible $\gamma\gamma$ background with a
signal to background  ratio of $\lsim 1/20$.
The instrumental requirements on the ECAL are very demanding, a
$\gamma\gamma$ effective mass resolution better than $\simeq$1 GeV at
$m_{h} \simeq$ 100 GeV \cite{seez}. In the associated production modes Wh
and $t\bar{t}$h with  h~$\rightarrow$ $\gamma\gamma$, the $S/B$ ratio is
of order 1 and calorimeter performance is less demanding:
but these channels require $\geq$ 10$^{5}$ pb$^{-1}$ \cite{associated}.
Evaluations of ways to exploit the h~$\rightarrow$ $b\bar{b}$ decay mode
in the Wh or $t\bar{t}$h final states leave little hope $-$ no way
to get a really significant signal has been found up to now \cite{wh}.

It is well known that the MSSM h can be abundantly produced in the
decays of charginos and neutralinos (primarily $\tilde{\chi}_{2}^{0}$).
In turn, the $\tilde{\chi}_{2}^{0}$ is a typical decay product of
squarks and gluinos (see Figs.~2.5a,b) which are also produced
abundantly. The idea is thus to use strongly produced 
$\tilde{q}$/$\tilde{g}$ and to look for the dominant decay mode 
h~$\rightarrow$ $b\bar{b}$, if $\tilde{\chi}_{i}^{0}$ $\rightarrow$
$\tilde{\chi}_{j}^{0}$h is kinematically allowed, and require
$E_{T}^{miss}$ and jet multiplicity to suppress the background
\cite{htobb, atlashbb}.

In this study we fix only one of the mSUGRA parameters, $A_{0}$ =
0, and vary the other four: $m_{0}$, $m_{1/2}$, tan$\beta$, $sign(\mu)$.
The most pronounced variation of physics quantities (masses, branching
ratios, cross-section) is with $m_{0}$ and $m_{1/2}$, thus
the investigation is done in the $m_{0}$, $m_{1/2}$ parameter plane
for several representative values of tan$\beta$ and $sign(\mu)$.
When $M_{\tilde{\chi}_{2}^{0}} - M_{\tilde{\chi}_{1}^{0}} > m_{h}$
the main source of Higgs is the decay
$\tilde{\chi}_{2}^{0}$ $\rightarrow$ $\tilde{\chi}_{1}^{0}$h.
This decay has a significant ($\simeq$20-90\%) branching ratio in a
large portion of the parameter space, see Fig.~2.17.
In regions of $m_{1/2} \lsim 200$-300 GeV
(depending on tan$\beta$) the three body decays
$\tilde{\chi}_{2}^{0}$ $\rightarrow$ $\tilde{\chi}_{1}^{0}$ $q\bar{q}$,
$\tilde{\chi}_{2}^{0}$ $\rightarrow$ $\tilde{\chi}_{1}^{0}$
$l^{+}l^{-}$ take over because the
$\tilde{\chi}_{2}^{0}$ $\rightarrow$ $\tilde{\chi}_{1}^{0}$h
decays are not kinematically allowed.
The boundary at $m_{0}$ $\sim 200$-300 GeV, on the other hand,
corresponds to the opening of two-body cascade modes, e.g.
 $\tilde{\chi}_{2}^{0}$ $\rightarrow$ $\tilde{l}_{L,R}^{\pm}l^{\mp}$
$\rightarrow$ $\tilde{\chi}_{1}^{0}$ $l^{+}l^{-}$.
Figure 2.17 also shows
that the tan$\beta$ = 2, $\mu$ $<$ 0 case looks the most promising   one,
whereas  the tan$\beta$ = 10, 30, $\mu$ $<$ 0 seem the least
favorable among the possibilities considered, the
$\tilde{\chi}_{2}^{0}$ $\rightarrow$ $\tilde{\chi}_{1}^{0}$h
branching ratio being the smallest on average in these latter cases.
Depending on tan$\beta$,
there is potentially a gap between the upper
limit explored up to now or foreseeable at LEP2 and Tevatron
at $m_{1/2}$ $\simeq$ 100 GeV ($m_h \lsim$ 100 GeV) and the lower
limit at $m_{1/2}$ $\simeq$ 200-250 GeV where $\tilde{\chi}_{2}^{0}$
$\rightarrow$
$\tilde{\chi}_{1}^{0}$h opens up. This comes about because within
mSUGRA there are simple mass relations:
$M_{\tilde{\chi}_{2}^{0}} \simeq 2M_{\tilde{\chi}_{1}^{0}} \simeq
m_{1/2}$, and the LSP mass is typically $\gsim$~100 GeV in the regions
where $\tilde{\chi}_{2}^{0}$ $\rightarrow$ $\tilde{\chi}_{1}^{0}$h is
allowed. The range of the h mass varies from
$\simeq$~80 to $\simeq$~120 GeV and the exact upper limit is determined
by the order to which the radiative corrections have been calculated.
 
With increasing $m_{0}$ (at fixed $m_{1/2}$) squarks become more massive
than gluino at some point (see, e.g. Fig.~2.16) and thus gluino
decays into squarks are not allowed anymore, decreasing significantly
the yield of \chnb \ from squark decays. Furthermore, the
B($\tilde{q}$ $\rightarrow$ $\tilde{\chi}_{2}^{0}q$) also decreases with
increasing $m_{0}$ because the decay  $\tilde{q}$ $\rightarrow$ 
$\tilde{g} q$ plays an increasingly important role. These factors, with
the variation of the total production cross-section (shown in Fig.~2.15)
and the variation of the kinematics of the decay chains,
require that for a quantitative evaluation of the
regions where h~$\rightarrow$ $b\bar{b}$ could be observed
one needs to investigate the parameter space point by point and optimise
cuts accordingly \cite{htobb}.

We considered the following SM background processes:
QCD 2 $\rightarrow$ 2 (including $b\bar{b}$), $t\bar{t}$ and W$tb$. The
$\hat{p}_{T}$ range of all these processes is subdivided into several
intervals to facilitate  accumulation of statistics in the
high-$\hat{p}_{T}$ region: 100-200 GeV (except for QCD), 200-400 GeV,
400-800 GeV and  $>$ 800 GeV. The accumulated statistics for all
channels correspond to 100 fb$^{-1}$, except for the QCD jet
(instrumental mismeasurement) background where the statistics correspond
to 0.2 - 50 fb$^{-1}$ depending on the $\hat{p}_{T}$ interval.  
It is very time consuming to produce a representative sample of QCD jets
in the low-$p_{T}$ range, since the cross section is huge and
we need extreme kinematical fluctuations and non-Gaussian tail
instrumental contributions for this type of background to be
within the signal selection cuts. On the other hand, our simulation of
instrumentally induced missing transverse energy is not yet fully
reliable as it strongly depends on the still evolving estimates of dead
areas, volumes due to services, non-Gaussian calorimeter response, etc.
So we cannot go confidently below $\simeq$100 GeV with the cut on
$E_{T}^{miss}$, where the QCD jet background becomes the dominant
contribution.

Initial requirements for all the SM and SUSY samples are the following:
\begin{itemize}
 \item at least 4 jets with $E_{T}^{jet}$ $>$ 20 GeV in 
 $\mid\eta^{jet}\mid$ $<$ 4.5,
 \item $E_{T}^{miss}$ $>$ 100 GeV,
 \item $Circularity$ $>$ 0.1.
\end{itemize}
No specific requirements are put on leptons.
If there are isolated muons or electrons with $p_{T}^{\mu, e}$ $>$ 10 GeV
within the acceptance they are also recorded.
The term ``isolated lepton''  here means satisfying simultaneously the 
following two requirements:
\begin{itemize} 
\item no charged particle with $p_{T}$ $>$  2 GeV in a cone $R$ = 0.3
around the direction of the lepton,
\item $\Sigma E_{T}^{calo}$ in the annular region 0.1 $< R <$ 0.3
around the lepton impact point has to be less than 10\% of the lepton
transverse energy.
\end{itemize}
At a later stage we investigate whether  selection or vetoing on 
isolated leptons can be used to help suppress the SM 
or internal SUSY backgrounds.

In this study we used an evaluation of the $b$-tagging performance
of CMS from impact parameter measurement in the tracker \cite{microv}.
An example is shown in Fig.~4.11. The $b$-tagging efficiency and
mistagging probability for jets is evaluated as a function of $E_{T}$
and jet pseudorapidity. It is obtained from a parameterized Gaussian
impact parameter resolution, dependent on the pixel point resolution,
radial position of the pixel layers and the effects of multiple scattering
on intervening materials, with the addition of a non-Gaussian impact
parameter measurement tail based on CDF data. Although there are some
specific assumptions made concerning the tails of the
impact parameter distributions extending beyond the parameterized Gaussian
parts in \cite{microv}, the presently on-going study of the expected
impact parameter resolution, with full pattern recognition and track
finding in CMS \cite{sasha} is in a very good agreement with the results
used as an input in the present study.
 
The $b$-jet tagging efficiency is parameterized as a function of
$E_{T}^{jet}$ in 3 intervals of $\eta$: 0 - 1, 1 - 1.75
and 1.75 - 2.4. A typical $b$-tagging efficiency at $\eta$ $\simeq$ 0
for $E_{T}^{jet}$ = 40 GeV is $\simeq$ 30\%, reaching
a maximal efficiency of $\simeq$ 60\% for high-$E_{T}$ jets.
Charm-jets have a typical tagging efficiency of
about 10\%, and for light quarks and gluons the $b$-tagging (mistagging)  
efficiency reaches a maximum of about 3\% for high-$E_{T}$ jets.
At the analysis stage the jets produced by the event generator are
``$b$-tagged'' according to the tagging efficiency for $b$-jets
and mistagging probability for quark and gluon jets.
If not specially mentioned otherwise,
h~$\rightarrow$ $b\bar{b}$ decay jets are tagged only
in the barrel $\mid\eta\mid$ $<$ 1.75 interval
and in all di-jet mass distributions
we take only one pair of $b$-jets per event, those closest
to each other in $\eta - \varphi$ space. No attention is paid to
jet ``charge'' or ``leading charge''.
More details can be found in \cite{htobb}.

As an example let us consider a representative mSUGRA point
$m_0 = m_{1/2} = 500$ GeV, tan$\beta$ = 2, $\mu$ $<$ 0.
Here the masses of relevant sparticles are:
$m_{\tilde{g}}$ = 1224 GeV,
$m_{\tilde{u}_{L}}$ = 1170 GeV, $m_{\tilde{\chi}_{2}^{0}}$ = 427 GeV,
$m_{\tilde{\chi}_{1}^{0}}$ = 217 GeV, $m_{h}$ = 89.7 GeV.
In Fig.~9.1 we compare
kinematical distributions for this point
with the SM background for $L_{int} = 100$ fb$^{-1}$.
The $E_{T}^{miss}$ distributions show
the most pronounced difference between signal and background, 
this then being the most important variable for the background
suppression. A significant difference exists also in the
jet multiplicity distributions.
A specific optimisation of cuts on $E_{T}^{miss}$,
$E_{T}^{jet}$, etc., is however performed in various regions of
mSUGRA parameter space.

Figure 9.2 shows the $b$-$b$ (jet-jet) mass distribution
for the sum of the SUSY and SM events assuming an ideal
$b$-tagging performance, i.e. 100\% $b$-tagging efficiency with
0\% mistagging probability for jets. The cuts are 
$E_{T}^{miss}$ $>$ 400 GeV and $E_{T}^{jet}$ $>$ 40 GeV.
The SM background shown in Fig.~9.2 separately as a shaded
histogram is small compared to the internal SUSY background.
The $b\bar{b}$ mass distribution is fitted with
a sum of a Gaussian and a quadratic polynomial.
The width of the signal peak is determined entirely by the
jet-jet effective mass resolution of the
detector (the intrinsic h width varies
from 3.2 to 4.3 MeV over the entire mSUGRA parameter space).
The position of the peak is shifted to lower mass since energy
losses in jets (finite cone size, neutrinos generated in hadronic
showers, etc.) are not corrected for.
The three distributions in the lower half in Fig.~9.2
are for events in the $M_{bb}$ mass window 70 - 100 GeV to
illustrate the kinematics of the h~$\rightarrow$ $b\bar{b}$ decay.
The average $b$-jet transverse energy is $\sim$110 GeV, and the $b$-jets
are very central as they result from the decay of massive ($\sim$1200
GeV) squarks and gluinos which are centrally produced. Clearly,
$b$-tagging beyond $\mid\eta^{jet}\mid$ $\simeq$ 1.5 is not very helpful
in this search.

If one now applies the nominal $b$-tagging performance expectations
described previously, we obtain the $b\bar{b}$ mass distribution given
in Fig.~9.3a. Under these nominal conditions and with 100 fb$^{-1}$ the
expected signal significance calculated as $S/\sqrt{B}$
within a $\pm 1\sigma$ interval around the peak value is 18.3.
The background is dominated by the internal SUSY background
which is largely irreducible, originating mainly from additional real
$b$-jets in the event, rather than from mistagged jets.
An improvement could possibly be obtained by taking into account ``jet
charge'' and thus enriching the sample in $b\bar{b}$ pairs as compared to
$b\bar{b}$ + $bb/\bar{b}\bar{b}$. This however requires track
reconstruction and charge determination
for fast tracks in jets, a task still under study in CMS
\cite{sasha}.

Figures 9.3-9.5 illustrate the expected sensitivity of the
h~$\rightarrow$ $b\bar{b}$ signal to various instrumental factors
\cite{htobb}. In Fig.~9.3b all $bb$-combinations per event are
included in the distribution (instead of just the one with the closest
distance in $\eta - \varphi$ space) and the signal significance is now
worse by $\simeq$12\%. Figures 9.3c and 9.3d
show the effect of increasing, or decreasing, the $b$-tagging efficiency
by 15\% in absolute value, whilst keeping mistagging at the nominal
level. The dependence on the $b$-tagging efficiency is very pronounced,
the signal significance varying by respectively $\pm$25\%.
This is not surprising as the $b$-tagging efficiency enters
quadratically in the number of signal events.
Figures 9.4a and 9.4b illustrate the effect
of the $\eta$ acceptance of the silicon pixel detector.   
As can be expected on the basis of Fig.~9.2,
increased acceptance beyond $\mid\eta^{jet}\mid$ $\simeq$ 1.5 does not
bring a significant improvement as the signal $b$-jets are very central.
With tagging acceptance increasing from $\mid\eta^{jet}\mid = 1.0$ to 
1.75 and 2.4, signal significance improves from 17.2 to 18.3 and 18.8,
respectively. Figures 9.4c and 9.4d
illustrate the dependence on tagging purity, where
one keeps the nominal $b$-tagging efficiency unchanged, but varies the
mistagging probability for all non-$b$-jets.   
Increasing the mistagging probability by a factor of 3 degrades
signal significance by $\simeq$25\%.
 
Since the observed width of the h~$\rightarrow$ $b\bar{b}$ peak
is entirely determined by the calorimetric jet-jet effective mass
resolution, one can expect
dependence of the signal observability on the energy resolution.
This is illustrated in Fig.~9.5.
The upper plot is for the nominal HCAL performance, corresponding to
$\sigma_{E}/E = 82\% /\sqrt{E} \oplus 6.5\%$ at $\eta$ = 0 and
the lower plot corresponds to the HCAL energy resolution
degraded to $120\% /\sqrt{E} \oplus 10\%$.
The resolution of the Gaussian fit to the Higgs peak is 7.6 GeV in   
the first case and 11 GeV in the second. The difference in $S/B$ between
Figs.~9.5a and 9.5b is significant, but not dramatic. This is due to the
fact that the h~$\rightarrow$ $b\bar{b}$ pair has a significant boost,
$p^{b\bar{b}}_{T}$ $\sim$ 200 GeV (Fig.~9.2) and the jet-jet opening
angle, whose measurement precision is determined by calorimeter
granularity plays an important role in the effective mass
resolution. The mass resolution also contains irreducible contributions
from final state gluon radiation, jet cone size and fragmentation
effects.

Figure 9.6 illustrates the h~$\rightarrow$ $b\bar{b}$ signal
observability as a function of $m_{h}$ over the allowed range of
$\sim$80 to 120 GeV. For this we select the points
all at the same $m_{0} = m_{1/2}$ = 500 GeV, but different values of
tan$\beta$ and $\mu$, with nominal $b$-tagging performance and
$L_{int}=100$ fb$^{-1}$. One can clearly see that the signal becomes
broader with increasing Higgs mass since the jet $E_{T}$ $>$ 40 GeV
cut-off does not play such a role as for a low-mass.

It is worth mentioning, that a significant part of the
mSUGRA parameter space
can be explored already at a very low luminosities.
For example, Fig.~9.7 shows the expected Higgs signal
with $L_{int} = 3$ fb$^{-1}$ for a few points.
All the distributions in this Figure are obtained with
the cuts: $E_{T}^{miss}$ $>$ 200 GeV and $E_{T}^{jet}$ $>$ 40 GeV.
For comparison, searches in the h~$\rightarrow \gamma \gamma$
channel at the corresponding mSUGRA points would require $\sim$10
times higher integrated luminosity \cite{htobb}.
 
Let us turn to the main problem,
how general is the possibility to observe this h~$\rightarrow$ $b\bar{b}$
signal in $\tilde{q}$, $\tilde{g}$ decays.
Figure 9.8
shows the domain of parameter space where the h~$\rightarrow$ $b\bar{b}$
signal is visible with $S/\sqrt{B} > 5$
for the case tan$\beta = 2$, $A_{0} = 0$,
$\mu < 0$ with nominal $b$-tagging
performance of CMS, for 10 and 100 fb$^{-1}$ integrated luminosities.
The isomass curves for the CP-even (h) and CP-odd (A) Higgses
are also shown in Fig.~9.8 by the dash-dotted lines. The bold broken
line denotes the region where B($\tilde{\chi}_{2}^{0}$ $\rightarrow$
$\tilde{\chi}_{1}^{0}$h) = 50\%.
The shaded regions along the axes denote the present theoretically (TH) or
experimentally (EX) excluded regions of  parameter space
not yet including LEP 96/97 results.
The LEP2 and Tevatron (with 1 fb$^{-1}$) sparticle reaches are also
shown by solid lines. With $L_{int} \leq 1$ fb$^{-1}$
the threshold of visibility of h~$\rightarrow$ $b\bar{b}$
at lowest $m_{1/2} = 170$-180 GeV
corresponds to $m_{\tilde{g},\tilde{q}}$ $\geq$ 400-450 GeV,
i.e. it begins just where the Tevatron searches will stop with $\sim$5
fb$^{-1}$.
With 100 fb$^{-1}$ the reach extends up to $m_h \simeq 90$ GeV,
$m_A \lsim 1500$ GeV.
Figures 9.9 and 9.10 show the domains of h signal visibility
for tan$\beta = 10$, $\mu < 0$ and tan$\beta = 30$, $\mu > 0$,
respectively.
In both cases there seems to be a significant observability gap
between the upper reach of LEP2 or the Tevatron and the
LHC/CMS low $m_h$ reach using
this method to look for h.
This has to be studied in more detail in the future.

We conclude that
the search for the h~$\rightarrow$ $b\bar{b}$ decay, when the
lightest Higgs
is produced in the cascade decays of the strongly interacting
sparticles,
seems to be a promising channel.
The large rejection factor needed to suppress backgrounds and achieve a $S/B$
of $\sim$1 is here provided by the $E_{T}^{miss}$ cut. Nothing similar
can be obtained in the search for the SM Higgs in H $\rightarrow$ $b\bar{b}$
nor for MSSM h~$\rightarrow$ $b\bar{b}$ in inclusive
h, Wh and $t\bar t$h final states.
mass in the $\simeq$450
to 700 GeV range for this search to be possible.
The study carried out here in the mSUGRA framework shows
that there is a significant domain of parameter space, just beyond
the $\tilde{g}$/$\tilde{q}$ mass reach of the Tevatron ($\sim$400-450
GeV), where observation of the h~$\rightarrow$ $b\bar{b}$ decay would be
possible with an integrated luminosity of only 1-3 fb$^{-1}$.
This parameter space domain can be significantly extended
with 100 fb$^{-1}$, where $\tilde{g}$, $\tilde{q}$ with masses in 1.5 TeV
range are probed. The h mass range from $\simeq$80 GeV up to
$\simeq$125 GeV can be covered.

Our investigations show that for the observation of the
h~$\rightarrow$ $b\bar{b}$ signal a calorimetric energy resolution of
$\simeq 100\% /\sqrt{E} \oplus 10\%$ is adequate,
but it should not be significantly worse than this.
The signal visibility depends most critically on the
$b$-tagging efficiency and to a lesser extent on the
mistagging probability; the acceptance of the $b$-tagging
pixel devices is not a real issue, provided the coverage is no smaller
than $\mid\eta\mid$ $\simeq$ 1.5. A significant improvement
in $b$-tagging efficiency could be obtained with a third pixel layer,
or by having pixel layer at a radius of 4 cm even for high luminosity
running, to be replaced every (few) year(s).
The mistagging probability depends on the non-Gaussian part of the impact
parameter measurement distribution, and thus on
the overall pattern recognition performance in the region close to the
beam, e.g. on the balance between Si and MSGC layers in the tracker.
Optimisation of this aspect might then involve
increasing the number of Si layers from 4 to 5, possibly
6, reducing correspondingly the MSGC layers and deserves a dedicated
study.

If this method is to become a viable alternative
to the h~$\rightarrow$ $\gamma\gamma$ search, it is important to evaluate
how general are the results of the present study,
i.e. what happens outside the mSUGRA scheme where
masses are not so constrained, for example in the MSSM.
Presumably what is found here in the framework of mSUGRA-MSSM is
valid as soon as the $\tilde{g}/\tilde{q}$
$\rightarrow$ $\tilde{\chi}^{0}_{2}$ $\rightarrow$
$\tilde{\chi}^{0}_{1}$h, or even more generally, the
$\tilde{g}/\tilde{q}$
$\rightarrow$ $\tilde{\chi}^{0}_{i}$ $\rightarrow$
$\tilde{\chi}^{0}_{j}$h
chains are kinematically allowed and the
$\tilde{\chi}^{0}_{i} \tilde{\chi}^{0}_{j}$h couplings non-vanishing.
A particular point of interest in this respect is the gap
at the lower  $\tilde{q} / \tilde{g}$
masses, between the domains where SUSY can be explored at the Tevatron with
5-6 fb$^{-1}$, the LEP2 reach in terms of $m_h$ and the lower mass reach
of this channel, if it exists in mSUGRA, can it be overcome in the MSSM
where mass relations are less rigid?

\newpage
 
\ \\
        
\vspace{25mm}
       
\begin{figure}[hbtp]
\begin{center}
\vspace{-30mm}
\hspace*{0mm}
\resizebox{16cm}{!}{\rotatebox{0}{\includegraphics{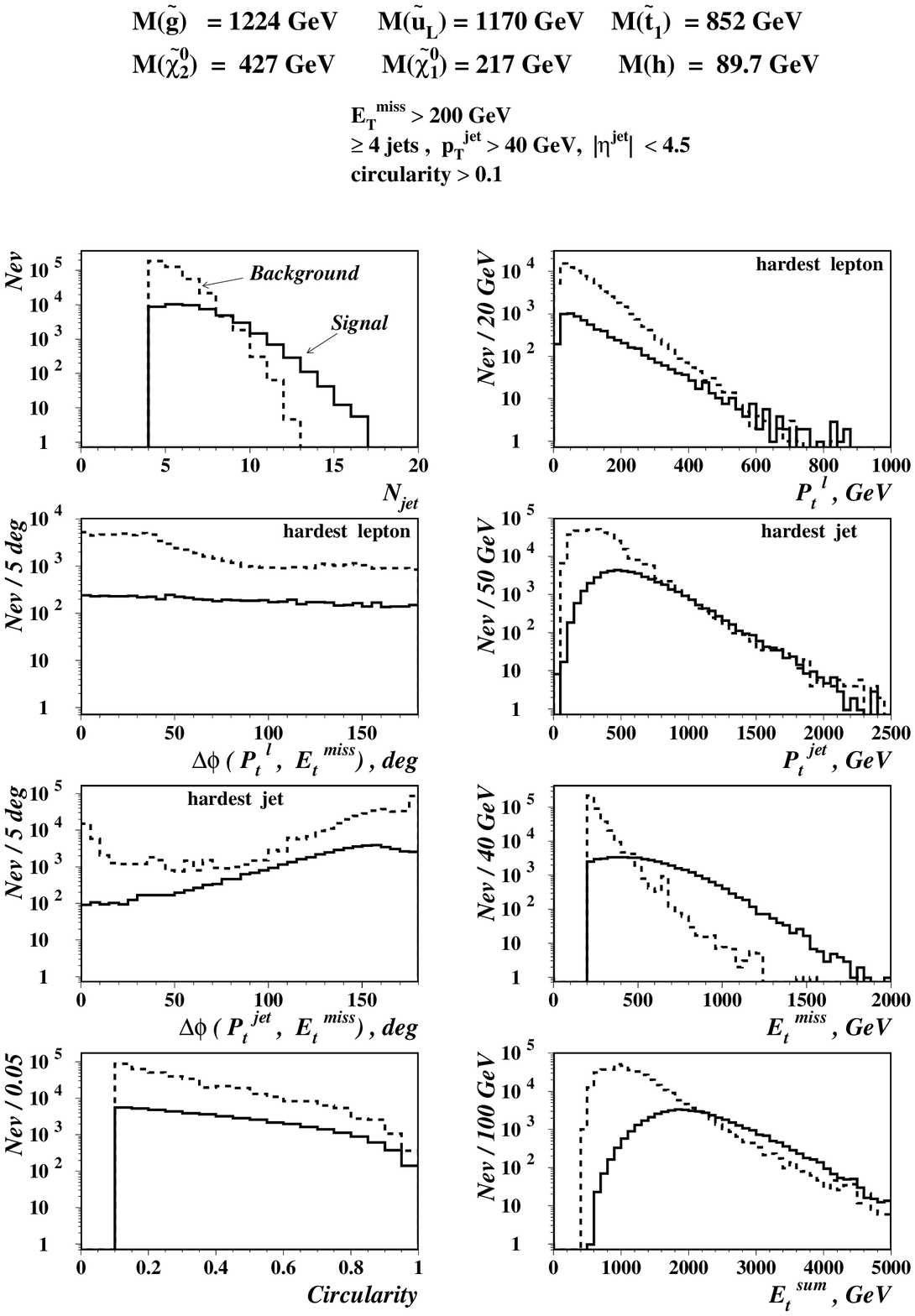}}}
\end{center}
\end{figure}

Figure 9.1:
Comparison of mSUGRA signal at $m_0=m_{1/2}=$500 GeV, tan$\beta$=2,
$\mu <$ 0. and SM background distributions.

\newpage
 
\ \\
 
\vspace{-10mm}

\begin{figure}[hbtp]
  \begin{center}
    \resizebox{15cm}{!}{\includegraphics{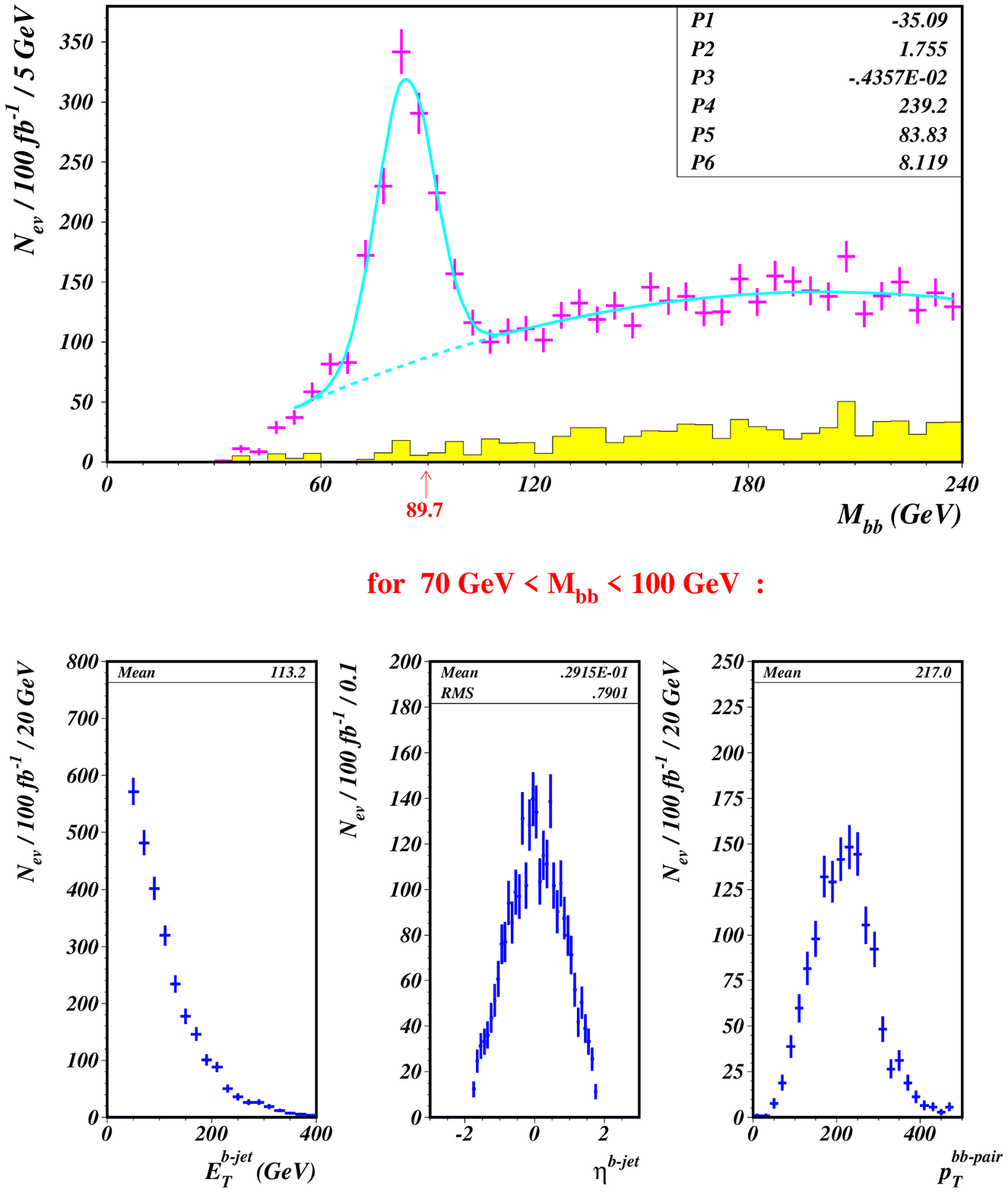}}
  \end{center}
\vspace{-10mm}
\end{figure}  

Figure 9.2: $b$-jet related distributions for the same point as in
Fig.~9.1 in case of ideal tagging performance.
The three distributions in the lower part
are for events in the $M_{bb}$ mass window 70 - 100 GeV.

\newpage
 
\ \\
 
\vspace{-20mm}

\begin{figure}[hbtp]
  \begin{center}
    \resizebox{14cm}{!}{\includegraphics{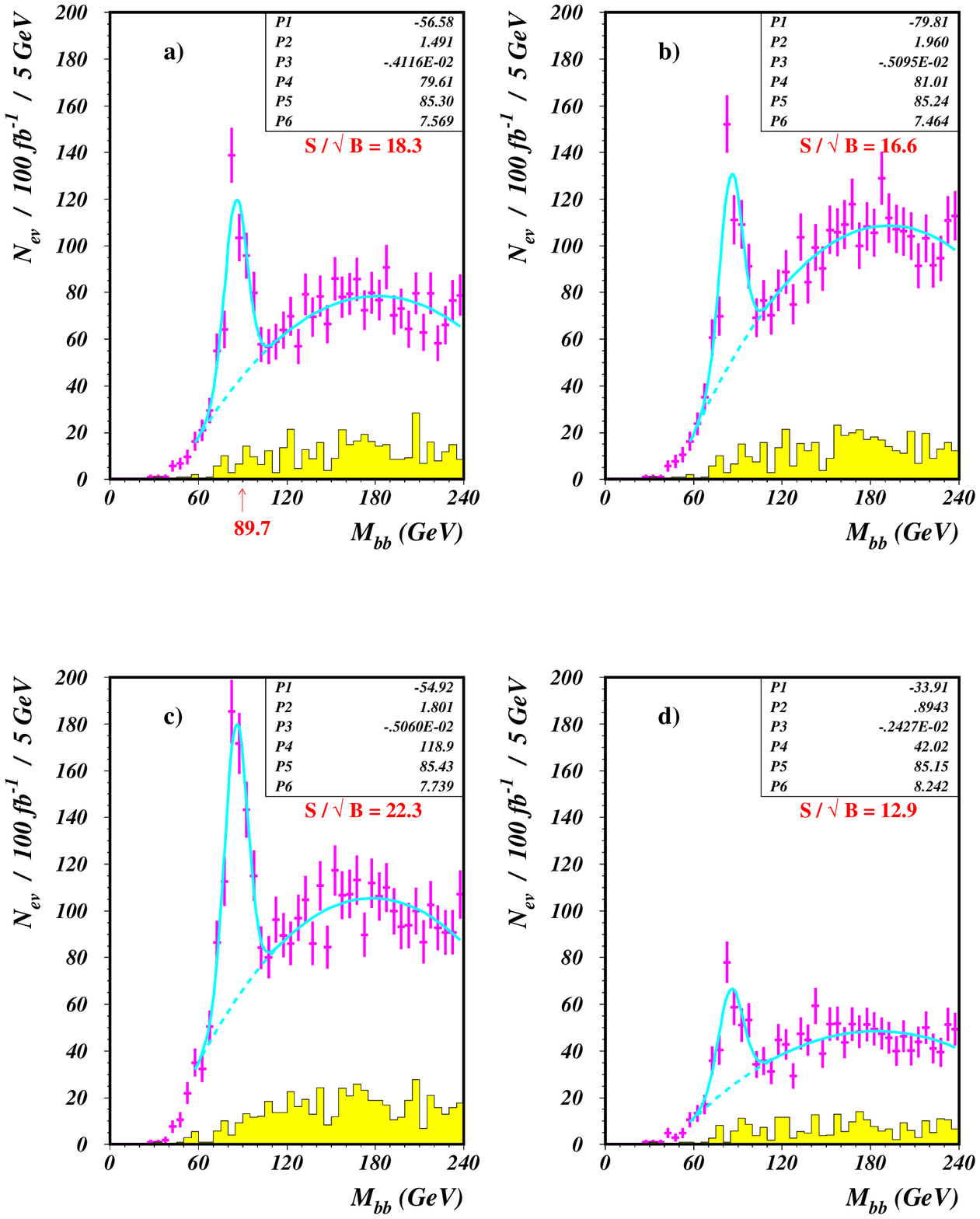}}
  \end{center}
\vspace{-10mm}
\end{figure}

Figure 9.3: Influence of the various instrumental factors
on signal observability in the same parameter space point as in Fig.~9.2
     with 100 fb$^{-1}$:
     a) with ``nominal'' $b$-tagging performance and data selection,
     b) all $bb$-combinations per event are included in histogram,
     c) $b$-tagging efficiency increased by 15\%  and
     d) $b$-tagging efficiency decreased by 15\%. The nominal mistagging
probability is assumed in all cases.

\newpage
 
\ \\
 
\vspace{-20mm}

\begin{figure}[hbtp]
  \begin{center}
    \resizebox{14cm}{!}{\includegraphics{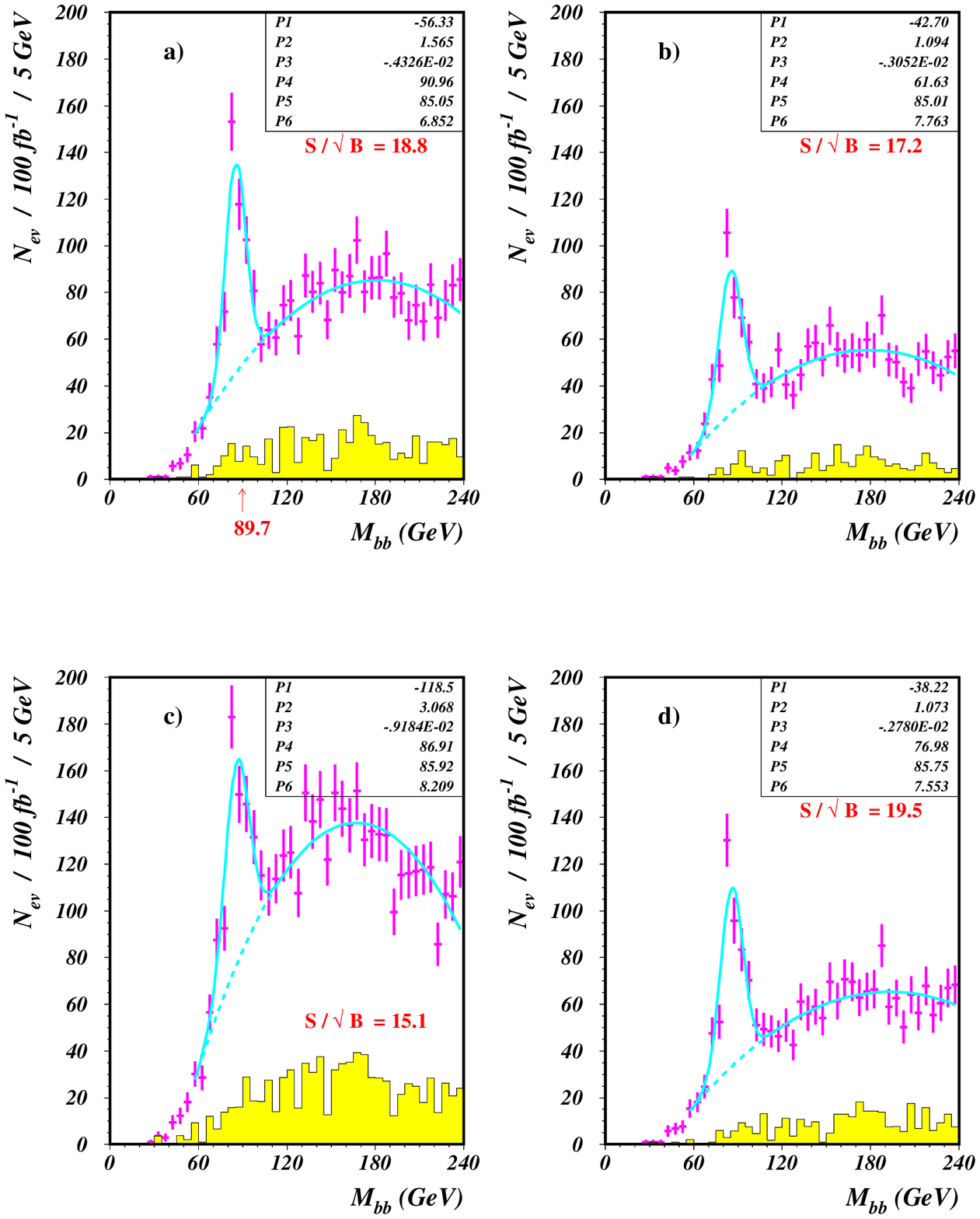}}
  \end{center}
\vspace{-10mm}
\end{figure}  

Figure 9.4:
Influence of various instrumental factors on
  h~$\rightarrow$ $b\bar{b}$ signal visibility in the same parameter
     space point as in Fig.~9.2:
    a) tagging acceptance increased compared to ``nominal''
     from $\mid\eta\mid$ $<$ 1.75 up to  $\mid\eta\mid$ $<$ 2.4,
    b) tagging acceptance decreased to $\mid\eta\mid$ $<$ 1.0,
    c) mistagging probability increased by a factor of 3 and  
    d) mistagging probability decreased by a factor of 2 relative to
nominal expectations.
 
\newpage
 
\ \\
    
\vspace{-20mm}

\begin{figure}[hbtp]
  \begin{center}
    \resizebox{14cm}{!}{\includegraphics{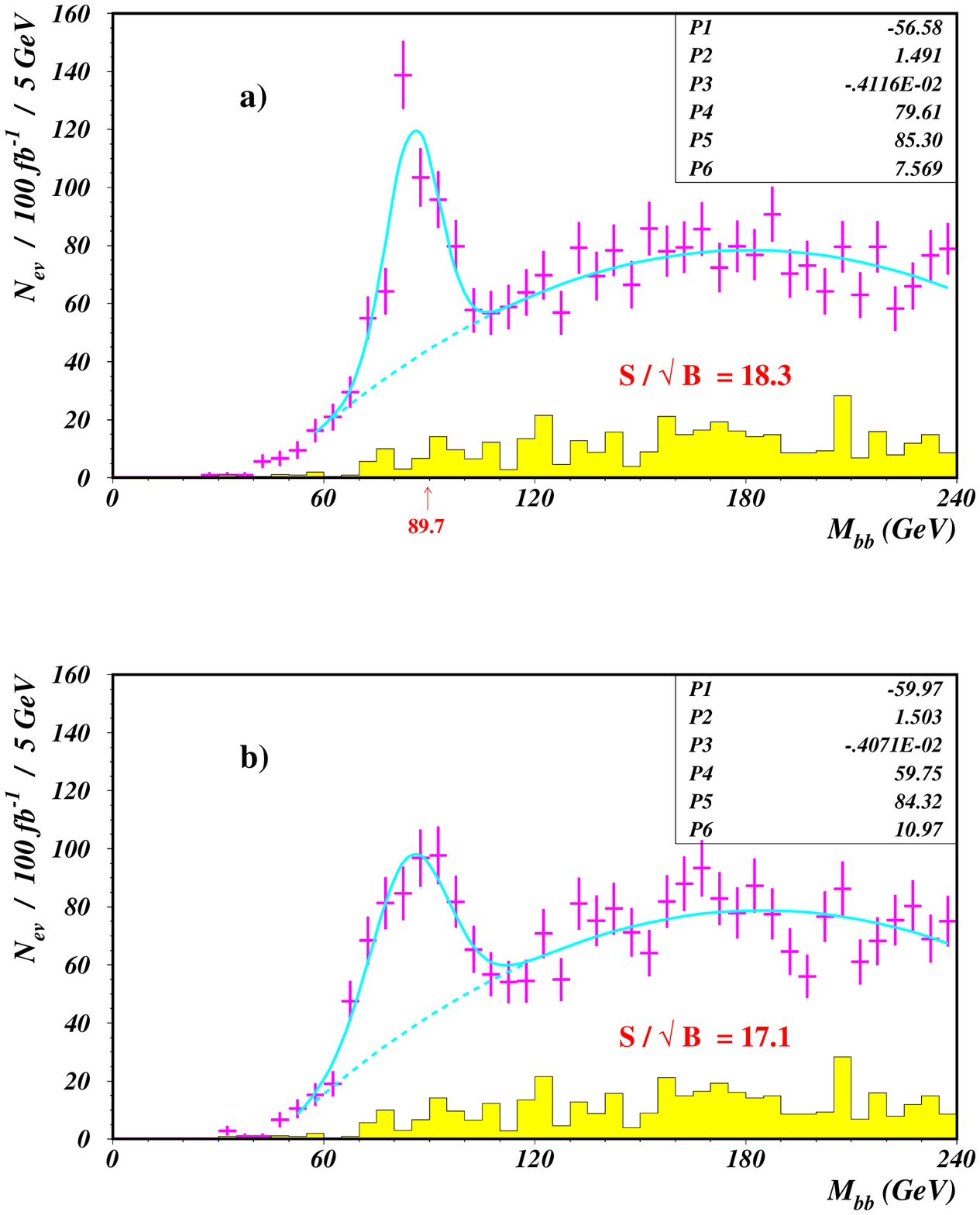}}
  \end{center}
\vspace{-10mm}
\end{figure}

Figure 9.5:
Effect of the assumed single hadron energy resolution on the
signal visibility: a) the nominal HCAL performance,
$\sigma_{E}/E = 82\% /\sqrt{E} \oplus 6.5\%$ at $\eta \sim 0$;
b) the HCAL energy resolution deteriorated to $120\% /\sqrt{E} \oplus
10\%$. mSUGRA point as in Figs.~9.2-9.4.

\newpage  
 
\ \\
 
\vspace{-20mm}
 
\begin{figure}[hbtp]
  \begin{center}
    \resizebox{14cm}{!}{\includegraphics{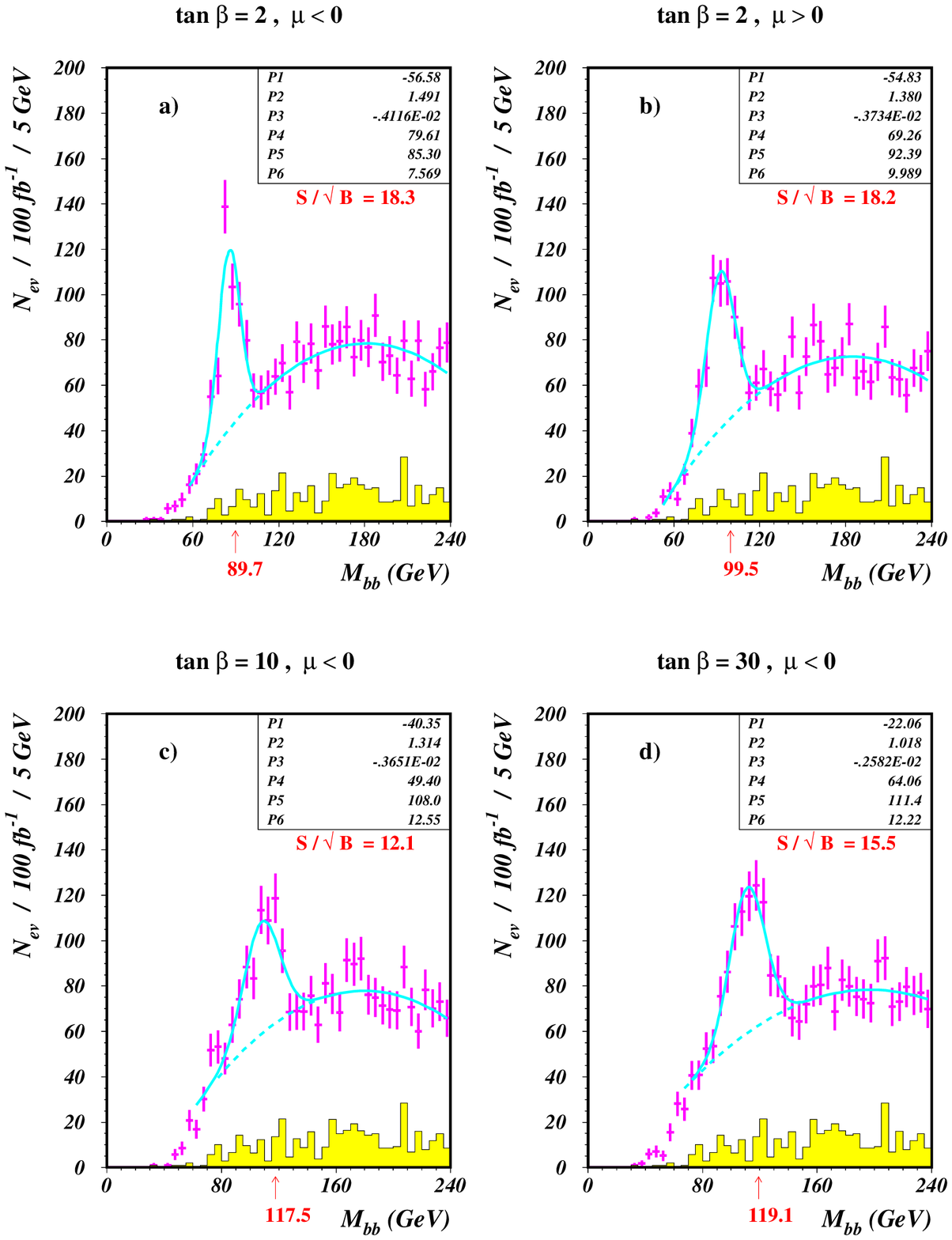}}
  \end{center}
\vspace{-10mm}
\end{figure}

Figure 9.6:
Dependence of the h~$\rightarrow$ $b\bar{b}$ signal
  visibility on $m_{h}$ over the allowed range $\sim$ 80 to 120 GeV,
  varying tan$\beta$ and $sign(\mu)$ at a fixed
  $m_{0} = m_{1/2}=500$ GeV. Nominal CMS instrumental
  performance, 100 fb$^{-1}$ integrated luminosity.

\newpage
 
\ \\
 
\vspace{-20mm}
 
\begin{figure}[hbtp]
  \begin{center}
    \resizebox{14cm}{!}{\includegraphics{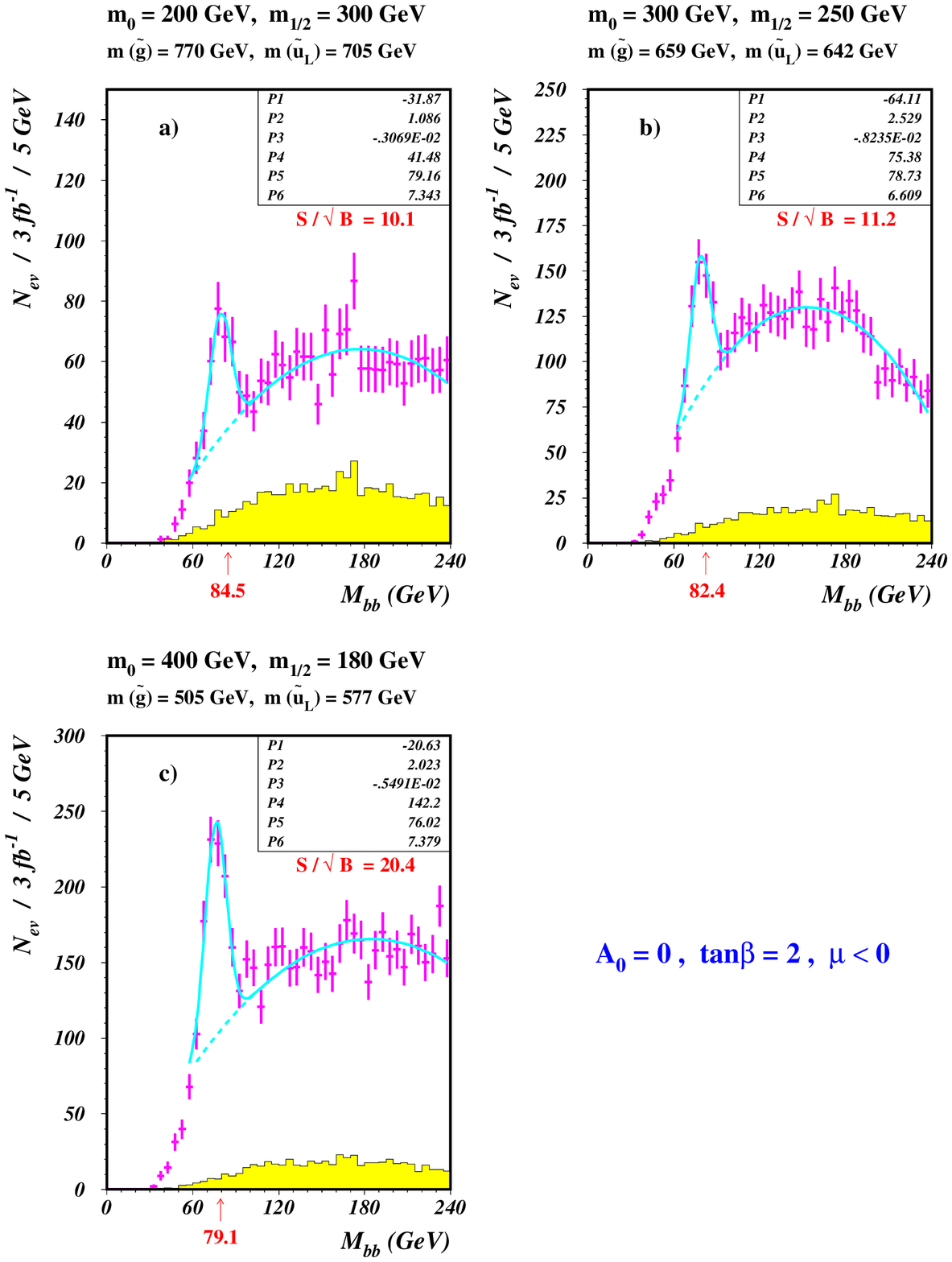}}
  \end{center}
\vspace{-10mm}
\end{figure}  
 
Figure 9.7:
Some points in parameter space accessible already with 3 fb$^{-1}$.
Nominal $b$-tagging performance of CMS is assumed.

\newpage
 
\ \\
 
\vspace{-20mm}

\begin{figure}[hbtp]
  \begin{center}
    \resizebox{12cm}{!}{\includegraphics{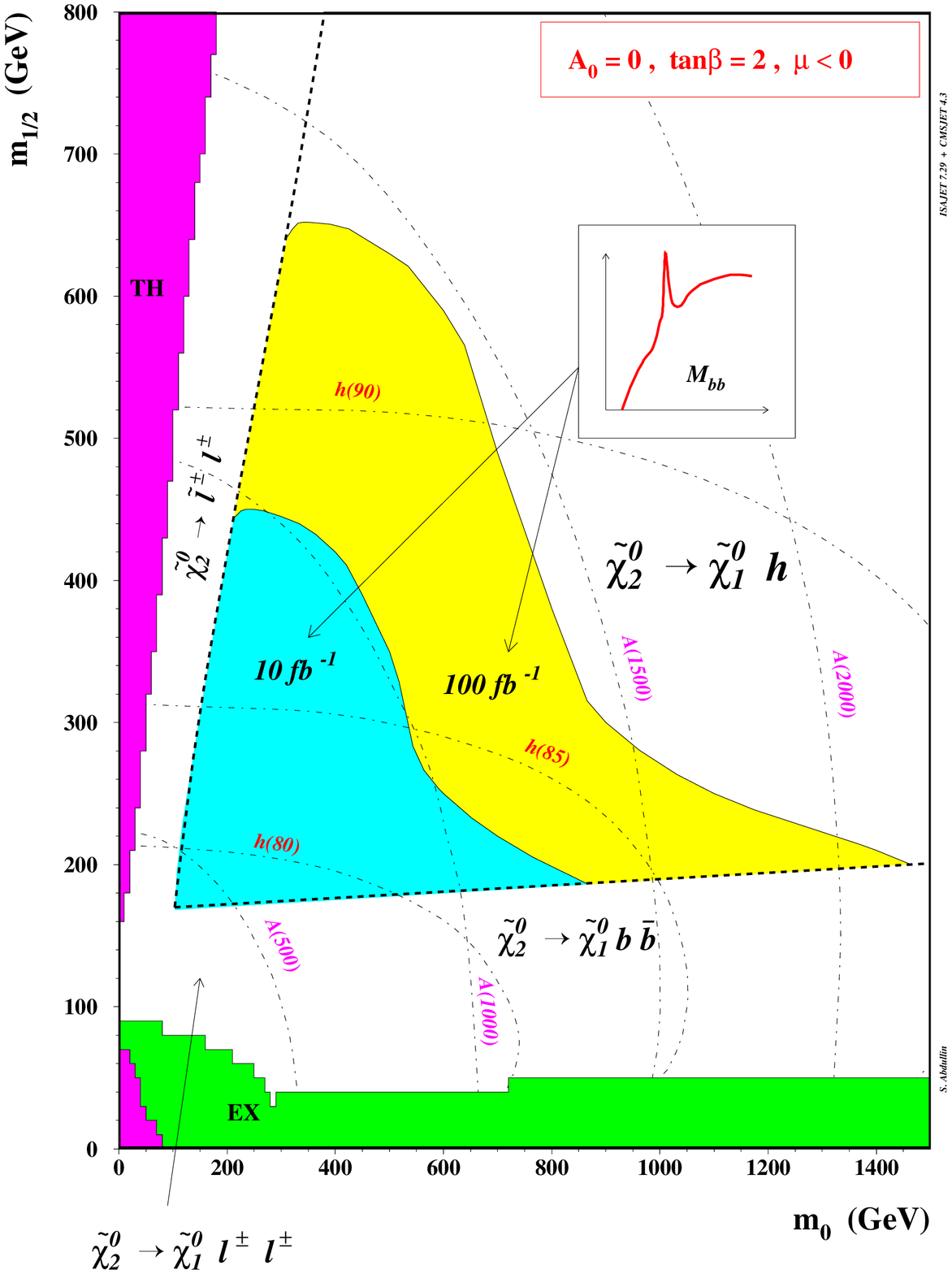}}
  \end{center}
\vspace{-5mm}
\end{figure}  

Figure 9.8:
5$\sigma$ visibility contours of h~$\rightarrow$ $b\bar{b}$
    for tan$\beta$ = 2, $A_{0}$ = 0 and $\mu$ $<$ 0 with 10 and 100
fb$^{-1}$. See also comments in text.

\newpage  
 
\ \\
 
\vspace{-20mm}
 
\begin{figure}[hbtp]
  \begin{center}
    \resizebox{12cm}{!}{\includegraphics{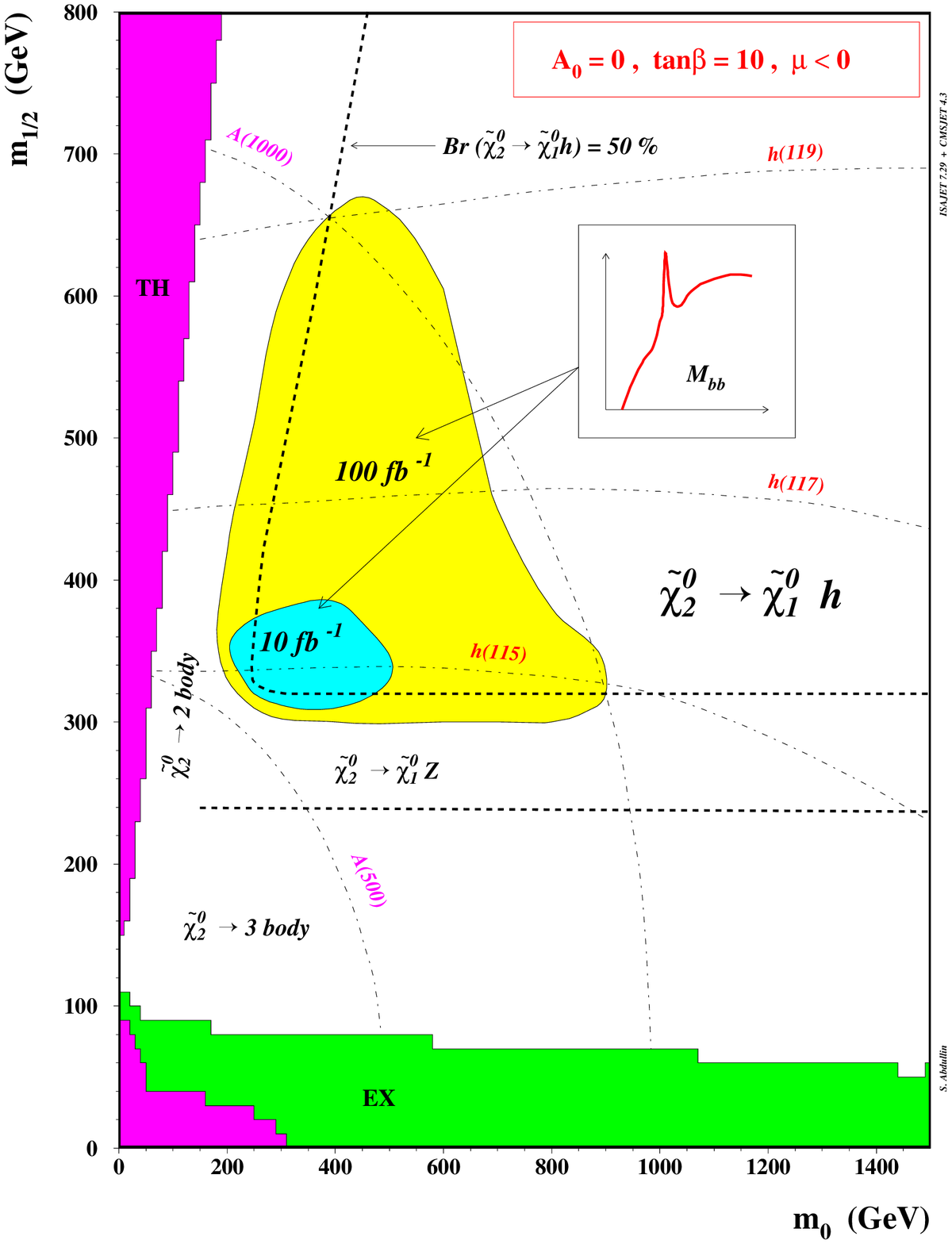}}
  \end{center}
\vspace{-5mm}
\end{figure}

Figure 9.9:  
5$\sigma$ visibility contours of h~$\rightarrow$ $b\bar{b}$
    for tan$\beta$ = 10, $A_{0}$ = 0 and $\mu$ $<$ 0 with 10 and 100
fb$^{-1}$. See also comments in text.
  
\newpage  
 
\ \\
 
\vspace{-20mm}
 
\begin{figure}[hbtp]
  \begin{center}
    \resizebox{12cm}{!}{\includegraphics{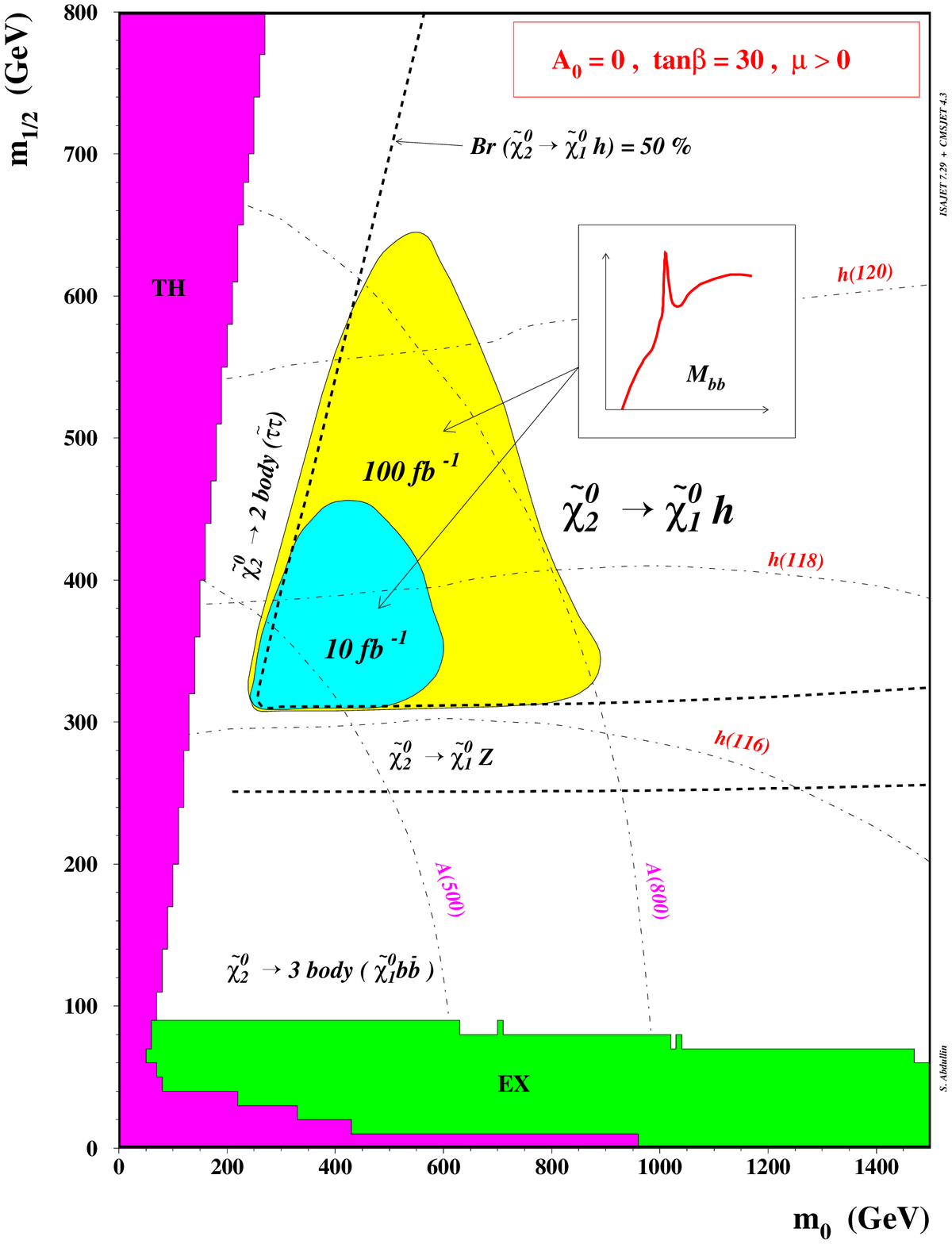}}
  \end{center}
\vspace{-5mm}
\end{figure}

Figure 9.10:
5$\sigma$ visibility contours of h~$\rightarrow$ $b\bar{b}$
    for tan$\beta$ = 30, $A_{0}$ = 0 and $\mu$ $<$ 0 with 10 and 100
fb$^{-1}$. See also comments in text.

\newpage


\section{Sensitivity to sparticle masses and \\
 model parameters}

\subsection{Establishing the SUSY mass scale}

Once a deviation from the SM expectations is established in
a number of final states indicating the presence of SUSY,
the determination of the ``SUSY mass scale'' would become the first
objective. The production
and decay of high mass strongly interacting particles, as would
be the case for gluinos and squarks, would lead to hard event
kinematics, with a number of hard central jets in the final state
accompanied by significant missing transverse energy due to escaping
LSPs and neutrinos. As suggested in \cite{paige1}, we can characterize
the ``hardness'' of an event by the scalar sum of transverse
energies of the four hardest jets and the missing transverse
energy:
\begin{equation}
E_{T}^{sum} = E_T^1 + E_T^2 + E_T^3 + E_T^4 + E_T^{miss}
\end{equation}
This variable is rather insensitive to the transverse energy flow
induced by event pile-up since it takes into account only the
calorimetric cells which are attached to a jet.

The following SM processes have been considered as a background to SUSY
strong production:
$t \bar{t}$, W/Z + jets and QCD jet production (PYTHIA
2 $\rightarrow$ 2 processes including $b\bar{b})$.
Events are selected by requiring:
\begin{itemize}
\item $E_T^{miss} >$ max(100 GeV, 0.2$E_{T}^{sum}$);
\item $\geq 4$ jets with $E_T > 50$ GeV, the hardest with $E_T > 100$ GeV
in $|\eta| < 5$;
\item Lepton veto: events are rejected if they contain
an isolated lepton with $p_T > 10$ GeV.
\end{itemize}
Figure 10.1 shows the $E_{T}^{sum}$ distributions for a representative
mSUGRA point (with $m_0 = 300$ GeV, $m_{1/2} = 150$ GeV, $A_0 = -600$
GeV, tan$\beta$ = 2, $\mu > 0$) and the SM background. The
background, which is dominated by $t \bar{t}$ and W/Z + jets productions,
is small compared to the SUSY signal.

The peak value of the $E_{T}^{sum}$ spectrum for the inclusive SUSY signal
provides a good estimate of the SUSY mass scale, defined
as in \cite{paige1}:
\begin{equation}
M_{SUSY} = {\mathrm min}(M_{\tilde{g}}, \ M_{\tilde{q}})
\end{equation}
and where $M_{\tilde{q}}$ is the mass of squarks from the first two
generations. Figure 10.2 shows the relationship between the peak
value of the $E_{T}^{sum}$ spectrum and $M_{SUSY}$ for
100 mSUGRA models chosen at random with $100 < m_{0} < 500$ GeV,
$100 < m_{1/2} < 500$ GeV, $-1000 < A_{0} < 1000$ GeV,
$1 <$ tan$\beta < 12$ and $sign(\mu) = \pm 1$ \cite{graham}.
At each point the peak value of $E_{T}^{sum}$ was found by a Gaussian fit.
The very strong correlation between the peak value of $E_{T}^{sum}$ and
$M_{SUSY}$ as can be seen in Fig.~10.2 could provide a first estimate of
the relevant SUSY mass scale.

\subsection{Constraints from \chnb \ leptonic decays}

Establishing the sparticle spectrum, the measurement of sparticle masses
and the model parameters will be the next step in uncovering SUSY, once
its existence and its mass scale have been established. In $R$-parity
conserving models undetectable LSP's  make this task nontrivial. 

As mentioned previously, some specific SUSY decays are particularly
useful for the determination of sparticle masses. The \chnb \ leptonic
decays may not only provide the first indication for SUSY in
significant regions of mSUGRA parameter space,
but also allow us to obtain information on some sparticle masses
by measuring the endpoint of the dilepton invariant mass spectrum
and exploiting its specific kinematics.
Before discussing how to exploit this feature in inclusive
two- and three-lepton final states, which is a recent development, let us
first review the $l^+l^-$ invariant mass spectrum in the exclusive
\chha \chnb $\rightarrow 3l^{\pm} + no~jets +$ \etm \ final state,
first described in \cite{edge1} for three-body decays
$\tilde{\chi}^{0}_{2} \rightarrow l^{+}l^{-} \tilde{\chi}^{0}_{1}$.
In this case the upper limit of dilepton invariant mass spectrum is:
\begin{equation}
M^{max}_{l^{+}l^{-}} = M_{\tilde{\chi}^{0}_{2}} - M_{\tilde{\chi}^{0}_{1}}
\end{equation}
Figure 10.3 shows the expected $l^{+}l^{-}$ mass spectrum
for mSUGRA ``Point 1'' (see Table 1.1) in the
exclusive \chha \chnb $\rightarrow 3l^{\pm} + no~jets +$ (\etm)
channel arising from direct (Electroweak) \chha \chnb \ production.
The SM and internal SUSY backgrounds are also shown, all for
$L_{int}  = 10^4$ pb$^{-1}$ \cite{chi12mass}. The edge is at 52 GeV and
with these statistics it could be measured with a precision better than
1 GeV. This measurement of $M^{max}_{l^{+}l^{-}} =
M_{\tilde{\chi}^{0}_{2}} - M_{\tilde{\chi}^{0}_{1}} = 52$ GeV
should allow, within mSUGRA, estimation of
$M_{\tilde{\chi}^{0}_{1}}, \ M_{\tilde{\chi}^{0}_{2}}, \
M_{\tilde{\chi}^{\pm}_{1}}, \ M_{\tilde{g}}$  
and $m_{1/2}$ through eqs. (2) -- (4) as discussed in section 2.1.

Even more information can be extracted from the \chnb \ two-body
(cascade) decays $\tilde{\chi}_2^{0} \rightarrow
l^{\pm}\tilde{l}^{\mp}_{L,R} \rightarrow l^+ l^- \tilde{\chi}_1^0$. The
kinematical upper limit in this case is:
\begin{equation}
M^{max}_{l^{+}l^{-}} = \frac{\sqrt{
(M^{2}_{\tilde{\chi}^{0}_{2}} - M^{2}_{\tilde{l}})
(M^{2}_{\tilde{l}} - M^{2}_{\tilde{\chi}^{0}_{1}}) }}
{M_{\tilde{l}} }
\end{equation}
i.e. it is sensitive also to the intermediate slepton mass.
In this case, however, the measurement of only the edge position does
not provide information on the masses of all the involved particles
unambiguously. Nonetheless, as discussed in the following, through
analyses of some kinematical distributions, the masses of \chna, \chnb \
and of the slepton can be determined with reasonable precision
\cite{inclchi2}. Therefore, with two-body decays one can constrain
$m_{1/2}$ and $m_0$ through eqs. (2), (3) and (5)/(6),
if the information on tan$\beta$ is available, for example from the
Higgs sector of the model.

Direct leptonic decays of the \chnb \ dominate
below $m_{1/2} \lsim 200$ GeV for nearly all values of
$m_{0}$; there is also a small region of high $m_{1/2}$
where direct three-body decays are open, as can be seen from Fig.~2.3a.
Cascade leptonic decays of \chnb \
via $\tilde{l}_{L}$ (Fig.~2.3b) or $\tilde{l}_{R}$ (Fig.~2.3c)
occur at almost any $m_{1/2}$ as soon as $m_{0} \lsim 0.5 m_{1/2}$,
where \chnb \ is heavier than the sleptons.

The expected position of the edge as a function of $m_{1/2}$ is
given in Fig.~10.4 for two different values of the
common scalar mass: $m_0 =$ 400 GeV and 100 GeV.
The box size in the Figure is proportional
to the branching ratio
B$(\tilde{\chi}^{0}_{2} \rightarrow l^{+}l^{-} + invisible)$.
At $m_0$ = 400 GeV the \chnb \
has only direct three-body decays for all values
of $m_{1/2}$ up to  $m_{1/2} \sim 180 $ GeV, where
``spoiler'' modes open up.
Measurement of $M^{max}_{l^{+}l^{-}}$ thus yields
the common gaugino mass parameter $m_{1/2}$
unambiguously in this region.
For $m_0=100$ GeV the situation is more complicated.
Here at different values of $m_{1/2}$ different leptonic
decay modes of $\tilde{\chi}^{0}_{2}$
dominate: direct decays up to
$m_{1/2} \lsim 130$ GeV;
cascade decays via $\tilde{l}_{R}$ at 130 GeV $ \lsim m_{1/2} \lsim 250$
GeV and, finally, cascade decays via $\tilde{l}_{L}$
at $m_{1/2} \gsim 250$ GeV.
In some regions
two $\tilde{\chi}^{0}_{2}$ decay channels coexist.

The essential issue, for the exploitation of an observed edge, 
is knowing whether the \chnb \ decay chain is three-body or two-body.
In general, the $p_T$ spectra of leptons from $\tilde{\chi}^{0}_{2}$
two-body decays are more asymmetric compared to the three-body decays. To
characterize the asymmetry we introduce the variable \cite{inclchi2}:  
\begin{equation}
{\mathcal A} = \frac{p_T^{max} - p_T^{min}} {p_T^{max} + p_T^{min}}
\end{equation}
where $p_T^{max}$ ($p_T^{min}$) corresponds to the lepton
of maximum (minimum) transverse momentum.

An example of decay type determination is discussed for the
mSUGRA point $m_0=100$ GeV, $m_{1/2}=150$ GeV. The sparticle masses
are $M_{\tilde{\chi}^{0}_{2}} = 135$ GeV,
$M_{\tilde{\chi}^{0}_{1}} = 65$ GeV
$M_{\tilde{l}_{R}} = 120$ GeV and \chnb \ decays
via
$\tilde{\chi}^{0}_{2} \rightarrow l^{\pm} \tilde{l}_{R}^{\mp} \rightarrow
l^+l^- + \tilde{\chi}^{0}_{1}$
with a probability of 0.54.
Figure 10.5 shows the dilepton mass spectrum
for this parameter space point in the inclusive $3~lepton$ events
for an integrated luminosity of
$L_{int} = 5 \times 10^3$ pb$^{-1}$.
The expected SM background is also shown.
Of the three
leptons with $p_T > 15 $ GeV only the two of opposite charge and same
flavor, which enter the
invariant mass distribution, are required to be isolated.
A very spectacular edge is situated at $\simeq 52$ GeV.
There is a second and much weaker edge at $\simeq$ 69 GeV which
is due to direct three-body decays with B$(\tilde{\chi}^{0}_{2}
\rightarrow l^+l^- + \tilde{\chi}^{0}_{1})=0.05$.
Thus at this mSUGRA point
there are two leptonic decay modes with very different
branching ratios. To identify the decay chain responsible
for the first edge, we look at the asymmetry
$\mathcal A$ distribution of
the lepton pairs with $M_{l^+l^-} < $ 52 GeV.
Figure 10.6 (full line) shows the $\mathcal A$ distribution for these
events. The pronounced asymmetry in $p_T$
indicates the cascade nature of decays. If we now pick up
the lepton pairs from the mass interval $55 < M_{l^+l^-} < 69$ GeV
below the second edge, then the corresponding $\mathcal A$ distribution
peaks near zero, as seen in Fig.~10.6 (dashed-dotted line), indicating
the three-body decay type of these events. 
To illustrate the generality of this approach, two more examples
of $\mathcal A$ distributions are also shown in Fig.~10.6
for mSUGRA parameter space points $m_0=100$ GeV, $m_{1/2}=400$ GeV
with pure 2-body decays (dashed line) and $m_0=400$ GeV, $m_{1/2}=150$ GeV
with pure 3-body decays (dotted line), respectively.

Once the decay type is determined,
the next step is to extract the masses of the particles
involved, the \chnb, \chna \ and $\tilde{l}$. In our example case of
two-body decay the $l^{+}l^{-}$ kinematical upper limit is
\begin{equation}
M^{max}_{l^{+}l^{-}} = \frac{\sqrt{
(M^{2}_{\tilde{\chi}^{0}_{2}} - M^{2}_{\tilde{l}}) 
(M^{2}_{\tilde{l}} - M^{2}_{\tilde{\chi}^{0}_{1}}) }}
{M_{\tilde{l}} } = 52 \; \mathrm{GeV}.
\end{equation}
This equation can be satisfied by an infinite
number of $M_{\tilde{\chi}^{0}_{1}}$, $M_{\tilde{\chi}^{0}_{2}}$
and $M_{\tilde{l}}$ mass combinations. To find a solution, we assume
$M_{\tilde{\chi}^{0}_{2}} = 2 M_{\tilde{\chi}^{0}_{1}}$ and generate
$\tilde{\chi}^{0}_{2} \rightarrow l^{\pm} \tilde{l}^{\mp}
\rightarrow l^{+}l^{-} \tilde{\chi}^{0}_{1}$
two-body sequential decays for various assumptions on
$M_{\tilde{l}}$ ($M_{\tilde{\chi}^{0}_{1}}$), e.g. from 70 GeV to 150 GeV
in 10 GeV steps, with the $\tilde{\chi}^{0}_{1}$ mass constrained
to provide the ``observed'' position of the edge.
For each value of $M_{\tilde{l}}$,
one obtains two series of solutions for equation (15).
Some of them are shown in Fig~10.7
in terms of dilepton invariant mass and $p_T$-asymmetry distributions,
in Figs.~10.7a,c for ``low mass'' solutions and Figs.~10.7b,d
for ``high mass'' solutions \cite{inclchi2}.
The solid lines in Figs.~10.7a-c
correspond to the ``observed'' spectra and the dotted  
histograms are the results obtained from eq. (15).
Clearly, the ``observed'' $l^{+}l^{-}$ invariant mass spectrum itself
and the $p_T$-asymmetry distributions
allow us to eliminate the ``high mass'' solutions of eq. (15).
To find the best combination of $M_{\tilde{\chi}^{0}_{1}}$,
$M_{\tilde{\chi}^{0}_{2}}$ and $M_{\tilde{l}}$
among the ``low mass'' solutions,
we perform a $\chi^2$-test on the $p_T$-asymmetry distributions,
taking into account only the difference in shapes,
but not the normalization. The result is shown
in Fig.~10.8. The horizontal line in this Figure corresponds to the
expected uncertainty of simulations due to the detector resolution,
background estimates, initial/final state radiations, etc.
We obtain the following precisions on masses:
$\delta M_{\tilde{\chi}^{0}_{1}} \lsim$ 5 GeV,
$\delta M_{\tilde{l}} \lsim$ 10 GeV.
The use of
the total number of observed events, as well as the dilepton
invariant mass spectrum itself in a combined
$\chi^2$-test would further improve these results.

As shown above for
the mSUGRA point under consideration, the second
edge in the dilepton mass spectrum at
$M^{max}_{l^{+}l^{-}} = 69$ GeV (see Fig.~10.6) is due to direct
three-body decays of \chnb. Using now two measured edge position values
and assuming again $M_{\tilde{\chi}^{0}_{2}} = 2
M_{\tilde{\chi}^{0}_{1}}$,
two solutions for the slepton mass can be directly obtained
$M_{\tilde{l}} =$ 120 GeV and $M_{\tilde{l}}=$ 77 GeV. The
corresponding dilepton mass spectra for these two solutions are
shown in Fig.~10.9. The clear difference in the predicted spectra
(the first edge in Fig.~10.5 which proceeds through $\tilde{l}$ vs.
Fig.~10.9) allows us to eliminate the $M_{\tilde{l}}=$ 77 GeV
solution. At this particular mSUGRA point the use of two observable
edges provides a precision of
$\delta M_{\tilde{\chi}^{0}_{1},\tilde{l}} \lsim$ 1 GeV
\cite{incl_lhcc, inclchi2}.

Another example of a double edge is given in Fig.~10.10
for the mSUGRA point (50 GeV, 125 GeV). Here
the sparticle masses are
$M_{\tilde{\chi}^{0}_{2}} = 116$ GeV,
$M_{\tilde{\chi}^{0}_{1}} = 55$ GeV,
$M_{\tilde{l}_{L}} = 110$ GeV,
$M_{\tilde{l}_{R}} = 78$ GeV. In this case two
two-body decays, via left and right
sleptons coexist with the comparable branching ratios
B($\tilde{\chi}^{0}_{2} \rightarrow l^{\pm} \tilde{l}_{L}^{\mp}
\rightarrow l^+  l^- \tilde{\chi}^{0}_{1}$) = 0.037
and
B($\tilde{\chi}^{0}_{2} \rightarrow l^{\pm} \tilde{l}_{R}^{\mp}
\rightarrow l^+  l^- \tilde{\chi}^{0}_{1}$) = 0.013
and their analysis could proceed as indicated above
providing a strong constraint on the underlying model.

\subsection{Determination of the squark mass}

The observation of an edge in the dilepton mass spectrum,
resulting from \chnb \ leptonic decays allows not only
the determination of \chna, \chnb, ($\tilde{l}$) masses, but also
enables the momenta to be fixed.
At the upper limit of the $l^+l^-$ spectrum in direct three-body decays
$\tilde{\chi}_2^{0} \rightarrow l^+l^- \tilde{\chi}_1^{0}$,
the $\tilde{\chi}_1^{0}$ is produced at rest in the
$\tilde{\chi}_2^{0}$ rest frame. Assuming knowledge of
the $\tilde{\chi}_1^{0}$ mass (or, equivalently, knowledge of
a relation between $\tilde{\chi}_1^{0}$ and $\tilde{\chi}_2^{0}$
masses), one can then, for events at the $l^+l^-$ edge, reconstruct the
$\tilde{\chi}_2^{0}$ momentum vector in the laboratory frame:
\begin{equation}
p_{\tilde{\chi}^{0}_{2}} = (1 + M_{\tilde{\chi}^{0}_{1}} /
M_{l^+l^-}) \times p_{l^+l^-}
\end{equation}
Once the four-momentum vector
of the $\tilde{\chi}_2^{0}$ is determined, one can search for a resonance
structure in, e.g. the $\tilde{\chi}_2^{0}$ + jet(s) invariant mass
distributions. This technique can be put to a good use to reconstruct
\g/\q \ masses \cite{paige1}
as the next-to-lightest neutralinos are
abundantly produced in gluino and squark decays,
see e.g. Figs.~2.5a,b.

An example of the \q \ mass reconstruction is given for
mSUGRA point with $m_0 = 300$ GeV, $m_{1/2} = 150$ GeV, $A_0 = -600$ GeV,
tan$\beta$ = 2 and $\mu > 0$, discussed in section 10.1. The
masses of relevant sparticles are
463 GeV, 495 GeV, 151 GeV, 425 GeV, 112 GeV and 57 GeV for
$\tilde{q}_{L,R}$ (squarks of the first two generations),
$\tilde{t}_2$, $\tilde{t}_1$,
\g, \chnb \ and \chna, respectively,
and the total SUSY cross-section (452 pb) is largely dominated
by $\sigma_{\tilde{t}_{1}\tilde{t}_{1}} = 195$ pb,
$\sigma_{\tilde{g}\tilde{q}_{L,R}} = 109$ pb,
$\sigma_{\tilde{g}\tilde{g}} = 77$ pb productions.
Gluinos are lighter than squarks of the first two generations,
but their decays to stop and sbottom are kinematically allowed,
giving several $b$-jets at the end of the decay chain.
Right squarks $\tilde{q}_{R}$ predominantly decay
to $\tilde{\chi}_1^0 + q$ or to \g$ + q$. Finally, the Left squarks
$\tilde{q}_{L}$
have the interesting for us decay mode to  \chnb + $q$
with a branching ratio of 0.27. To extract this latter decay chain
from internal SUSY and SM backgrounds we have adopted the
following procedure \cite{graham}.
\begin{itemize}
\item Require two isolated leptons with $p_T > 10$ GeV;
and an angular separation
$\Delta R_{l^{+}l^{-}} = \sqrt{\Delta \eta^2 + \Delta \phi^2} < 1$.
\item The scalar sum of calorimeter cell transverse energies and of muon
transverse momenta, $\Sigma E_{T}^{calo} + p_T^{\mu} > 600$ GeV to
ensure production of hard events. Require \etm \ $>$ 100 GeV.
\item The $l^+l^-$ invariant mass has been reconstructed
and the edge position at $\simeq$54 GeV has been observed.
The $p_T$-asymmetry distribution of dilepton pairs indicates
it is due to a the three-body decay of type
$\tilde{\chi}^{0}_{2} \rightarrow l^+l^- \tilde{\chi}^{0}_{1}$.
\item Events with a dilepton mass close to the kinematical limit,
i.e. 49 GeV $< M_{l^+l^-} <$ 54 GeV, have been selected for further
analysis;
the \chnb \ momentum has been determined using eq. (16).
\item The invariant mass of the \chnb \ and each non-$b$-jet with
$E_T^{jet} > 300$ GeV has then been reconstructed. Such a hard jet
is unlikely to originate from a QCD radiation. Furthermore,
due to the large mass difference
$M_{\tilde{q}_L} - M_{\tilde{\chi}^{0}_{2}} = 351$ GeV the
$E_T$-spectrum of quark-jets from the $\tilde{q}_L$ decay is harder
than both the inclusive SUSY and SM jet spectra. Hence this requirement
enhances the $\tilde{q}_L$ signal.
\end{itemize}

Figure 10.11 shows the reconstructed $\tilde{\chi}^{0}_{2}$ + jet
mass spectrum for SUSY and SM events for an integrated luminosity
of $L_{int} = 10^4$ pb$^{-1}$.
As a source of the SM backgrounds we have considered
$t \bar{t}$, W/Z + jets, WW, ZZ, ZW and $b\bar{b}$ processes.
Their contribution is negligible compared to the internal
SUSY background. A peak at $\sim$450 GeV is clearly visible and
a Gaussian plus polynomial fit yields a value of 447 $\pm$ 4 GeV
(statistical error) for the reconstructed $\tilde{q}_{L}$ mass.
Significant systematic uncertainties are associated with this
measurement due to the jet reconstruction algorithm, energy scale,
approximations in determining the $\tilde{\chi}_2^{0}$ four-momentum,
in particular the dilepton mass interval which has been chosen
for analysis, etc. The overall systematic uncertainty is estimated
to be about $\pm 5 \%$. A much more detailed analysis is needed to 
understand fully the achievable precision \cite{graham}.

As a concluding remark, the
observation of a resonance peak in the $\tilde{\chi}_2^{0}$ + jet
invariant mass spectrum would be interpreted as a squark undergoing
a two-body decay to \chnb \ and a quark. The technique described here
is also applicable for gluino mass reconstruction, for example
in $\tilde{g} \rightarrow q \bar{q}' \tilde{\chi}_2^{0}$ with subsequent
decay of $\tilde{\chi}_2^{0} \rightarrow \tilde{\chi}_1^{0}$.

\subsection{Sensitivity to model parameters in $3l + no \ jets +
E_T^{miss}$ final states}

Since information on $m_{1/2}$ can be obtained from the $l^+l^-$
edge position in regions of parameter space where it is visible,
one can then attempt to constrain the
$m_0$ parameter from the event rates. Let us first discuss the case of
$3l + no \ jets + E_T^{miss}$ events \cite{chi12mass}.
For a fixed $m_{1/2}$ the $\sigma \cdot$B(\chha\chnb~$\rightarrow 3l
+invisible$) decreases monotonically with increasing $m_0$ in the region
where \chnb \ has direct three-body decays.
Table 10.1 gives the expected SUSY event rates for a series
of points with different $m_0$ but with fixed $m_{1/2} = 100$ GeV, i.e.
corresponding to
approximately the same $M_{l^+l^-}^{max}$. The remaining parameters are
tan$\beta = 2$, $A_0 = 0$, $\mu < 0$. All expected SUSY sources of such
events are included.
The applied cuts are (Set 1 in section 8): 

\begin{table}
\begin{center}

Table 10.1: SUSY cross-sections (in fb) after selection. Events are
counted below ``observed'' edge at 54 GeV.

\vspace*{3mm}

\begin{tabular}{|r||l||l|l|l|l|}  
\hline
\hline
Point     & \chha\chnb &
$\tilde{g} \tilde{g}$ $\tilde{g} \tilde{q}$ $\tilde{q} \tilde{q}$ &
$\tilde{g}\tilde{\chi}\  \tilde{q}\tilde{\chi}$ &
$\tilde{\chi}\tilde{\chi}$ &
$\tilde{l}\tilde{l}\  \tilde{l}\tilde{\nu}\  \tilde{\nu}\tilde{\nu}$
 \\ \hline\hline
(100,100) & 71.6 & 14.1 & 11.3 & 5.9 & 14.5 \\
\hline
(150,100) & 53.8 & 15.9 & 5.3 & 2.6 & 5.3 \\
\hline
(200,100) & 45.6 & 11.9 & 6.0 & 4.0 & 0.0 \\
\hline
(300,100) & 29.7 & 0.5 & 1.0 & 0.5 & 0.0 \\
\hline
(400,100) & 23.0 & 0.0 & 0.0 & 0.0 & 0.0 \\
\hline
(1000,100) & 15.7 & 0.0 & 0.0 & 0.0 & 0.0 \\
\hline
(1800,100) & 12.3 & 0.0 & 0.0 & 0.0 & 0.0 \\
\hline
\hline
\end{tabular}

\vspace*{5mm}

Table 10.2: SM cross-sections (in fb) after selection. Events are
counted below ``observed'' edge at 54 GeV.

\vspace*{3mm}

\begin{tabular}{|l|l|l|l|l|l|}
\hline
\hline
WZ  & ZZ  & $t\bar{t}$ & W$tb$ & Z$b\bar{b}$ & $bb$ \\
\hline
1.42  & 0.48  & 1.07       & 0.46  & 0.27        & $<$ 0.15 \\
\hline
\hline
\end{tabular}

\end{center}
\end{table}

\begin{table}[b]
\begin{center}
Table 10.3: statistical ($L_{int}= 10^4$ pb$^{-1}$)  and systematic
errors on $m_0$ measurement for $m_{1/2} \sim 100$ GeV.

\vspace*{3mm} 

\begin{tabular}{|r||c|c|}
\hline
\hline
Source & $m_0$ $\in$ [100 GeV; 300 GeV]  & $m_0$ $\in$ [300 GeV; 400 GeV]
\\
\hline
\hline
statistical error  & 8 $\div$ 20 GeV & 20 $\div$ 55 GeV \\
\hline
\hline
edge measurement precision & 4 $\div$ 8 GeV & 8 $\div$ 25 GeV \\
\hline
uncertainty on SM background & 2.5 $\div$ 9 GeV & 9 $\div$ 30 GeV \\
\hline
uncertainty on luminosity  & 27 $\div$ 35 GeV & 35 $\div$ 85  GeV \\
\hline
\hline
statistical $\oplus$ systematic error & 30 $\div$ 45 GeV & 45 $\div$ 105
GeV \\
\hline
\hline
\end{tabular}
\end{center}
\end{table}

\begin{itemize}
\item
$3 l$ with $p_T^l>15$ GeV in $|\eta^l|<2.4(2.5)$ for muons (electrons);
\item
Isolation:  no track with  $p_T>1.5$ GeV in a cone $R=0.3$ about the
lepton direction, for all leptons;
\item
Jet veto: no jet with $E_T^{jet}>$  25 GeV in $|\eta^{jet}|<3.5$;
\item
$Z$-mass cut:  no lepton pair  with $M_{l^{+}l^{-}} > 81$ GeV.
\end{itemize}

From the selected events we count the ones with
$M_{l^+l^-}<M_{l^+l^-}^{max}=54$ GeV.
Contributions from different
SUSY sources
-- \chha \chnb \ direct production, strong production, 
gluino/squark associated production, gaugino pair  and
slepton pair productions -- are given separately. The main contribution
is always due to \chha \chnb \ direct production, but below
$m_0 \sim 300$ GeV other SUSY processes also contribute
significantly.

The expected SM event rates after selection cuts are given in Table 10.2.
All SM processes together amount 3.85 fb, which is always
smaller than the \chha \chnb \ direct production rate at 
$m_{1/2}$ = 100 GeV, and much smaller than that for
$m_0 \lappeq $ 300 GeV (Table 10.1).

\begin{table}
\begin{center}
Table 10.4: SUSY cross-sections (in fb) after selection. Events are
counted below the ``observed'' edge at 72 GeV.

\vspace*{3mm}

\begin{tabular}{|l||l||l|l|l|l|}  
\hline
\hline
Point     & \chha\chnb &
$\tilde{g} \tilde{g}$ $\tilde{g} \tilde{q}$ $\tilde{q} \tilde{q}$ &
$\tilde{g}\tilde{\chi}\  \tilde{q}\tilde{\chi}$ &
$\tilde{\chi}\tilde{\chi}$ &
$\tilde{l}\tilde{l}\  \tilde{l}\tilde{\nu}\  \tilde{\nu}\tilde{\nu}$
 \\ \hline\hline
(150,150) & 26.3 & 0.3 & 1.5 & 1.3 & 2.9 \\
\hline
(200,150) & 19.9 & 0.7 & 0.8 & 0.8 & 0.8 \\
\hline
(300,150) & 10.7 & 0.5 & 0.1 & 0.1 & 0.0 \\
\hline
(400,150) & 6.4 & 0.0 & 0.0 & 0.1 & 0.0 \\
\hline
(1000,150) & 3.6 & 0.0 & 0.0 & 0.0 & 0.0 \\
\hline
(1800,150) & 3.3 & 0.0 & 0.0 & 0.0 & 0.0 \\
\hline
\hline
\end{tabular}

\vspace*{5mm}

Table 10.5: SM cross-sections (in fb) after selection. Events are
counted below the ``observed'' edge at 72 GeV.

\vspace*{3mm}

\begin{tabular}{|l|l|l|l|l|l|}
\hline
\hline
WZ  & ZZ  & $t\bar{t}$ & W$tb$ & Z$b\bar{b}$ & $bb$ \\
\hline
1.53  & 0.42  & 1.66       & 0.50  & 0.33    & $<$ 0.15 \\
\hline
\hline
\end{tabular}
\end{center}
\end{table}

\begin{table}[b]
\begin{center}
Table 10.6: statistical ($L_{int}= 10^5$ pb$^{-1}$)  and systematic
errors on $m_0$ measurement for $m_{1/2} \sim 150$ GeV.

\vspace*{3mm}

\begin{tabular}{|r||c|c|}
\hline
\hline
Source & $m_0$ $\in$ [150 GeV; 300 GeV]  & $m_0$ $\in$ [300 GeV; 400 GeV]
\\
\hline
\hline
statistical error  & 5 $\div$ 10 GeV & 10 $\div$ 18 GeV \\
\hline
\hline
edge measurement precision & 2 $\div$ 8 GeV & 5 $\div$ 8 GeV \\
\hline
uncertainty on SM background & 9 $\div$ 24 GeV & 24 $\div$ 55 GeV \\
\hline
uncertainty on luminosity  & 28 $\div$ 30 GeV & 30 $\div$ 40  GeV \\
\hline
\hline
statistical $\oplus$ systematic error & 32 $\div$ 40 GeV & 40 $\div$ 75 
GeV \\  
\hline
\hline
\end{tabular}
\end{center}
\end{table}

Table 10.1 shows that the data selection does not wash out
significantly the event rate sensitivity to $m_{0}$
expected from the behavior of $\sigma\cdot$B for \chha \chnb \ direct
production (Fig.~2.4).
Figure 10.12a shows $N_{SUSY}$ versus $m_{0}$ for
$m_{1/2} = 100$ GeV, where $N_{SUSY}$ is the 
SM-background-subtracted number of 
$3l + no \ jets + E_T^{miss}$ events surviving cuts
for an  integrated luminosity of
$L_{int}= 10^4 pb^{-1}$.  The
error band includes both statistical and systematic uncertainties; The
systematic uncertainties include effects due to
the edge measurement precision,
the uncertainty on the rate of SM processes and
a 10$\%$ uncertainty on the luminosity measurement.
The resulting error on $m_0$ is 30 to 45 GeV for
100 GeV $\lsim m_0 \lsim$ 300 GeV and 45 to 105 GeV for
300 GeV $\lsim m_0 \lsim$ 400 GeV. Above $m_0 \simeq 400 $ GeV
sensitivity to $m_0$ is lost (see also Fig.~2.4).
Table 10.3 gives the statistical and the various systematic uncertainty
contributions to the $m_0$ measurement error; the error on the integrated
luminosity is the dominant contribution \cite{chi12mass}.

For $m_{1/2}$ = 150 GeV the expected event rate is much lower than for 
$m_{1/2}$ = 100 GeV, and $L_{int}= 10^5$ pb$^{-1}$ is needed to obtain a
reasonable precision on $m_0$. The cuts applied are the same as
for the $m_{1/2}$ = 100 GeV points, except for a relaxed
jet veto requirement (Set 2 in section 8):
no jet with $E_T^{jet}>$  30 GeV in $|\eta^{jet}|<3.$
Selected events are counted below the $l^+l^-$
edge at 72 GeV. The event rates for the mSUGRA points
and the SM background processes are given 
in Tables 10.4 and 10.5. After cuts the sum of
SM processes is 4.6 fb. The SUSY/SM ratio exceeds 1.4 for $m_0 \lappeq$
400 GeV. Figure 10.12b shows $N_{SUSY}$ versus $m_{0}$ for
$m_{1/2} = 150$ GeV assuming $L_{int}= 10^5 pb^{-1}$. The
sensitivity to $m_0$ is
32 to 40 GeV for 150 GeV $\lsim m_0 \lsim$ 300 GeV and 
40 to 75 GeV for 300 GeV $\lsim m_0 \lsim$ 400 GeV. 
Contributions from the different sources of
uncertainty are given in Table 10.6
The error on the luminosity is still the
main source of uncertainty, although the
error on the SM background now has a significant influence 
due to the lower signal/background ratio
compared to the $m_0$=100 GeV case.


\subsection{Sensitivity to model parameters in inclusive $l^+l^-  +
E_T^{miss} + jets$ final states}

More generally, information on event rates from the observation of 
an edge in inclusive studies can also be used to constraint the model
parameters. Let us discuss the case of 
$l^+ l^- \,+ \, E_T^{miss}\, (+ \,jets)$ final states \cite{ll_incl}.
Within mSUGRA the expected edge position $M_{l^+l^-}^{max}$ in the dilepton 
mass distribution can be obtained from equations (12), (13).
Figure 10.13 shows the contours of expected 
values of $M_{l^+l^-}^{max}$ in the ($m_0,m_{1/2}$) parameter plane. 
Different lines with same values of $M_{l^+l^-}^{max}$ belong to
domains I, II and III  which 
correspond to the three possible decay modes of $\tilde{\chi}_2^0$ to
$l^+l^-\tilde{\chi}_1^0$ final states.
The regions of $M_{l^+l^-}^{max}$ accessible at LHC are:

\begin{tabbing} 
for 
$\hspace{2cm}$ \= $\tilde{\chi}_2^0 \rightarrow \tilde{\chi}_1^0  l^+ l^- 
\hspace{1cm}$ \= $- \hspace{1cm}$ 50 GeV $\lappeq$ \= $M_{l^+l^-}^{max} 
\lappeq$ 
90 GeV  \hspace{0.5cm} (I) \\
\> $\tilde{\chi}_2^0 \rightarrow \tilde{l}_R l$ \> $-$ \>
$M_{l^+l^-}^{max} \gappeq 10$ GeV \hspace{0.5cm} (II) \\
\> $\tilde{\chi}_2^0 \rightarrow \tilde{l}_L l$ \> $-$ \>
$M_{l^+l^-}^{max} \gappeq 20$ GeV \hspace{0.5cm} (III) \\ 
\end{tabbing}

\vspace{-0.3cm}
The first case is limited by the appearance of the spoiler modes  
$\tilde{\chi}_2^0 \rightarrow h^0 (Z^0)\tilde{\chi}_1^0$. In the last two 
cases the upper limit on accessible $M_{l^+l^-}^{max}$ is determined 
only by the available statistics. A measurement of $M_{l^+l^-}^{max}$ in
the inclusive  dilepton mass distribution, with a single edge, thus
constrains the model parameters in general to three lines in the
($m_0,m_{1/2}$) parameter plane. In the case of $M_{l^+l^-}^{max} \gappeq
90$ GeV the constraint is stronger, as there are just two possible
lines. The most favorable case is when
the measured $M_{l^+l^-}^{max}$ value is large, $M_{l^+l^-}^{max} \gappeq 
180$ GeV.
Then one is left with a single line in the ($m_0,m_{1/2}$) parameter plane.
The observation of two edges due to the simultaneous presence of two types of
$\tilde{\chi}_2^0$ decay modes to  $l^+l^-\tilde{\chi}_1^0$ would
give even stronger constraints,
see Refs. \cite{ll_incl, inclchi2, incl_lhcc}.

The proper line in the ($m_0,m_{1/2}$) plane 
can be determined by the method discussed in \cite{ll_incl}. 
When the correct $M_{l^+l^-}^{max}$ line is identified, 
the next step is to find 
the point ($m_0,m_{1/2}$) on this line. In general the cross section 
decreases with increasing $m_{1/2}$ and $m_0$, thus the study of the event 
rate along the corresponding $M_{l^+l^-}^{max}$ lines 
will determine the point in parameter space.
As an example of the general situation let us discuss 
the case of $M_{l^+l^-}^{max}=74 \pm 1$ GeV, with three lines 
corresponding to domains I,II and III, respectively.
The $(m_0,m_{1/2})$ points analysed are given in tables 10.7-10.9. 

\begin{table}[t]
\begin{center}

Table 10.7: $M_{l^+l^-}^{max}$ values (in GeV) at the investigated 
$(m_0,m_{1/2})$ points from domain I,
$\tilde{\chi}_2^0 \rightarrow \tilde{\chi}_1^0 + l^+ + l^-$.

\vspace*{3mm}

\begin{tabular}{| c |
@{\extracolsep{0.5mm}}c@{\extracolsep{2mm}}|
@{\extracolsep{0.5mm}}c@{\extracolsep{2mm}}|
@{\extracolsep{0.5mm}}c@{\extracolsep{2mm}}|
@{\extracolsep{0.5mm}}c@{\extracolsep{2mm}}|
@{\extracolsep{0.5mm}}c@{\extracolsep{2mm}}|
@{\extracolsep{0.5mm}}c@{\extracolsep{2mm}}|
@{\extracolsep{0.5mm}}c@{\extracolsep{2mm}}|
@{\extracolsep{0.5mm}}c@{\extracolsep{2mm}}|}
\hline 
&(120,160)&(130,160)&(180,160)&(200,160)&(220,160)&(240,160)&
(290,160)&(350,160)\\
\hline
\hline
\hline
$M_{l^+l^-}^{max}$ & 74 & 74 & 74 & 74 & 74& 74& 74&  73.7\\
\hline
\end{tabular}

\vspace{5mm}

Table 10.8: $M_{l^+l^-}^{max}$ values (in GeV) at the investigated 
$(m_0,m_{1/2})$ points from domain II,
$\tilde{\chi}_2^0 \rightarrow \tilde{l}_R^{\pm} + l^{\mp}
\rightarrow \tilde{\chi}_1^0 + l^+ + l^-$.

\vspace*{3mm}

\begin{tabular}{|c|c|c|c|c|c|} 
\hline 
$(m_0,m_{1/2})\rightarrow$ &(80,162)&(90,170)&(105,180)&(110,187)&(120,195) \\
\hline
\hline
\hline 
$M_{l^+l^-}^{max}$ & 74 & 74 & 73 & 75 & 73 \\
\hline
\end{tabular}

\vspace{5mm}

Table 10.9: $M_{l^+l^-}^{max}$ values (in GeV) at the investigated 
$(m_0,m_{1/2})$ points from domain III,
$\tilde{\chi}_2^0 \rightarrow \tilde{l}_L^{\pm} + l^{\mp} 
\rightarrow \tilde{\chi}_1^0 + l^+ + l^-$.

\vspace*{3mm}

\begin{tabular}{|c|c|c|c|c|} 
\hline 
$(m_0,m_{1/2})\rightarrow$  &(20,195)&(40,210)&(60,230)&(80,255) \\
\hline
\hline
\hline
$M_{l^+l^-}^{max}$ & 73 & 73 & 73 &  73 \\
\hline
\end{tabular}
\end{center} 
\end{table}

As this is an inclusive analysis, we assume a low luminosity
of $L_{int}=10^3$ pb$^{-1}$.
We discuss first domain III, where the situation is simplest. 
To minimize the uncertainties due to background,
we require  $p_T^{l_{1,2}}>15$ GeV and $E_T^{miss}>130$ GeV. 
The dependence of the resulting event rate on $m_0$ is shown in
Fig.~10.14. The errors are calculated taking 
into account the statistical error and assuming a
systematic error of 30 $\%$ for the background uncertainty; a systematic 
error due to the precision of the edge position measurement is also taken
into account. From the observed event rate $m_0$ can be determined 
with good precision, $\delta m_0 \simeq 4$ GeV. 
The parameter $m_{1/2}$ is then given by the $M_{l^+l^-}^{max}$-line in
the 
$(m_0,m_{1/2})$ plane (Fig.~10.13). The precision obtained in such a way 
is $\delta m_{1/2} \simeq 4$~GeV.

In domain II, the event rate along a line of definite $M_{l^+l^-}^{max}$ 
first increases and then decreases with increasing
$m_0$, Fig.~10.15. This is mainly 
due to the change in the branching ratios. 
The event rate thus does not determine
$m_0$ uniquely: a given event rate corresponds 
in general to two $m_0$ values. The ambiguity can, however, 
be solved at high luminosity $L_{int}=10^5$ pb$^{-1}$, when
two edges in the $M_{l^+l^-}$ distribution can be observed
\cite{ll_incl}. 

For domain I, the $m_0$ dependence of the event rate is shown in
Fig.~10.16a,
again for $M_{l^+l^-}^{max}\simeq 74 \pm 1$ GeV. There is a steep increase of 
the rate at $m_0 \simeq 120-130$ GeV. This is due to the decay  
channel $\tilde{\chi}_2^0 \rightarrow l^+ l^- \tilde{\chi}_1^0$ just
opening up in this region. As can be seen from Fig.~10.16a, there is an
ambiguity  in the determination of $m_0$ if the event rate is in the
region $3700\lappeq N_{EV} \lappeq 5600$  
or 120 GeV $\lappeq m_0 \lappeq $ 240~GeV. 
Additional information can however be obtained from 
the average number of jets $<$$N_{jet}$$>$ in the events.
Figure 10.16b shows $<$$N_{jet}$$>$ as a function of $m_0$.
$<$$N_{jet}$$>$ is 
increasing with $m_0$ as more jets are produced with increasing 
squark-mass. 
With the measured $<$$N_{jet}$$>$ we can resolve the ambiguity in the 
region 120 GeV $\lappeq m_0 \lappeq $ 240 GeV and determine $m_0$
with $\delta m_0 \simeq $7-3~GeV.     


\newpage
 
\begin{figure}[hbtp]
\vspace{-2mm}
\hspace*{25mm}
\resizebox{11cm}{83mm}{\rotatebox{0}{\includegraphics{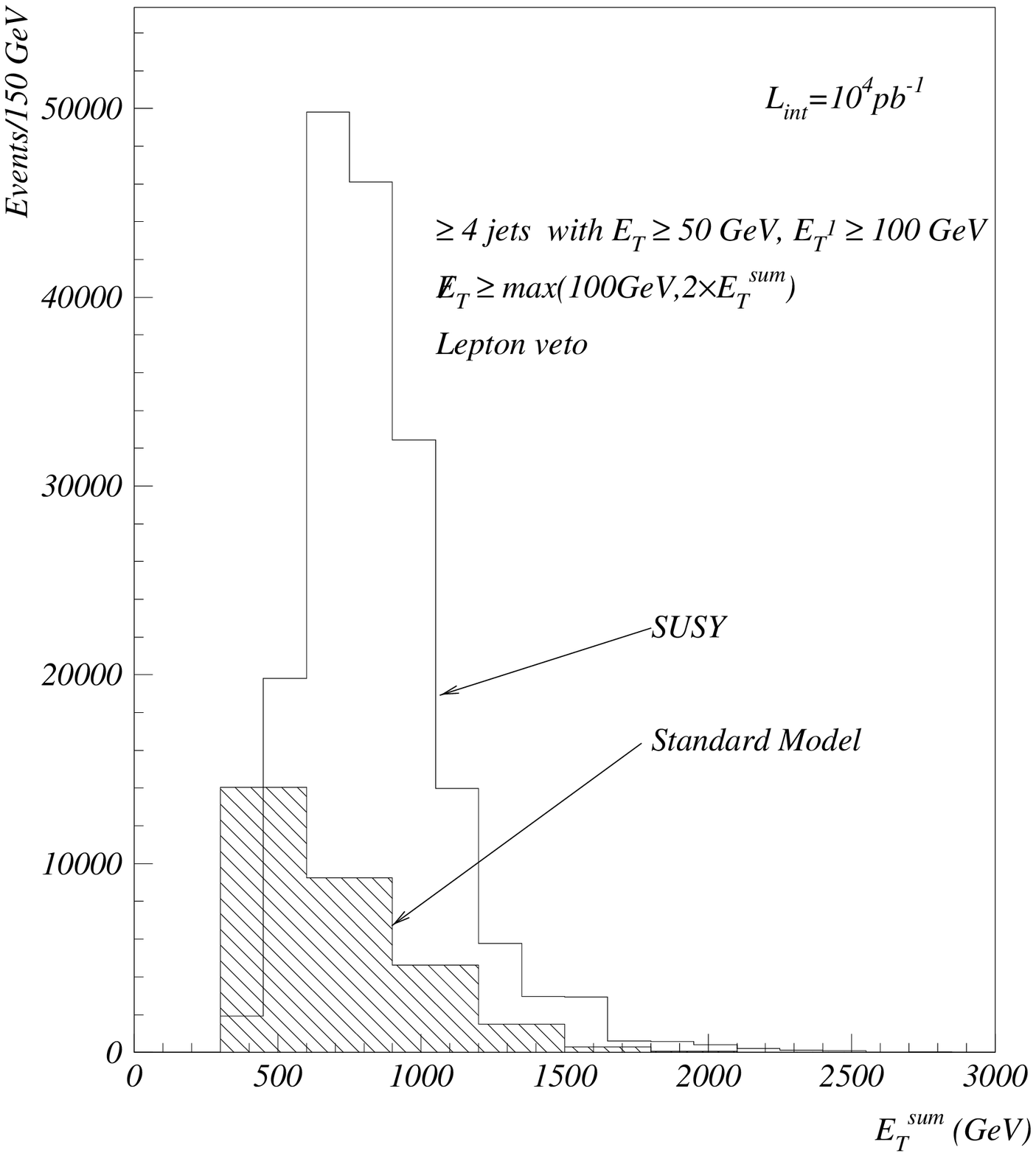}}}

\vspace{3mm}

Figure 10.1: $E_T^{sum}$ distribution for both the inclusive SUSY and the
SM backgrounds after the event selection cuts have been applied.
\end{figure}

\vspace{10mm}
 
\begin{figure}[h]
\hspace*{40mm}
\resizebox{76mm}{!}{\rotatebox{0}{\includegraphics{D_Denegri_1099n.ill}}}
 
\vspace{3mm}
 
Figure 10.2: The relationship between the peak value of the $E_T^{sum}$
distribution and the SUSY mass scale, as defined in the text.
\end{figure}

\newpage
 
\ \ \\
 
 \vspace{30mm}
 
\begin{figure}[hbtp]
\vspace{-80mm}
\hspace*{-20mm}
\resizebox{18cm}{!}{\rotatebox{0}{\includegraphics{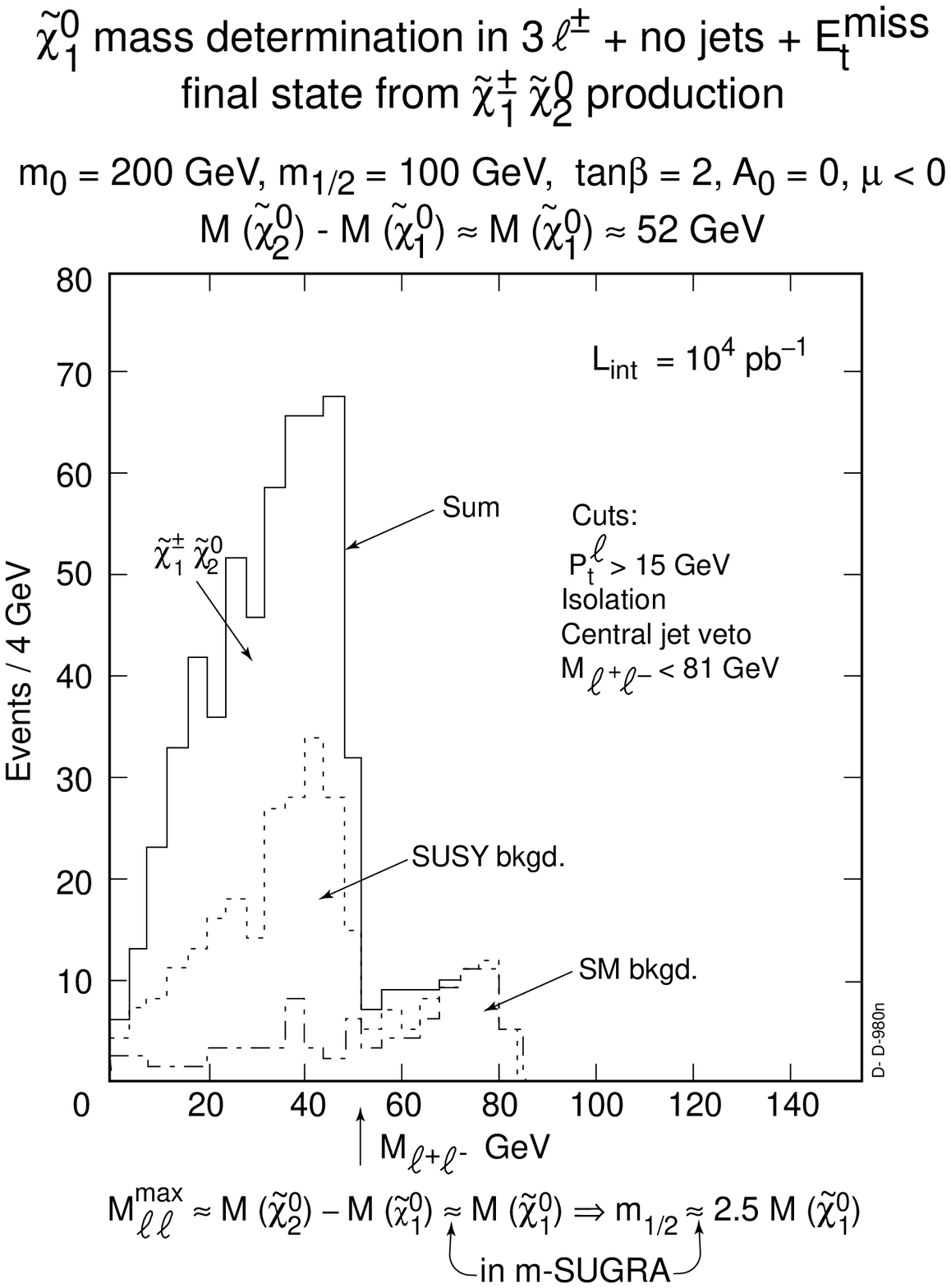}}}
 
\vspace{10mm}
 
Figure 10.3: Dilepton invariant masses distribution for mSUGRA point
($m_0$=200 GeV, $m_{1/2}$=100 GeV) with $L_{int}= 10^4$ pb$^{-1}$ in the
$3l + no \ jets + (E_T^{miss})$ events. Contributions from
SM and SUSY backgrounds are also shown.
\end{figure} 

\newpage
 
\ \ \\
 
 \vspace{30mm}

\begin{figure}[hbtp]
\vspace{-30mm}
\hspace*{10mm}
\resizebox{18cm}{!}{\rotatebox{0}{\includegraphics{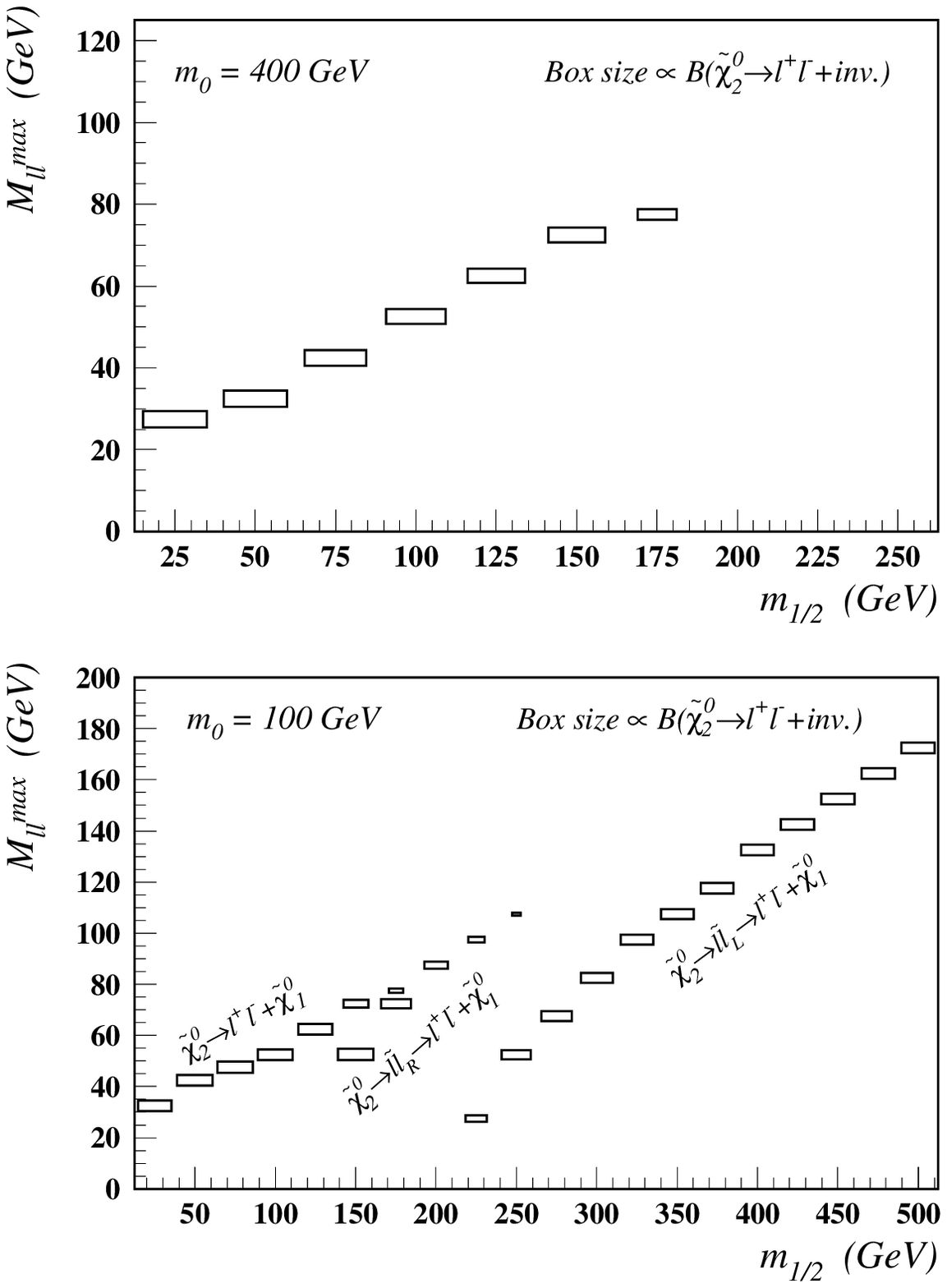}}}

\vspace{10mm}

Figure 10.4:
Correlation between $M^{max}_{l^{+}l^{-}}$
and $m_{1/2}$ for fixed  $m_{0}= 400$ GeV and $m_{0}= 100$ GeV.
The other mSUGRA parameters are:
tan$\beta$=2, $A_{0} = 0$ and $\mu < 0$.

\end{figure}
 
\newpage

\ \ \\
 
\begin{figure}[hbtp]
\begin{center}
\resizebox{12cm}{!}{\rotatebox{0}{\includegraphics{D_Denegri_0996n.ill}}}
\end{center}
\end{figure}

\vspace*{5mm}

Figure 10.5:
$l^+l^-$ mass distribution for mSUGRA point
($m_0 =100$ GeV, $m_{1/2} =150$ GeV);
the other mSUGRA parameters are:
tan$\beta$=2, $A_{0} = 0$ and $\mu < 0$.
The shaded histogram corresponds to the SM background.

\newpage

\ \ \\
 
\begin{figure}[hbtp]
\vspace{0mm}
\begin{center}
\resizebox{14cm}{!}{\rotatebox{0}{\includegraphics{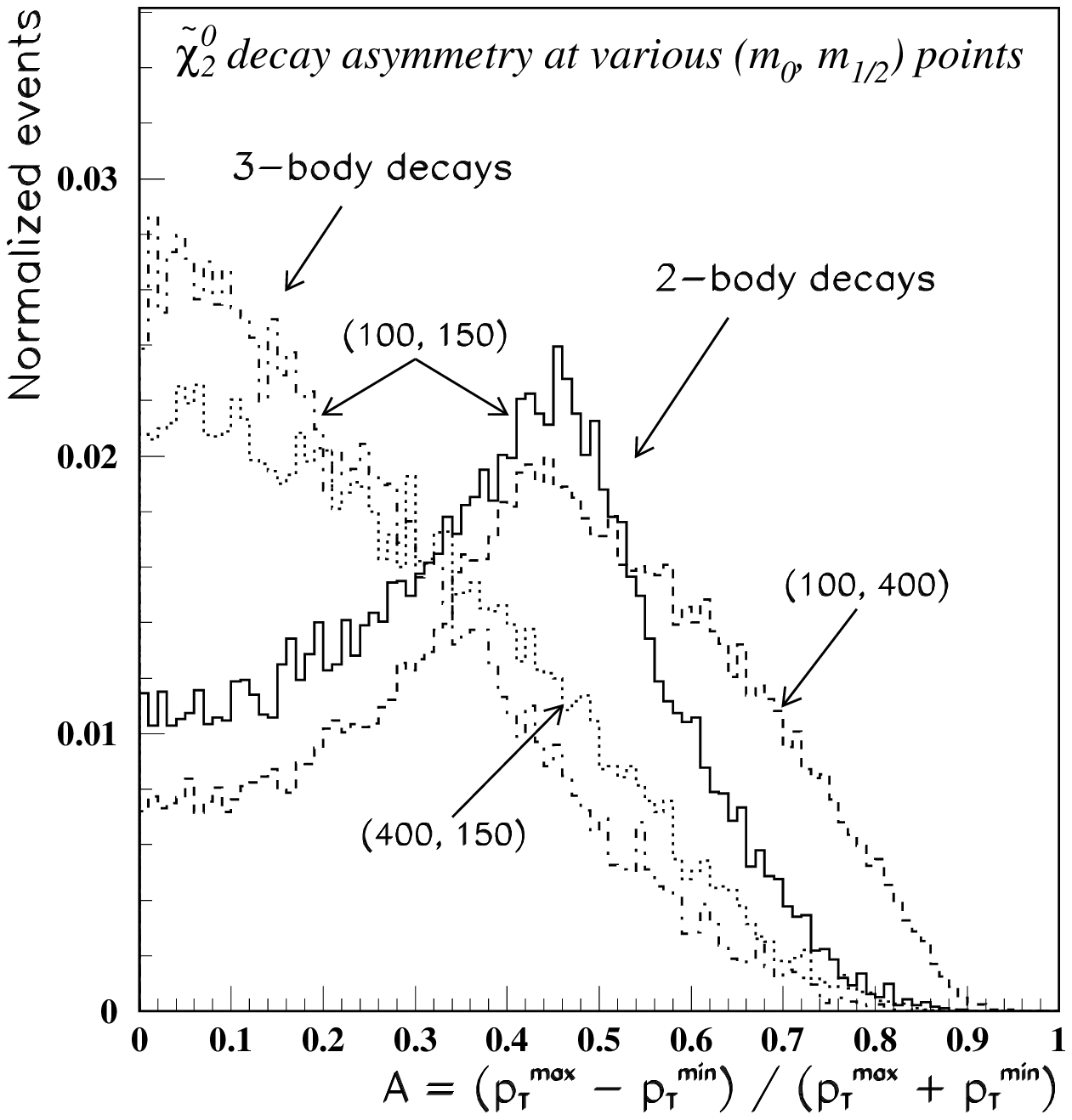}}}
\end{center}
\end{figure}
 
Figure 10.6:
Lepton transverse momentum asymmetry distributions in
$\tilde{\chi}^{0}_{2}$ decays. mSUGRA parameters $(m_{0}, m_{1/2})$
are the following: 100 GeV, 150 GeV (full and dashed-dotted lines);
100 GeV, 400 GeV (dashed line) and 400 GeV, 150 GeV (dotted line).

\newpage  

\ \ \\
 
 \vspace{20mm}

\begin{figure}[hbtp]
\vspace{10mm}
\hspace*{-10mm} 
\resizebox{17cm}{!}{\rotatebox{0}{\includegraphics{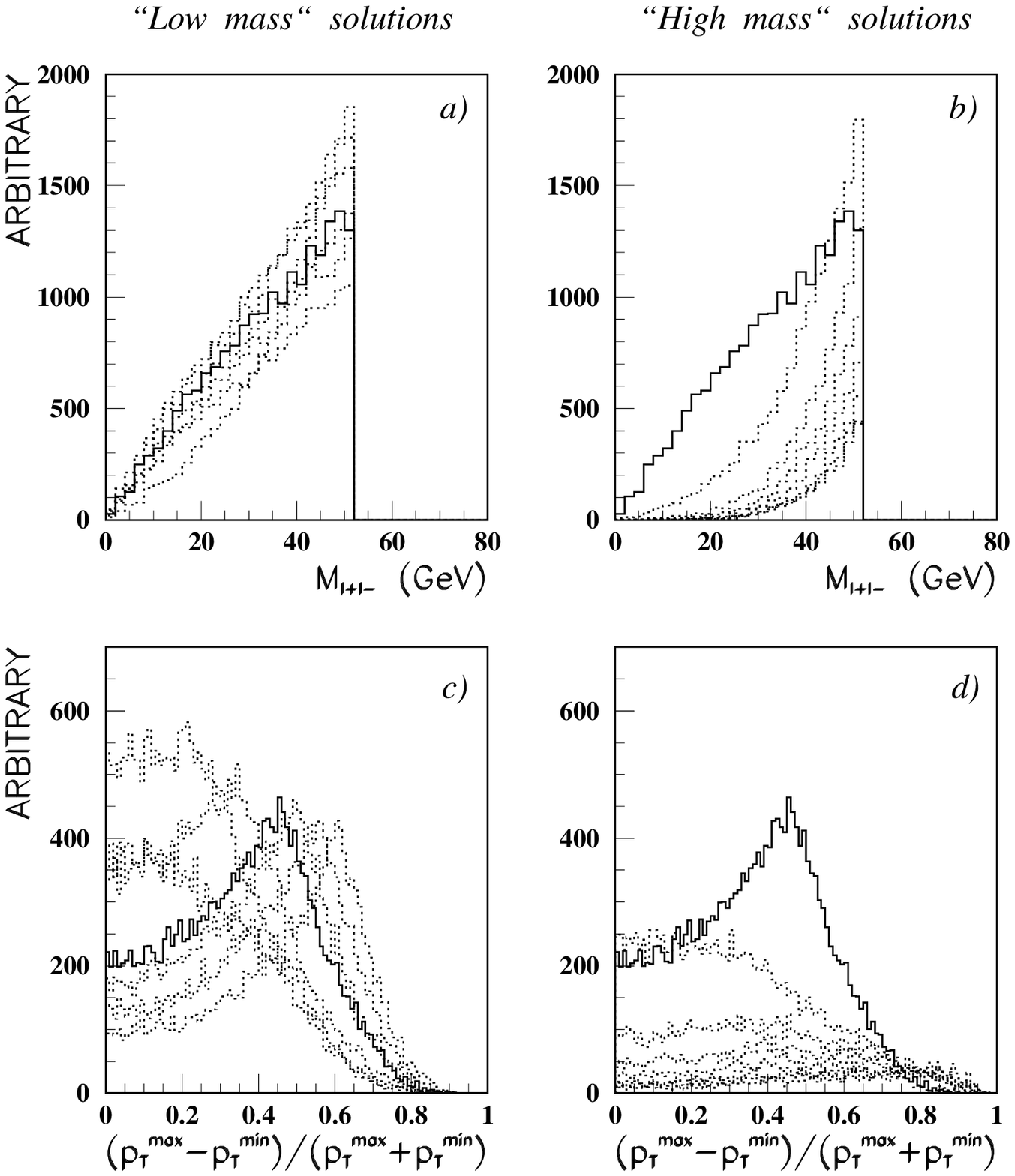}}}
\end{figure}
 
 \vspace{-20mm}
 
Figure 10.7:
The dotted lines correspond to the invariant mass spectra   
and transverse momentum asymmetry
distributions in $\tilde{\chi}^{0}_{2}$ two-body decays for several
``low mass'' and ``high mass'' solutions of eq. (10). The full line
corresponds to the initial (observed) spectrum.

\newpage  

\ \ \\
 
 \vspace{20mm}
 
\begin{figure}[hbtp]
\vspace{0mm}
\hspace*{-10mm}
\resizebox{17cm}{!}{\rotatebox{0}{\includegraphics{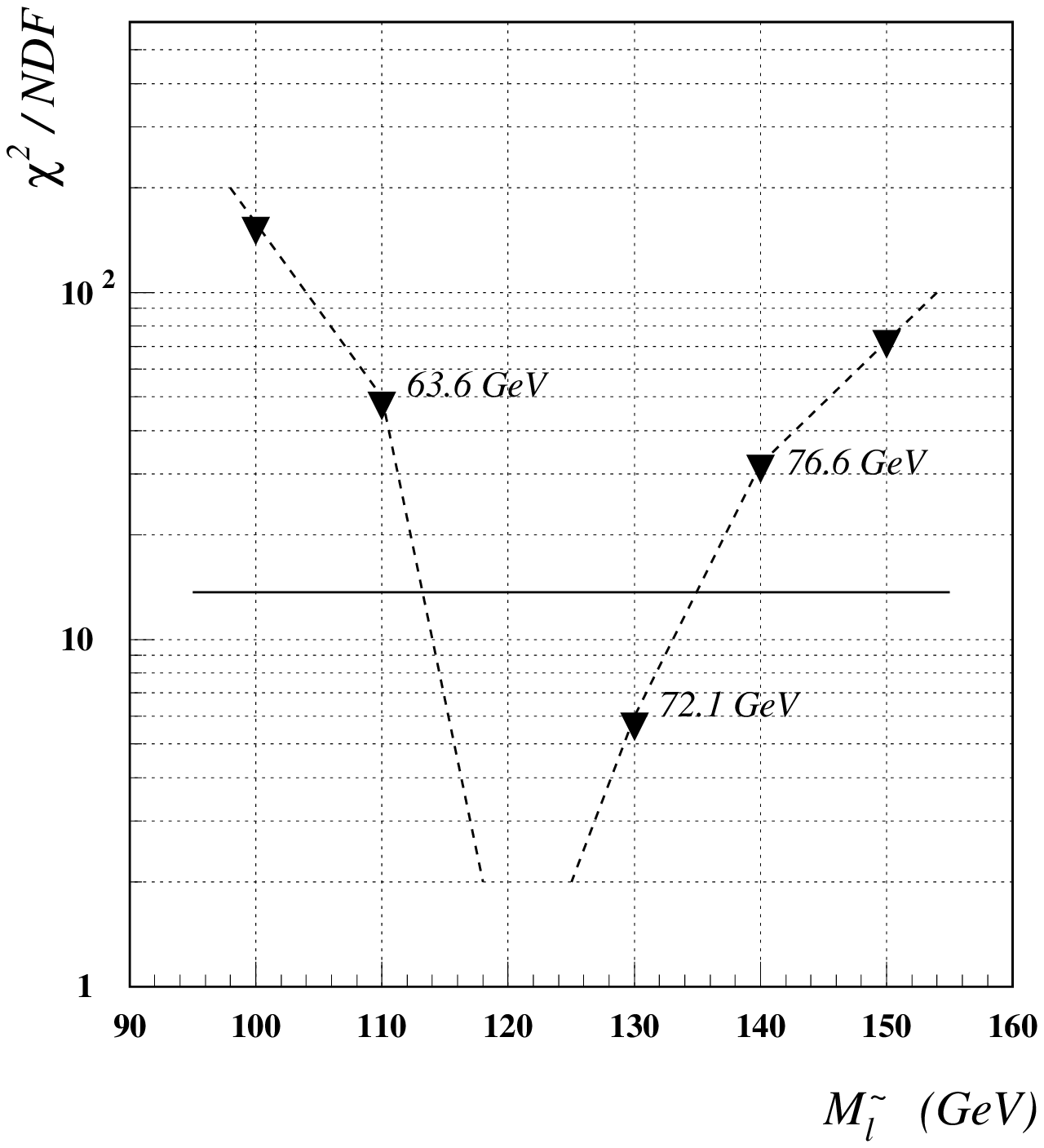}}}
\end{figure}
 
 \vspace{-30mm}
 
Figure 10.8:
Results of the $\chi^2$ test of the shape of $p_T$-asymmetry
distributions as a function of slepton mass. The corresponding values of
$M_{\tilde{\chi}^{0}_{1}}$ are also indicated.

\newpage
 
\ \ \\  
 
 \vspace{20mm}

\begin{figure}[hbtp]
\vspace{10mm}
\hspace*{-10mm}
\resizebox{17cm}{!}{\rotatebox{0}{\includegraphics{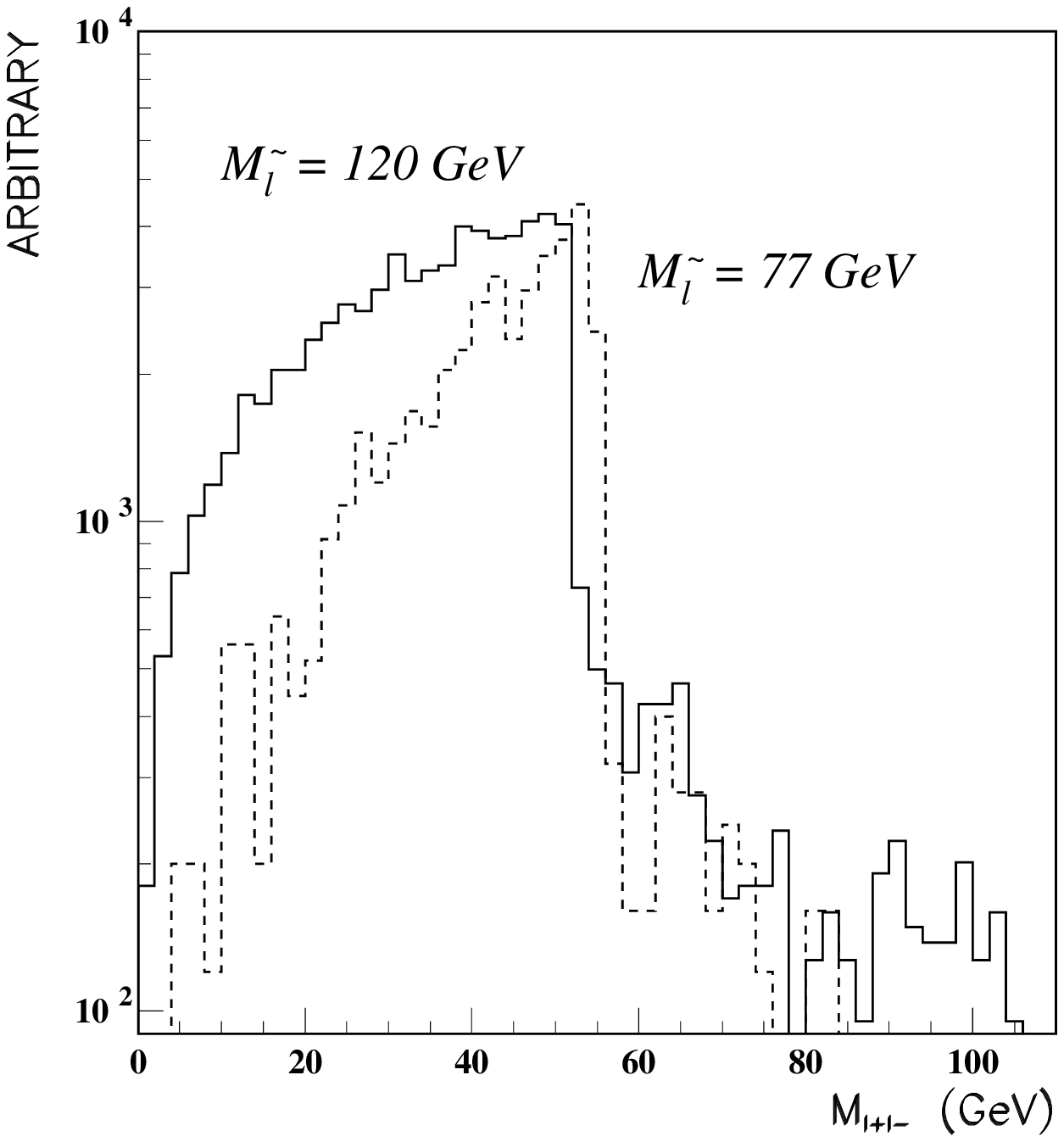}}}
\end{figure}
 
 \vspace{-25mm}
 
Figure 10.9: Predicted $l^+l^-$ invariant mass spectra
for two values of slepton masses: $M_{\tilde{l}} = 120$ GeV (full
line) and $M_{\tilde{l}} = 77$ GeV (dashed line);
mSUGRA parameters as in Fig.~10.6.

\newpage
 
\begin{figure}[hbtp]
\vspace{10mm}
\hspace*{20mm}
\resizebox{17cm}{!}{\rotatebox{0}{\includegraphics{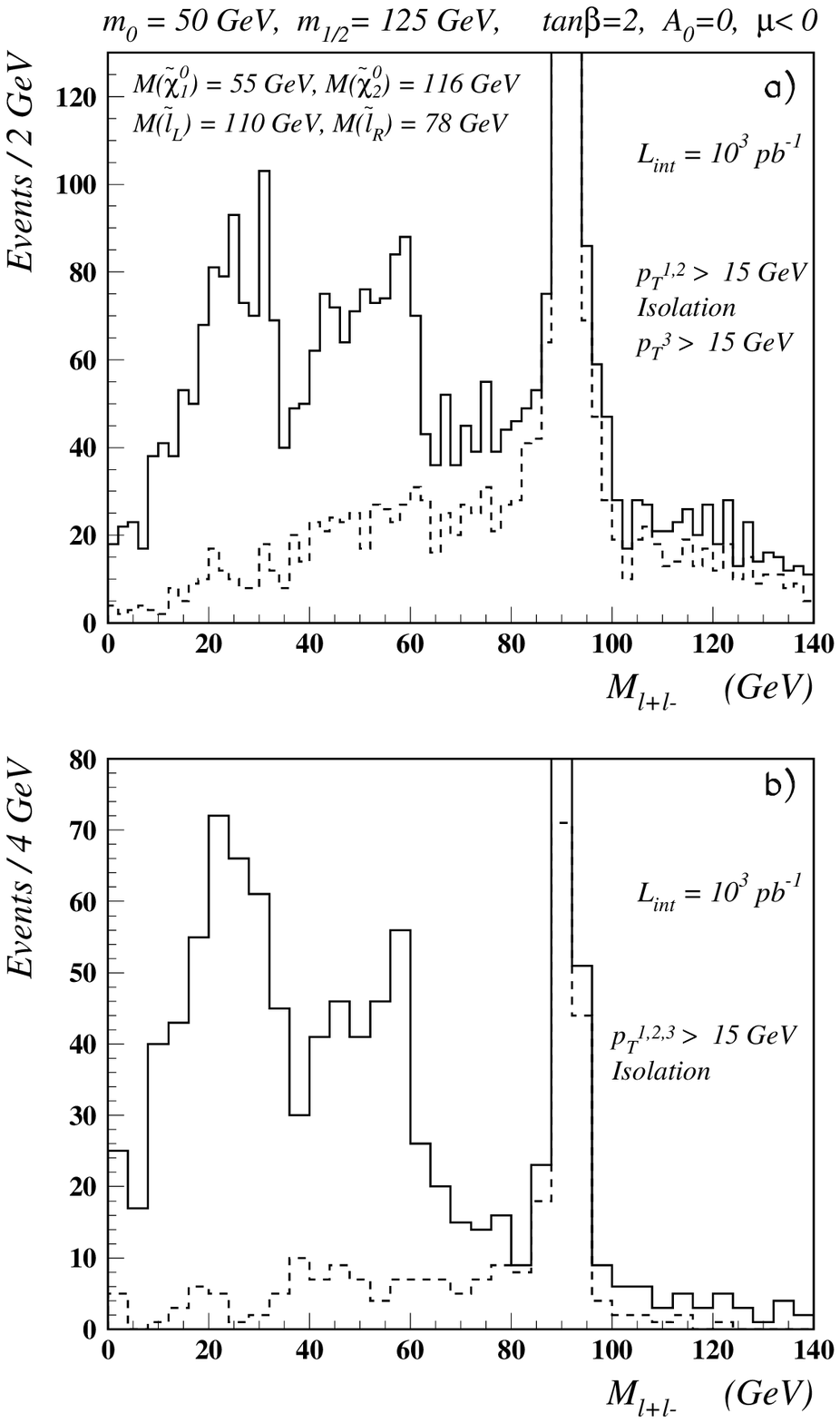}}}
\end{figure}
 
\vspace{10mm}
 
Figure 10.10: 
Expected $l^+l^-$ mass spectrum for mSUGRA Point:
$m_0 =50$ GeV, $m_{1/2} =125$ GeV,
tan$\beta$=2, $A_{0} =0$ and $\mu < 0$.
a) Leptons with $p_T>15$ GeV in $|\eta|<2.4$ are considered.
Two leptons which enter invariant
mass spectrum are required to be isolated;
b) as in previous case, but all three leptons are isolated.
The dotted histogram corresponds to the SM background.

\newpage
 
\ \ \\
 
 \vspace{30mm}
 
\begin{figure}[hbtp]
\vspace{-40mm}
\resizebox{15cm}{!}{\rotatebox{0}{\includegraphics{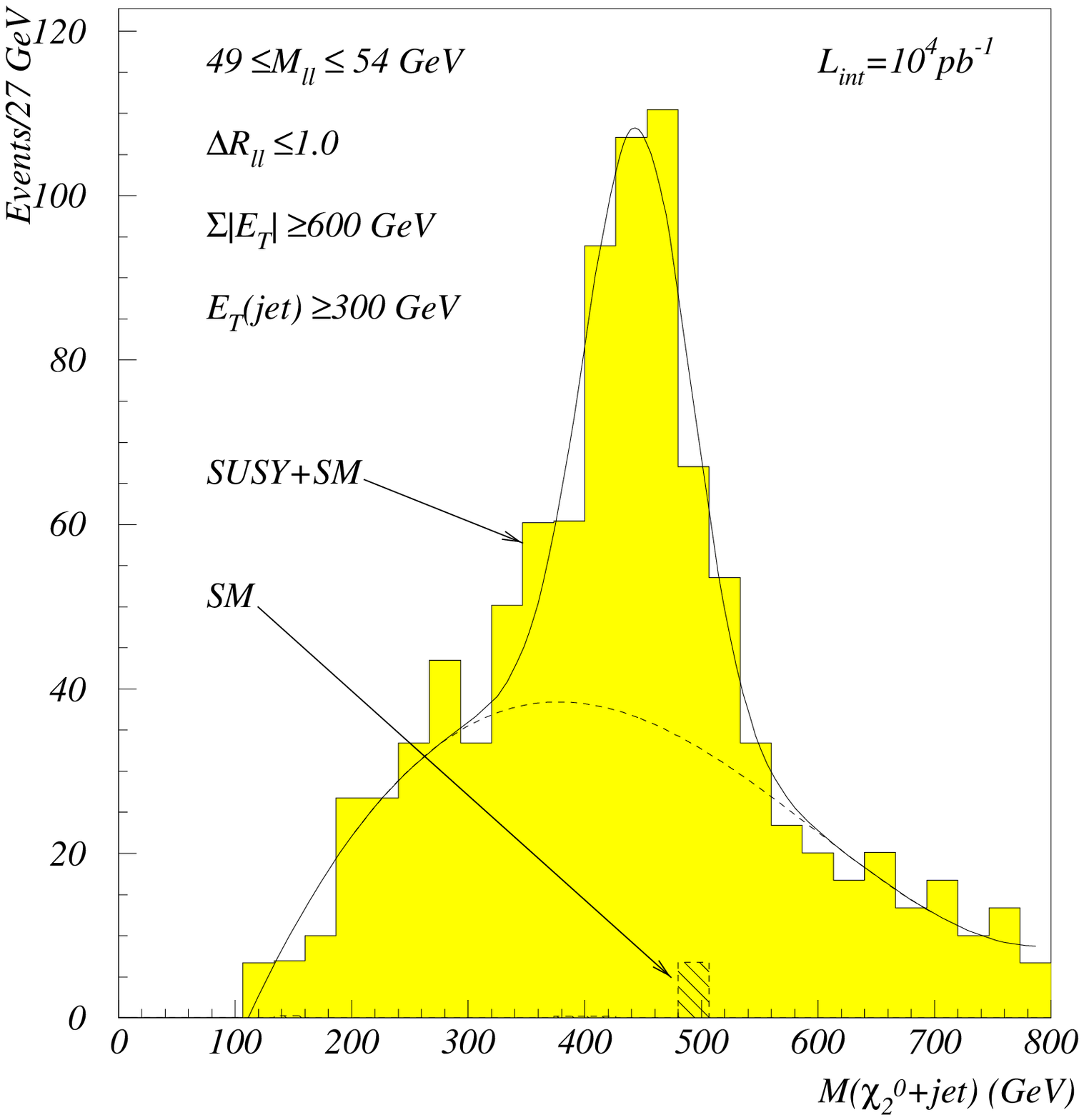}}}
 
\vspace{10mm}
 
Figure 10.11: The reconstructed \chnb+jet mass spectrum for
the SUSY + SM events.
\end{figure}

\newpage

\ \\

\begin{figure}[hbtp]
\hspace*{20mm}
\resizebox{17cm}{!}{\rotatebox{0}{\includegraphics{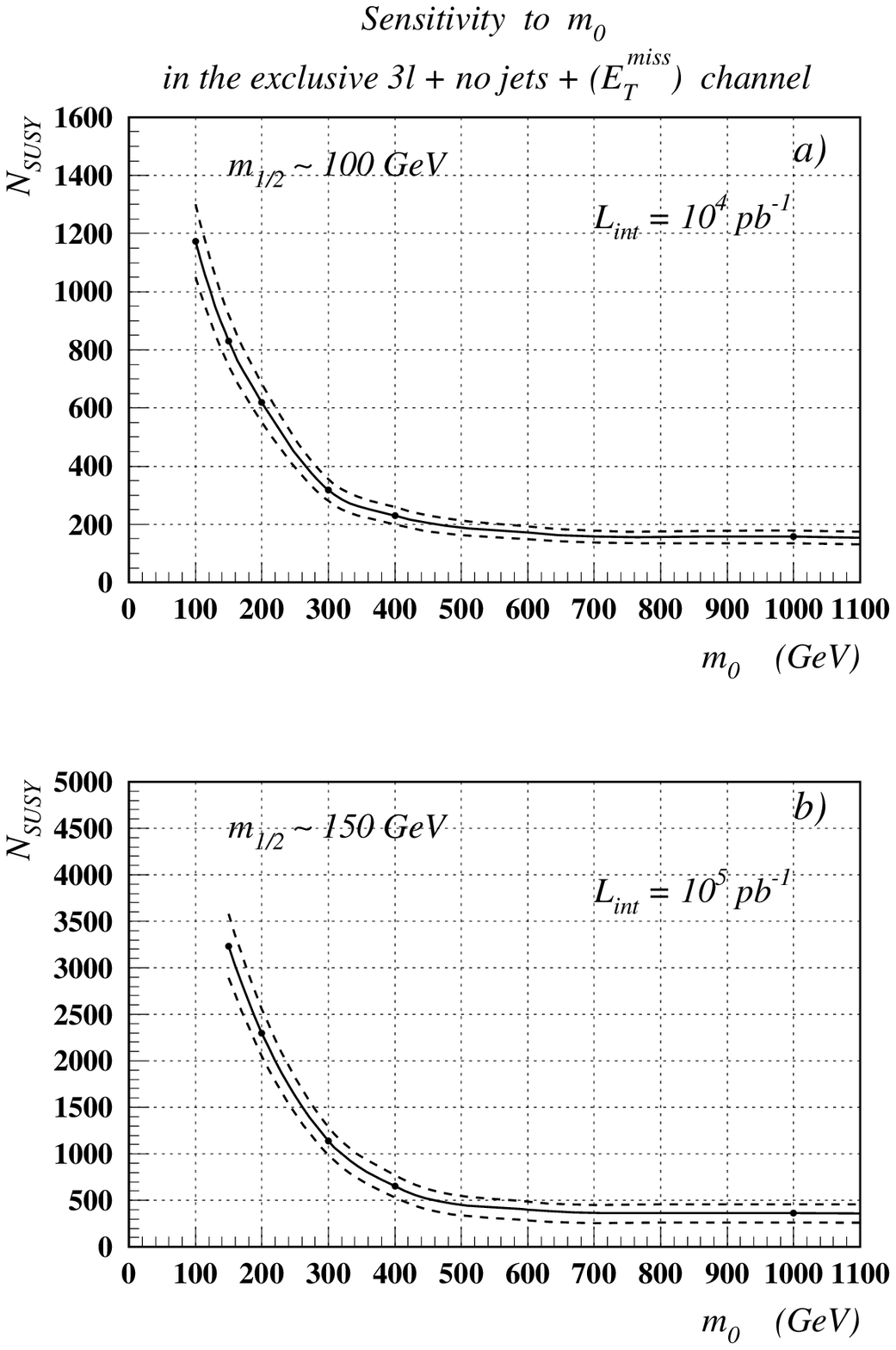}}}
\end{figure}

 \vspace{10mm}

Figure 10.12: a) Number of SUSY $3l + no \ jets + (E_T^{miss})$ events
with
$L_{int}= 10^4 pb^{-1}$
 as a function of
common scalar mass $m_0$ for $m_{1/2} \sim$ 100 GeV.
b) Number of SUSY $3l + no \ jets + (E_T^{miss})$ events with
$L_{int}= 10^5 pb^{-1}$
 as a function of
common scalar mass $m_0$ for $m_{1/2} \sim$ 150 GeV.  

\newpage
 
\ \ \\
 
\begin{figure}
\vspace{10mm}
\begin{center}
\resizebox{130mm}{130mm}
{\includegraphics{D_Denegri_1066n.ill}}
\vspace{0.2cm}
\end{center}
\end{figure}

Figure 10.13: Contour lines of expected $M_{l^+l^-}^{max}$ values (in GeV)
in the invariant dilepton mass distribution corresponding to the
three different $\tilde{\chi}_2^0$ decay modes
in the region of the ($m_0,m_{1/2}$) parameter plane accessible with
$10^5$ pb$^{-1}$.

\newpage

\begin{figure}
\begin{center}
   \epsfig{file=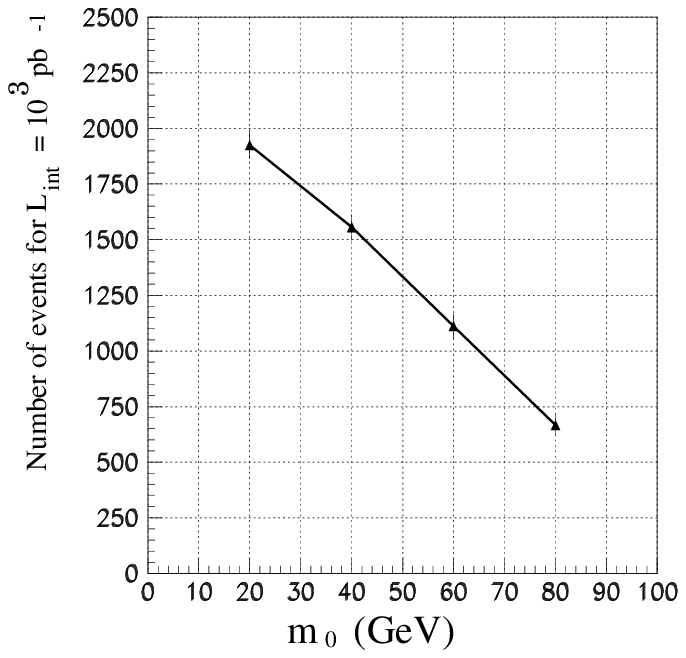,width=8.5cm,height=7.5cm}
\vspace{-0.4cm}
\end{center}
\end{figure}

Figure 10.14:
Expected $2 l+E_T^{miss}$ event rate with $L_{int}=10^3$
pb$^{-1}$
along $M_{l^+l^-}^{max}=74\pm 1$ GeV contour line
in domain III ($\tilde{\chi}_2^0\rightarrow \tilde{l}_L^{\pm} l^{\mp}
\rightarrow \tilde{\chi}_1^0 l^+ l^-$) as a function of $m_0$.
Event selection criteria are:
$p_T^{l_{1,2}}>15$ GeV, $E_T^{miss}>130$ GeV and
$M_{l^+l^-}<M_{l^+l^-}^{max}$
at corresponding points.

\begin{figure}[h]
\vspace{10mm}
\begin{center}
   \epsfig{file=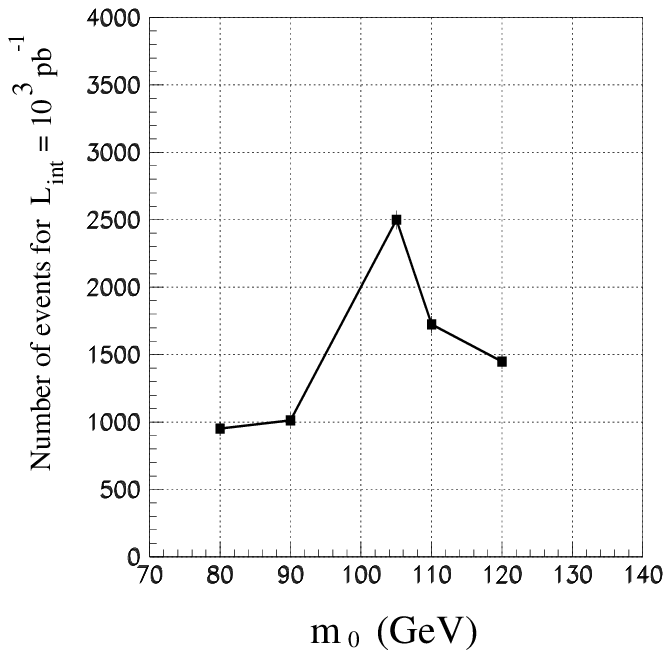,width=8.5cm,height=7.5cm}
\vspace{-0.4cm}
\end{center}
\end{figure}

Figure 10.15:
Expected  $2 l+E_T^{miss}$ event rate with $L_{int}=10^3$
pb$^{-1}$
along $M_{l^+l^-}^{max}=74\pm 1$ GeV contour line in domain II
($\tilde{\chi}_2^0\rightarrow \tilde{l}_R^{\pm} l^{\mp}
\rightarrow \tilde{\chi}_1^0 l^+ l^-$) as a function of $m_0$.
Event selection criteria are:
$p_T^{l_{1,2}}>15$ GeV, $E_T^{miss}>130$ GeV and
$M_{l^+l^-}<M_{l^+l^-}^{max}$
at corresponding points.

\newpage
 
\begin{figure}
\begin{center}
\epsfig{file=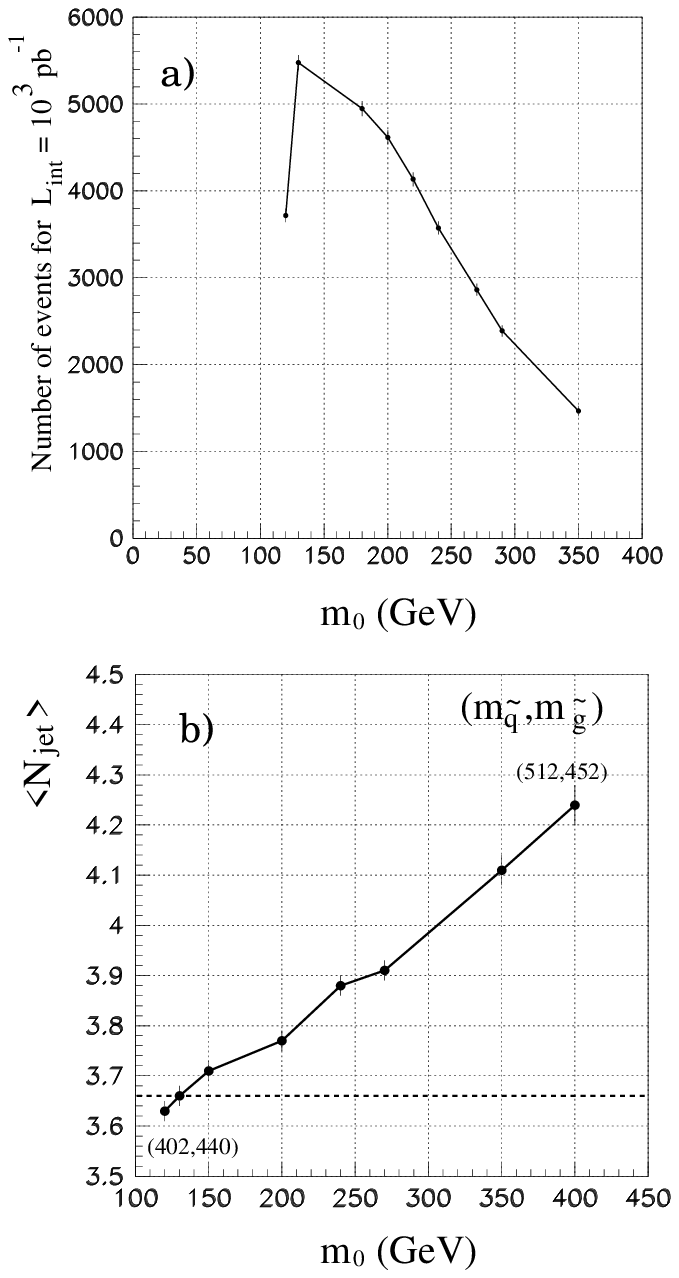,width=10cm,height=15cm}
\end{center}
\end{figure}

Figure 10.16: a) Expected  $2 l+E_T^{miss}$ event rate with $L_{int}=10^3$
pb$^{-1}$ along $M_{l^+l^-}^{max}=74\pm 1$ GeV contour line in domain I
($\tilde{\chi}_2^0\rightarrow \tilde{\chi}_1^0 l^+ l^-$)
as a function of $m_0$.
Event selection criteria are:
$p_T^{l_{1,2}}>15$ GeV, $E_T^{miss}>130$ GeV and
$M_{l^+l^-}<M_{l^+l^-}^{max}$
at corresponding points;
b) Average number of jets ($E_T^{jet}>30$ GeV, $\mid \eta_{jet} \mid <3$)
at investigated points from domain I. The numbers in parenthesis show the
masses of squarks and gluinos (in GeV) at corresponding points.

\newpage


\section{Summary and conclusions}

We have investigated the ability of the CMS detector at
the Large Hadron Collider to discover and characterise
supersymmetry.  As a benchmark model, we used the minimal
supergravity-inspired supersymmetric standard model
(mSUGRA). In the investigated scenarios the lightest supersymmetric
particle (LSP) is stable and is the \chna.

Discovery of supersymmetry at the LHC will be relatively 
straightforward. It would be revealed by large excesses
of events over the standard model expectations in a
number of characteristic signatures -- for example missing
$E_T$ plus jets, with one or more isolated leptons; an
excess of trilepton events; a very characteristic signature in the
$l^+ l^-$ invariant mass distribution.

Because of the importance of missing $E_T$ signatures,
considerable effort has been expended in
optimising the performance of the CMS detector for $E_T^{miss}$
measurements. In particular we have tried to minimize sources
of non-Gaussian tails in the calorimeter response due to cracks
and dead materials.

For the cases we have studied in detail and for $10^5$~pb$^{-1}$
integrated luminosity,

\begin{itemize}

\item
Squarks and gluinos can be discovered up to masses well in
excess of 2~TeV, i.e. for a parameter space region $m_0 \lsim 2-3$~TeV
and $m_{1/2} \lsim 1$~TeV (see Fig.~11.1).
This covers the entire parameter space over which supersymmetry
can plausibly be relevant to electroweak symmetry breaking
without excessive fine-tuning. Over the lifetime of LHC experiments
these searches would probe the $\simeq$~3~TeV region.

\item
Charginos and neutralinos can be discovered from an
excess of events in dilepton or trilepton final
states.  Inclusive searches could give early indications
from their copious production in squark and gluino cascade decays
just from the $l^+ l^-$ spectrum shapes.
With $10^5$~pb$^{-1}$ the
dilepton edge due to the $\tilde\chi^0_2$ leptonic
decays can be seen up to $m_{1/2} \sim$~900~GeV
for $m_0 \lsim 400$~GeV (see Fig.~11.2).
Applying isolation requirements and a jet veto would
enable direct chargino/neutralino production to be
selected; the mass reach in this case covers up to
$\sim 170$~GeV for $\tilde\chi^{\pm}_1$, $\tilde\chi^0_2$ and
$\sim 90$~GeV for $\tilde\chi^0_1$.  The region of
sensitivity (Fig.~11.1)
is entirely determined by the production cross
section and decay branching ratios rather than by any
detector limitations.

\item
Directly-produced sleptons can be detected up to masses
of $\sim 350$~GeV using dilepton signatures. Again, the    
use of a jet veto is crucial, and the lepton opening angle
can be used to discriminate between the slepton signal
and other SUSY backgrounds.  The region of sensitivity
(Fig.~11.1) is once again largely determined by the production cross
section and decay branching ratios.
It is also interesting to remark that the domain of parameter space
explorable through direct slepton production searches matches closely
the \chna \ cosmological dark matter favored region of parameter
space. This is due to relic \chna \ densities being largely determined
by slepton masses. Observation of a dilepton edge up to
$m_{1/2} \sim$~900~GeV gives indirect access to slepton
production for masses up to $M_{\tilde{l}} \sim 700$~GeV.

\item
Squark and gluino production may in addition be a
copious source of Higgs bosons through the cascade decays
of the SUSY intermediate particles. In this case it will be relatively
simple to reconstruct the lightest SUSY Higgs from its
decay h~$\to b\overline b$ by requiring two $b$-tagged jets.
Unlike the case of the Standard Model Higgs in this  
decay mode, good signal to background ratios of order 1 can be obtained
selecting SUSY events using a hard cut on $E_T^{miss}$ thanks to
the escaping LSP's (Fig.~11.2).
\end{itemize}

Once supersymmetry has been discovered, the challenge
will be to determine the sparticle spectrum and to pin down model
parameters. Over much of the
parameter space, we have shown that, within this model,
neutralino masses
can be determined using the dilepton invariant mass
distribution. Details of the \chnb \ decay scheme may be extracted
from the lepton $p_T$-asymmetry, for example.
What can exactly be measured and with what precision depends on the
region in $m_0, m_{1/2}$ space Nature has chosen, and what
topology/final state we use. As visible from Fig.~11.1 some regions of
parameter space can be explored through several channels
allowing for example simultaneous observation of \q, \g \ and
of \chha, \chnb \ and $\tilde{l}$ direct production
giving thus access to a larger spectrum and
providing more constraints on the supersymmetric models.

One of the attractive features of $R$-parity conserving SUSY models
is that the \chna \ -- lightest supersymmetric particle is a plausible
candidate for the cosmological dark matter.
The region of parameter space where
this is true for tan$\beta = 2$ is shown in Fig.~11.1; as already
mentioned, it is covered by both the squark and gluino searches
and the slepton search, and overlaps the
region where inclusive charginos and neutralino signals
are visible (Fig.~11.2).
Specifically, in \q/\g \ searches even with $10^4$~pb$^{-1}$
the \chna \ (WIMP) is probed up to
$\sim 350$ GeV for any $m_0$, up to $\sim 450$ GeV for $m_0 \lsim 400$
GeV, i.e. well beyond the cosmologically plausible domain.
The cosmologically favored region (hatched area in Fig.~11.1)
is embedded deeply within
the explorable domain of parameter space, implying that, if SUSY is
responsible for cosmological dark matter, this could hardly escape 
discovery at the LHC, at least for tan$\beta$ values investigated.
This result is probably valid much more generally than
implied by this specific model since there is a large safety margin,
and a variety of experimental signatures/final states with different  
sensitivities, backgrounds and systematics. In this respect
a detailed investigation of large tan$\beta$ values would certainly
be of interest.

We have not yet explored other supersymmetry scenarios
in detail, but some general comments may be made. 
Generator-level studies \cite{baer2} have shown that   
even if the lightest neutralino decays hadronically, the
missing $E_T$ plus jets and isolated leptons signature
will remain sensitive to squark and gluino production.     
We also expect that the inclusive
multilepton signature would be robust against changes
in the SUSY decays. Gauge-mediated SUSY breaking models, which
predict decays of the lightest neutralino into a
photon plus a gravitino, would give in our detector striking signatures
of diphotons plus missing $E_T$ that would be easily
detected \cite{GMSB}.

We therefore believe that it will be very hard for electroweak-scale
supersymmetry, if it exists, to escape detection
at the LHC.  If supersymmetry is still not found
after the LHC has accumulated few times $10^{5}$~pb$^{-1}$
of data then it is essentially excluded at the
electroweak scale, and some other new physics is presumably at work.
On the other hand if supersymmetry is discovered at
LEP or the Tevatron, there will be a rich program
of SUSY studies to be carried out at the LHC since 
so much more of the Higgs and sparticle spectrum will be accessible.

\newpage

\ \\

\vspace{25mm}

\begin{figure}[hbtp]
\vspace{-30mm}
\hspace*{5mm}
\resizebox{15cm}{!}{\rotatebox{0}{\includegraphics{Avto925.ill}}}
\end{figure}

\vspace{0mm}

Figure 11.1: $5\sigma$ significance contours at $L_{int}= 10^5$ pb$^{-1}$
for different SUSY channels.

\newpage

\ \\
 
\vspace{25mm}

\begin{figure}[hbtp]
\vspace{-20mm}
\hspace*{0mm}
\begin{center}
\resizebox{15cm}{!}{\rotatebox{0}{\includegraphics{D_Denegri_1103n.ill}}}
\end{center}
\end{figure}

Figure 11.2: mSUGRA parameter reach in h~$\rightarrow$ $b\bar{b}$
decays at $10^4$ pb$^{-1}$ and $10^5$ pb$^{-1}$ integrated luminosities.
The reach in \chnb \ via observation of edge(s) in dilepton mass distribution
is also shown.


\newpage
 
{\Large {\bf Acknowledgements}}

\vspace{10mm}

Over the past few years we have benefited from numerous discussions
with our colleagues, from their suggestions and criticism.
It is a pleasure for us to thank in first place H.~Baer who has
spent many hours in discussions with us and in following our studies.
We are also grateful to A.~Bartl, W.~Majerotto, F.~Pauss, C.~Seez,
M.~Spira, T.~Virdee for useful discussions and comments.

\newpage


\end{document}